\documentclass[usenatbib,onecolumn]{mnras}

\usepackage{newtxtext,newtxmath}
\usepackage{soul}
\usepackage[T1]{fontenc}
\usepackage{cuted}
\usepackage{subcaption}

\DeclareRobustCommand{\VAN}[3]{#2}
\let\VANthebibliography\thebibliography
\def\thebibliography{\DeclareRobustCommand{\VAN}[3]{##3}\VANthebibliography}

\usepackage{graphicx}
\usepackage{amsmath}
\usepackage{tabularx}
\usepackage{booktabs}
\usepackage{multirow}
\usepackage{multicol}
\usepackage{xcolor}
\usepackage{listings}
\usepackage{needspace}

\usepackage{longtable}
\hypersetup{hypertexnames=false}
\renewcommand{\theHfigure}{main.\arabic{figure}}
\renewcommand{\theHtable}{main.\arabic{table}}
\lstdefinestyle{pyhermescode}{
    language=Python,
    basicstyle=\ttfamily\small,
    keywordstyle=\color{blue},
    commentstyle=\color{gray},
    stringstyle=\color{purple},
    showstringspaces=false,
    breaklines=true,
    breakatwhitespace=false,
    keepspaces=true,
    columns=fullflexible,
    frame=single,
    rulecolor=\color{gray!40},
    aboveskip=0.6\baselineskip,
    belowskip=0.6\baselineskip
}
\AddToHook{env/lstlisting/before}{\par\Needspace{6\baselineskip}\noindent}

\title{\textsf{Hermes} - Towards an Optimal High-Performance Algorithm for Cosmic Statistics of Large Data Sets}

\author[Feng, Xu \& Luan et al.]{
Long-long Feng,$^{1}$\thanks{flonglong@mail.sysu.edu.cn}
Tengpeng Xu,$^{1}$\thanks{xutp5@mail.sysu.edu.cn}
Tian-Cheng Luan,$^{1}$\thanks{dingdluan@gmail.com}
Jiawei Li,$^{1}$
Xin Sun,$^{1}$
Wenjie Ju,$^{1}$
\newauthor
Zhuoyang Li,$^{2}$
Shiyu Yue,$^{3,4}$
Weishan Zhu,$^{1}$
and
Yan-Chuan Cai$^{5}$
\\
$^{1}$School of Physics and Astronomy, Sun Yat-Sen University, Zhuhai 519082, China \\
$^{2}$Department of Astronomy, Tsinghua University, Beijing 100084, China \\
$^{3}$ Department of Physics, The Chinese University of Hong Kong, Shatin, New Territories, Hong Kong SAR, China \\
$^{4}$ JC STEM Lab of Astronomical Instrumentation, The Chinese University of Hong Kong, Shatin, New Territories, Hong Kong SAR,
China \\
$^{5}$Institute for Astronomy, University of Edinburgh, Blackford Hill, Edinburgh, EH9 3HJ, UK
}

\date{Accepted 2026 November 16. Received 2026 November 16; in original form 2026 August 21}

\pubyear{2026}

\begin{document}
\raggedbottom
\label{firstpage}
\pagerange{\pageref{firstpage}--\pageref{lastpage}}
\maketitle

\begin{abstract}
We present \textsf{Hermes}, an \textit{in situ} multiresolution framework for efficient
and flexible measurements of cosmic large-scale-structure statistics from
discrete catalogues. \textsf{Hermes} replaces
conventional \textit{ex situ} counting of particle tuples with algebraic
operations among window-filtered continuous fields. A discrete point
distribution is first reconstructed as a continuous density field in a compact
scaling-function representation, in which two- and higher-point statistics are
formulated as spatial products of locally filtered fields. Within this
\textit{in situ} formulation, conventional binning schemes for counts-in-cells
statistics, two- and higher-point correlation functions are unified as specific
choices of window functions, while generalised statistics can be constructed by
modifying the window kernels without redesigning the underlying estimator. We further introduce \textsf{PyHermes}, an open-source \textsf{Python} implementation of this framework, combining multiresolution field reconstruction with FFT-based window
convolution, MPI/thread parallelism, and GPU acceleration for high-order field
contractions. The framework supports a broad range of measurements, including
isotropic and anisotropic two-point statistics, marked
correlations, standard and multipole three-point functions, filtered
statistics, and Fourier-space differential operators for constructing derived
physical fields. Using halo catalogues from cosmological \(N\)-body simulations, we demonstrate \textsf{PyHermes} through a range of clustering statistics and perform
performance benchmarks to quantify its computational efficiency and scalability
for large-scale data sets. By separating field representation from statistical windows, \textsf{Hermes} provides a unified computational framework in which a single reconstructed field can be reused for a wide range of standard and customised cosmic statistics. This field-window architecture enables scalable analysis of the large data sets expected from current and future galaxy surveys.
\end{abstract}


\section{Introduction}

In the concordance cosmological model, the formation of cosmic structures is
described by the hierarchical clustering scenario, in which cold dark matter
dominates the matter content and gravitational instability drives the growth
of density fluctuations. The
primordial density perturbations are generally believed to originate from
quantum fluctuations generated during inflation, which subsequently evolve through gravitational
growth as the Universe expands. At early times, these fluctuations are well
approximated by a Gaussian random field, whose statistical properties are fully characterized by
second-order statistics: the power spectrum in Fourier space or equivalently
the two-point correlation function (2PCF) in configuration space.

The 2PCF therefore provides the fundamental statistic for characterising the
clustering of cosmic density fields and has been extensively applied to galaxy
surveys. It also forms the basis for precision measurements of
cosmological parameters from large-scale-structure observations
(e.g., \citealt{Eisenstein2005,Tegmark2006,Percival2011,Anderson2014,Alam2017,Ivanov2020,
DESI2024VI_BAO,DESI2024VII_cosmo}). Most recently, DESI DR2 baryon acoustic
oscillation measurements using galaxies, quasars, and the Ly$\alpha$ forest
have provided stringent constraints on the expansion history and dark
energy \citep{DESIDR2Cosmo}. These measurements were supported by
dedicated validation analyses
\citep{DESIDR2BAOValidation,DESIDR2LyAValidation} and improved absorber catalogues
\citep{DESIDR2DLACatalog}, while
companion studies explored extended dark-energy models and neutrino physics
\citep{DESIDR2DarkEnergy,DESIDR2Neutrinos}.

However, gravitational evolution inevitably drives the density field into a
nonlinear and non-Gaussian regime. The nonlinear collapse of matter
fluctuations generates mode coupling across scales, leading to enhanced
variance and non-Gaussian covariance. As a result, second-order statistics
no longer capture all of the available cosmological information
(e.g., \citealt{Scoccimarro1999,Cooray2002,Rimes2005,Neyrinck2006}).
This motivates the development of higher-order statistics, including
polyspectra and higher-order correlation functions.

The three-point correlation function (3PCF) and its Fourier counterpart,
the bispectrum, are the lowest-order connected probes of non-Gaussianity.
They test non-linear gravitational evolution in cold dark matter models
\citep{Gaztanaga1994,Jing1998,Scoccimarro2001,Jing2004},
with measurements across galaxy redshift surveys
\citep{Pan2005,Nichol2006,Slepian2017}.
Their applications include tests of gravity \citep{Sugiyama2023},
constraints on primordial non-Gaussianity and inflation
\citep{Desjacques2010,Achucarro2022}, and measurements of non-linear
galaxy bias \citep{Fry1993,Desjacques2018}. Furthermore, joint power-spectrum
and bispectrum analyses can tighten constraints on cosmic growth, geometry,
and cosmological parameters
\citep{Gagrani2017,Agarwal2021,Samushia2021}.
Such analyses have been validated using simulations
\citep{Novell-Masot2023} and applied to DESI DR1 observations
\citep{DESIBispectrum2025}.

Nevertheless, an important question remains: is extending the hierarchy
of correlation functions to arbitrarily high orders sufficient to recover
all the information contained in nonlinear density fields? The lognormal (LN)
field provides an important example that highlights a fundamental limitation of
moment-based statistics. Even a complete hierarchy of moments may fail to
uniquely determine the underlying probability distribution, because the LN
distribution is moment-indeterminate
\citep{carron2011,CarronNeyrinck2012}. The LN model has nevertheless been shown to provide a successful approximation to the one- and two-point PDFs of
nonlinear matter density fields evolved from Gaussian initial conditions
and has therefore been widely used to describe non-Gaussian cosmological fields
\citep{coles1991,Kayo2001}.

These limitations motivate statistical summaries beyond low-order correlation functions,
such as counts-in-cells (CIC) distributions of discrete tracers and one-point
PDFs of smoothed continuous fields. By probing both the bulk and tails,
these statistics can capture non-Gaussian information missed by a finite
set of moments. For example, the void probability function depends on
correlations of all orders \citep{White1979}, while some distributions,
including the lognormal, are not uniquely specified even by their complete
moment sequence \citep{Aitchison1957,coles1991}. Applications include
galaxy surveys \citep{Wild2005,Hurtado2017,Repp2020},
weak-lensing PDFs and cumulative distribution functions
\citep{Clerkin2016,Anbajagane2023}, and density-split lensing
\citep{Gruen2018,Burger2023}.
Related PDF-based approaches also probe primordial non-Gaussianity
\citep{Matarrese2000,Valageas2002,Uhlemann2018,Friedrich2020},
modified gravity and dark energy \citep{Cataneo2022},
and neutral hydrogen distributions for 21\,cm intensity mapping
\citep{Leicht2018}.

A complementary direction is to incorporate information carried by galaxy
properties and their environments. The nonlinear clustering of matter is
affected not only by gravitational evolution but also by baryonic processes
on small scales \citep{vanDaalen2011} and environmental effects associated
with the cosmic web.
The dependence of galaxy clustering on intrinsic properties such as luminosity,
colour, and morphology, known as the \textit{segregation effect}, has been
extensively studied
(e.g., \citealt{Beisbart2000,Wang2007,zehavi2011,Guo2014}). Similarly,
\textit{density-split statistics}, obtained by dividing samples according to
their local environment, have demonstrated additional sensitivity to
non-Gaussian information beyond the conventional 2PCF and nonlinear power spectrum
(e.g., \citealt{Gruen2018,Paillas2021,Paillas2023a,Paillas2023b}).

Marked clustering statistics (MCS) provide another flexible approach for
exploiting information beyond the conventional correlation functions.
In this approach,
each tracer is assigned a weight determined by intrinsic or environmental
properties. Such marks modify the effective density field and enable the
correlation between clustering and additional physical properties to be
measured
(e.g., \citealt{Sheth2005,Skibba2006,White2009,Skibba2013,White2016,Philcox2020}).
Marked power spectra can recover additional cosmological information relative
to the standard power spectrum \citep{Massara2021,Massara2023}.
Recent work has further shown how optimising the mark function can
substantially enhance the cosmological information recovered by marked
power spectra \citep{Cowell2024}. From this perspective,
CIC, density-split, and marked statistics can all be interpreted as
measurements of appropriately filtered density fields, supplemented where
necessary by density selection or nonlinear weighting.

With the advent of ongoing and forthcoming deep and wide galaxy surveys,
including Euclid \citep{laureijs2011}, the Dark Energy Spectroscopic
Instrument (DESI) survey \citep{DESI2016}, the Nancy Grace Roman Space
Telescope \citep{Akeson2019}, the Vera C. Rubin Observatory \citep{Ivezic2019},
and the China Space Station Telescope (CSST)
\citep{2026SCPMA.6939501C,2025SCPMA.6880402G},
large-scale structure observations are entering a new era
of unprecedented volume and statistical precision. These surveys will provide
increasingly detailed maps of cosmic density fields and enable stringent tests
of cosmological models and structure formation physics at the percent level.

Achieving this precision, however, requires not only accurate theoretical
modelling of nonlinear structure formation
(e.g., \citealt{Bernardeau2002,Cooray2002,baldauf2020}), but also scalable
algorithms capable of analysing rapidly growing galaxy catalogues and large
ensembles of simulations. The computational cost of clustering measurements
remains a major bottleneck. Even for the simplest two-point correlation
function, direct pair counting requires
$\mathscr{O}(N_{\mathrm{g}}^{2})$ operations for a catalogue containing
$N_{\mathrm{g}}$ galaxies. More generally, brute-force estimation of the
$N$-point correlation function requires enumeration of all possible
$N$-tuples, resulting in a scaling of
$\mathscr{O}(N_{\mathrm{g}}^{N})$. Such approaches become impractical for modern
surveys containing millions to billions of objects.

Building on standard correlation-function estimators
\citep{Landy1993,Szapudi1998}, substantial efforts have been devoted
to developing fast $N$-point correlation-function (NPCF) algorithms suited
to modern high-performance computing architectures.
For the 2PCF, hierarchical tree methods accelerate pair counting through
spatial partitioning and efficient neighbour searches
\citep[e.g.,][]{Moore2001,Zhang2005,Rohin2018,Zhao2023}, while linked-list
and skip-list methods provide efficient alternatives \citep{Donoso2019}.
For higher-order statistics, multi-tree traversal and graph-based methods
reduce the cost of configuration counting
\citep{Gray2004,March2012,Sabiu2019}, while multipole and separable-basis
expansions exploit angular separability to avoid explicit enumeration of
higher-order particle tuples
\citep{Szapudi2004,Slepian2015,Slepian2018,Philcox2022}.
These developments have also enabled massively parallel implementations
for billion-object catalogues \citep{Friesen2017}.
For gridded density fields, convolution and FFT techniques provide an
efficient means of evaluating the required filtered fields and angular
moments \citep{Slepian2016,Portillo2018,Sunseri2023}.
At fixed radial binning and angular truncation, such methods can achieve
$\mathscr{O}(N_{\mathrm{c}}\log N_{\mathrm{c}})$ scaling, where
$N_{\mathrm{c}}$ is the number of grid cells \citep{Sunseri2023}.
The convolution stage of ConKer instead has a reported scaling of
$\mathscr{O}(N_{\mathrm{c}}^{4/3}\log N_{\mathrm{c}})$ when the radial
sampling is refined together with the spatial grid \citep{Brown2022}.

Recent field-based formulations have further recast pair and triplet
counting in terms of window-filtered density fields, providing a unified
treatment of binning and smoothing \citep{Yue2024}.
Combined with multiresolution field reconstruction, this perspective has
led to an optimised \textit{in situ} multipole algorithm for the isotropic
3PCF \citep{Ju2026}.
Complementary advances in the treatment of survey geometry include
window-deconvolved power-spectrum and bispectrum estimators with
minimum-variance weighting \citep{PolyBin3D}, as well as efficient
survey-window convolution of theoretical bispectrum models
\citep{Wang2025Window}.

Several mature software packages implementing these approaches are now publicly
available, including
\textsf{POWMES} \citep{Colombi2009}, \textsf{CUTE} \citep{Alonso2012},
\textsf{CosmoBolognaLib} \citep{Marulli2016}, \textsf{Halotools}
\citep{Hearin2017}, \textsf{nbodykit} \citep{Hand2018},
\textsf{Corrfunc} \citep{Sinha2020}, \textsf{Triumvirate}
\citep{Wang2023Triumvirate}, \textsf{PolyBin3D} \citep{PolyBin3D},
\textsf{HIPSTER} \citep{Philcox2021a}, and \textsf{Encore}
\citep{Philcox2021b}. Despite these advances, existing packages differ in
their supported statistics, estimators, and computational representations.
A more general framework that treats diverse cosmic statistics within a
common formulation, e.g., as measurements of suitably filtered density fields,
remains highly desirable.

Furthermore, cosmological inference requires accurate predictions of
clustering statistics and reliable covariance estimates for likelihood
construction. Bayesian analyses typically involve repeated likelihood
evaluations using MCMC or related sampling techniques, while mock-based
covariance estimation requires repeated measurements of the statistics
across large simulation ensembles. Noise in covariance and precision
matrices estimated from finite mock samples propagates into cosmological
parameter uncertainties \citep{Hartlap2007,SellentinHeavens2016}.
Depending on the dimensionality of the data vector
and the required precision, limiting the resulting degradation of parameter
constraints to the percent level can require thousands to tens of thousands
of independent realisations \citep{Taylor2013,Dodelson2013}.
To reduce this computational burden, various approaches have been
developed, including eigenmode truncation \citep{Gaztanaga2005},
linear and non-linear covariance shrinkage \citep{pope2008,Joachimi2017},
covariance tapering \citep{paz2015}, sparse precision-matrix estimation
\citep{padmanabhan2016}, mock-calibrated jackknife methods
\citep{Trusov2024}, semi-analytical covariance modelling
\citep{Connell2016,Connell2019,Philcox2019,Rashkovetskyi2023},
and analytical covariance calculations in the Gaussian approximation
\citep{Hou2022,Chellino_2026}.
Nevertheless, efficient evaluation of clustering statistics across large
simulation suites remains a fundamental computational challenge for
precision cosmology.

Motivated by this challenge, we develop \textsf{Hermes}
(HypER-speed MultirEsolution cosmic Statistics), an open-source,
parallel and GPU-accelerated framework for efficient cosmic statistics implemented in
\textsf{Python}. The core algorithm of \textsf{Hermes} is based on the
\textsf{MRACS} (Multi-Resolution Analysis--Cosmic Statistics) scheme
\citep{Feng2007}, which represents density fields and statistical windows in a
common multiresolution space. By exploiting the completeness and
orthogonality of compactly supported basis functions, \textsf{Hermes} replaces
explicit object-level counting with efficient algebraic operations on
reconstructed field coefficients.

Beyond the development of a high-performance implementation, this work
introduces a generalised formulation of cosmic statistics based on window
operators. In this framework, each statistical estimator is defined by a window function that specifies the filtering, binning, or
projection operation applied to the density field. Different statistics can
therefore be constructed by changing the window kernel rather than designing a separate counting algorithm for each estimator. The window representation also provides a direct
connection between measurements and theoretical predictions, enabling a
consistent treatment of finite binning, smoothing effects, and higher-order
statistics. This work does not attempt to determine a
universally optimal window for cosmological applications; instead, optimising
the signal-to-noise ratio of specific measurements is left to future studies
and practical survey analyses.

The main contributions of this work are fourfold:
\begin{enumerate}
\item We formulate a broad class of cosmic statistics, including
counts-in-cells, two-point and three-point correlation functions, multipole
statistics, filtered variances, and operator-based measurements involving
derivatives and gravitational potentials, as window operations on
multiresolution density fields.

\item We demonstrate that conventional 2PCF and 3PCF estimators correspond to
specific choices of window operators within this unified framework. The same
formalism naturally extends to generalised statistics, including weighted,
marked, and higher-order correlation functions, through appropriate choices of
window kernels.

\item We develop \textsf{PyHermes}, a public \textsf{Python}
implementation of this framework, providing reusable reconstructed fields,
analytic and user-defined window functions, task-level configurations,
MPI/thread parallelism, and CUDA acceleration for computationally intensive
operations such as 3PCF multipole summations.

\item Using cosmological simulations, including the
\textsf{Quijote} \citep{villaescusa2020quijote} and \textsf{Kun} \citep{Jiutian, Chen2025EmulatorI} halo catalogues, we demonstrate the practical applications of the framework, including multiresolution density reconstruction, generalised redshift-space binning windows, standard and multipole 3PCFs, weighted fields,
differential and Newtonian-potential operators, as well as representative runtime and
memory scaling.
\end{enumerate}

This paper is organised as follows. Section~\ref{sec:concept-formulation} introduces the mathematical framework of \textsf{Hermes}, including the multiresolution field representation, window formalism, and the field-based formulation of statistical estimators. Section~\ref{sec:algorithm-demonstration} develops the corresponding algorithms for window-filtered fields, higher-order correlation functions, and multipole decompositions. Section~\ref{sec:implementation-performance} presents the \textsf{PyHermes} implementation, representative applications and validation tests using cosmological simulations, and the computational performance and parallelisation strategy. Finally, Section~\ref{sec:Conclusions} summarises the main results and discusses future extensions of \textsf{PyHermes}.

\section{\textsf{Hermes} - Concept and Formulation}
\label{sec:concept-formulation}

This section introduces the mathematical framework underlying the
\textsf{Hermes} approach. We first discuss one-point statistics, exemplified by
the counts-in-cells statistic, which can be formulated as measurements of a
spatial point process through a window function. We then extend this viewpoint
to two-point statistics and review the essential perspective of transforming the
conventional \textit{ex situ} pair-counting formulation into an
\textit{in situ} field-product representation introduced in \citet{Yue2024}.
This reformulation leads to a generalised definition of the 2PCF and provides a
unified description of conventional 2PCF estimators through window functions.
The same framework naturally extends to higher-order correlation functions.
By combining the \textit{in situ} view of spatial clustering with the
window-function formalism, we highlight the fundamental role of window
operators in constructing flexible cosmic statistics.

We adopt the following notation throughout this paper. The number-density field
is denoted by \(n(\mathbf{x})\), and the density contrast is defined as
\[
    \delta(\mathbf{x})=\frac{n(\mathbf{x})}{\bar n}-1.
\]
The symbols \(D\) and \(R\) denote data and random catalogues, respectively,
while \(D^mR^n\) represents a generic combination of data and random fields
appearing in pair-counting estimators. A generic window function is denoted by
\(W_{\mathcal{P}}\), where \(\mathcal{P}\) represents the associated set of
window parameters.

Unless otherwise specified, Fourier transformation pair follows the non-unitary
convention
\[
    \hat f(\mathbf{k}) =
    \int f(\mathbf{x})\mathrm{e}^{-\mathrm{i}\mathbf{k}\cdot\mathbf{x}}d^3\mathbf{x}, \quad f(\mathbf{x})
    = \frac{1}{(2\pi)^3}
    \int \hat f(\mathbf{k})
    \mathrm{e}^{\mathrm{i}\mathbf{k}\cdot\mathbf{x}}d^3\mathbf{k}.
\]

For two generic fields \(f(\mathbf{x})\) and \(g(\mathbf{x})\), we define their
inner product in function space as
\begin{equation}
    \langle f,g\rangle
    \equiv
   \int d^3\mathbf{x}\,
    f(\mathbf{x})g(\mathbf{x})
    \equiv
    \langle f(\mathbf{x})g(\mathbf{x})\rangle_{\mathbf{x}},
\end{equation}
where the final expression emphasises that the inner product is equivalent to
the spatial average of the product field \(f(\mathbf{x})g(\mathbf{x})\).

\subsection{Spatial point process and the one-point statistics}

In cosmology, a discrete galaxy distribution
$\{\mathbf{x}_i\}_{i=1}^{N}$ is commonly regarded as a Poisson sampling of an
underlying continuous matter or tracer density field. The corresponding number
density field can be represented as a weighted sum of Dirac delta functions,
\begin{equation}\label{eq:dendis}
n(\mathbf{x})=\sum_{i=1}^N w_i
\delta_{\mathrm{D}}^3(\mathbf{x}-\mathbf{x}_i),
\end{equation}
where $w_i$ denotes the weight assigned to the $i$th object. In practical
applications of cosmic statistics, these weights may account for catalogue
completeness or selection corrections (e.g. \citealt{Singh2021,Karim2023}),
marks associated with intrinsic galaxy properties or local environments, or represent
physical quantities used to construct weighted fields, such as mass, velocity,
or momentum.

For compactness, we adopt normalised weights satisfying
$\sum_i w_i=1$ unless otherwise specified. This convention fixes the overall
normalisation of the represented tracer measure and leaves dimensionless
normalised multi-point statistics unchanged. When physical fields are
constructed, however, the corresponding physical normalisation of the weights
should be retained. For example, assigning $w_i=v_i^x$, where $v_i^x$ is the
$x$-component of the velocity of the $i$th particle, Eq.(\ref{eq:dendis}) represents the $x$-component of a velocity-weighted density field (or, for equal-mass particles, multiplying by the particle mass gives the momentum density field), rather than a number-density field.

The fundamental operation underlying many cosmic-statistics estimators is the
counting of objects within a prescribed geometric region, which is the basic
idea of counts-in-cells (CIC) statistics for large-scale structure. Spatial
correlation measurements can also be interpreted within this framework. For
example, estimating the two-point correlation function requires counting
pairs of objects within a set of separation bins, such as spherical shells.
Such counting operations can be formulated as convolutions of the density field
with a window function $W_{\mathcal{P}}(\mathbf{x})$ that specifies the geometry
and other properties of the counting region. Here $\mathcal{P}$ denotes the set
of parameters defining the window, including its scale, shape, and orientation.

Explicitly, the window-filtered density field can be written as
\begin{equation}\label{eq:1pt_CIC}
n_W(\mathbf{x}) =
\int W_{\mathcal{P}}\left(\mathbf{x}-\mathbf{x}^{\prime}\right)
n\left(\mathbf{x}^{\prime}\right)d^3\mathbf{x}^{\prime}
=
\sum_{i=1}^{N}w_i
W_{\mathcal{P}}(\mathbf{x}-\mathbf{x}_i),
\end{equation}
where the window function is normalised as
\begin{equation}
\int W_{\mathcal{P}}(\mathbf{x})d^3\mathbf{x}=1 .
\end{equation}
With this normalisation, $n_W(\mathbf{x})$ represents a locally averaged
number density (or a weighted density for general weights). For example, when
$W_{\mathcal{P}}$ is chosen as a spherical top-hat window of radius $R$, the
above expression gives the mean number density within the spherical cell
centred at $\mathbf{x}$. If the window is instead left unnormalised, the same
operation directly gives the object count inside the corresponding geometric
volume.

\subsection{The two-point statistics: from \textit{ex situ} to \textit{in situ} perspective}\label{sebsec:2pcf}

According to the \textit{in situ} perspective of the two-point correlation
function (2PCF) \citep{Yue2024}, a general second-order statistic of the
density field can be expressed as
\begin{equation}
\label{eq:CFdef}
    \xi_{\mathcal P}
    =
    \left\langle
    \delta(\mathbf{x})
    \delta_{\mathcal P}(\mathbf{x})
    \right\rangle ,
\end{equation}
where the filtered field is defined as
\begin{equation}
    \delta_{\mathcal P}(\mathbf{x})
    =(W_{\mathcal P}\circ\delta)(\mathbf{x})\equiv \int W_{\mathcal P}({\bf x}-{\bf y})\delta({\bf y})d^3{\bf y}.
\end{equation}
Here, \(W_{\mathcal P}\) denotes a window kernel characterised by the parameter
set \(\mathcal P\), and \(\circ\) represents convolution. Therefore, different
choices of \(W_{\mathcal P}\) define different second-order statistics.

With this notation, theoretical correlation functions are defined for the continuous density-contrast field $\delta$, whereas catalogue-based estimators
introduced later are constructed from the discrete number-density field $n$. For a Poisson-sampled density field, the cross-correlation between the original fluctuation field and its window-filtered counterpart is related to the power spectrum through
\begin{equation}
\label{eq:CFpow}
\xi_{\mathcal P}
=
\int\frac{d^3\mathbf{k}}{(2\pi)^3}
\hat W_{\mathcal P}^*(\mathbf{k})
P(\mathbf{k})
+
\frac{1}{\bar n}W_{\mathcal P}(\mathbf{0}),
\end{equation}
where $\hat W_{\mathcal P}$ is the Fourier transform of the window kernel and the power spectrum is defined by \(
\left\langle
\hat\delta(\mathbf{k})
\hat\delta^*(\mathbf{k}')
\right\rangle
= (2\pi)^3
\delta_{\mathrm{D}}^{3}
(\mathbf{k}-\mathbf{k}')
P(\mathbf{k})
\). The second term in Eq.~\eqref{eq:CFpow} represents the Poisson shot-noise contribution. It vanishes for separation windows that exclude the origin, such as spherical-shell bins.

For the conventional two-point correlation function, which measures the excess probability of finding pairs separated by a vector $\mathbf r$, \(\xi(\mathbf r)
    =\left\langle
    \delta(\mathbf x)
    \delta(\mathbf x+\mathbf r)
    \right\rangle \)
the parameter set reduces to $\mathcal P=\{\mathbf r\}$. The translation field of $\delta({\bf x})$ by a displacement ${\mathbf r}$ can be written as
\begin{equation}
    \delta(\mathbf x+\mathbf r) \equiv \delta_{\bf r}({\bf x}) = (W_{\mathbf r}\circ\delta)(\mathbf x)
\end{equation}
where the translation kernel and its Fourier transformation are given by 
\begin{equation}
\label{eq:translation-window}
    W_{\mathbf r}(\mathbf x)
    =\delta_{\mathrm{D}}^{3}(\mathbf x+\mathbf r),\quad \hat W_{\mathbf r}(\mathbf k)
    =
    e^{i\mathbf k\cdot\mathbf r}.
\end{equation}

In practical measurements, the separation vector is not fixed but averaged
within a finite-volume bin $\mathcal{V}_{\mathrm{bin}}$ with a volume $V_{\mathrm{bin}}$ defined in separation space. The
corresponding binned field is therefore

\begin{equation}
\label{eq:bin-field}
    \bar\delta_{\mathrm{bin}}(\mathbf x)
    =
    \frac{1}{V_{\mathrm{bin}}}
    \int_{\mathcal{V}_{\mathrm{bin}}}
    \delta(\mathbf x+\mathbf r)
    d^3\mathbf r =(W_{\mathrm{bin}}\circ\delta)(\mathbf x),
\end{equation}
where the effective binning kernel is
\begin{equation}
\label{eq:window-r-space}
    W_{\mathrm{bin}}(\mathbf x)
    =
    \frac{1}{V_{\mathrm{bin}}}
    \int_{\mathcal{V}_{\mathrm{bin}}}
    W_{\mathbf r}(\mathbf x)
    d^3\mathbf r =
    \frac{1}{V_{\mathrm{bin}}}
    \int_{\mathcal{V}_{\mathrm{bin}}}
    \delta_{\mathrm{D}}^{3}
    (\mathbf x+\mathbf r)
    d^3\mathbf r ,
\end{equation}
in which the translation kernel Eq.~\eqref{eq:translation-window} was used. 

Introducing the variable transformation \( \mathbf r \rightarrow \mathbf x+\mathbf r \), the integration domain is shifted from $V_{\mathrm{bin}}$ to $\mathbf x+V_{\mathrm{bin}}$, yielding
\begin{equation}
    W_{\mathrm{bin}}(\mathbf x)
    =
    \frac{1}{V_{\mathrm{bin}}}
    \int_{\mathbf x+\mathcal{V}_{\mathrm{bin}}}
    \delta_{\mathrm{D}}^{3}(\mathbf y)
    d^3\mathbf y .
\end{equation}

We define the three-dimensional indicator function of a volume $\mathcal{V}$ as

\begin{equation}
\label{eq:indicator}
    \Theta_{\mathcal V}(\mathbf x)
    =
    \begin{cases}
    1,&\mathbf x\in \mathcal{V},\\
    0,&\mathbf x\notin \mathcal{V} .
    \end{cases}
\end{equation}
Since the Dirac delta selects whether the integration domain contains the
origin, the binning kernel can be written as
\begin{equation}
\label{eq:window-top-hat}
    W_{\mathrm{bin}}(\mathbf x)
    =
    \frac{1}{V_{\mathrm{bin}}}
    \Theta_{\mathcal{V}_{\mathrm{bin}}}(-\mathbf x).
\end{equation}
Thus, the binning kernel is a normalised top-hat function in separation space,
obtained by averaging the elementary translation operators over the bin volume.
For inversion-symmetric bins, such as spherical shells, 
$\mathcal{V}_{\mathrm{bin}}=-\mathcal{V}_{\mathrm{bin}}$ and the minus sign in
Eq.~\eqref{eq:window-top-hat} becomes irrelevant.

In Fourier space, the binning window is
\begin{equation}
\label{eq:window-k-space}
    \hat W_{\mathrm{bin}}(\mathbf k)
    =
    \frac{1}{V_{\mathrm{bin}}}
    \int_{\mathcal{V}_{\mathrm{bin}}}
    \hat W_{\mathbf r}(\mathbf k)
    d^3\mathbf r =
    \frac{1}{V_{\mathrm{bin}}}
    \int_{\mathcal{V}_{\mathrm{bin}}}
    e^{i\mathbf k\cdot\mathbf r}
    d^3\mathbf r
    =
    \left\langle
    e^{i\mathbf k\cdot\mathbf r}
    \right\rangle_{\mathbf r}.
\end{equation}
Therefore, a finite-bin two-point statistic can be interpreted as the
correlation measured with an effective window obtained by averaging the
translation operator over the adopted separation domain. This interpretation
naturally generalises to anisotropic bins, multipole projections, and higher
order statistics within the unified window-function framework of
\textsf{PyHermes}.

\subsubsection{Real Space: Spherical Shell}
\label{subsec:Real Space: Spherical Shell}

\begin{figure}
    \centering
    \includegraphics[width=0.80\textwidth]{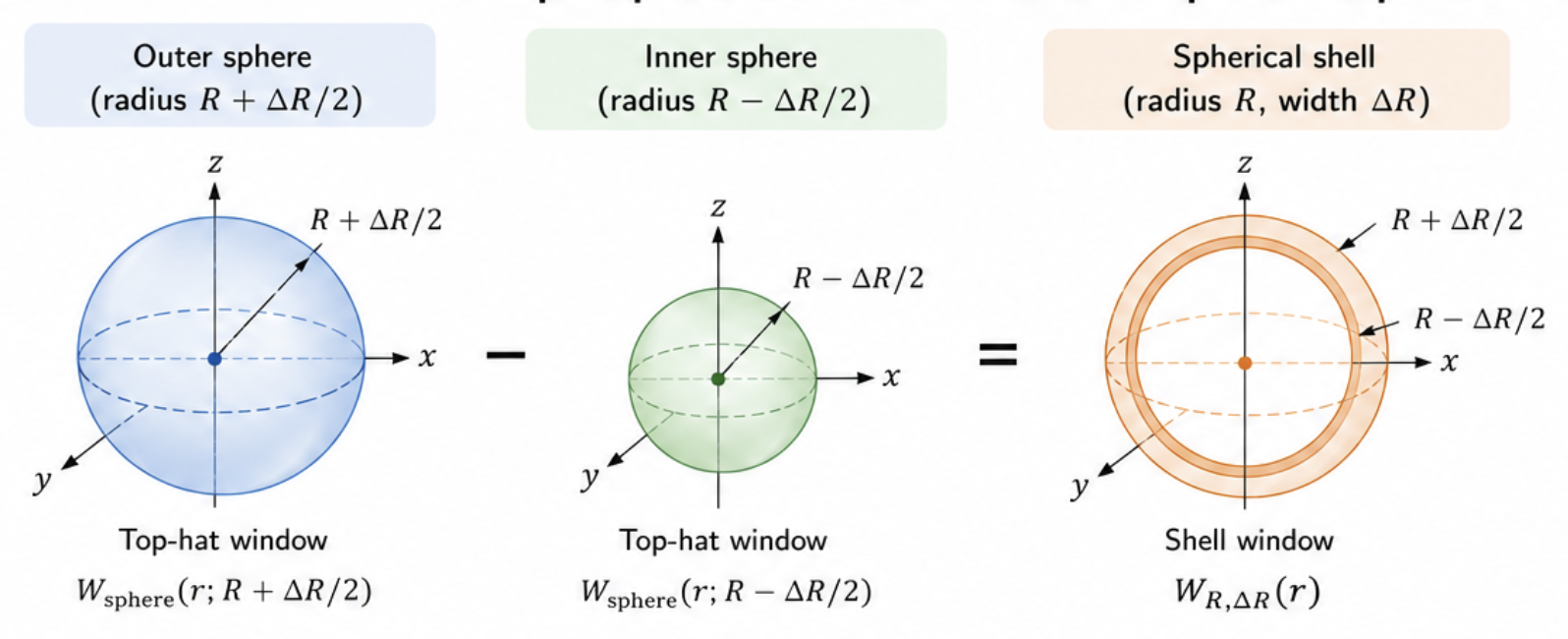}
    \caption{Real-space construction of a spherical-shell window as the subtraction of two spherical top-hat windows. The outer top-hat has radius $R+\Delta R/2$, while the inner top-hat has radius $R-\Delta R/2$. Their difference leaves a shell of mean radius $R$ and width $\Delta R$, corresponding to the normalised shell window $W_{R,\Delta R}(r)$. }
    \label{fig:sphere-shell}
\end{figure}
  
The conventional binning scheme estimates correlation functions by counting pairs within a prescribed binning volume, which can be represented by a sharp-edged top-hat window function. Such a discontinuous window is natural for direct arithmetic counting in discrete point samples, where each pair either belongs to a given bin or not. However, binning always involves a trade-off between resolution and statistical uncertainty: narrow bins preserve scale-dependent features and reduce bin-averaging effects but suffer from increased variance due to fewer pairs, whereas wider bins suppress statistical fluctuations at the cost of smoothing the underlying signal and introducing finite-volume averaging effects.

For a homogeneous and isotropic point process, the two-point correlation function can be estimated by counting pairs within a spherical shell centred on a reference point. A finite-width radial bin centred at separation $R$ is described by an inner and outer radius, $R_\pm=R\pm{\Delta R}/2$, and the corresponding window function is a spherical-shell top-hat specified by two parameters: the central separation scale $R$ and the shell width $\Delta R$, as illustrated by Fig.~\ref{fig:sphere-shell}. Following Eq.~(\ref{eq:window-r-space}), the spherical-shell window is
\begin{equation}\label{eq:Wshell}
 W_{R, \Delta R}(\mathbf{x}) = \displaystyle{\frac{1}{V_{R_+} - V_{R_-}}}\int_{{\mathcal V}_{R,\Delta R}}\delta_{\mathrm{D}}^3(\mathbf{x}+\mathbf{r})d^3\mathbf{r}
 = \frac{\Theta(R_+-|\mathbf{x}|) - \Theta(R_--|\mathbf{x}|)}{V_{R_+} - V_{R_-}},
\end{equation}
where ${\mathcal V}_{R,\Delta R}$ denotes the spherical shell bounded by the inner and outer radii $R_{\pm}$. The corresponding sphere volumes are
$V_{R_{\pm}}=(4\pi/3)R_{\pm}^{3}$, and the shell volume is
${\mathcal V}_{R,\Delta R}=V_{R_+}-V_{R_-}$. The argument $|\mathbf{x}|$ is the radial distance from the shell centre, while $\mathbf{r}$ is the elementary translation vector being averaged over the bin volume.
It is the subtraction of two adjacent spherical top-hat windows. Its Fourier transformation reads
\begin{equation}\label{eq:Wshell_k}
\hat{W}_{R, \Delta R}(k)= \frac
{V_{R_+}\hat{W}_{\mathrm{sphere}}(k,R_+) - V_{R_-}\hat{W}_{\mathrm{sphere}}(k,R_-)}{V_{R_+}-V_{R_-}},
\end{equation}

In the infinitesimally thin-shell limit $\Delta R \rightarrow 0$, Eqs.~(\ref{eq:Wshell}) and (\ref{eq:Wshell_k}) reduce to
\begin{equation}
W_{R, \Delta R}(\mathbf{x}) \rightarrow {W}_{\mathrm{shell}}(\mathbf{x};R) = \displaystyle{\frac{1}{4\pi R^2}}\delta_{\mathrm{D}}(|\mathbf{x}|-R), \quad
\hat{W}_{R, \Delta R}(\mathbf{k}) \rightarrow \hat{W}_{\mathrm{shell}}(k,R) =\frac{\sin(kR)}{kR}.
\end{equation}
The above equations recover the conventional shell-averaged expression of the two-point correlation function. The thin-shell window also follows from
\begin{equation}
    \frac{d}{dV_R}\Bigl[V_R W_{\mathrm{sphere}}(\cdot,R)\Bigr] = W_{\mathrm{shell}}(\cdot,R),
\end{equation}
and thus
\begin{equation}
    \xi_{\Delta R}(R)
    =\frac{1}{V_{R,\Delta R}}\int_{V_{R,\Delta R}}\xi(r)\,dV_r
    =\frac{1}{V_{R,\Delta R}}\int_{R_-}^{R_+}\xi(r)\,4\pi r^2\,dr.
\end{equation}

\subsubsection{The Filtered Two-Point Correlation Function}
\label{subsec: Filtered_2PCF}

\begin{figure}
    \centering
    \includegraphics[width=1.0\textwidth]{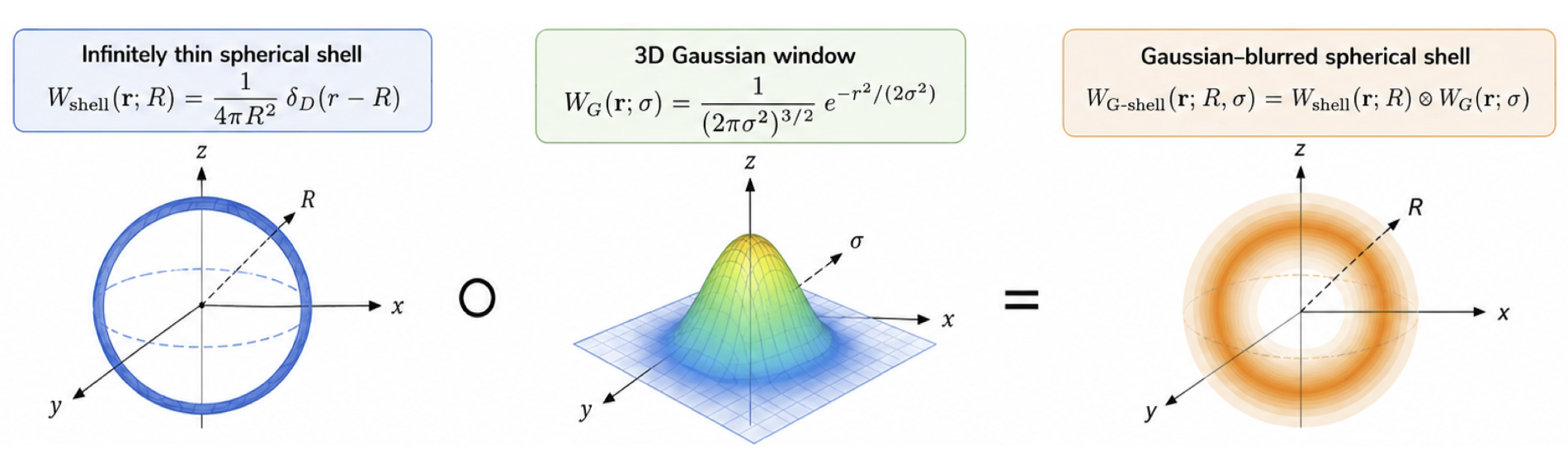}
    \caption{Real-space construction of a Gaussian-blurred spherical-shell window - An infinitesimally thin spherical shell of radius $R$ is convolved with a three-dimensional Gaussian smoothing kernel of width $\sigma$. The convolution preserves spherical symmetry and normalisation while broadening the delta-function shell into a smooth shell of finite radial width around $r=R$.}
    \label{fig:gaussian-shell}
\end{figure}

Following the \textit{in situ} perspective of generalised 2PCF \citep{Yue2024}, non-sharp-edged binning window functions are allowed. Generally, we consider two homogeneous random fields $n_A$ and $n_B$, whose cross-correlation function is conventionally defined by
\begin{equation}\label{eq:CFdef-filtered}
    \xi_{AB}(\mathbf{r}) = \langle \delta_{A}(\mathbf{x}) \delta_{B}(\mathbf{x}+\mathbf{r}) \rangle,
\end{equation}
where $\delta_A=(n_A-\bar n_A)/\bar n_A$ and $\delta_B=(n_B-\bar n_B)/\bar n_B$ are the corresponding density contrasts. If the two fields are identical,, $n_A=n_B$, Eq.~(\ref{eq:CFdef-filtered}) becomes an autocorrelation function.

As a special case, let $n_A$ and $n_B$ be obtained from a single field $n(\mathbf{x})$ filtered by the window functions $W_A$ and $W_B$, respectively:
\begin{equation}\label{eq:pairden}
\delta_A(\mathbf{x}) = W_A(\mathbf{x}) \circ \delta(\mathbf{x}), \quad \delta_B(\mathbf{x}+\mathbf{r}) = W_B(\mathbf{x})\circ W_{\mathbf{r}}(\mathbf{x}) \circ \delta(\mathbf{x}).
\end{equation}
The 2PCF of $\delta_A$ and $\delta_B$ is
\begin{equation}
    \xi_{AB}(\mathbf{r}) =\langle W_A \circ \delta,  W_B\circ W_{\mathbf{r}} \circ \delta\rangle
    =\langle \delta,  W_A^{\dagger} \circ W_B\circ W_{\mathbf{r}} \circ \delta\rangle.
\end{equation}

We further assume that $n(\mathbf{x})$ is an $N$-galaxy distribution $\{\mathbf{x}_i\}_{i=1}^{N}$, which is a Poisson sampling of the underlying density field $n(\mathbf{x}) = \bar{n}(1+\delta(\mathbf{x}))$ with power spectrum $P(\mathbf{k})$. Accordingly, $\xi_{AB}(\mathbf{r})$ is related to $P(\mathbf{k})$ via
\begin{equation}\label{eq:CFpow-filtered}
\xi_{AB}(\mathbf{r}) = \displaystyle{\frac{1}{(2\pi)^3}}\int \hat{W}^*_A(\mathbf{k})\hat{W}_B(\mathbf{k})\Bigl[P(\mathbf{k}) + \frac{1}{\bar{n}}\Bigr]\mathrm{e}^{\mathrm{i}\mathbf{k}\cdot\mathbf{r}}d^3\mathbf{k},
\end{equation}
where the term $1/\bar{n}$ on the right-hand side is the Poisson shot-noise contribution and must be subtracted. For notational simplicity, we omit this term below.

Consider filtering the density field by the same window function $W(\mathbf{x})$ for both fields. Taking $W_A(\mathbf{x}) = W_B(\mathbf{x}) = W(\mathbf{x})$ in Eq.~(\ref{eq:CFpow-filtered}), the two-point autocorrelation function becomes
\begin{equation}\label{eq:sACF}
    \xi_W(\mathbf{r}) = \langle\delta_W(\mathbf{x})\delta_W(\mathbf{x}+\mathbf{r})\rangle
    = \frac{1}{(2\pi)^3}\int P(\mathbf{k}) |\hat{W}(\mathbf{k})|^2 \mathrm{e}^{\mathrm{i}\mathbf{k}\cdot\mathbf{r}} d^3\mathbf{k}.
\end{equation}
If the particle distribution can be modelled by a homogeneous and isotropic random field, $\xi_W(\mathbf{r})$ will depend on the distance $r=|\mathbf{r}|$ only and can be given by taking a surface average of radius $r$ around a reference point, i.e.,
\begin{equation}
     \xi_W(r) =  \frac{1}{4\pi r^2} \int \xi_W(\mathbf{r}) r^2 d\Omega.
\end{equation}
It can be shown that the 2PCF measured from a filtered density field $\delta_W=W\circ\delta$ can be written as
\begin{equation}
    \xi_W(r) = \langle \delta_W,W_{\mathrm{shell}}\circ\delta_W\rangle = \frac{1}{2\pi^2}\int_0^{\infty} P(k) |\hat{W}(k)|^2 \hat{W}_{\mathrm{shell}}(k,r) k^2\,dk = \langle  \delta,W^{\dagger}\circ W\circ W_{\mathrm{shell}}\circ\delta\rangle.
\end{equation}
Therefore, at the level of the two-point statistic, smoothing the original density-field via $W$ is mathematically equivalent to replacing the spherical-shell binning window with the effective window \( W_{\mathrm{eff}}
\equiv W^\dagger\circ W\circ W_{\mathrm{shell}}\). 

For the Gaussian kernel,
\(\widehat{W}_{\mathrm{gauss}}(k)
=\exp(-k^2a^2/2)\), the real-space kernel
\(W_{\mathrm{gauss}}\) is real and inversion symmetric, so that
\(W_{\mathrm{gauss}}^\dagger=W_{\mathrm{gauss}}\). The convolution
\(W_{\mathrm{gauss}}^\dagger\circ W_{\mathrm{gauss}}\) is therefore
another Gaussian kernel with variance \(2a^2\). Hence, the
autocorrelation of the Gaussian-smoothed field can equivalently be
evaluated using the original unsmoothed field together with a
Gaussian-blurred spherical-shell binning window. In configuration
space, the effective window is
\[
W_{\mathrm{eff}}
=
W_{\mathrm{gauss}}^\dagger
\circ W_{\mathrm{gauss}}
\circ W_{\mathrm{shell}}
=
W_{\mathrm{G\text{-}shell}}(r;R,\sigma),
\qquad
\sigma=\sqrt{2}\,a.
\]
Explicitly, this window is given by \citep{Yue2024}
\begin{equation}
W_{\mathrm{G\text{-}shell}}(r;R,\sigma)
=
\frac{1}{(2\pi\sigma^2)^{3/2}}
\exp\!\left[-\frac{r^2+R^2}{2\sigma^2}\right]
\frac{\sinh(rR/\sigma^2)}{rR/\sigma^2}.
\end{equation}
In the narrow-shell limit, \(\sigma\ll R\) and \(r\simeq R\), using
\(\sinh x\simeq e^x/2\) for \(x\gg1\) gives
\begin{equation}
\label{eq:thin_GaussianShell}
W_{\mathrm{G\text{-}shell}}(r;R,\sigma)
\simeq
\frac{1}{(2\pi\sigma^2)^{3/2}}
\frac{\sigma^2}{2rR}
\exp\!\left[-\frac{(r-R)^2}{2\sigma^2}\right]
\simeq
\frac{1}{(2\pi\sigma^2)^{3/2}}
\frac{\sigma^2}{2R^2}
\exp\!\left[-\frac{(r-R)^2}{2\sigma^2}\right].
\end{equation}
Compared with a finite-width spherical-shell top-hat, the effective
Gaussian-shell window replaces the sharp radial boundaries by a smooth
profile centred on \(r=R\), with characteristic Gaussian radial width
\(\sigma\). Figure~\ref{fig:gaussian-shell} illustrates how convolution
of the infinitesimally thin spherical-shell window with the effective
Gaussian kernel produces this smoothly broadened radial selection.

For Gaussian smoothing windows with different smoothing radii, the factor
$|\hat{W}_{\mathrm{Gauss}}(k)|^2$ acts as a low-pass filter, progressively
suppressing small-scale modes as the smoothing scale increases. The resulting
2PCF is therefore not a different clustering statistic, but the correlation
function measured through a modified effective window. This interpretation
provides a direct framework for quantifying the impact of finite smoothing and
binning effects when comparing measurements with theoretical models.

Similarly, the sphere-averaged 2PCF can be given by
\begin{equation}
		\bar{\xi}_W(R) = \frac{3}{4\pi R^3}\int_{0}^{R} \xi_W(r)\cdot 4\pi r^2\,dr = \langle\delta_W,W_{\mathrm{sphere}}\circ \delta_W\rangle,
\end{equation}
and the density variance,
\begin{equation}\label{eq:varw}
	\sigma_W^{2}(R)=\int[W_{\mathrm{sphere}}\circ\delta_W ]^2(\mathbf{x})\,d^3\mathbf{x}
    =\langle W_{\mathrm{sphere}}\circ\delta_W, W_{\mathrm{sphere}}\circ\delta_W\rangle =\langle\delta_W, W_{\mathrm{sphere}} \circ W_{\mathrm{sphere}} \circ \delta_W \rangle.
\end{equation}
The corresponding expressions for $\bar{\xi}_W$ and $\sigma_W^{2}$ can be obtained from Eq.~(\ref{eq:sACF}) by replacing $W_{\mathrm{shell}}$ with $W_{\mathrm{sphere}}$ and by using the additional factor $|\hat{W}_{\mathrm{sphere}}|^2$, respectively.

\subsubsection{Plane-parallel Redshift Space: Double-ring and Gaussian-blurred Ring Windows}
\label{subsec:RSD Circular_Ring}

\begin{figure*}
    \centering
    \includegraphics[width=1.0\textwidth]{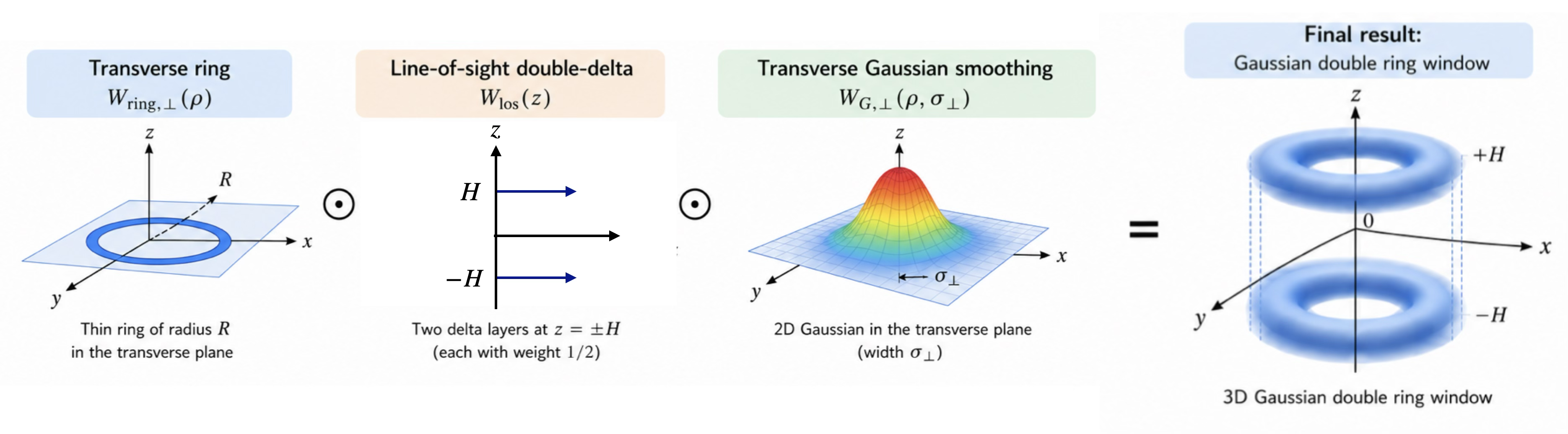}
    \caption{Real-space construction of the Gaussian-blurred double-ring window - The elementary double-ring window is the product of the transverse circular-ring
kernel \(W_{\mathrm{ring},\perp}(\rho;r_\perp)\) and the line-of-sight kernel
\(W_{\mathrm{los}}(z;r_\parallel)
=\{\delta_{\mathrm D}(z-r_\parallel)+
\delta_{\mathrm D}(z+r_\parallel)\}/2\).
It therefore consists of two infinitesimally thin circular rings centred at
\(z=\pm r_\parallel\), with a centre-to-centre separation \(2r_\parallel\).
Convolution with the transverse Gaussian kernel
\(W_{\mathrm{G},\perp}(\rho;\sigma_\perp)\) broadens the radial profile of each
ring while leaving its line-of-sight position unchanged. The resulting window
is consequently a pair of Gaussian-blurred rings.}
\label{fig:gauss_ring}
\end{figure*}

In redshift space, clustering statistics become anisotropic due to the contribution of peculiar velocities. Besides the separation scale, this anisotropy can be characterised by the dependence of clustering statistics on the orientation of the separation vector relative to the line of sight. Under the plane-parallel approximation, we parameterise the two-point correlation function by the transverse and line-of-sight separations, $r_\perp$ and $r_\parallel$.

Corresponding to the cylindrical symmetry of redshift-space clustering, the conventional spherical-shell translation window is replaced by an axisymmetric ring window, which performs an angular average over a circular locus at fixed $(r_\perp,r_\parallel)$. The anisotropic 2PCF can therefore be written as
\begin{equation}
    \xi(r_{\parallel}, r_{\perp}) = \langle \delta(\mathbf{x}),\delta(\mathbf{x}) \circ W_{\mathrm{ring}}(r_{\parallel}, r_{\perp}) \rangle
\end{equation}
where $W_{\mathrm{ring}}(r_\parallel,r_\perp)$ denotes the translation kernel associated with the cylindrical coordinate binning scheme. In cylindrical coordinates, the infinitesimally thin ring window is given by
\begin{equation}\label{eq:window-ring-realspace}
W_{\rm ring}(\rho,z)=\frac{\delta_{\mathrm{D}}(\rho-r_\perp)}{2\pi r_\perp}\frac{\delta_{\mathrm{D}}(|z|-r_\parallel)}{2},
\end{equation}
where we take a symmetric double ring along the line-of-sight axis at $z=\pm r_{\parallel}$. Taking the azimuthal average in the transverse plane, its Fourier-space representation becomes
\begin{equation}\label{eq:window-ring-kspace}
     \hat{W}_{\rm ring}(k_\parallel,k_{\perp})=\langle \mathrm{e}^{i\mathbf{k}\cdot\mathbf{r}}\rangle_{\mathrm{ring}}=\cos(k_{\parallel}r_{\parallel})J_0(k_{\perp}r_{\perp}),
\end{equation}
where $J_0$ is the zeroth-order Bessel function.

For realistic measurements, finite bin widths must be taken into account.
Under the plane-parallel approximation, a bin centred at the transverse and
line-of-sight separations \((r_\perp,r_\parallel)\) consists of an annulus of
width \(\Delta r_\perp\) in the transverse plane and two line-of-sight slabs
of width \(\Delta r_\parallel\), centred at \(z=\pm r_\parallel\). Defining
\(\rho_\pm=r_\perp\pm{\Delta r_\perp}/2,\)
the corresponding normalised window in configuration space is
\begin{equation}
W_{\mathrm{ann}}(\rho,z)
=
\frac{
\Theta(\rho-\rho_-)\Theta(\rho_+-\rho)
}{
\pi(\rho_+^2-\rho_-^2)
}
\frac{
\Theta\!\left(\displaystyle\frac{\Delta r_\parallel}{2}
              -|z-r_\parallel|\right)
+
\Theta\!\left(\displaystyle\frac{\Delta r_\parallel}{2}
              -|z+r_\parallel|\right)
}{
2\Delta r_\parallel
}.
\end{equation}
where \(\rho\) and \(z\) denote the transverse and line-of-sight coordinates,
respectively. Its Fourier-space representation is
\begin{equation}
\widehat W_{\mathrm{ann}}(k_\perp,k_\parallel)
=
\frac{
\sin(k_\parallel\Delta r_\parallel/2)
}{
k_\parallel\Delta r_\parallel/2
}
\cos(k_\parallel r_\parallel)
\left\langle
J_0(k_\perp\rho)
\right\rangle_{\Delta r_\perp},
\end{equation}
with the transverse annular average defined by
\begin{equation}
\left\langle J_0(k_\perp\rho)\right\rangle_{\Delta r_\perp}
=
\frac{2}{\rho_+^2-\rho_-^2}
\int_{\rho_-}^{\rho_+}
J_0(k_\perp\rho)\,\rho\,d\rho
=
\frac{1}{r_\perp\Delta r_\perp}
\int_{r_\perp-\Delta r_\perp/2}^{r_\perp+\Delta r_\perp/2}
J_0(k_\perp\rho)\,\rho\,d\rho .
\end{equation}
In the limit
\(\Delta r_\perp,\Delta r_\parallel\rightarrow0\), this finite annular-bin window reduces to the double circular-ring window Eq.(\ref{eq:window-ring-realspace}) and Eq.(\ref{eq:window-ring-kspace}). 

A Gaussian-blurred ring is obtained simply by multiplying the transverse factor by a Gaussian filter,
\begin{equation}
    \widehat W_{\mathrm{G\text{-}ring}}(k_\perp,k_\parallel;r_\perp,r_{\parallel},\sigma_\perp)
    =\underbrace{\cos(k_\parallel r_\parallel)}_{\widehat W_{\mathrm{los}}}\underbrace{J_0(k_\perp r_{\perp})}_{\widehat W_{\mathrm{ring},\perp}}
    \underbrace{\exp\!\left[-\frac12 k_\perp^2\sigma_\perp^2\right]}_{\widehat W_{\mathrm{G},\perp}}
    .
\end{equation}
In the transverse plane, the inverse Fourier transform is
\begin{equation}
W_{\mathrm{G\text{-}ring}}^{\perp}(\rho,\sigma_{\perp})=\mathrm{iFT}\left[\mathrm{e}^{-k_{\perp}^2\sigma_{\perp}^2/2}J_0(k_{\perp}r_\perp)\right] =
\frac{1}{2\pi \sigma_{\perp}^2}
\mathrm{e}^{-(\rho^2+r_\perp^2)/(2\sigma_{\perp}^2)}
I_0\!\left(\frac{\rho r_\perp}{\sigma_{\perp}^2}\right),
\end{equation}
where $I_0$ is the zeroth-order modified Bessel function.
Similar to the spherical Gaussian shell Eq.~(\ref{eq:thin_GaussianShell}), in the narrow-ring limit, $\sigma_{\perp} \ll r_\perp$ and $\rho \sim r_\perp$, the asymptotic form $I_0(x)\approx {\mathrm{e}^x}/{\sqrt{2\pi x}}$ for $x\gg 1$ gives
\begin{equation}
W_{\mathrm{G\text{-}ring}}^{\perp}(\rho,\sigma_{\perp}) \approx \frac{1}{(2\pi)^{3/2}r_\perp\sigma_{\perp}}
\exp\!\left[
-\frac{(\rho-r_\perp)^2}{2\sigma_{\perp}^2}
\right].
\end{equation}
Thus, the inverse Fourier transform corresponds to a normalised Gaussian-ring window, whose radial profile is centred at $\rho=r_\perp$ with characteristic width $\sigma_{\perp}$. Accordingly, convolution with the transverse Gaussian kernel broadens each circular ring in the radial direction, producing a pair of Gaussian-blurred rings while leaving their line-of-sight locations unchanged. Figure~\ref{fig:gauss_ring} provides a schematic illustration of the resulting double Gaussian-ring window in redshift space.

An alternative to the ring-based binning scheme is to employ other axisymmetric windows, such as the two-dimensional disk and cylindrical-surface windows (corresponding to items (g) and (h) in Table~A1). As discussed by Li et al. (2026, in preparation), combining different axisymmetric windows can provide complementary sensitivity to redshift-space distortions and improve cosmological parameter constraints from galaxy surveys.

The axisymmetric windows - including the cylinder, cylindrical shell, disk, and ring windows - differ in the geometrical domains over which the averaging is performed: a three-dimensional volume, a cylindrical side surface, a two-dimensional transverse disk, or a one-dimensional circular ring. Consequently, they introduce distinct scale- and orientation-dependent filters in Fourier space, leading to complementary sensitivities to the Kaiser effect, Fingers-of-God damping, and higher-order redshift-space anisotropies. Our alternative approach recover perfectly the same results as with the normal pair-counting, but the flexible window basis of the new approach provides a promising route toward constructing optimised estimators for cosmic statistics.

\subsection{Compensated filters and band-averaged power spectra}

Low-pass and compensated filters provide complementary descriptions of
the density field. A normalised low-pass window,
\(W_{\mathcal P}(\mathbf{x})\), satisfies
\begin{equation}
    \int W_{\mathcal P}(\mathbf{x})\,d^3\mathbf{x}=1,
\end{equation}
and therefore preserves a spatially constant field. Its Fourier response
is concentrated near \(k=0\), so the corresponding filtered variance
accumulates power over modes below a characteristic smoothing scale.
Such a filter is naturally suited to describing the coarse-grained
density or the mean-square fluctuation within a smoothing region.

A compensated filter $\mathcal{W}_{\rm comp}$ instead satisfies
\begin{equation}
    \int \mathcal{W}_{\rm comp}(\mathbf{x})\,d^3\mathbf{x}=0,
    \qquad
    \widehat{\mathcal{W}}_{\rm comp}(\mathbf{0})=0.
\end{equation}
It removes the constant mode and responds only to spatial variations.
Besides reducing sensitivity to an additive mean-density offset, this
condition suppresses modes far below the characteristic filtering scale
and allows the filter to isolate fluctuations within a more localised
range of wavenumbers.

A general way to construct a compensated kernel is to differentiate a
normalised low-pass window family with respect to one of its scale
parameters. For a scale parameter \(p\in\mathcal P\), we define the
unnormalised kernel
\begin{equation}\label{eq:hpDeriv}
    {\mathcal{W}}_{p}(\mathbf{x})
    \equiv
    \frac{\partial W_{\mathcal P}(\mathbf{x})}
         {\partial\ln p}.
\end{equation}
Because the normalisation of \(W_{\mathcal P}\) is independent of \(p\),
\begin{equation}
    \int {\mathcal{W}}_{p}(\mathbf{x})\,d^3\mathbf{x}
    =
    \frac{\partial}{\partial\ln p}
    \int W_{\mathcal P}(\mathbf{x})\,d^3\mathbf{x}
    =0.
\end{equation}
The scale derivative therefore converts the low-pass window into a
compensated wavelet filter with a scale-localised band-pass response.
A particular example is the Gaussian-derivative wavelet (GDW), introduced
by \citet{Wangyun2022A,Wangyun2022B} to characterise the scale dependence
of matter clustering and its environmental dependence. The compensated
wavelet filters currently implemented in \textsf{PyHermes} are summarised
in rows (j)--(l) of Table~\ref{tab:kernel} in
Appendix~\ref{app:window_functions}.

For comparisons between different scales, we further normalise the compensated kernel to have unit \(L^2\) norm. Under the Fourier convention introduced in Section~\ref{sec:concept-formulation}, Parseval's theorem gives
\begin{equation}\label{eq:hpNorm}
    \int
    \left|{\mathcal{W}}_{p}(\mathbf{x})\right|^2
    d^3\mathbf{x}
    =
    \frac{1}{(2\pi)^3}
    \int
    \left|\widehat{\mathcal{W}}_{p}(\mathbf{k})\right|^2
    d^3\mathbf{k}
    =1.
\end{equation}
The variance of the filtered density field is consequently
\begin{equation}\label{eq:band_averaged_power}
    \sigma_{\mathcal{W}}^2(p)
    =
    \frac{1}{(2\pi)^3}
    \int
    P(\mathbf{k})
    \left|\widehat{\mathcal{W}}_{p}(\mathbf{k})\right|^2
    d^3\mathbf{k}.
\end{equation}
Because
\(\widehat{\mathcal{W}}_{p}(\mathbf{0})=0\) and the Fourier band-pass response of
a scale derivative is typically localised around a characteristic
wavenumber, Eq.~(\ref{eq:band_averaged_power}) represents a weighted,
band-averaged power spectrum. This differs from the variance obtained
with a conventional low-pass smoothing window, which measures power
accumulated below the smoothing scale.

This distinction also has a direct interpretation in the MRA framework.
Scaling functions provide the low-pass approximation of the field and
retain its coarse-grained density, whereas wavelet functions have zero
mean and encode the detail added between adjacent resolution levels.
The \(L^1\) normalisation of the low-pass window therefore preserves the
mean number or mass density, while the \(L^2\) normalisation of the
compensated filter provides an energy-based convention for comparing
fluctuation amplitudes across scales. The resulting statistic is closely
related to the wavelet power spectrum used in discrete wavelet analyses
\citep{fang2000,yang2001a,yang2001b,yang2002}.

\subsection{The N-Point correlation functions}

The two-point correlation function can be naturally generalised to arbitrary
higher-order correlation functions. For a spatial point process, consider
$N$ infinitesimal disjoint volume elements
$d^3{\bf r}_1,\ldots,d^3{\bf r}_N$. The joint probability of finding one point in each volume element is
\begin{equation}
dP =\lambda_N({\bf r}_1,\ldots,{\bf r}_N) \,
d^3{\bf r}_1\cdots d^3{\bf r}_N ,
\end{equation}
where $\lambda_N$ denotes the $N$-point product density of the point process. The reduced $N$-point correlation functions are defined as the connected moments of the density contrast field,
\begin{equation}
\xi^{(N)}({\bf r}_1,\ldots,{\bf r}_N)
=
\left\langle
\delta({\bf r}_1)
\cdots
\delta({\bf r}_N)
\right\rangle_c .
\end{equation}

For a statistically homogeneous field, the correlation functions are invariant under translations and therefore depend only on the relative separations between points rather than on the absolute coordinate origin.

In practical measurements, however, the idealized point configuration must be replaced by finite bins. Following the {\it in situ} formulation of the filtered 2PCF, we introduce a filtered density field \(\delta_W({\bf x})=
(W\circ\delta)({\bf x}),\) where $W$ represents a smoothing or binning window. The corresponding filtered NPCF can then be written as
\begin{equation}
\xi_W^{(N)}
({\bf r}_1,\ldots,{\bf r}_{N-1})
=
\int d^3{\bf s}\,
\left\langle
\delta_W({\bf s})
\prod_{i=1}^{N-1}
\delta_W({\bf s}+{\bf r}_i)
\right\rangle_c =
\int d^3{\bf s}\,
\left\langle
(W\circ\delta)({\bf s})
\prod_{i=1}^{N-1}
\left[
W\circ W_{{\bf r}_i}\circ\delta
\right]({\bf s})
\right\rangle_c,
\end{equation}
where $W_{{\bf r}_i}$ denotes the translation operator associated with the
relative separation ${\bf r}_i$. Under the ergodic assumption, the ensemble
average can be replaced by a spatial average over a sufficiently large
sampling volume.

The introduction of filtered density fields serves three purposes: it provides
a natural description of finite binning effects, enables efficient NPCF
estimation through operations on filtered fields, and can be naturally
implemented within the \textsf{MRA} framework for high-performance
computations. Analogous to the relation between the filtered 2PCF and the power spectrum, the connection between the filtered NPCF and the corresponding polyspectrum involves a window correction. In Fourier space, each density leg contributes a window factor, giving the overall correction $\prod_{i=1}^N W(\mathbf{k}_i)\big\vert_{\sum_i\mathbf{k}_i=0}$. The filtered 3PCF and its Fourier counterpart, the bispectrum, for spherical top-hat and Gaussian smoothing windows have been derived explicitly by \citet{Yue2024}.

\section{\textsf{Hermes} - Algorithm and Demonstration}
\label{sec:algorithm-demonstration}

Section~2 introduced the statistical role of window functions in defining
cosmic statistics. In this section, we describe their computational
realisation within the multiresolution-analysis (MRA) framework.
We first introduce the reconstruction of continuous density fields from
discrete point distributions by projecting the fields onto an MRA
space spanned by compactly supported basis functions. The corresponding window
functions are represented and manipulated in the same MRA space through
a bilinear expansion \citep{Feng2007}. Taking advantage of the completeness and orthogonality of the basis functions, together with the \textit{in situ} formulation of spatial
statistics, we demonstrate how two-point statistics can be efficiently
evaluated through field--window operations. Furthermore, by combining the MRA
representation with the multipole decomposition of isotropic window kernels,
the same framework can be naturally extended to higher-order correlation
functions.

The key computational advantage of the MRA framework comes from
replacing particle-based algorithms with algebraic operations among
MRA coefficients. These operations, together with the flexible
representation of window functions in the same MRA space, constitute the
computational foundation of \textsf{Hermes} and enable a unified and efficient treatment of
different cosmic-statistics estimators.

Meanwhile, we demonstrate the statistical applications of \textsf{Hermes} using numerical simulations. The main set of simulations used for demonstrations is taken from the \textsf{Quijote} project \citep{villaescusa2020quijote}, which provides a suite of more than 82,000 full $N$-body simulations designed to quantify the information content of cosmological observables. The \textsf{Quijote} simulations include a large number of independent realisations in boxes of side length $1000\,h^{-1}\mathrm{Mpc}$, each containing $512^3$ dark-matter particles and approximately $4\times 10^5$ haloes.
Specifically, we use the fiducial cosmology adopted by the Quijote simulations, with parameters $(h,\Omega_{\mathrm{m}},\Omega_{\mathrm{b}},n_{\mathrm{s}},\sigma_8)=(0.6711,0.3175,0.049,0.9624,0.834)$.

The \textsf{Kun} simulations used in this work are the emulator-run component
of the broader \textsf{Jiutian} simulation suite developed for CSST
extragalactic surveys \citep{Jiutian,Chen2025EmulatorI}. 
The \textsf{Kun} suite contains 129 high-resolution \(N\)-body simulations sampling an extended cosmological
parameter space, including variations in \(\Omega_{\mathrm{b}}\),
\(\Omega_{\mathrm{cb}}\), \(H_0\), \(n_{\mathrm{s}}\), \(A_{\mathrm{s}}\), \(w_0\), \(w_a\),
and \(\sum m_\nu\). The parameter ranges are listed in
Table~\ref{tab:kun_param_ranges}. Each \textsf{Kun} simulation evolves \(3072^3\) cold-dark-matter particles in a
periodic box of side length \(1\,h^{-1}\mathrm{Gpc}\). For each cosmological model,
we construct a fixed-number-density halo sample by ranking haloes by mass and
selecting the most massive objects until the target number density
\(\bar n=5\times10^{-3}\,h^3\mathrm{Mpc}^{-3}\) is reached. For a simulation box of
side length \(1\,h^{-1}\mathrm{Gpc}\), this corresponds to \(5\times10^6\) haloes
in each cosmological model. This fixed-number-density selection provides a
consistent tracer definition across different cosmologies and is used for the
emulator-based clustering measurements shown here.

\begin{table*}
\centering
\caption{Cosmological parameter ranges of the \textsf{Kun} simulations.
The first eight quantities are the sampled parameters of the simulation suite.
The quantities \(\Omega_{\mathrm{m}}\) and \(\sigma_8\) are included because they are
used in the emulator training in this work.}
\label{tab:kun_param_ranges}
\small
\setlength{\tabcolsep}{7pt}
\renewcommand{\arraystretch}{1.18}
\begin{tabular}{@{}cccccc@{}}
\hline
Parameter &
\(\Omega_{\mathrm{b}}\) &
\(\Omega_{\mathrm{cb}}\) &
\(H_0\,(\mathrm{km\,s^{-1}\,Mpc^{-1}})\) &
\(n_{\mathrm{s}}\) &
\(A_{\mathrm{s}}\) \\
\hline
Range &
\([0.04,\,0.06]\) &
\([0.24,\,0.40]\) &
\([60,\,80]\) &
\([0.92,\,1.00]\) &
\([1.70,\,2.50]\times10^{-9}\) \\
\hline
Parameter &
\(w_0\) &
\(w_a\) &
\(\sum m_\nu\,(\mathrm{eV})\) &
\(\Omega_{\mathrm{m}}\) &
\(\sigma_8\) \\
\hline
Range &
\([-1.30,\,-0.70]\) &
\([-0.50,\,0.50]\) &
\([0.00,\,0.30]\) &
\([0.2400,\,0.4089]\) &
\([0.5445,\,1.1829]\) \\
\hline
\end{tabular}
\end{table*}

Following the emulator construction strategy adopted in the CSST analysis, we
measure the corresponding statistic for all 129 cosmological models and train
an independent Gaussian-process emulator to learn its dependence on cosmology. In our baseline
implementation, 100 cosmological models are randomly selected as the training
set, while the remaining 29 models are used as an independent test set to
assess the interpolation accuracy.
Related CSST emulator studies based on the same \textsf{Kun} simulation suite
have also been carried out for halo mass functions and biased tracer
clustering \citep{Chen2025EmulatorII,Zhou2025EmulatorIII}.

\subsection{Filtered field reconstruction based on multiresolution analysis}

The microscopic density field associated with a discrete point
distribution is a weighted sum of Dirac delta distributions and is
therefore not an ordinary function that can be evaluated pointwise. The
central step of our algorithm is to represent its finite-resolution
projection in a complete basis. Specifically, we project the singular
density field in Eq.~(\ref{eq:dendis}) onto a space spanned by compactly
supported orthonormal scaling functions from multiresolution analysis \citep{Daubechies1992,Fang1998}.

Let \(\phi(x)\) denote the one-dimensional father scaling function, where
\(x\) is a dimensionless coordinate rescaled to the unit computational
domain. At resolution level \(J\), dilation by \(2^J\) and translation
by the integer index \(\mathbf{l}\) give the \(\mathcal{D}\)-dimensional tensor-product
basis
\begin{equation}
\phi_{J\mathbf{l}}(\mathbf{x})
=
\prod_{i=1}^{\mathcal D}\phi_{J,l_i}(x_i),
\qquad
\phi_{J,l}(x)
=
2^{J/2}\phi(2^J x-l),
\qquad
l\in\mathbb{Z},
\end{equation}
which satisfy the orthogonality condition 
\begin{equation}
\left\langle
\phi_{J\mathbf{l}},
\phi_{J\mathbf{m}}
\right\rangle
=
\delta_{\mathbf{l}\mathbf{m}}.
\end{equation}
Here we have adopted the same resolution level \(J\) in every spatial
dimension. 

\begin{figure*}
    \centering
\includegraphics[width=0.95\textwidth]{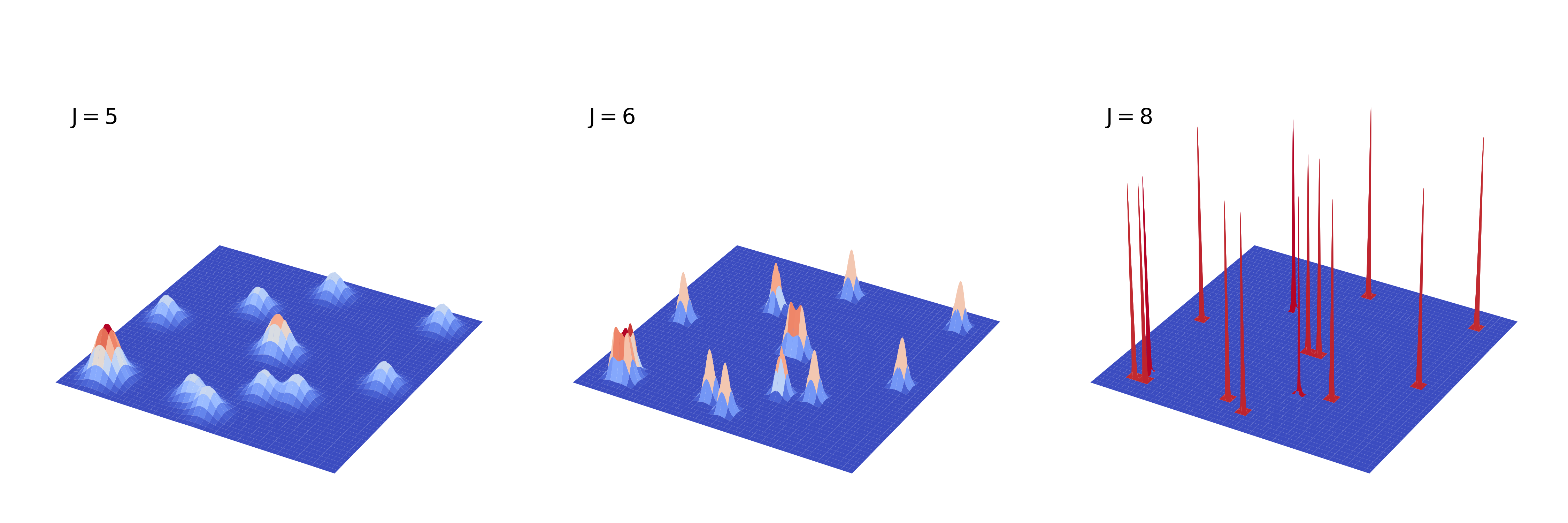}
    \caption{Schematic reconstruction of the same point distribution in the
    multiresolution scaling-function basis at three resolution levels. At low
    resolution, each particle is represented by a broad compact-support basis
    response, producing a smooth coarse-grained field. As the resolution level
    \(J\) increases, the reconstructed peaks become more localised, illustrating
    how the scaling-function representation approaches the underlying singular
    point field.}
    \label{fig:field_resolution}
\end{figure*}

\begin{figure*}
    \centering
    \includegraphics[width=0.90\textwidth]{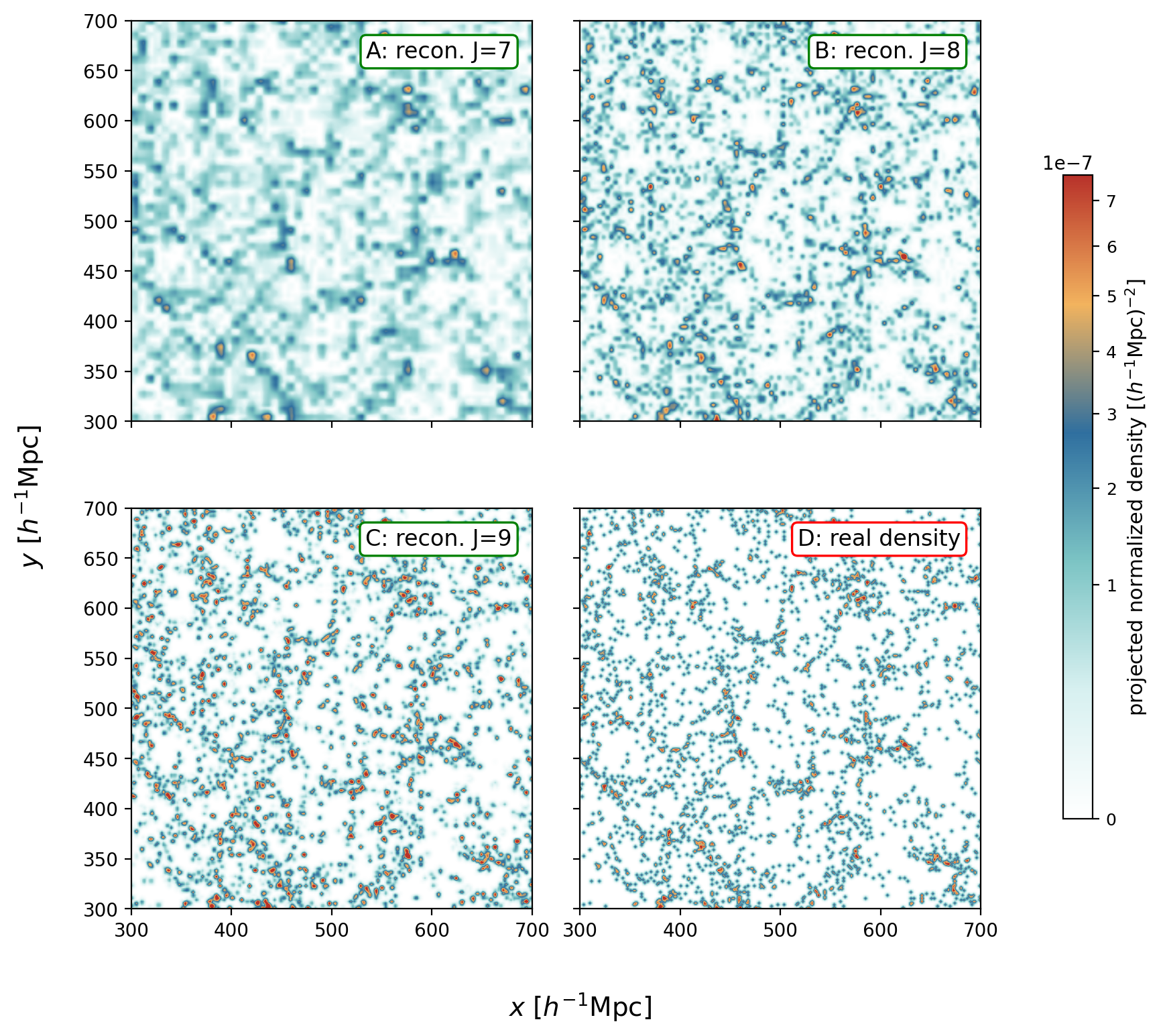}
    \caption{Demonstration of continuous density-field reconstruction from a discrete
\textsf{Quijote} halo catalogue using the MRA representation in
\textsf{PyHermes}. The projected density fields are reconstructed at three
resolution levels, $J=7$, $J=8$, and $J=9$ (panels A--C), and compared with the
direct particle projection (panel D) in the same spatial slab. Increasing the
MRA resolution gradually restores small-scale halo structures, while lower
resolution levels provide smoother coarse-grained representations. This
resolution dependence illustrates the central role of the MRA basis in
controlling the trade-off between localisation and smoothness during the
conversion from discrete catalogues to continuous physical fields.}
    \label{fig:sfc_field_resolution}
\end{figure*}

For the density field $n({\bf x})$, its finite-resolution projection via the operation $P_J$ at the scale-resolution $J$ can be reconstructed as
\begin{equation}
n(\mathbf{x}) \rightarrow n_J(\mathbf{x}) \equiv \hat{P}_J n(\mathbf{x})=\sum_{\mathbf{l}} \epsilon_{J\mathbf{l}}\phi_{J\mathbf{l}}(\mathbf{x}),
\end{equation}
with
\begin{equation}\label{eq:phi_projection}
\epsilon_{J\mathbf{l}} = \langle \phi_{J\mathbf{l}},n\rangle=\int n(\mathbf{x})\phi_{J\mathbf{l}}(\mathbf{x}) d^3\mathbf{x} = \sum_{i=1}^{N}w_i\phi_{J\mathbf{l}}(\mathbf{x}_i).
\end{equation}

In a conventional mass-assignment scheme, particle masses are partitioned onto a regular mesh using an assignment kernel \citep{Hockney1988}, such as the nearest-grid-point (NGP), cloud-in-cell (CIC), or triangular-shaped-cloud (TSC) kernel. The resulting density values are associated with grid cells or grid nodes, and the density at an arbitrary spatial position, when required, is obtained by interpolating these gridded values. This sampling procedure introduces kernel-dependent smoothing and can fold unresolved small-scale modes into the resolved Fourier modes, giving rise to the familiar mass-assignment aliasing. Subsequent interpolation or deconvolution of the assignment window does not, in general, completely remove this effect.


In the MRA construction, the discrete particle distribution is projected
directly onto compactly supported scaling functions. Although the resulting
coefficients are stored on a regular translation lattice, hereafter, referred to the MRA lattice, they are basis
projections rather than pointwise density samples. Their scaling-function
expansion defines a continuous finite-resolution field that can be evaluated
at arbitrary spatial positions. Using the same basis consistently for
analysis and reconstruction strongly suppresses the assignment-window
distortions and aliasing associated with conventional mesh assignment and
interpolation, thereby improving Fourier-space density and power-spectrum
recovery \citep{Cui2008,YangYB_2009}. The remaining resolution dependence
is a controlled projection effect encoded by the scaling-function response,
rather than an uncontrolled folding of unresolved modes into the reconstructed field.

For a periodic box of side length \(L\), the computational resolution at the scale \(J\) restricts the translation index to
\(\mathbf{l}\in\{0,\ldots,2^J-1\}^{\mathcal D}\). This index labels a regular spatial
translation lattice with associated positions
$ \mathbf{x}_{\mathbf{l}}=({L}/{2^J})\mathbf{l}$. 
The lattice spacing is therefore
\((\Delta L)_J=L/2^J\). The corresponding scaling coefficients form a
\(\mathcal{D}\)-dimensional regular array containing $N_{\mathrm{grid}}=2^{\mathcal{D}J}$ elements; in three dimensions, this is a
\(2^J\times2^J\times2^J\) coefficient array. These coefficients are basis
projections, rather than pointwise samples of the density field at the lattice positions.

Because the basis functions have compact support, the sum runs only over particles within that support and neighbouring cells. The precision of the reconstructed field $n_J(\mathbf{x})$ depends on the dilation level $J$, as indicated by the completeness relation of the basis functions in the MRA space \citep{Yue2024},
\begin{equation}
    \sum_{\mathbf{l}} \phi_{J\mathbf{l}}(\mathbf{x}) \phi_{J\mathbf{l}}\left(\mathbf{x}^{\prime}\right) \,= \,\Delta_J\left(\mathbf{x}, \mathbf{x}^{\prime}\right) \xrightarrow[J \rightarrow \infty]{} \delta_{\mathrm{D}}^3\left(\mathbf{x}-\mathbf{x}^{\prime}\right).
\end{equation}
The asymptotic convergence of the MRA reconstruction with increasing
\(J\), as implied by completeness, is illustrated in
Fig.~\ref{fig:field_resolution}. The same point distribution is
reconstructed at progressively higher resolution levels. As \(J\)
increases, the reconstructed field becomes increasingly localised, and
the smooth, coarse-grained peaks converge in the distributional sense
towards the original sum of Dirac delta functions. Throughout this
paper, we use the Daubechies \(D4\) scaling-function basis
(\texttt{db2} in \textsf{PyWavelets}\footnote{\url{https://github.com/PyWavelets/pywt}}) at various MRA resolution levels
\(J\).

\begin{figure*}
    \centering
    \includegraphics[width=0.90\textwidth]{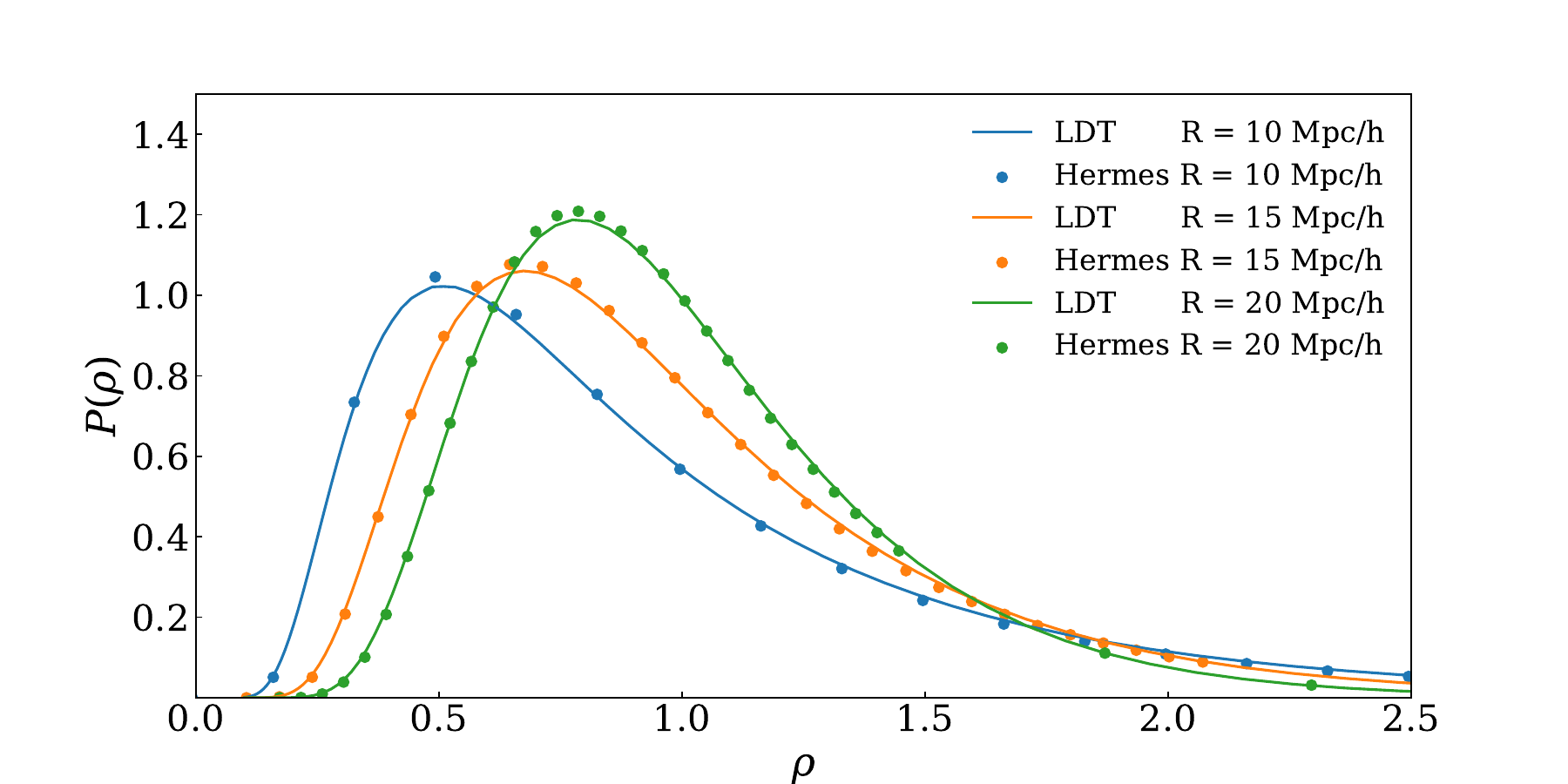}
    \caption{PDFs of the normalised matter density field smoothed with spherical windows of
    radii $R=10$, $15$, and $20\,h^{-1}\mathrm{Mpc}$ at $z=0$.
   Symbols show the \textsf{Hermes} measurements from the \textsf{Quijote} dark-matter
    field, while solid curves give the theoretical predictions from
    \textsf{pyLDT-cosmo} \citep{Cataneo2022}. The PDFs are measured from $10^8$ randomly sampled spherical windows, with all densities normalised by the mean density.}
    \label{fig:PDF-CIC-sphere}
\end{figure*}

Fig.~\ref{fig:sfc_field_resolution} shows the projected density reconstructed from the same \textsf{Quijote} halo catalogue at $J=7$, $J=8$, and $J=9$, together with a direct halo-binned reference in the same slab. Increasing $J$ progressively recovers smaller-scale structures of the cosmic web, while lower resolutions provide smoother coarse-grained representations. This illustrates the role of the MRA resolution in controlling the balance between spatial localisation and field smoothness.

Similarly, in the scaling-function basis, the kernel at scale $J$ can be written in bilinear form,
\begin{equation}\label{eq:w_bilinear}
W(\mathbf{x}, \mathbf{y})\rightarrow W_J(\mathbf{x}, \mathbf{y})\equiv P_JW(\mathbf{x}, \mathbf{y})=\sum_{\mathbf{l},\mathbf{m}} w^J_{\mathbf{l},\mathbf{m}} \phi_{J, \mathbf{l}}(\mathbf{x}) \phi_{J, \mathbf{m}}(\mathbf{y}),
\end{equation}
where
\begin{equation}\label{eq:w_bilinear_coff}
w^J_{\mathbf{l},\mathbf{m}}=\int W(\mathbf{x},\mathbf{y}) \phi_{J, \mathbf{l}}(\mathbf{x}) \phi_{J, \mathbf{m}}(\mathbf{y}) d^3\mathbf{x}\,d^3\mathbf{y}.
\end{equation}
It is easy to show that the first-order statistic in Eq.~(\ref{eq:1pt_CIC}) can be written as
\begin{equation}\label{eq:filtered-density-readout}
n_W(\mathbf{x}) \rightarrow n_W^J(\mathbf{x}) \equiv P_Jn_W(\mathbf{x})=\sum_{\mathbf{l}} \tilde{\epsilon}_{J\mathbf{l}} \phi_{J,\mathbf{l}}(\mathbf{x}),
\end{equation}
in which
\begin{equation}\label{eq:filtered-SFC-multiplication}
\tilde{\epsilon}_{J\mathbf{l}} = \sum_{\mathbf{m}} w^{J}_{\mathbf{l},\mathbf{m}} {\epsilon}_{J\mathbf{m}}.
\end{equation}

For a homogeneous, translation-invariant kernel \(W\), the matrix
elements depend only on the difference between the two translation
indices, $w^{J}_{\mathbf{l},\mathbf{m}}= w^{J}_{\mathbf{l}-\mathbf{m}}$. 
The operator therefore has a multidimensional Toeplitz structure, and
Eq.~(\ref{eq:filtered-SFC-multiplication}) is a discrete convolution on
the regular MRA lattice. For the periodic domains considered here, the
index difference is understood modulo \(2^J\) in each dimension, so the
Toeplitz operator becomes circulant and is diagonalised by the discrete
Fourier transform. The filtered coefficients can consequently be
evaluated as
\begin{equation}
    \widetilde{\boldsymbol{\epsilon}}_J
    =
    \operatorname{IFFT}
    \left[
        \operatorname{FFT}(\mathbf{w}_J)
        \odot
        \operatorname{FFT}(\boldsymbol{\epsilon}_J)
    \right],
\end{equation}
where \(\odot\) denotes element-wise multiplication. The FFT is taken over the discrete translation index \(\mathbf{l}\),
which labels positions on the MRA lattice; the kernel array
\(\mathbf{w}_J\) is indexed by the corresponding displacement
\(\mathbf{l}-\mathbf{m}\). This is therefore a spatial FFT of the MRA
coefficient arrays, rather than an FFT applied directly to the original
particle catalogue or to pointwise samples of the reconstructed
continuous field. At fixed resolution \(J\), the convolution has time
complexity \(O\!\left(N_{\rm grid}\log N_{\rm grid}\right),\). This convolution cost is independent of the number of galaxies once the MRA
coefficients have been constructed; the preceding projection of the
particle catalogue onto the MRA basis scales separately as
\(O(N_{\rm gal})\).

This construction provides an efficient counts-in-cells algorithm for
arbitrary window geometries. Once the filtered MRA coefficients have been
constructed, the readout in Eq.~(\ref{eq:filtered-density-readout}) involves only those scaling functions whose compact supports overlap the sampling position. For a fixed scaling-function family, their number is bounded, so each readout has \(O(1)\) complexity, independent of the total particle number and, apart from the one-time window-convolution step, of the window geometry. A measurement over \(N_{\rm samp}\) randomly selected centres therefore scales as \(O(N_{\rm samp})\). This linear sampling cost makes counts-in-cells and higher-order clustering measurements computationally feasible for large data sets.

As a demonstration, we measure counts-in-cells PDFs from the
\textsf{Quijote} fiducial dark-matter realisation 10000 at \(z=0\).
Using the filtered-field readout described above, we estimate the one-point PDFs of the spherically smoothed density field from \(10^8\) cell centres sampled uniformly throughout the simulation volume. The resulting distributions are shown in Fig.~\ref{fig:PDF-CIC-sphere}. 

For comparison, we compute the corresponding theoretical matter PDFs
using the large-deviation framework
\citep[e.g.][]{BernardeauReimberg2016,Uhlemann2016,Friedrich2020,
Cataneo2022,McCarthyGould2024}. This framework predicts the one-point
PDF of the density field smoothed with spherical top-hat windows in the
mildly nonlinear regime. Starting from Gaussian initial conditions, it
combines spherical-collapse dynamics with mass conservation to construct
the nonlinear rate function of the late-time density. The PDF is then
obtained numerically either by an inverse Laplace transform of the
cumulant-generating function or through a saddle-point approximation in
the log-density \citep{Bernardeau2014,Uhlemann2016,Cataneo2022}. We evaluate the predictions for the \textsf{Quijote} cosmology and the same smoothing windows using \textsf{pyLDT cosmo}\footnote{\url{https://github.com/mcataneo/pyLDT-cosmo}}.
The \textsf{PyHermes} measurements agree well with the large-deviation
predictions, demonstrating that the MRA field--window formulation
accurately recovers counts-in-cells statistics while making the
evaluation of \(10^8\) randomly sampled cells computationally feasible.
Any residual differences in the extreme tails may be affected by the
finite number of rare cells, the simulation resolution, and the
approximations entering the mildly nonlinear theoretical model.

\begin{figure*}
    \centering
    \includegraphics[width=\textwidth]{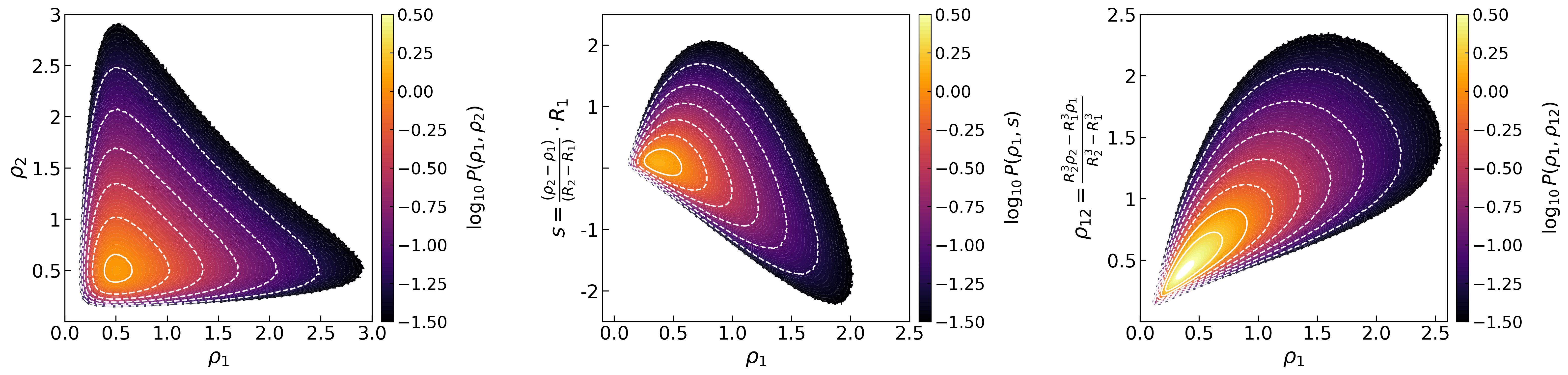}
    \caption{
    Joint PDFs of the matter densities measured with \textsf{Hermes} from the same data set as in Fig.~\ref{fig:PDF-CIC-sphere}.
    All panels are obtained by randomly sampling \(10^9\) cell pairs.
    The left panel shows the joint PDF of the matter densities \((\rho_1,\rho_2)\) in two spherical cells of radius \(10\,h^{-1}\mathrm{Mpc}\), separated by \(50\,h^{-1}\mathrm{Mpc}\).
    The middle and right panels use two concentric spherical cells with radii \(R_1=10\,h^{-1}\mathrm{Mpc}\) and \(R_2=11\,h^{-1}\mathrm{Mpc}\), where \(\rho_1\) and \(\rho_2\) denote the matter densities within the inner and outer spheres, respectively.
    The middle panel shows the joint PDF of \(\rho_1\) and the density slope \(s=(\rho_2-\rho_1)R_1/(R_2-R_1)\), while the right panel shows the joint PDF of \(\rho_1\) and the matter density in the spherical shell between \(R_1\) and \(R_2\), defined as \(\rho_{12}=(R_2^3\rho_2-R_1^3\rho_1)/(R_2^3-R_1^3)\). All densities are normalised by the mean density.
    The colour scale represents the logarithmic probability density, and the white lines correspond to \(\log_{10}P=-1.5,-1.25,\ldots\).}   
    \label{fig:2D-PDF}
\end{figure*}

Furthermore, Figure~\ref{fig:2D-PDF} presents joint two-cell PDFs obtained by randomly
sampling \(10^9\) cell pairs. The left panel quantifies the statistical
dependence between the densities in two separated spherical cells, thereby
retaining information about spatial correlations that is absent from the
corresponding one-point PDFs. For the concentric cells, the middle panel
recasts the joint density distribution in terms of the inner density and
the local radial density slope, directly characterising variations of the
density profile across the cell boundary. The right panel instead relates
the inner density to the density in the surrounding spherical shell,
providing a natural measure of the connection between a region and its
immediate environment. These joint distributions can therefore be used to
construct conditional density statistics and environment-dependent
clustering observables. The smooth sampling of these two-dimensional PDFs
from \(10^9\) cell pairs also illustrates the computational efficiency of
the MRA readout for high-dimensional counts-in-cells measurements.

\subsection{Statistical Estimator}

For measuring the 2PCF in a catalogue of $N$ objects, several edge-corrected estimators have been proposed. The most widely used is the \textsf{Landy--Szalay} estimator \citep[hereafter LS;][]{Landy1993}, whose symbolic expression is
\begin{equation}
\hat{\xi}_{\mathrm{LS}} = \frac{DD-2DR+RR}{RR},
\end{equation}
where $D$ denotes a catalogue of data points, and $R$ is a random sample from
the exact same survey volume. $DD$, $RR$, and $DR$ denote the pair
products formed from the corresponding data and random number-density fields.
In the \textsf{Hermes} field language the LS numerator is more naturally evaluated as a
single product of the difference field $\Delta \equiv D-R$, namely, for any binning window \(W_{\mathcal{P}}\),
\begin{equation}
    \Delta DD[W_{\mathcal{P}}]
    \equiv
    \left\langle
    \Delta(\mathbf{x})\,(W_{\mathcal{P}}\circ\Delta)(\mathbf{x})
    \right\rangle
    =DD[W_{\mathcal{P}}]-DR[W_{\mathcal{P}}]
    -RD[W_{\mathcal{P}}]+RR[W_{\mathcal{P}}].
\end{equation}
For a symmetric binning window, \(DR=RD\), so the dimensionless
correlation associated with the window is
\begin{equation}
    \xi[W_{\mathcal{P}}]
    =\frac{\Delta DD[W_{\mathcal{P}}]}{RR[W_{\mathcal{P}}]}.
    \label{eq:xi-deltaDD-over-RR}
\end{equation}
Thus \textsf{Hermes} does not need to regard \(DD\), \(DR\), \(RD\), and \(RR\)
as four independent catalogue loops; the numerator can be obtained from the
reconstructed difference field. The summation over scaling-function translations
$\mathbf{l}$ below is an inner-product contraction, not an average over Fourier
modes. Using the symbolic notation, the NPCF can be written in a compact form
without confusion \cite{Szapudi1998}
\begin{equation}
\hat{\xi}_{\mathrm{N}} = \frac{\prod_{i=1}^{N}(D_i-R_i)}{\prod_{i=1}^{N}R_i} = \frac{(D-R)^N}{R^N}.
\end{equation}
In conventional pair counting, the $D$ and $R$ catalogues are mapped to binned count products using a chosen separation binning scheme. The bin width introduces a familiar trade-off between resolution and variance: broad bins erase scale-dependent information, whereas narrow bins generally have larger variances.

In the \textsf{Hermes} toolkit, the central concept is an \textit{in situ}
perspective on $N$-point clustering statistics. Conventional pair counting in
separation bins is reinterpreted as a counts-in-cells operation over the
corresponding binning volume, which can be evaluated by convolving the
reconstructed density field with a window function specified by the binning
scheme. This viewpoint provides a natural algebraic extension of conventional
correlation-function estimators. In particular, the sharp-edged binning windows
used in discrete pair counting can be generalised to arbitrary kernels,
including smooth, compensated, low-pass, high-pass, and band-pass filters.

For higher-order correlation functions, the computational efficiency can be
further improved when the geometry of the binning window is independent of the
orientation of the underlying spatial configuration. In this case, the same
window kernel can be reused under arbitrary translations and rotations of the
$N$-point configuration, avoiding explicit sampling over individual orientations.
A natural choice is therefore a rotationally symmetric window, such as a
spherical top-hat or a spherical Gaussian filter, as demonstrated in
\citet{Yue2024}. Such symmetric windows provide an efficient route to
constructing generalised N-point estimators while preserving the
\textit{in situ} field-product formulation.

\subsection{Fast algorithm based on field reconstruction}

\subsubsection{Pair counting and the 2PCF in real space}\label{sec:pair-counting}

According to our strategy, the count-level pair product at separation $R=|\mathbf{r}|$,
$DD(\mathbf{r})=\langle n(\mathbf{x})n(\mathbf{x}+\mathbf{r})\rangle$,
can be naturally embedded within a generalised framework of \textit{in situ} correlation functions. As discussed in Section~\ref{sebsec:2pcf}, the dimensionless 2PCF $\xi$ is defined from the fluctuation field $\delta$, while the catalogue estimator is built from count products such as $DD$, $DR$, and $RR$. The pair product $DD(\mathbf{r})$ may be equivalently reformulated as the cross-correlation of two distinct fields evaluated at the same spatial location, namely the original field $n(\mathbf{x})$ and the filtered field $n_{\mathbf{r}}(\mathbf{x}) \equiv W_{\mathbf{r}}(\mathbf{x}) \circ n(\mathbf{x})$, where $W_{\mathbf{r}}$ is the translational binning window function defined in Eq.~(\ref{eq:window-r-space}).

This \textit{in situ} perspective leads to a simple numerical algorithm for 2PCF estimation. According to \cite{Feng2007}, for object $i$ at $\mathbf{x}_i$ in the data, the binned pair count can be expressed as $n_{\mathbf{r}}(\mathbf{x}_i) = \sum_{\mathbf{l}} \tilde{\epsilon}_{j\mathbf{l}}(\mathbf{r})\phi_{j\mathbf{l}}(\mathbf{x}_i)$.
Summing over all particles in the sample, the total pair count is
\begin{equation}\label{eq:pair-counting}
    DD = \sum_{i=1}^N w_i n_{\mathbf{r}}(\mathbf{x}_i) = \sum_{i=1}^N w_i\sum_{\mathbf{l}} \tilde{\epsilon}_{j\mathbf{l}}(\mathbf{r})\phi_{j\mathbf{l}}(\mathbf{x}_i) =\sum_{\mathbf{l}}\tilde\epsilon_{j\mathbf{l}}(\mathbf{r})\int d^3\mathbf{x}\,\phi_{j\mathbf{l}}(\mathbf{x})n(\mathbf{x})
    =\sum_{\mathbf{l}}\tilde\epsilon_{j\mathbf{l}}(\mathbf{r})\epsilon_{j\mathbf{l}}.
\end{equation}
Here $\{\epsilon_{j\mathbf{l}}\}$ are the coefficients of the original field in the multiresolution basis, and $\{\tilde\epsilon_{j\mathbf{l}}(\mathbf{r})\}$ are the coefficients after applying the binning window $W_{\mathbf{r}}$.

For the conventional 2PCF, Eqs.~(\ref{eq:w_bilinear})--(\ref{eq:w_bilinear_coff}) imply that the transformation matrix associated with a spatial translation, $W_{\mathbf{r}}(\mathbf{x})=\delta_{\mathrm{D}}^3(\mathbf{x}+\mathbf{r})$, takes the form
\begin{equation}\label{eq:translationCF}
    w^j_{\mathbf{l},\mathbf{m}}=w^j_{\mathbf{l}-\mathbf{m}}=\Phi_{\mathbf{l}-\mathbf{m}}(\mathbf{r}),
\end{equation}
where ${\Phi}_{\mathbf{l}-\mathbf{m}}(\mathbf{r}) = \Phi(\mathbf{r}+\mathbf{l}-\mathbf{m})$, and $\Phi(\mathbf{r})$ is the autocorrelation of the scaling function,
\begin{equation}\label{eq:cf-basis}
    \Phi(\mathbf{r})=\langle \phi(\mathbf{s}),(W_{\mathbf{r}}\circ\phi)(\mathbf{s})\rangle.
\end{equation}
Equation~(\ref{eq:translationCF}) then leads to the following transformation in multiresolution space,
\begin{equation}\label{eq:shift-scaling_coeff}
    \tilde\epsilon_{j\mathbf{l}}(\mathbf{r})= \langle\phi_{j,\mathbf{l}},W_{\mathbf{r}}\circ n\rangle
    = \sum_{\Delta \mathbf{l}}{\epsilon}_{j\mathbf{l}-\Delta \mathbf{l}}{\Phi}_{\Delta\mathbf{l}}(\mathbf{r}).
\end{equation}
Since Eq.~(\ref{eq:shift-scaling_coeff}) is a convolution, it can be efficiently evaluated using FFTs in a periodic volume.

It is worth emphasising that no assumption regarding spatial symmetry has been imposed on the density field. Consequently, the above formulation applies to arbitrary vector separations. Nevertheless, Eq.~(\ref{eq:shift-scaling_coeff}) reveals that the basis autocorrelation $\Phi(\mathbf{r})$ acts as an effective window function. When the underlying field possesses rotational symmetry, one may perform an appropriate rotational average, $\langle \Phi(\mathbf{r})\rangle_{\mathrm{rot}}$, over the corresponding geometric coordinates. Transforming Eq.~(\ref{eq:shift-scaling_coeff}) into Fourier space and averaging over the symmetry group yields
\begin{equation}\label{eq:shift-scaling_coeff-kspace}
   \langle\tilde\epsilon_{j\mathbf{l}}(\mathbf{r})\rangle_{\mathrm{rot}} = \sum_{\mathbf{n}}\hat{\epsilon}_{j\mathbf{n}}\hat\Phi_{\mathbf{n}}\hat{W}_{\mathrm{bin}}(\mathbf{k}_{\mathbf{n}}) \mathrm{e}^{\mathrm{i}\mathbf{k}_{\mathbf{n}}\cdot\mathbf{l}},
\end{equation}
where $\hat{W}_{\mathrm{bin}}(\mathbf{k}_{\mathbf{n}})=\langle \mathrm{e}^{\mathrm{i}\mathbf{k}_{\mathbf{n}}\cdot\mathbf{r}}\rangle_{\mathrm{rot}}$ is the binning function introduced in Eq.~(\ref{eq:window-k-space}), $\mathbf{k}_{\mathbf{n}} = (2\pi/L)\mathbf{n}$, and a hat denotes the Fourier counterpart of a quantity.

The above formalism extends naturally to arbitrary two-point statistics, including marked correlation functions. For two filtered fields, $n_{A}=W_A\circ n$ and $n_{B}=W_B\circ n$, their correlation at the same spatial location can be expressed as
\begin{equation}\label{eq:2rdEs}
\begin{aligned}
\langle n_{A}(\mathbf{x})n_{B}(\mathbf{x}) \rangle_{\mathbf{x}}
  =  \displaystyle{\sum_{\mathbf{l},\mathbf{m}}}\tilde{\epsilon}^A_{j\mathbf{l}}\tilde{\epsilon}^B_{j\mathbf{m}}
\langle\phi_{j\mathbf{l}},\phi_{j\mathbf{m}}\rangle
=\sum_{\mathbf{l}}\tilde{\epsilon}^A_{j\mathbf{l}}\,\tilde{\epsilon}^B_{j\mathbf{l}}.
\end{aligned}
\end{equation}

In deriving Eq.~(\ref{eq:2rdEs}), we have invoked the ergodic hypothesis, replacing ensemble averages with spatial averages over a sufficiently large survey volume, and exploited the orthogonality of the basis functions. As a result, the correlation reduces to a simple scalar product of the coefficient arrays associated with the two filtered fields.

Equation~(\ref{eq:2rdEs}) provides a generalised form of pair counting. In practical applications, random catalogues are required to account for shot noise and survey geometry. The same filtering procedure is therefore applied to both data and random samples, and shot noise can be removed directly at the coefficient level, $\Delta \tilde\epsilon_{j\mathbf{l}} =\tilde\epsilon^D_{j\mathbf{l}}-\tilde\epsilon^R_{j\mathbf{l}}$, where $D$ and $R$ denote the data and random catalogues. The desired second-order statistic is then obtained through Eq.~(\ref{eq:2rdEs}).

Finally, we note that the scalar product in Eq.~(\ref{eq:pair-counting}) is written in configuration space. By Parseval’s theorem, the same quantity may equivalently be evaluated in Fourier space
\begin{equation}\label{eq:pair-counting-kspace}
    DD = \sum_{\mathbf{l}}\epsilon_{j\mathbf{l}}\tilde{\epsilon}_{j\mathbf{l}}(\mathbf{r}) =\sum_{\mathbf{n}} \hat{\epsilon}_{j\mathbf{n}}^*\hat{\tilde\epsilon}_{j\mathbf{n}}(\mathbf{r}).
\end{equation}

\begin{figure*}
    \centering
    \vskip0.5cm
    \includegraphics[width=0.9\textwidth]{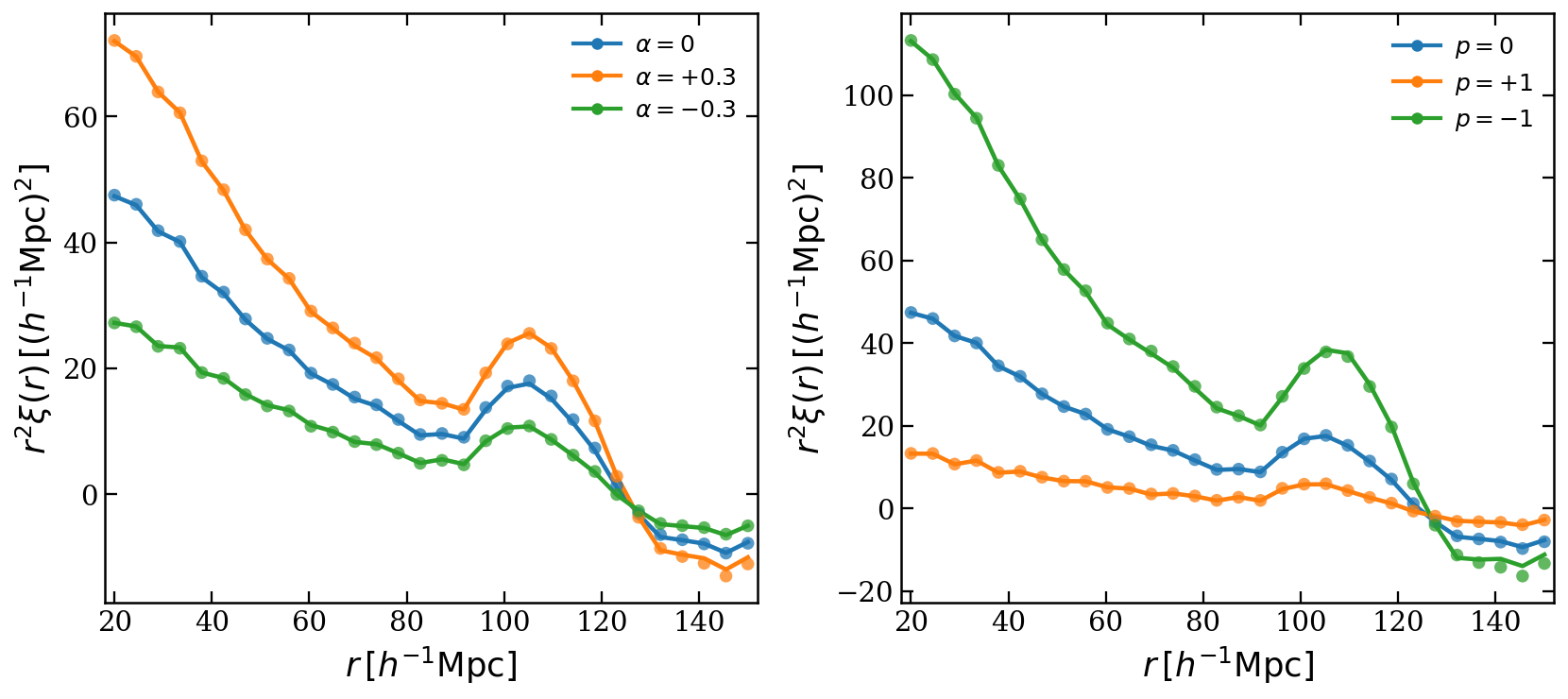}
    \caption{Examples of marked two-point correlation functions measured with two different
density-dependent marks. The left panel shows the power-law density mark defined in Eq.~\eqref{eq:power-law-mark}, with $\alpha=0$, $+0.3$, and $-0.3$. The case $\alpha=0$ corresponds to the standard unmarked two-point correlation function, while positive and negative values of $\alpha$ up-weight haloes in high- and low-density environments, respectively. The right panel shows the inverse-density mark of Eq.~\eqref{eq:inverse-density-mark}, with $p=0$, $+1$, and $-1$, where $p=0$ gives the unmarked result. }
    \label{fig:mark_2pcf}
\end{figure*}

Based on the density reconstruction in the MRA framework, we extend conventional discrete pair counting algebraically to general continuous fields. In this formulation, the radial binning function is no longer restricted to a sharp-edged top-hat bin, but can be replaced by a more general window function, including both sharp-edged and smooth non-sharp-edged filters. Generalised 2PCFs with several typical smooth binning windows have been demonstrated in \cite{Yue2024}. Here, we present two further examples based on low-pass and high-pass filters. The low-pass case is the marked 2PCF, whereas the high-pass case corresponds to a configuration-space estimator of the band-averaged power spectrum.

The marked 2PCF is a generalisation of the ordinary 2PCF in which each object, such as a galaxy, halo, or simulation particle, is assigned a weight, or \textit{mark}, according to some physical or environmental property. A common and physically useful choice is to define the mark as a function of the local smoothed density field $\delta_R(\mathbf{x})$ filtered on scale $R$: $m(\mathbf{x})=f[\delta_R(\mathbf{x})]$. The smoothing scale $R$ controls which environment is probed. A small $R$ makes the mark sensitive to non-linear halo-scale or filamentary environments, whereas a larger $R$ probes the large-scale density environment. Two particularly useful density-based choices are the power-law density mark and the inverse-density mark.
\begin{itemize}
    \item Power-law density mark (e.g. \citealt{Xiao2026,Yang2020,Xiao2022}, \citealt{xiao2025_AP}): it is taken to be
\begin{equation}\label{eq:power-law-mark}
m^{\mathrm{PL}}(\mathbf{x})=[1+\delta_R(\mathbf{x})+\epsilon]^\alpha,
\end{equation}
where $\epsilon$ is a small regularisation parameter introduced to avoid numerical problems in very low-density regions. The parameter $\alpha$ controls how strongly the mark responds to the local density environment. For $\alpha>0$, high-density regions receive larger weights, so the marked statistic becomes more sensitive to clusters, massive haloes, and filaments. For $\alpha<0$, low-density regions are weighted more strongly, and the statistic becomes sensitive to voids and underdense environments. For $\alpha=0$, the marked 2PCF reduces to the ordinary one.
    \item Inverse-density mark (\citealt{White2016}): it is designed specifically to enhance low-density regions. A commonly used form is
\begin{equation}\label{eq:inverse-density-mark}
    m^{\mathrm{ID}}(\mathbf{x})=\left[\frac{\rho_*+\bar\rho}{\rho_*+\rho_R(\mathbf{x})}\right]^p,
\end{equation}
where $p>0$ controls the strength of the weighting and $\rho_*$ is an offset parameter that prevents the mark from diverging when $\rho_R(\mathbf{x})\rightarrow 0$. The inverse-density mark gives larger weights to galaxies or haloes in underdense environments. By up-weighting low-density regions, the inverse-density mark increases the sensitivity of the two-point statistic to physical effects that are most prominent in such environments.
\end{itemize}

Utilising the pair-counting estimator in Eq.~\eqref{eq:2rdEs} or Eq.~\eqref{eq:pair-counting-kspace}, the marked 2PCF can be measured straightforwardly within the framework of \textsf{PyHermes}. As is conventional, we first normalise the marks by their weighted mean,
\begin{equation}
\bar m=\frac{\sum_i w_i m_i}{\sum_i w_i},
\qquad
\tilde m_i=\frac{m_i}{\bar m},
\end{equation}
where $w_i$ denotes the weight of object $i$. The detailed implementation is described later in Sec.~\ref{subsec:weighted_marked_fields}. We then measure the marked 2PCFs for a halo sample from the \textsf{Kun} simulation suite, using both the power-law and inverse-density marks. In both cases, the smoothing scale is set to $R=8\,h^{-1}\mathrm{Mpc}$. For the power-law density mark we adopt $\epsilon=0.1$, while for the inverse-density mark we take $\rho_{*}=1$. Figure~\ref{fig:mark_2pcf} illustrates the dependence of these two marks on the power-law indices $\alpha$ and $p$. The left panel shows the power-law mark defined in Eq.~\eqref{eq:power-law-mark}, whereas the right panel shows the inverse-density mark defined in Eq.~\eqref{eq:inverse-density-mark}. The solid curves denote the emulator predictions obtained by training on the \textsf{Kun} simulations. Since a power-law mark with a negative exponent is formally equivalent to an inverse-density weighting, the correspondence between the two classes of marks can be clearly seen by comparing the left and right panels of Fig.~\ref{fig:mark_2pcf}.

As a further demonstration, we consider two compensated, scale-localised wavelet filters: the three-dimensional isotropic Gaussian-derivative wavelet (GDW) and the spherical cosine wavelet (CWS) \citep{Wangyun2022A,Wangyun2022B}. The underlying continuous wavelet transform represents the density field jointly in position and scale and is therefore well suited to characterising spatially localised, multiscale cosmic structures, including haloes, filaments, walls, and void boundaries. For the GDW kernel (item $l$ in Table~\ref{tab:kernel}), its Fourier counterpart has a particularly simple form,
$\propto k^2R^2 \exp\left(-(kR)^2/2\right)$. The CWS (item $k$ in Table~\ref{tab:kernel}) is likewise a compensated band-pass filter, but its Fourier response behaves as \(\propto(kR)^4\) for \(kR\ll1\), resulting in stronger suppression of long-wavelength modes than for the GDW. At \(kR\gg1\), both kernels are exponentially damped by their Gaussian cutoff. For either kernel, the variance of the filtered density field is related to the matter power spectrum through Eq.(\ref{eq:band_averaged_power}), and provides a weighted, band-averaged measure of clustering power around a characteristic scale \(k\sim R^{-1}\). In this sense, both wavelet filters can be interpreted as configuration-space realisations of band-power measurements.

\begin{figure*}
\centering
    \includegraphics[width=\textwidth]{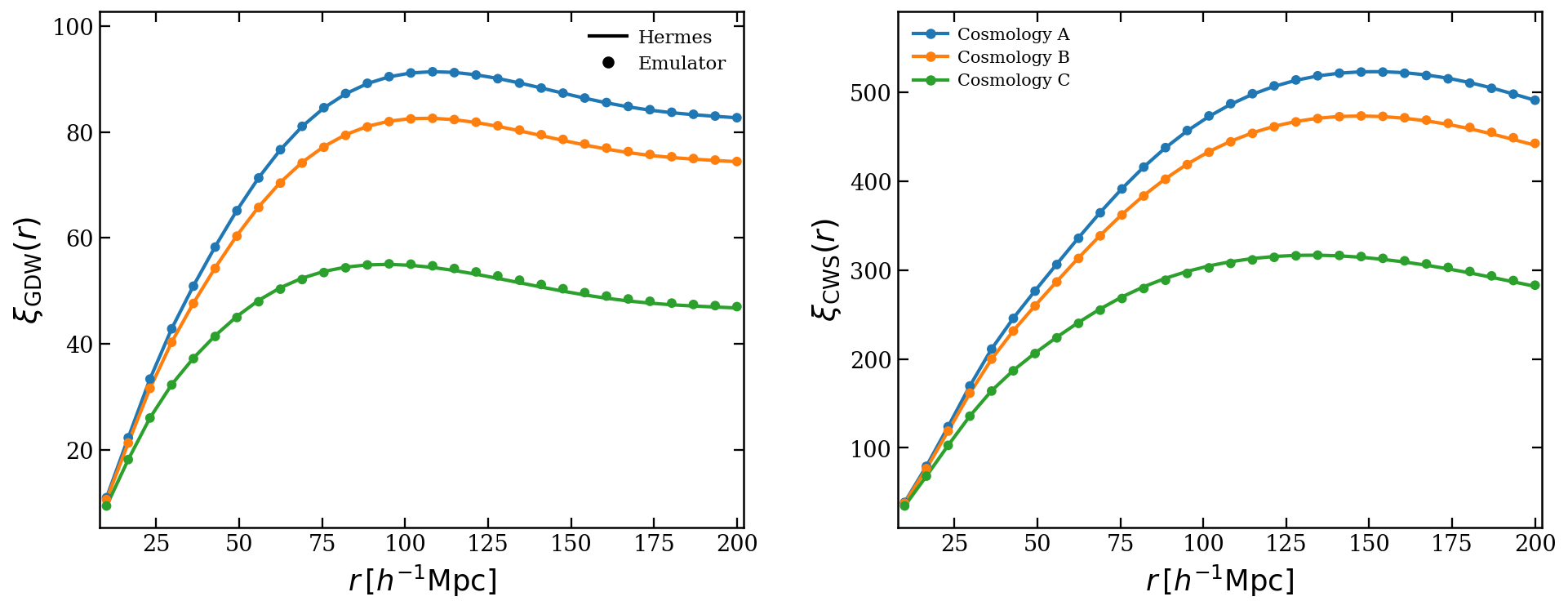}
    \caption{Examples of the GDW and CWS windowed statistics for three different cosmologies. The left and right panels show $\xi_{\mathrm{GDW}}(r)$ and $\xi_{\mathrm{CWS}}(r)$, respectively. Solid lines show the \textsf{Hermes} results, while filled circles show the emulator predictions.
    The cosmological parameters are ordered as $(\Omega_{\mathrm{b}},H_0,n_{\mathrm{s}},w,w_{\mathrm{a}},M_\nu,\Omega_{\mathrm{m}},\sigma_8)$:
    A = $(0.0490,67.66,0.9665,-1,0,0.0600,0.3111,0.8101)$,
    B = $(0.0500,70.00,0.9600,-1,0,0.1500,0.3233,0.8356)$,
    and C = $(0.0469,74.38,0.9525,-0.8688,-0.2188,0.1781,0.3585,0.8704)$, where $H_0$ and $M_\nu$ are expressed in units of
\(\mathrm{km\,s^{-1}\,Mpc^{-1}}\) and \(\mathrm{eV}\), respectively.}
    \label{fig:GDW_CWS}
\end{figure*}

Figure~\ref{fig:GDW_CWS} shows the two-point statistics of the wavelet-filtered density fields obtained with the GDW (left panel) and CWS (right panel) kernels. The solid curves show direct \textsf{PyHermes} measurements for three cosmological samples from the \textsf{Kun} simulation suite, while the filled circles show the corresponding predictions from Gaussian-process emulators.

We emphasise that sufficiently accurate theoretical predictions throughout the fully nonlinear regime are not currently available for the complete set of clustering observables considered here, particularly for marked and higher-order statistics. A practical alternative is therefore to construct a cosmology-dependent emulator tailored to each observable. We use the \textsf{Kun} simulation suite to build such simulation-based emulators. These emulators are measurement specific: the prediction for each statistic incorporates its particular mark definition, filtering kernel, radial binning, estimator normalisation, and numerical resolution. The emulator thus predicts the measured, forward-modelled observable, including its measurement response, rather than an estimator-independent continuum statistic.

The simulations used for emulator training and testing are disjoint. Figures~\ref{fig:mark_2pcf} and~\ref{fig:GDW_CWS} compare emulator predictions with direct \textsf{PyHermes} measurements from simulations that were not included in the training set. Their agreement therefore demonstrates the predictive performance of the emulators for previously unseen simulations and their ability to reproduce the cosmological dependence of these nonlinear clustering statistics. These comparisons are intended to illustrate the application of the \textsf{HERMES} framework to emulator-based cosmological analyses, rather than to provide an independent validation of the underlying algorithms. The numerical accuracy of the field reconstruction and correlation-function estimators has been established in our previous validation studies \citep{Feng2007,Yue2024,Ju2026}. Further validation through comparisons with external codes is presented in Section~\ref{sec:implementation-performance}.

\subsubsection{Generalising to 3PCF and NPCF monopoles}

The \textit{in situ} formulation can be most clearly illustrated through the estimation of the monopole moments of higher-order correlation functions. We begin with the simplest non-trivial example, namely the monopole component of the three-point correlation function (3PCF). After averaging over all angular configurations, the monopole 3PCF measures the excess probability of finding two neighbours within spherical shells of radii $R_1$ and $R_2$ around a common central point. In practical applications, these shells are represented by finite-width radial bins. For brevity, we denote the corresponding shell window functions simply by $W_{R_i}$, suppressing the additional parameters that specify the shell geometry.

Within the \textit{in situ} framework, the triplet count can be expressed as
\begin{equation}\label{eq:3pcfddd}
DDD = \langle n(\mathbf{x}){n}_{R_1}(\mathbf{x}){n}_{R_2}(\mathbf{x})\rangle =\langle n(\mathbf{x}),{n}_{R_1,R_2}^{\mathrm{2\text{-}fold}}(\mathbf{x})\rangle,
\end{equation}
where $n_{R_i}(\mathbf{x})=(W_{R_i}\circ n)(\mathbf{x}),
\quad i=1,2,$ denotes the shell-filtered density field. Expanding the fields on the scaling-function basis $\phi_{j\mathbf{l}}$, $n(\mathbf{x})= \sum_{\mathbf{l}}\epsilon_{j\mathbf{l}}
\phi_{j\mathbf{l}}(\mathbf{x}), \quad n_{R_i}(\mathbf{x}) = \sum_{\mathbf{l}}
\tilde{\epsilon}_{j\mathbf{l}}(R_i)
\phi_{j\mathbf{l}}(\mathbf{x})$,

we define the product of the two shell fields as a \textit{two-fold density field},
\begin{equation}
\label{eq:2-fold-density}
n^{\mathrm{2\text{-}fold}}_{R_1,R_2}(\mathbf{x})
\equiv
n_{R_1}(\mathbf{x})n_{R_2}(\mathbf{x}) =
\sum_{\mathbf{l}}
\tilde{\epsilon}_{j\mathbf{l}}(R_1,R_2)
\phi_{j\mathbf{l}}(\mathbf{x}),
\end{equation}
whose decomposition coefficients are
\begin{equation}
\tilde{\epsilon}_{j\mathbf{l}}(R_1,R_2)=
\left\langle
\phi_{j\mathbf{l}},
n_{R_1}n_{R_2}
\right\rangle.
\end{equation}

Substituting the basis expansions of the shell-filtered fields yields
\begin{equation}\label{eq:3PCF-ddd-monopole}
     \tilde{\epsilon}_{j\mathbf{l}}({R_1,R_2})
    =\sum_{\mathbf{l}_1}\sum_{\mathbf{l}_2}
    \tilde{\epsilon}_{j\mathbf{l}_1}(R_1)\tilde{\epsilon}_{j\mathbf{l}_2}(
    R_2)
    \langle \phi_{j\mathbf{l}}(\mathbf{x}),
    \phi_{j\mathbf{l}_1}(\mathbf{x})\phi_{j\mathbf{l}_2}(\mathbf{x})\rangle
    =
    \sum_{\mathbf{a}} \tilde\epsilon_{j,\mathbf{l}+\mathbf{a}}(R_1)
    \sum_{\mathbf{b}} \Gamma_{\mathbf{a},\mathbf{b}}
    \tilde\epsilon_{j,\mathbf{l}+\mathbf{b}}(R_2).
\end{equation}
Here we have introduced the relative offsets $\mathbf{a}=\mathbf{l}_1-\mathbf{l}$ and
$\mathbf{b}=\mathbf{l}_2-\mathbf{l}$. Because the overlap of translated scaling functions depends only on relative separations, the resulting coefficients are independent of the absolute location $\mathbf{l}$. The quantities
\begin{equation}\label{eq:con-coeff}
  \Gamma_{\mathbf{a},\mathbf{b}}
  = \langle
  \phi_{j\mathbf{l}}(\mathbf{x}),\phi_{j,\mathbf{l}+\mathbf{a}}(\mathbf{x})\phi_{j,\mathbf{l}+\mathbf{b}}(\mathbf{x})\rangle =\frac{1}{N_j}\int \phi(\mathbf{x})\phi(\mathbf{x}-\mathbf{a})\phi(\mathbf{x}-\mathbf{b})\,d^3\mathbf{x},
\end{equation}
are the \textit{triple connection coefficients}, with $N_j=2^{3j}$ the total number of scaling cells at resolution level $j$. Equation~(\ref{eq:3PCF-ddd-monopole}) is the central result of the construction. It replaces explicit triplet counting by a local stencil contraction involving only neighbouring scaling coefficients and a precomputed table of connection coefficients.

The sums over \(\mathbf{a}\) and \(\mathbf{b}\) extend only over the
compact support of the scaling functions. Let \(C_{\rm s}\) denote the
extent of the one-dimensional support in units of the MRA lattice
spacing. Each shift index can take \(2C_{\rm s}-1\) relative values.
After excluding the non-overlapping combinations, the number of
potentially non-zero shift pairs per dimension is bounded by
$N_{\rm ov}=(2C_{\rm s}-1)^2-C_{\rm s}(C_{\rm s}-1)$. 
The fully expanded local contraction in
Eq.~(\ref{eq:3PCF-ddd-monopole}) therefore has the computational
complexity $\mathscr{O}\!\left(
N_{\rm ov}^{\mathcal{D}}N_{\rm grid}
\right), N_{\rm grid}=2^{\mathcal{D}j}$, where \(N_{\rm grid}\) is the total number of MRA lattice elements and
\(\mathcal{D}\) is the spatial dimension. For the Daubechies \(D4\) scaling function adopted in \textsf{PyHermes}, the
four-tap refinement filter gives \(C_{\rm s}=3\). Consequently,
\(N_{\rm ov}=19\), and in three dimensions the fully expanded
contraction contains at most
\(19^3=6859\) potentially non-zero terms per lattice element. At fixed
basis, the computational cost therefore scales linearly with
\(N_{\rm grid}\).

In practice, the multidimensional connection coefficients are evaluated from separable one-dimensional overlap tables. The resulting operation becomes a compact, grid-local stencil contraction that is free of inter-cell communication and can be efficiently parallelised on GPUs. In \textsf{PyHermes}, each contraction is executed independently by CUDA threads, making the construction of the two-fold field computationally inexpensive.

Once the two-fold density field has been constructed, the triplet count reduces to an \textit{in situ} correlation between the original density field and the two-fold field. Exploiting the orthogonality of the scaling-function basis, the monopole component of the 3PCF can be evaluated as
\begin{equation}\label{eq:DDD-vector-product}
    DDD(R_1,R_2) = \sum_{\mathbf{l}}\epsilon_{j\mathbf{l}} \tilde{\epsilon}_{j\mathbf{l}}({R_1,R_2}).
\end{equation}
Equation~(\ref{eq:DDD-vector-product}) is the direct higher-order
counterpart of Eq.~(\ref{eq:2rdEs}). The key point is that an explicit
enumeration of particle triplets is entirely avoided. Once the particle
field has been projected onto the MRA basis and the shell-filtered
coefficient fields have been constructed, their contraction through the
connection coefficients reduces the 3PCF estimator to an inner product
in the MRA space. At fixed basis, the cost of this contraction depends
on the size of the MRA lattice and the compact-support stencil, rather
than on the number of particles.

The above construction naturally generalises to the monopole component of the $N$-point correlation function. Let
\begin{equation}
D^N(\mathcal{R})
\equiv \left\langle n(\mathbf{x}) n_{R_1}(\mathbf{x}) \cdots n_{R_{N-1}}(\mathbf{x})\right\rangle_{\mathbf{x}} =\left\langle n(\mathbf{x}),n^{(N-1)\text{-}\mathrm{fold}}_{\mathcal{R}}(\mathbf{x}) \right\rangle,
\end{equation}
where $\mathcal{R}=\{R_1,\ldots,R_{N-1}\}$, $n_{R_i}(\mathbf{x})=(W_{R_i}\circ n)(\mathbf{x})$
denotes the shell-filtered density field at radius $R_i$. Analogous to the two-fold field introduced for the 3PCF, we define the $(N-1)$-fold density field
\begin{equation}
n^{(N-1)\text{-}\mathrm{fold}}_{\mathcal{R}}(\mathbf{x})
=n_{R_1}(\mathbf{x})\cdots n_{R_{N-1}}(\mathbf{x})
=\sum_{\mathbf{l}}\tilde\epsilon_{j\mathbf{l}}({\mathcal{R}})\phi_{j\mathbf{l}}(\mathbf{x}),
\end{equation}
with decomposition coefficients
\begin{equation}
\label{eq:DN_general}
\tilde{\epsilon}_{j\mathbf{l}}(\mathcal{R})=
\left\langle
\phi_{j\mathbf{l}}, n^{(N-1)\text{-}\mathrm{fold}}_{\mathcal{R}}
\right\rangle.
\end{equation}

Substituting the basis expansions of the shell-filtered fields yields
\begin{equation}
\tilde\epsilon_{j\mathbf{l}}({\mathcal{R}})=
\sum_{\mathbf{l}_1,\dots,\mathbf{l}_{N-1}}
\Gamma^{(N)}_{\mathbf{l}_1,\dots,\mathbf{l}_{N-1}}
\tilde\epsilon_{j,\mathbf{l}+\mathbf{l}_1}(R_1)\cdots \tilde\epsilon_{j,\mathbf{l}+\mathbf{l}_{N-1}}(R_{N-1}),
\end{equation}
where
\begin{equation}
\Gamma^{(N)}_{\mathbf{l}_1,\dots,\mathbf{l}_{N-1}}=
\langle\phi_{j\mathbf{l}}, \phi_{j,\mathbf{l}+\mathbf{l}_1}\cdots \phi_{j,\mathbf{l}+\mathbf{l}_{N-1}}\rangle,
\end{equation}
is the corresponding higher-order connection coefficient.

A key property of the multiresolution basis is that all higher-order connection coefficients can be reduced recursively to the triple connection coefficients. Since
\begin{equation}
\phi_{j\mathbf{l}_1}(\mathbf{x})\phi_{j\mathbf{l}_2}(\mathbf{x})
=\sum_{\mathbf{l}}
\Gamma_{\mathbf{l}_1-\mathbf{l},\,\mathbf{l}_2-\mathbf{l}}\phi_{j\mathbf{l}}(\mathbf{x}),
\end{equation}
repeated application of this contraction identity gives
\begin{equation}
\Gamma^{(N)}_{\mathbf{l}_1,\dots,\mathbf{l}_{N-1}}
=\sum_{\mathbf{m}_1,\dots,\mathbf{m}_{N-3}}
\Gamma_{\mathbf{l}_1-\mathbf{m}_1,\;\mathbf{l}_2-\mathbf{m}_1}
\Gamma_{\mathbf{m}_1-\mathbf{m}_2,\;\mathbf{l}_3-\mathbf{m}_2}
\cdots \Gamma_{\mathbf{m}_{N-4}-\mathbf{m}_{N-3},\;\mathbf{l}_{N-2}-\mathbf{m}_{N-3}}
\Gamma_{\mathbf{m}_{N-3},\;\mathbf{l}_{N-1}}.
\end{equation}

Consequently, the coefficients of the $(N-1)$-fold field can be constructed entirely from nested contractions involving the precomputed triple connection coefficients,
\begin{equation}\label{eq:DN_gamma_recursive}
\tilde\epsilon_{j\mathbf{l}}({\mathcal{R}})=
\sum_{\mathbf{l}_1,\dots,\mathbf{l}_{N-1}}
\sum_{\mathbf{m}_1,\dots,\mathbf{m}_{N-3}}
\Big[\Gamma_{\mathbf{l}_1-\mathbf{m}_1,\;\mathbf{l}_2-\mathbf{m}_1}
\Gamma_{\mathbf{m}_1-\mathbf{m}_2,\;\mathbf{l}_3-\mathbf{m}_2}
\cdots
\Gamma_{\mathbf{m}_{N-4}-\mathbf{m}_{N-3},\;\mathbf{l}_{N-2}-\mathbf{m}_{N-3}}
\Gamma_{\mathbf{m}_{N-3},\;\mathbf{l}_{N-1}}
\Big]
\tilde\epsilon_{j,\mathbf{l}+\mathbf{l}_1}(R_1)\cdots \tilde\epsilon_{j,\mathbf{l}+\mathbf{l}_{N-1}}(R_{N-1}),
\end{equation}
where the bracketed factor represents the chain of triple-connection contractions in the preceding equation.

Once the $(N-1)$-fold field has been constructed, the monopole estimator of
the $N$-point correlation function reduces to a simple inner product in the MRA
coefficient space,
\begin{equation}\label{eq:NPCF-inner-product}
    D^N(\mathcal{R})
    =\sum_{\mathbf{l}}
    \epsilon_{j\mathbf{l}}
    \tilde{\epsilon}_{j\mathbf{l}}(\mathcal{R}).
\end{equation}
Equation~(\ref{eq:NPCF-inner-product}) provides the natural generalisation of
the 2PCF and 3PCF estimators derived above. The key feature of the
\textit{in situ} formulation is that explicit enumeration of $N$-tuples is
completely avoided. Instead, the construction of the $(N-1)$-fold density field
is achieved through a sequence of local stencil contractions determined by the
compact support of the connection coefficients, followed by a single
field--field inner product in the multiresolution coefficient space.

For a basis with compact support, only a finite number of relative shifts contribute to each contraction. The computational complexity of the fully expanded contraction is therefore $\mathscr{O}\!\left(
C_{\mathrm{s}}^{(2N-4)\mathcal{D}}N_{\mathrm{grid}}\right)$. In practice, however, the recursive construction is evaluated sequentially through repeated triple contractions, so that the computational cost scales linearly with $N_{\mathrm{grid}}$ and grows only polynomially with the order of the correlation function.

\begin{figure*}
  \centering
      \includegraphics[width=\textwidth]{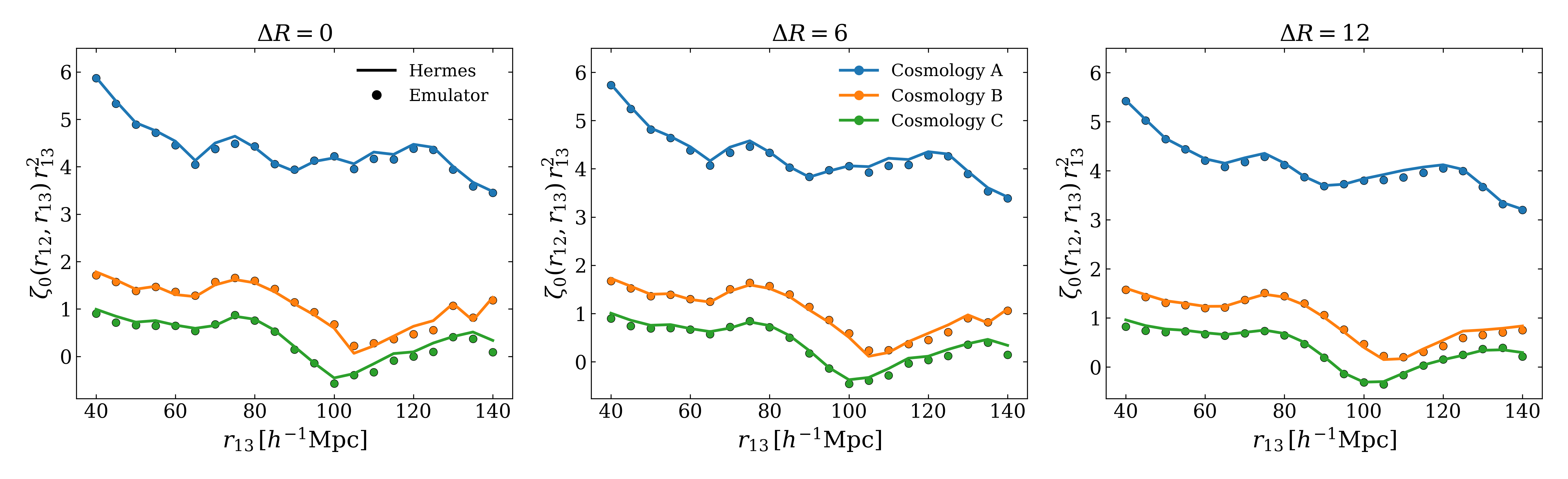}
      \caption{Examples of the 3PCF monopole for three different cosmologies. The first side length is fixed at $r_{12}=20\,h^{-1}\mathrm{Mpc}$, while the second
      side length varies over $40 \le r_{13} \le 140\,h^{-1}\mathrm{Mpc}$. From left to right, the panels show shell thicknesses $\Delta R=0$, $6$, and $12\,h^{-1}
      \mathrm{Mpc}$, respectively. The plotted quantity is $\zeta_0(r_{12},r_{13})\,r_{13}^{2}$. Solid lines show the \textsf{Hermes} measurements, while filled
      circles show the emulator predictions.
      The cosmological parameters are ordered as $(\Omega_{\mathrm{b}},H_0,n_{\mathrm{s}},w,w_{\mathrm{a}},M_\nu,\Omega_{\mathrm{m}},\sigma_8)$:
      A = $(0.0470,64.22,0.9831,-0.7422,-0.0391,0.1617,0.2555,0.5670)$,
      B = $(0.0523,63.28,0.9994,-1.2766,0.1328,0.1289,0.3422,0.8039)$,
      and C = $(0.0439,69.84,0.9856,-1.2859,-0.1953,0.2648,0.4021,1.1138)$, where $H_0$ and $M_\nu$ are expressed in units of
\(\mathrm{km\,s^{-1}\,Mpc^{-1}}\) and \(\mathrm{eV}\), respectively.}
      \label{fig:3pcf_monopole_shell_thickness}
  \end{figure*}

Figure~\ref{fig:3pcf_monopole_shell_thickness} illustrates the impact of radial binning on the measured 3PCF monopole using \textsf{PyHermes} and compares it with emulator predictions obtained using the same training framework as in Fig.~\ref{fig:GDW_CWS}. We fix the first side length at \(r_{12}=20\,h^{-1}\mathrm{Mpc}\) and vary the second side length over \(40\leq r_{13}\leq140\,h^{-1}\mathrm{Mpc}\), while adopting three different shell widths, \(\Delta R=0\), \(6\), and \(12\,h^{-1}\mathrm{Mpc}\). Increasing the bin width corresponds to applying a broader radial window, which averages over neighbouring triangle configurations. As a result, small-scale fluctuations and high-frequency features in the monopole are gradually suppressed, leading to smoother and more stable measurements. Importantly, the overall scale dependence and the relative differences among the three cosmologies remain unchanged, demonstrating that finite binning primarily modifies the resolution of the measurement rather than the underlying cosmological information. The excellent agreement between the \textsf{Hermes} measurements and emulator predictions for all binning choices further validates the implementation of the windowed 3PCF estimator. The results also highlight the flexibility of the window formalism: the binning scale can be adjusted to balance statistical precision, nonlinear sensitivity, and measurement resolution.

\subsubsection{The binning function and multipole decomposition of NPCFs}\label{sec:multipole-decomposition}

As discussed in Section~\ref{sebsec:2pcf}, the binning window function in
Fourier space can be interpreted as the rotational average of the spatial
translation operator over the binning volume [Eq.~(\ref{eq:window-k-space})].
Using the Rayleigh plane-wave expansion, it can be written as
\begin{equation}
\hat{W}_{\mathrm{bin}}(\mathbf{k})
=
\left\langle e^{i\mathbf{k}\cdot\mathbf{r}}\right\rangle_{\mathrm{rot}}
=
4\pi\sum_{\ell=0}^{\infty}\sum_{m=-\ell}^{\ell}
i^\ell j_\ell(kr)
Y_{\ell m}(\Omega_{\hat{\mathbf{k}}})
\left\langle
Y_{\ell m}^{*}(\Omega_{\hat{\mathbf{r}}})
\right\rangle_{\mathrm{rot}},
\end{equation}
where $j_\ell$ is the spherical Bessel function,
$Y_{\ell m}$ denotes the spherical harmonic function, and
$\langle\cdot\rangle_{\mathrm{rot}}$ represents the rotational average over the
symmetry group relevant to the chosen clustering statistic. Here, the radial
selection is assumed to be an infinitesimally thin spherical shell, such that no radial averaging over the bin width is performed.

For the anisotropic two-point correlation function in redshift space, under
the plane-parallel approximation, the rotational average is performed over the
azimuthal angle around the fixed line of sight (LOS). The anisotropic
correlation function can therefore be expanded in Legendre multipoles,
\begin{equation}
\xi(r,\mu)=\sum_{\ell}\xi_{\ell}(r)\mathcal{L}_{\ell}(\mu),
\end{equation}
where $\mathcal{L}_{\ell}$ denotes the Legendre polynomial and the multipole
moment is defined as
\begin{equation}
\xi_{\ell}(r)=\frac{2\ell+1}{2}
\int_{-1}^{1}\xi(r,\mu)\mathcal{L}_{\ell}(\mu)d\mu .
\end{equation}

Using Eq.~(\ref{eq:shift-scaling_coeff-kspace}), the multipole component of the
pair-count field can be directly evaluated as \citep{Ju2026}
\begin{equation}
DD_{\ell}
=
\sum_{\mathbf n}
\left|\hat\epsilon_{j,\mathbf n}\right|^2
\hat{\Phi}_{\mathbf n}
\hat W_{\mathrm{bin}}^{\ell}(k_{\mathbf n}),
\end{equation}
where the corresponding multipole window kernel is
\begin{equation}
\hat W_{\mathrm{bin}}^{\ell}(\mathbf k)
=
(2\ell+1)i^\ell
j_\ell(kr)
\mathcal{L}_{\ell}(\cos\theta_{\mathbf k}).
\end{equation}
Here the radial dependence is written in the thin-shell limit, while a
finite-width radial bin can be incorporated by replacing
$j_\ell(kr)$ with its average over the corresponding radial window.

The above construction can be generalised to the multipole decomposition of higher-order correlation functions. However, the notion of multipoles for the NPCF differs fundamentally from that for the 2PCF. In the latter case, the rotational average is taken about a preferred line of sight, and the resulting multipoles characterise anisotropic clustering in redshift space. In contrast, the NPCF in real space is rotationally invariant. The relevant average is therefore performed over all possible orientations of a fixed polyhedral configuration in three-dimensional space.

An $N$-point configuration may be specified by choosing a primary vertex and $N-1$ displacement vectors,
$\{\mathbf{r}_1,\ldots,\mathbf{r}_{N-1}\}$. The corresponding secondary-vertex fields are $n_{\mathbf{r}_i}(\mathbf{x}) = W_{\mathbf{r}_i}(\mathbf{x})\circ n(\mathbf{x})$
where $W_{\mathbf{r}_i}$ denotes the translation operator associated with the displacement $\mathbf{r}_i$. The $N$-tuple count for a fixed polyhedral configuration can then be written as
\begin{equation}\label{eq:N-tuple-counting}
D^N(\mathbf{r}_1,\ldots,\mathbf{r}_{N-1}) = \int d^3\mathbf{x}
\left\langle n(\mathbf{x}) n_{\mathbf{r}_1}(\mathbf{x})\cdots n_{\mathbf{r}_{N-1}}(\mathbf{x})\right\rangle_{\mathrm{rot}}=\int d^3\mathbf{x}\,n(\mathbf{x})n^{(N-1)\text{-}\mathrm{fold}}(\mathbf{x},\mathbf{r}_1,\ldots,\mathbf{r}_{N-1}),
\end{equation}
where two distinct averages are involved. The first, denoted by $\langle\cdots\rangle_{\mathrm{rot}}$, is the rotational average over all orientations of the polyhedral configuration. The second is the spatial average over the survey volume, which replaces the ensemble average under the ergodic hypothesis.

Here we have introduced the rotationally averaged $(N-1)$-fold density field,
\begin{equation}\label{eq:N1-counting-field}
    n^{(N-1)\text{-}\mathrm{fold}}(\mathbf{x},\mathbf{r}_1,\ldots,\mathbf{r}_{N-1}) =
    \left\langle  n_{\mathbf{r}_1}(\mathbf{x})\cdots n_{\mathbf{r}_{N-1}}(\mathbf{x})\right\rangle_{\mathrm{rot}},
\end{equation}
which plays the role of the generalised counting field for the NPCF. As will be shown below, the multipole decomposition of the NPCF can be formulated entirely in terms of the harmonic expansion of this $(N-1)$-fold field, thereby extending the \textit{in situ} framework developed for the 2PCF and 3PCF to arbitrary order.

In real space, the NPCF is invariant under a global rotation of the entire polyhedral configuration. Denoting the set of displacement vectors by
$\mathbf{R}=\{\mathbf{r}_1,\ldots,\mathbf{r}_{N-1}\}$ and a rotation operator by $\mathcal{O}\in\mathrm{SO}(3)$, rotational invariance implies
$\zeta(\mathbf{x},\mathcal{O}\mathbf{R})=\zeta(\mathbf{x},\mathbf{R})$. Consequently, the $(N-1)$-fold counting field can be expanded in a complete basis of isotropic functions
\begin{equation}\label{eq:N-counting-MultipoleExp}
   n^{(N-1)\text{-}\mathrm{fold}}(\mathbf{x},\mathbf{R}) =\sum_{\Lambda}n^{(N-1)\text{-}\mathrm{fold}}_{\Lambda}(\mathbf{x},\mathcal{R})\mathcal{P}_{\Lambda}(\hat{\mathbf{R}}),
\end{equation}
where $\mathcal{R}=\{R_1,\ldots,R_{N-1}\}$, with $R_i=|\mathbf{r}_i|$, denotes the radial configuration, and
$\hat{\mathbf{R}}=\{\hat{\mathbf{r}}_1,\ldots,\hat{\mathbf{r}}_{N-1}\}$
collects the angular degrees of freedom. The functions
$\mathcal{P}_{\Lambda}(\hat{\mathbf{R}})$ form a basis for the rotationally invariant subspace of the tensor product of $(N-1)$ spherical-harmonic spaces. Following \citet{Philcox2021b}, they may be written as
\begin{equation}
    \mathcal{P}_{\Lambda}(\hat{\mathbf{R}})=\sum_{m_1 \ldots m_{N-1}} C_M^{\Lambda} Y_{\ell_1 m_1}\left(\hat{\mathbf{r}}_1\right) \ldots Y_{\ell_{N-1} m_{N-1}}\left(\hat{\mathbf{r}}_{N-1}\right),
\end{equation}
where $C_M^{\Lambda}$ denotes the coupling coefficients that combine the individual angular momenta into a rotational scalar (see Eq.~(5) of \citealt{Philcox2021b}). A detailed mathematical construction of these basis functions may be found in \citet{Cahn_2023}.

Substituting Eq.~(\ref{eq:N-counting-MultipoleExp}) into Eq.~(\ref{eq:N-tuple-counting}) yields
\begin{equation}
D^N(\mathbf{R})=
\sum_{\Lambda} D_\Lambda^N(\mathcal{R})\mathcal{P}_{\Lambda}(\hat{\mathbf{R}}),
\end{equation}
with
\begin{equation}
D_\Lambda^N(\mathcal{R}) = \int d^3\mathbf{x}\,n(\mathbf{x})\,n^{(N-1)\text{-}\mathrm{fold}}_\Lambda (\mathbf{x},\mathcal{R}).
\end{equation}
The coefficients $D_\Lambda^N(\mathcal{R})$ are precisely the multipole moments of the NPCF.

Following \citet{Ju2026}, the plane-wave factor
$\mathrm{e}^{\mathrm{i}\mathbf{k}\cdot\mathbf{r}}$ can be interpreted as the
Fourier-space representation of the spatial translation operator. The
rotational average over the joint translations of $(N-1)$ separation vectors
can then be expanded in a basis of isotropic angular functions,
\begin{equation}\label{eq:joint-translation-kspace}
\left\langle
\mathrm{e}^{\mathrm{i}\mathbf{k}_1\cdot\mathbf{r}_1}
\cdots
\mathrm{e}^{\mathrm{i}\mathbf{k}_{N-1}\cdot\mathbf{r}_{N-1}}
\right\rangle_{\mathrm{rot}}
=
\sum_{\Lambda}\mathcal{E}(\Lambda)
\left[
\sum_{\mathcal{M}}C_{\mathcal{M}}^{\Lambda}
\hat{W}^{\ell_1m_1}_{r_1}(\mathbf{k}_1)
\cdots
\hat{W}^{\ell_{N-1}m_{N-1}}_{r_{N-1}}(\mathbf{k}_{N-1})
\right]\times
\mathcal{P}_{\Lambda}
(\hat{\mathbf r}_1,\ldots,\hat{\mathbf r}_{N-1}),
\end{equation}
where
\begin{equation}\label{eq:multipole-window}
    \hat{W}_{r}^{\ell m}(\mathbf{k})
    =
    4\pi i^\ell j_\ell(kr)Y_\ell^m(\hat{\mathbf k})
\end{equation}
is the harmonic decomposition of the spherical-shell translation window with
radius $r$.

Equation~(\ref{eq:joint-translation-kspace}) establishes the connection between
rotational averaging and multipole decomposition. It demonstrates that the
isotropic angular projection of an arbitrary $N$-point configuration can be
factorised into products of single-vertex multipole window functions, with the
coupling between different angular momenta fully described by the coefficients
$C_{\mathcal M}^{\Lambda}$.

Within the MRA framework, the shell-filtered fields associated with each multipole mode can be constructed independently via inverse FFTs. For a given harmonic mode $(\ell,m)$ and shell radius $R_i$, the corresponding scaling-function coefficients (SFCs) are
\begin{equation}\label{eq:iFFT-multipoleSFC}
\epsilon_{j\mathbf{l}}^{\ell m}(R_i)=\mathrm{iFFT}\left[\hat{\epsilon}_{j\mathbf{k}}\hat\Phi_{\mathbf{k}}G(\mathbf{k})\hat{W}_{R_i}^{\ell m}(\mathbf{k})\right], \quad i=1, \ldots, N-1,
\end{equation}
where $G(\mathbf{k})$ denotes the additional filters applied to the original density field. These multipole-filtered coefficient fields constitute the fundamental building blocks for constructing the NPCF multipoles through the connection-coefficient formalism developed above.

For a given multipole order $\Lambda$, Eq.~(\ref{eq:iFFT-multipoleSFC}) yields a complete set of multipole-filtered scaling-function coefficients $\{\epsilon^{\ell m}_{j\mathbf{l}}(R_i)\}$ for all harmonic indices
specified by $\Lambda$ and $\mathcal{M}$, and for each of the
$N-1$ vertices $R_i$ ($i=1,\ldots,N-1$). These coefficient fields constitute the fundamental building blocks for the construction of the NPCF multipoles.

Furthermore, the recursive contraction rule derived in Eq.~(\ref{eq:DN_gamma_recursive}) for the monopole component of the $(N-1)$-fold density field can be generalised straightforwardly to each harmonic channel specified by the index set $\{\Lambda,\mathcal{M}\}$. This yields the corresponding multipole-filtered coefficient field $\tilde\epsilon_{j\mathbf{l}}^{\Lambda \mathcal{M}}(\mathcal{R})$.

The SFCs of the $(N-1)$-fold density field at multipole order $\Lambda$ are then obtained by coupling all $\mathcal{M}$ indices through the isotropic basis coefficients,
\begin{equation}
\tilde{\epsilon}_{j\mathbf{l}}^{\Lambda}(\mathcal{R}) = \langle \phi_{j\mathbf{l}},n^{(N-1)\text{-}\mathrm{fold}}_{\Lambda}(\mathbf{x},\mathcal{R})\rangle=\mathcal{E}(\Lambda)\sum_{\mathcal{M}}C_{\mathcal{M}}^{\Lambda} \tilde{\epsilon}^{\Lambda,\mathcal{M}}_{j,\mathbf{l}}(\mathcal{R}),
\end{equation}
The corresponding multipole component of the $(N-1)$-fold density field is therefore reconstructed as
\begin{equation}
    n^{(N-1)\text{-}\mathrm{fold}}_{\Lambda}(\mathbf{x},\mathcal{R})=\sum_{\mathbf{l}}\tilde{\epsilon}_{j\mathbf{l}}^{\Lambda}(\mathcal{R})\phi_{j\mathbf{l}}(\mathbf{x}).
\end{equation}

Under this construction, the multipole moment of the NPCF reduces to an \textit{in situ} two-point correlation between the original density field and the reconstructed $(N-1)$-fold density field,
\begin{equation}\label{eq:NPCF_multipole_definition}
    D_\Lambda^N(\mathcal{R})=\int d^3\mathbf{x}\,n(\mathbf{x})n^{(N-1)\text{-}\mathrm{fold}}_{\Lambda}(\mathbf{x},\mathcal{R})=\langle n(\mathbf{x}), n^{(N-1)\text{-}\mathrm{fold}}_{\Lambda}(\mathbf{x},\mathcal{R}) \rangle,
\end{equation}
Exploiting the orthogonality of the multiresolution basis, Eq.~(\ref{eq:NPCF_multipole_definition}) immediately reduces to a scalar product in coefficient space,
\begin{equation}\label{eq:NPCF_innerproduct}
D^N_{\Lambda}(\mathcal{R}) = \sum_{\mathbf{l}}\langle n(\mathbf{x}),\phi_{j\mathbf{l}}(\mathbf{x})\rangle \langle \phi_{j\mathbf{l}}(\mathbf{x}),n^{(N-1)\text{-}\mathrm{fold}}_{\Lambda}(\mathbf{x},\mathcal{R}) \rangle=\sum_{\mathbf{l}}\epsilon_{j\mathbf{l}}\tilde\epsilon^{\Lambda}_{j\mathbf{l}}(\mathcal{R}).
\end{equation}
Equation~(\ref{eq:NPCF_innerproduct}) is the central result of the \textit{in situ} formalism. For an $N$-point correlation function of arbitrary order, the estimator ultimately reduces to a simple inner product between the coefficient arrays of the original density field and those of the reconstructed, multipole-filtered $(N-1)$-fold field.

For the special case of the three-point correlation function ($N=3$), the multipole index becomes
$\Lambda=\{\ell_1,\ell_2\}$. The coupling coefficients reduce to \citep{Philcox2021b}
\begin{equation}
    C_M^{\Lambda}=C_{m_1 m_2}^{\ell_1 \ell_2}=\frac{(-1)^{\ell_1-m_1}}{\sqrt{2 \ell_1+1}} \delta_{\ell_1 \ell_2}^{\mathrm{K}} \delta_{m_1,-m_2}^{\mathrm{K}},
\end{equation}
and the corresponding isotropic basis functions become
\begin{equation}
    \mathcal{P}_{\ell}\left(\hat{\mathbf{r}}_1, \hat{\mathbf{r}}_2\right)=\frac{\sqrt{2 \ell+1}}{4 \pi}(-1)^{\ell} \mathcal{L}_{\ell}\left(\hat{\mathbf{r}}_1 \cdot \hat{\mathbf{r}}_2\right),
\end{equation}
Accordingly, Eq.~(\ref{eq:joint-translation-kspace}) simplifies to
\begin{equation}\label{eq:2-fold-translation}
    \left\langle\mathrm{e}^{\mathrm{i} \mathbf{k}_1 \cdot \mathbf{r}_1}\mathrm{e}^{\mathrm{i} \mathbf{k}_2 \cdot \mathbf{r}_2}\right\rangle_{\mathrm{rot}}=\frac{1}{4\pi}\sum_{\ell=0}^{\infty}\left(\sum_{m=-\ell}^\ell \hat{W}_{r_1}^{\ell m *}\left(\mathbf{k}_1\right) \hat{W}_{r_2}^{\ell m}\left(\mathbf{k}_2\right)\right) \mathcal{L}_\ell\left(\hat{\mathbf{r}}_1 \cdot \hat{\mathbf{r}}_2\right).
\end{equation}
The normalisation adopted in Eq.~(\ref{eq:multipole-window}) differs slightly from that used in \citet{Ju2026}, leading to the additional prefactor $1/(4\pi)$ in Eq.~(\ref{eq:2-fold-translation}). Apart from this convention-dependent normalisation, the present formulation reproduces the 3PCF multipole estimator derived in \citet{Ju2026}.

\begin{figure*}
    \centering
    \includegraphics[width=0.98\textwidth]{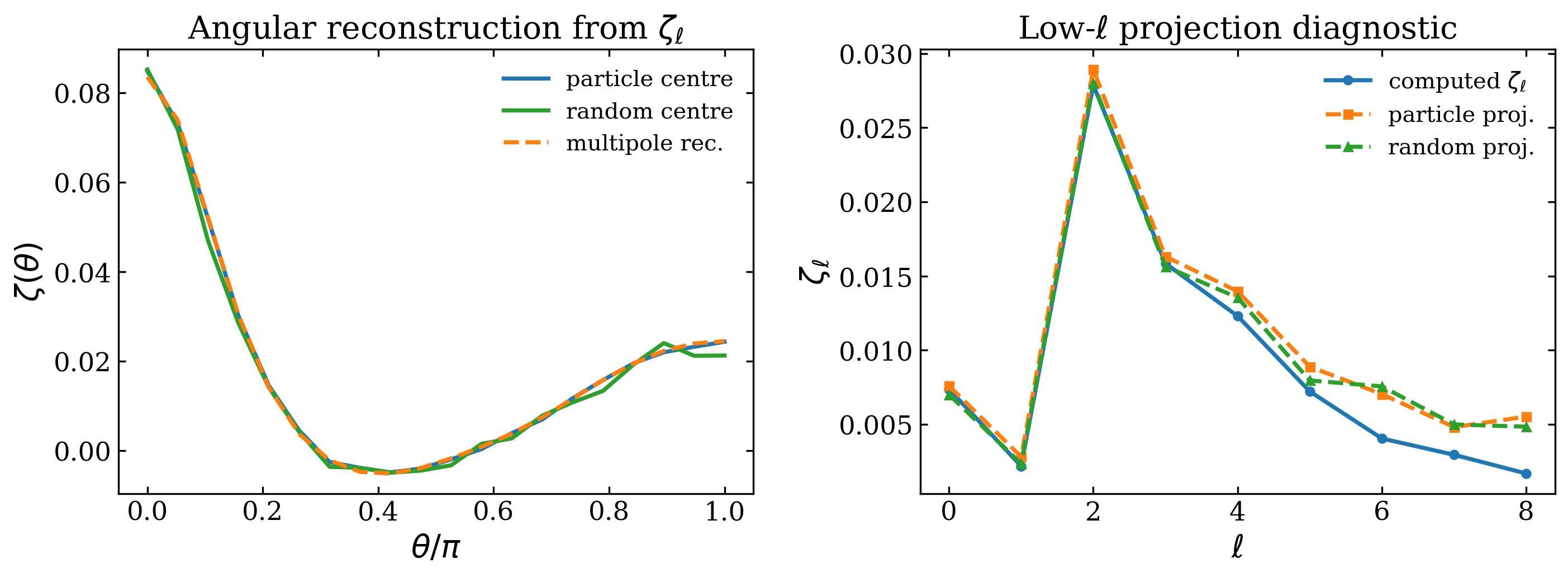}
    \caption{Connection between the standard angular 3PCF and its multipole representation at \((r_{12},r_{13})=(20,40) \,h^{-1}\mathrm{Mpc}\), computed with \textsf{PyHermes} for the \textsf{Quijote} halo example. The first vertex of each triangle is designated as the primary centre. In the particle-centred measurement, \(n_{\rm part}=406,728\) primary centres are sampled from the halo positions, with \(n_{\rm rot}=10^3\) random triangle orientations about each centre. In the box-random-centred measurement, \(n_{\rm rand}=8\times10^6\) primary centres are sampled uniformly from the simulation volume, with \(n_{\rm rot}=200\) orientations per centre. Left: direct angular 3PCF measurements from the two centring schemes, compared with \(\zeta(\theta)\) reconstructed from the directly measured multipole coefficients \(\zeta_\ell\). Right: multipole coefficients obtained by projecting the direct angular measurements onto the Legendre basis, compared with those evaluated directly using the multipole estimator.}
    \label{fig:threepcf_multipole_consistency}
\end{figure*}

\subsection{Particle-based summation}\label{sec:particle-based-summation}

The fast algorithm described above evaluates the estimator through local contractions of scaling-function coefficients followed by a deterministic sum over the MRA translation lattice defined in Section 3.1. We refer to this procedure as the grid-based formulation. The term “grid” therefore denotes the regular lattice on which the MRA coefficients are indexed and stored. An alternative is to retain the same reconstructed and window-filtered fields but evaluate the final spatial average directly at the original particle positions, which we refer to as the particle-based formulation. In essence, this approach corresponds to the conventional pair- or tuple-counting strategy used in correlation-function measurements, but applied to the filtered fields constructed within the \textit{in situ} framework.

The equivalence between the grid-based and particle-based formulations is already evident for the 2PCF. From Eq.~(\ref{eq:pair-counting}), the pair count may be written either as a scalar product of scaling-function coefficients or as a summation over particles. Given an $N_{\mathrm{p}}$-particle sample specified by Eq.~(\ref{eq:dendis}), the spatial average of an arbitrary function $F(\mathbf{x})$ becomes
\begin{equation}\label{eq:particle-based-sum}
    \int d^3\mathbf{x}\, n(\mathbf{x})F(\mathbf{x}) = \sum_{i=1}^{N_{\mathrm{p}}} w_i F(\mathbf{x}_i).
\end{equation}

Equation~(\ref{eq:particle-based-sum}) may be regarded as a Monte Carlo evaluation of the spatial integral, where the particle positions sample the underlying density field. Here we have assumed that the primary vertex is \textit{naked}, i.e.\ no filter is applied to the central density field. If instead the primary field is smoothed by a window function $W$, $n_W(\mathbf{x}) =
(W\circ n)(\mathbf{x})$, the particle representation can be recovered by transferring the filter to the test function,
\begin{equation}
  \langle n_W(\mathbf{x}) F(\mathbf{x})\rangle_{\mathbf{x}}=\langle (W\circ n)(\mathbf{x}), F(\mathbf{x})\rangle = \langle n(\mathbf{x}), (W^{\dagger}\circ F)(\mathbf{x})\rangle = \sum_{i=1}^{N_{\mathrm{p}}} w_i (W^{\dagger}\circ F)(\mathbf{x}_i),
\end{equation}
where $W^\dagger$ denotes the adjoint operator of $W$.

For the isotropic NPCF, the function $F(\mathbf{x})$ is identified with the multipole-filtered $(N-1)$-fold density field, $F(\mathbf{x}) = n^{(N-1)\text{-}\mathrm{fold}}_{\Lambda}(\mathbf{x},\mathcal{R})$. Equation~(\ref{eq:particle-based-sum}) therefore provides a direct particle-based estimator for the NPCF multipoles. However, within the grid-based formulation the construction of
$n^{(N-1)\text{-}\mathrm{fold}}_\Lambda$ requires the recursive contraction of scaling-function coefficients through the connection-coefficient relation Eq.~(\ref{eq:DN_gamma_recursive}). The fully expanded form scales as $\mathscr{O}\!\left(C_{\mathrm{s}}^{\, (2N-4)\mathcal{D}} N_{\mathrm{grid}}\right)$. Although the compact support keeps $C_{\mathrm{s}}$ small, the cost still grows rapidly with increasing $N$.

An alternative strategy follows from the factorised form of the rotationally averaged translation operator in Eq.~(\ref{eq:joint-translation-kspace}). For each harmonic mode $(\ell,m)$ and shell radius $R_i$, we first construct the multipole-filtered density field
\begin{equation}
    n^{\ell m}(\mathbf{x},R_i) =\sum_{\mathbf{l}}\epsilon_{j\mathbf{l}}^{\ell m}(R_i)\phi_{j\mathbf{l}}(\mathbf{x}),
\end{equation}
using the coefficients obtained from Eq.~(\ref{eq:iFFT-multipoleSFC}). The multipole component of the $(N-1)$-fold density field can then be assembled directly in configuration space,
\begin{equation}\label{eq:particle_nfold}
n^{(N-1)\text{-}\mathrm{fold}}_\Lambda\left(\mathbf{x},\mathcal{R}\right)
    =\varepsilon(\Lambda)\sum_{m_1, \ldots m_{N-1}}  C_M^{\Lambda}
    n^{\ell_1 m_1}\left(\mathbf{x};R_1\right)
      n^{\ell_2 m_2}\left(\mathbf{x};R_2\right)
    \ldots n^{\ell_{N-1} m_{N-1}}\left(\mathbf{x};R_{N-1}\right).
\end{equation}

Compared with the recursive connection-coefficient approach, Eq.~(\ref{eq:particle_nfold}) avoids the explicit construction of higher-order contraction kernels. Once the multipole-filtered fields have been generated by FFTs, the $(N-1)$-fold field is obtained through local products in configuration space, followed by a summation over particles according to Eq.~(\ref{eq:particle-based-sum}).

This particle-based implementation can be advantageous when the particle number is significantly smaller than the number of MRA grid cells, i.e. $N_{\mathrm{p}} \ll N_{\mathrm{grid}}$, since the final spatial average scales with the number of particles rather than the number of grid cells. In this regime, the particle-based formulation provides a practical alternative to the grid-based contraction algorithm while preserving exactly the same multipole decomposition and estimator definition.

On the other hand, the conventional particle-based approach to
$N$-point correlation-function (NPCF) estimation relies on brute-force
$N$-tuple counting, which can also be interpreted as a two-step averaging
procedure. For a given $N$-point polyhedral configuration, (i) one vertex is
selected as the primary vertex of each data point, and the corresponding
configuration is averaged over all possible orientations while keeping the
primary vertex fixed; (ii) the spatial average is obtained by summing over all
possible choices of primary vertices in the catalogue. This procedure can be 
viewed as a Monte Carlo integration over the translational
and rotational degrees of freedom of the $N$-point configuration.

In practice, such calculations can be substantially accelerated using the
multiresolution-analysis (MRA) framework implemented in \textsf{Hermes}. A
concrete application to 3PCF measurements was presented by
\citet{Yue2024}. Since bin counting can be formulated as counts-in-cells
estimation, a key step towards efficient $N$-tuple evaluation is to associate
each secondary vertex with a binning window that represents the corresponding
geometric selection. In \citet{Yue2024}, a triple-sphere binning scheme,
implemented using either spherical top-hat or Gaussian windows, was employed
for numerical tests. The same construction naturally extends to general NPCF
measurements.

To illustrate the connection between the angular and multipole representations of the 3PCF, we consider triangles with \((r_{12},r_{13})=(20,40) \,h^{-1}\mathrm{Mpc}\). We evaluate the angular 3PCF using two primary-vertex choices: halo positions (particle-centred, with \(n_{\rm part}\) sampled vertices) and uniformly sampled positions (random-centred, with \(n_{\rm rand}\) vertices). For each vertex, \(n_{\rm rot}\) random orientations approximate the rotational average. These are distinct centre- and orientation-sampling operations. As shown in the left panel of Figure~\ref{fig:threepcf_multipole_consistency}, reconstructing \(\zeta(\theta)\) from the measured multipoles \(\zeta_\ell\) reproduces both angular measurements, confirming consistency between the centering schemes. The right panel projects the angular measurements onto the Legendre basis. The low-order multipoles agree with the direct estimator, validating the two representations. Differences increase at high \(\ell\) because rapidly oscillating Legendre polynomials amplify effects from angular binning, finite orientation sampling, primary-vertex noise, and truncation at \(\ell_{\max}\). Thus, the direct estimator is preferable for accurate high-\(\ell\) coefficients, while angular reconstruction is useful for visualization and consistency checks.

\section{\textsf{Hermes} -- Implementation and Performance}
\label{sec:implementation-performance}

\textsf{PyHermes} is an open-source \textsf{Python} toolkit built upon the
algorithmic framework introduced above. It provides a high-performance and
flexible platform for reconstructing physical fields within the MRA framework
and for performing a broad range of cosmic-statistics measurements. The main
features of \textsf{PyHermes} are summarised as follows:

\begin{itemize}

\item \textbf{Speed}:
By exploiting the field-based MRA formulation, \textsf{PyHermes} replaces direct
particle-pair or particle-tuple enumeration with algebraic operations among MRA
coefficients. The computational cost is therefore primarily determined by the
number of MRA grid coefficients, scaling approximately as
\(\mathscr{O}(N_{\mathrm{grid}}\log N_{\mathrm{grid}})\), rather than directly by the number of objects in the catalogue. This enables efficient reconstruction of physical
fields and evaluation of clustering statistics for extremely large data sets.
Performance-critical kernels are further accelerated through \textsf{Numba}'s just-in-time
(JIT) compilation\footnote{\url{https://numba.pydata.org}},
which translates numerical \textsf{Python} routines into optimised machine code while
preserving the flexibility of the \textsf{Python} interface.

\item \textbf{Parallelism}:
\textsf{PyHermes} combines MPI-based distributed parallelism with
thread-level acceleration to efficiently utilise multi-core and
high-performance computing systems. Computational tasks, including field
construction, sampling, and higher-order correlation measurements, can be
distributed among MPI processes, while local numerical kernels are accelerated
through multi-threading.

\item \textbf{Cross-platform compatibility}:
\textsf{PyHermes} can be installed through standard \textsf{Python} package managers,
including \textsf{pip} and \textsf{conda}. Its lightweight implementation
minimises platform-dependent compiled components and provides consistent
execution across both personal workstations and HPC environments.

\item \textbf{Flexibility}:
\textsf{PyHermes} employs \textsf{PyWavelets}
for wavelet transformations, allowing users to select from a range of wavelet bases
\citep{Lee2019PyWaveletsAP}. In addition,
\textsf{PyHermes} provides a library of commonly used window functions and
supports user-defined windows, enabling customised estimators for a wide range of
statistical applications.

\end{itemize}

\begin{figure*}
    \centering
    \includegraphics[width=1.0\textwidth]{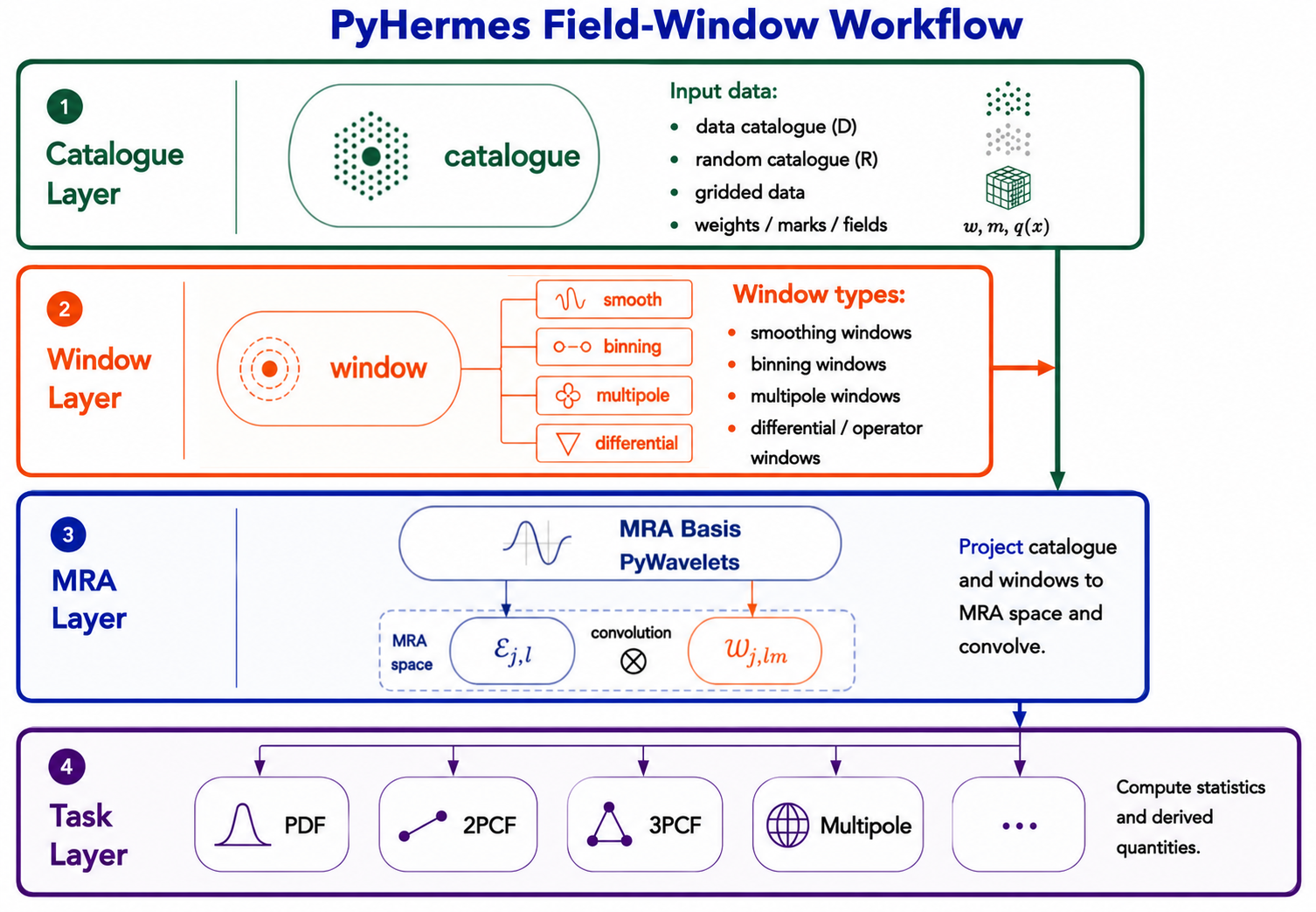}
    \caption{Schematic overview of the field--window workflow in \textsf{PyHermes}. The framework is organised into four layers: catalogue, window, MRA, and task.
The catalogue layer supplies input data, including data and random catalogues,
gridded fields, and physical weights or marks. The window layer specifies the
field operations required by different applications, such as smoothing,
binning, multipole decomposition, and differential operators. These inputs are
projected onto the MRA basis in the MRA layer, where the reconstructed fields and
window kernels are combined through convolution. The resulting
filtered fields are then processed by the task layer to compute a variety of
statistics and derived quantities, including PDFs, 2PCF, 3PCF, multipoles, and physical-field diagnostics. This unified architecture separates the field representation from the statistical measurement, allowing different estimators to be constructed through flexible choices of window operators.}
    \label{fig:pipeline}
\end{figure*}

The numerical demonstrations and performance benchmarks presented in this
section use the \textsf{Quijote} fiducial-cosmology halo catalogue from
realisation 8000 at snapshot 004 (\(z=0\)). The corresponding \textsf{FoF} halo sample
contains 406,728 haloes in a periodic cubic box with side length
\(1000\,h^{-1}\mathrm{Mpc}\). For redshift-space examples, the same halo catalogue
is transformed using the plane-parallel redshift-space displacement along the
specified line of sight. 

These examples are intended as representative validation tests of the
algorithms and implementations rather than as a complete analysis pipeline for
observational surveys. In particular, all demonstrations assume periodic
boundary conditions and do not account for survey masks, angular selection
functions, or spatially varying random catalogues. Realistic survey geometries
and non-uniform random samples can be incorporated within the same framework.
Task-oriented examples, parameter files, and additional diagnostic plots are
provided in the public documentation.

By integrating an efficient MRA-based field representation, hybrid MPI/thread
parallelism, GPU acceleration, and a flexible window-function framework,
\textsf{PyHermes} provides a unified computational environment for a broad range of cosmic
statistics. The combination of compiled performance through JIT acceleration
and the accessibility of \textsf{Python} allows the framework to operate efficiently
on both local workstations and large-scale HPC platforms.

\subsection{\textsf{PyHermes} workflow and task architecture}\label{subsec:pyhermes_engine}

The \textsf{PyHermes} framework is organised into four functional layers:
Catalogue, MRA field, Window, and Task, as illustrated in Fig.~\ref{fig:pipeline}.

\begin{itemize}

\item \textbf{Catalogue layer:}
The catalogue layer defines the input data to be analysed. Particle positions
specify the tracer distribution, while optional attributes, including weights,
marks, and particle-carried physical quantities, can be attached to each
object.

\textsf{PyHermes} supports both in-memory particle arrays and catalogue files in
raw binary, \textsf{NumPy} NPZ, \textsf{Gadget} binary, HDF5, and \textsf{FoF} group formats. It can
therefore directly process both \textsf{Quijote} \textsf{FoF} halo catalogues and
dark-matter particle snapshots, with user-defined weights, physical quantities,
and unit conversions specified through reader options.

\item \textbf{MRA-field layer:}
The \texttt{SFCProjection} task projects particle catalogues onto the scaling
function basis and computes the coefficients
\(\epsilon_{j\mathbf{l}}\) defined in Eq.~(\ref{eq:phi_projection}).
The resulting \texttt{SFCField} object represents the reconstructed density or
physical field in the MRA basis and stores the corresponding box geometry,
resolution level \(J\), scaling-function basis, normalisation, and grid
metadata.

The wavelet transformation module is implemented through the
\textsf{PyWavelets} library, allowing flexible choices of compact-support
wavelet bases. The default configuration adopts the Daubechies D4 wavelet,
while other built-in filters or user-defined wavelets can be employed for
specialised applications.

\item \textbf{Window layer:}
A \texttt{WindowFunc} object represents a real- or Fourier-space kernel together
with its associated parameters. A window is compatible with an
\texttt{SFCField} when they share identical box geometry, resolution level,
and scaling-function basis.

Within this unified interface, window operators provide the mathematical
building blocks for a broad range of operations, including smoothing,
separation binning, counts-in-cells statistics, angular multipole projections,
Fourier-space differential operators, and inverse-Laplacian operations for
constructing Newtonian potential fields.

\item \textbf{Task layer:}
High-level analysis tasks, including \texttt{Counting},
\texttt{Corr\_2PCF}, \texttt{Corr\_3PCF}, and
\texttt{Corr\_3PCF\_Multipole}, combine MRA-represented fields and window
operators according to the estimators introduced in
Section~\ref{sec:algorithm-demonstration}. Each task produces
estimator-specific output objects while sharing the same underlying field and
window infrastructure.

\end{itemize}

\par\noindent
\begin{lstlisting}[style=pyhermescode,basicstyle=\ttfamily]
# Project a catalogue into a reusable SFCField.
SFCProjection(read_param(config_path="./configs/param_sfc_projection.yaml")).run()

# Measure 2PCF, standard 3PCF, and 3PCF multipoles from configuration files.
Corr_2PCF(read_param(config_path="./configs/param_2pcf.yaml")).run()
Corr_3PCF(read_param(config_path="./configs/param_3pcf_pcenter_nrot20.yaml")).run()
Corr_3PCF_Multipole(read_param(config_path="./configs/param_3pcf_multipole_lmax14.yaml")).run()
\end{lstlisting}

The current \textsf{PyHermes} release implements counting statistics, isotropic
and anisotropic two-point correlation functions (2PCFs), standard three-point
correlation functions (3PCFs), 3PCF multipoles, marked statistics, weighted and
differential fields, and Newtonian-potential reconstruction. More general
statistics, including direct 2PCF multipoles and higher-order NPCF estimators,
can be constructed within the same framework by combining additional window
convolutions and products of filtered fields.

A central design principle of \textsf{PyHermes} is the separation between
catalogue projection and statistical measurement. Once an
\texttt{SFCField} object has been constructed, it can be reused for
different measurements, including one-point statistics, two-point products,
three-point products, multipole estimators, and smoothed or weighted-field
analyses, without reprocessing the original catalogue. This design avoids
redundant projections and provides an efficient and reproducible workflow for
large-scale structure statistics.

\begin{table*}
    \centering
    \caption{Main roles of window operators in \textsf{PyHermes}. The same
\texttt{WindowFunc} interface is used throughout the framework; the role of
each operator is determined by how its kernel is applied in the field--window
operations of the estimator.}
    \label{tab:window_roles}
    \begin{tabular}{p{0.22\textwidth}p{0.31\textwidth}p{0.39\textwidth}}
\toprule
Window classification
 & Operators & Functions \\
\midrule
     Smoothing operators &
    \texttt{sphere}(b), \texttt{cubic}(e), \texttt{cylinder}(f),
     \texttt{gaussian}(c) &
     Low-pass filters for smoothing reconstructed fields and defining finite-volume
     counts-in-cells measurements. They can be combined with separation-binning
    windows to construct generalised spatial kernels. \\

    Continuous wavelet transforms &
    \texttt{cw}(j), \texttt{cws}(k), \texttt{gdw}(l) &
    Band-pass filters for measuring localised spatial fluctuations and real-space density variances corresponding to the band-averaged power spectrum. \\

    Binning window functions&
    \texttt{shell}(a), \texttt{ring}(i), \texttt{disk}(h),
    \texttt{cylshell}(g) &
    The infinitesimal-width binning windows used for isotropic and anisotropic
    NPCF measurements can be combined with the \texttt{gaussian}(c) smoothing
    window to introduce an additional Gaussian smoothing scale, while preserving the same estimator structure.\\

    Multipole windows &
    \texttt{legendre\_multipole} &
    Construct spherical-harmonic projection windows and the corresponding
    multipole-decomposed fields for NPCF measurements. \\

    Differential and potential operators &
    \texttt{directional\_derivative}(m), \texttt{laplacian}(n),
    Newtonian gravitational potential (o) &
    Fourier-space representations of differential operators and the Newtonian gravity kernel associated with the inverse Laplacian operator. Combinations of these operators construct derived physical fields, including divergence,
    curl/vorticity, tidal tensors, gravitational potential, and acceleration fields.\\

    Custom and composite operators &
    user-defined kernel functions or algebraic combinations of built-in operators & Enable customised statistics and new estimators through flexible combinations of window kernels and field operations. \\
    \bottomrule
     \end{tabular}
     \vspace{0.2cm}
    \begin{minipage}{0.95\textwidth}
    \vspace{0.15cm}
    \footnotesize
    Notes: The letters in parentheses correspond to the window definitions
    listed in Table~\ref{tab:kernel} in Appendix~\ref{app:window_functions}.
    \end{minipage}
\end{table*}

\subsection{Multiresolution field construction and window operations}
\label{sec:MRA_window}

The lower-level interface of \textsf{PyHermes} separates catalogue projection
from subsequent statistical operations. The \texttt{SFCProjection} task
projects a discrete catalogue onto the MRA basis and returns a reusable
\texttt{SFCField}:

\par\noindent
\begin{lstlisting}[style=pyhermescode,basicstyle=\ttfamily]
task = SFCProjection()       # catalogue-to-field projection task
task.fin = {                 # public Quijote FoF halo catalogue
    "path": "https://pyhermes.astroslacker.com/downloads/group_tab_004.0",
    "download_path": "./data/quijote_halos/8000/groups_004/group_tab_004.0",
    "format": "fof",
}
# task.particle_pos = pos     # optional in-memory particle positions
# task.catalog_weight = wg    # optional catalogue or selection weights
# task.field_value = x        # optional physical values, e.g. velocity or mass
task.J = 8                   # field resolution, with L = 2**J grid cells per axis
task.wavelet_mode = "db2"    # scaling-function family used for the MRA basis
D = task.run()               # reusable SFCField containing epsilon coefficients
\end{lstlisting}

The \texttt{fin} interface accepts local or \texttt{HTTP(S)} sources in generic
binary, NumPy archive, Gadget binary/HDF5, and friends-of-friends formats
(\texttt{bin}, \texttt{npz}, \texttt{gadget}, \texttt{gadget\_hdf5},
\texttt{gadget-fof}, and \texttt{fof}). Remote files are cached at
\texttt{download\_path}; format-specific \texttt{reader\_params} can select the
particle type, position units, weights, and physical field values. Users may
instead provide these quantities directly through \texttt{particle\_pos},
\texttt{catalog\_weight}, and \texttt{field\_value}. In MPI runs, the catalogue
is read once and distributed across ranks.

The options \texttt{wavelet\_mode}, \texttt{wavelet\_level}, and
\texttt{phi\_resolution} specify the scaling-function basis and its numerical
representation. The resulting field \texttt{D} stores the coefficients
$\epsilon_{j\mathbf{l}}$ of Eq.~(\ref{eq:phi_projection}), while
\texttt{sfc\_info} records the box geometry, resolution, and basis needed to
construct compatible windows.

Once the field is constructed, different measurements are specified through
window operations. In the current implementation, normalised windows satisfy
$\widehat W(0)=1$ under the Fourier convention adopted in the code and
Table~\ref{tab:kernel} in Appendix~\ref{app:window_functions}. Length arguments are given in physical box
units and are internally rescaled to the $L=2^J$ grid. The basic object-level
algebra is correspondingly compact:
\vskip0.2cm
\par\noindent\begin{lstlisting}[style=pyhermescode,basicstyle=\ttfamily]
smooth = WindowFunc({"type": "sphere", "len_args": {"R": 5}}, D.sfc_info, threads=8)  # top-hat smoothing window
D_smooth = D @ smooth                         # Fourier-space window convolution
DeltaField = D_smooth / D_smooth.field_mean_density() - 1  # smoothed density contrast
\end{lstlisting}

The window application in the second line is evaluated in Fourier space. For a
general window, the translation phase factor in
Eq.~(\ref{eq:shift-scaling_coeff-kspace}) is replaced by the Fourier-space
window kernel $\widehat W(\mathbf{k})$, and \textsf{PyHermes} constructs the discrete
kernel
\begin{equation}\label{eq:discrete-window-kernel}
K_W(\mathbf{k})=\widehat W(\mathbf{k})\,\widehat\Phi(\mathbf{k}),
\qquad
\epsilon_W=\mathrm{iFT}\left[\mathrm{FT}(\epsilon)\,K_W(\mathbf{k})\right],
\end{equation}
where $\widehat{\Phi}$ is the Fourier transform of the scaling-function
autocorrelation $\Phi$ defined in Eq.~(\ref{eq:cf-basis}); it accounts for the
finite scaling-function representation. The method
\texttt{field\_mean\_density()} supplies the mean density used to form the
contrast, and arithmetic on \texttt{SFCField} objects constructs differences
and products of windowed fields. The main window roles are summarised in
Table~\ref{tab:window_roles}; their analytic forms are listed in the appendix
and online documentation.

\subsection{Composing windows and weights in \textsf{PyHermes}}

Within the \textsf{PyHermes} framework, weighting and spatial filtering are treated in a unified manner: object weights modify the field being represented, whereas
window functions define how the reconstructed field is sampled or combined in
a statistical estimator. Consequently, weighted correlations, marked
statistics, and cross-correlations between different physical fields can be
constructed using the same field--window operations without introducing
separate pair-counting schemes.

\subsubsection{Window function composition in 2PCF estimation}

For 2PCF measurements, \textsf{PyHermes} supports both the convolution
expression \texttt{(A @ W) * B}, followed by a grid mean, and the equivalent
Parseval sum accessed through \texttt{A.pair\_with(B).evaluate([W, ...])}. The latter is particularly suited to evaluating multiple windows in a batch.
For two real fields on the same MRA grid and basis, applying
Parseval's theorem [Eq.~(\ref{eq:pair-counting-kspace})] to the
window relation in Eq.~(\ref{eq:discrete-window-kernel}) gives
\begin{equation}\label{eq:pair-product-parseval}
DD_{AB}(\mathbf r)
=\sum_{\mathbf l}\epsilon^{B}_{j\mathbf l}
  \widetilde\epsilon^{A}_{j\mathbf l}(\mathbf r)
=\sum_{\mathbf n}\bigl(\widehat\epsilon^{B}_{j\mathbf n}\bigr)^*
  \widehat{\widetilde\epsilon}^{A}_{j\mathbf n}(\mathbf r)
=\sum_{\mathbf n}\bigl(\widehat\epsilon^{B}_{j\mathbf n}\bigr)^*
  \widehat\epsilon^{A}_{j\mathbf n}
  K_{W_{\mathbf r}}(\mathbf k_{\mathbf n}).
\end{equation}
For $A=B$, the cross-power reduces to
$\lvert\widehat\epsilon_{j\mathbf n}\rvert^2$.
The Fourier-space formulation reuses source FFTs and cross-power across
windows, substantially accelerating measurements at multiple separations.

The flexibility in choosing window functions substantially broadens the scope of clustering measurements within a unified computational framework. Rather than being restricted to predefined geometric bins, the window function can be designed to isolate, suppress, or enhance specific spatial scales and environmental dependencies according to the scientific objective. Different statistics are therefore realised by modifying the window operators, while the underlying field representation, window convolution, and products of filtered fields remain unchanged.

In general, a window function is specified by a set of parameters describing its spatial translation and orientation, the geometry of the binning volume, the characteristic scale or smoothing length, and other filter-specific properties. In the \textsf{PyHermes} task interface, window functions can be selected from the built-in window library or defined by the user in \textsf{Python}. The built-in library provides a collection of commonly used geometric and filtering kernels, including \texttt{shell}, \texttt{sphere}, \texttt{gaussian}, \texttt{gaussian\_shell}, \texttt{ring}, \texttt{disk}, \texttt{cylshell}, and \texttt{cylinder}, together with the smoothing, filtering, differential, and potential windows listed in Table~\ref{tab:window_roles}. They share the same \texttt{WindowFunc} interface, while each is used according to its smoothing, binning, multipole, differential, or potential role.

Window functions can also be combined algebraically through linear
superposition, multiplication in Fourier space, and convolution in real
space. These operations provide a flexible way to construct composite windows
from simpler building blocks for two-point correlation-function measurements
and other clustering statistics. In the following, we demonstrate their
implementation in \textsf{PyHermes} using three representative examples:
the finite-width spherical-shell window introduced in
Section~\ref{subsec:Real Space: Spherical Shell}, the Gaussian-smoothed spherical-shell window described in Section~\ref{subsec: Filtered_2PCF}, and the Gaussian-smoothed circular-ring window presented in
Section~\ref{subsec:RSD Circular_Ring}. 

The examples below use the direct convolution expression for individual
separations and the batched Parseval interface for an entire radial curve.

\begin{itemize}

\item \textbf{Spherical shell}: 
\vskip0.1cm 

The elementary isotropic binning window is an infinitesimally thin shell centred at a given separation. For example, a $10\,h^{-1}\mathrm{Mpc}$ shell can be created as
\vskip0.2cm
\par\noindent
\begin{lstlisting}[style=pyhermescode,
                   basicstyle=\ttfamily]
W_shell = WindowFunc({"type": "shell", "len_args": {"R": 10}}, DeltaField.sfc_info, threads=8)
xi_shell = ((DeltaField @ W_shell) * DeltaField).as_array().mean()
\end{lstlisting}
where \texttt{DeltaField.sfc\_info} carries the grid and basis metadata of the reconstructed field, including the resolution level $J$, box size, and scaling function, ensuring that the window kernel is constructed on the same multiresolution grid and in the same basis as \texttt{DeltaField}. The \texttt{threads} argument specifies the number of \textsf{Numba} threads used to construct the kernel. 

Window arithmetic is particularly useful when the desired selection or filtering kernel can be constructed more naturally from simpler analytic components. For example, a finite radial bin corresponds to a spherical shell of finite thickness and can be expressed as the difference between two spherical top-hat windows, as illustrated in Fig.~\ref{fig:sphere-shell}. In the \textsf{PyHermes} framework, it can be defined by

\vskip0.2cm
\par\noindent
\begin{lstlisting}[style=pyhermescode,
                   basicstyle=\ttfamily]
s, delta_R = 20, 5                      # radial separation and bin width
R_in = s - 0.5 * delta_R                # inner bin edge
R_out = s + 0.5 * delta_R               # outer bin edge

W_in = WindowFunc({"type": "sphere", "len_args": {"R": R_in}}, DeltaField.sfc_info, threads=8)
W_out = WindowFunc({"type": "sphere", "len_args": {"R": R_out}}, DeltaField.sfc_info, threads=8)

# Volume-normalised finite shell from two spherical top-hat windows.
W_thickshell = (R_out**3 * W_out - R_in**3 * W_in) / (R_out**3 - R_in**3)
xi_s = ((DeltaField @ W_thickshell) * DeltaField).as_array().mean()
\end{lstlisting}

For an entire radial curve, the built-in finite-shell representation also
enables automatic reuse of radial window work:
\vskip0.2cm
\par\noindent
\begin{lstlisting}[style=pyhermescode, basicstyle=\ttfamily]
radii = np.linspace(5.0, 150.0, 31)
shell_windows = [
    WindowFunc({"type": "thick_shell", "len_args": {"R": float(r), "delta_R": delta_R}}, DeltaField.sfc_info, threads=8) for r in radii]
with DeltaField.pair_with(DeltaField, threads=8) as pair:
    xi = pair.evaluate(shell_windows)
\end{lstlisting}
Passing a complete window list enables shared planning across samples;
\texttt{evaluate} returns the values in the same order. Explicit composite
windows remain valid, but may still require materialized kernels.

An alternative construction is obtained by convolving an infinitesimally thin
spherical-shell window with a Gaussian smoothing kernel. In Fourier space, this operation corresponds to multiplying the shell window by the Gaussian transfer function, while in configuration space it produces a
Gaussian-smoothed spherical-shell window, as illustrated in
Fig.~\ref{fig:gaussian-shell}. This example demonstrates how window
composition provides a flexible way to introduce controlled smoothing into the binning scheme. 

The left-hand panel of Fig.~\ref{fig:twopcf_binning_window_comparison}
compares 2PCFs measured from one \textsf{Quijote} simulation sample using thin-shell, finite-width-shell, and Gaussian-shell windows. The thin-shell window retains the highest radial resolution and consequently exhibits more pronounced bin-to-bin fluctuations. Increasing the shell width or the Gaussian smoothing scale averages the product of filtered density fields over a broader range of pair separations, progressively suppressing small-scale fluctuations and producing a smoother measurement. Despite this smoothing, the large-scale shape and the main features of the correlation signal remain broadly preserved. The results demonstrate that the window parameters provide a flexible means of controlling the effective radial resolution of the measurement within the same 2PCF estimator.

In the right panel of
Figure~\ref{fig:twopcf_binning_window_comparison}, we compare the isotropic
2PCF measured with \textsf{PyHermes} at different MRA resolution levels
with the results obtained using \textsf{pycorr}/\textsf{Corrfunc}\footnote{https://github.com/cosmodesi/pycorr}
\citep{Sinha2020} and \textsf{TreeCorr}\footnote{https://github.com/rmjarvis/treecorr} \citep{Jarvis2004}. For the
\(L=1000\,h^{-1}\mathrm{Mpc}\) simulation box, the nominal spatial
resolution of the \textsf{MRA} algorithm is $(\Delta L)_J={L}/{2^J}$,
which gives \((\Delta L)_J\simeq7.8\), \(3.9\), and
\(2.0\,h^{-1}\mathrm{Mpc}\) for \(J=7\), \(8\), and \(9\), respectively. All three codes use 31 equally spaced radial bins with edges
between \(5\) and \(150\,h^{-1}\mathrm{Mpc}\), giving
\(\Delta s=145/31\simeq4.68\,h^{-1}\mathrm{Mpc}\).

\begin{figure*}
    \centering
    \begin{minipage}{0.49\textwidth}
        \centering
        \includegraphics[width=\linewidth]{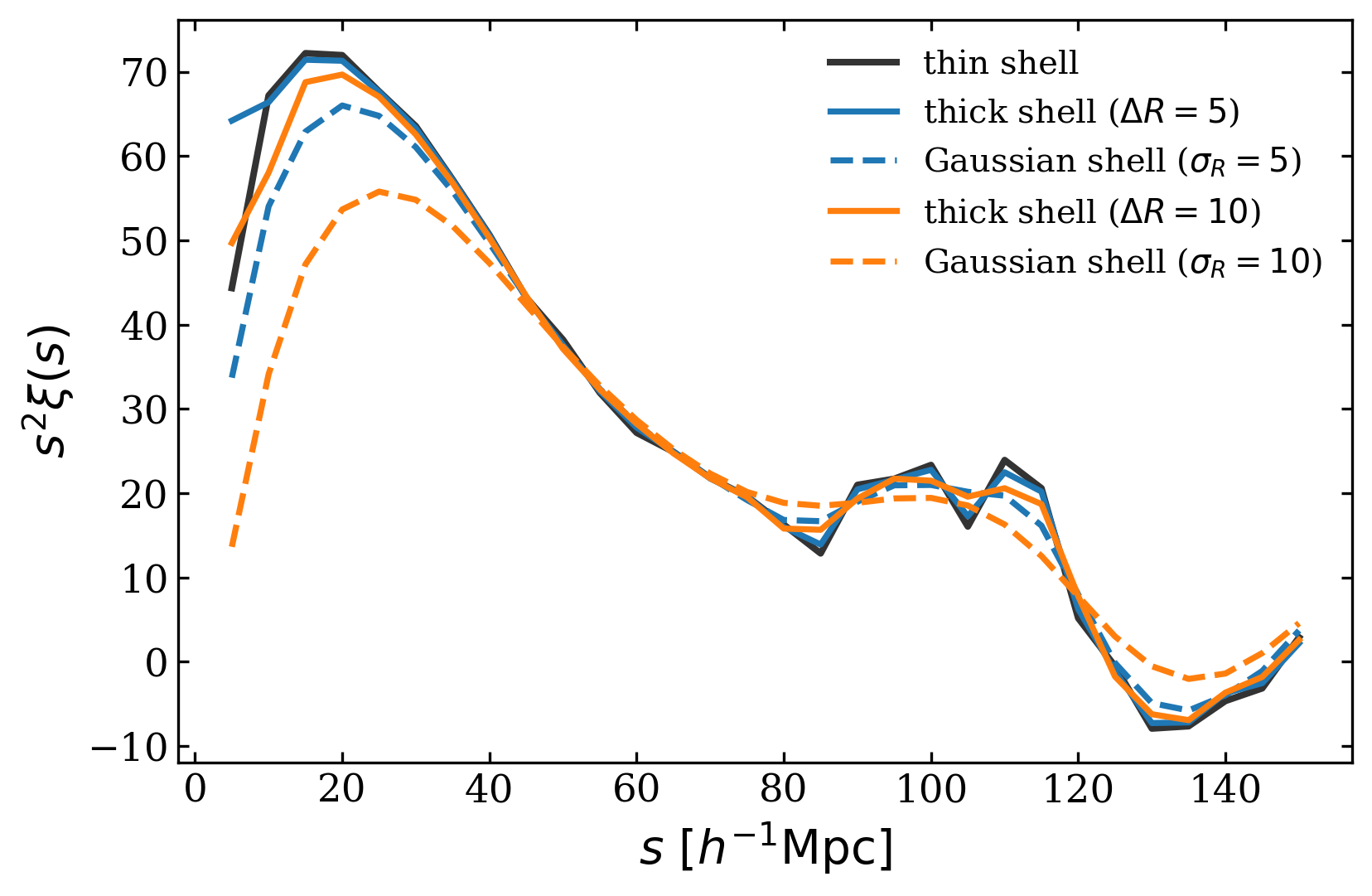}
    \end{minipage}
    \hfill
    \begin{minipage}{0.49\textwidth}
        \centering
        \includegraphics[width=\linewidth]{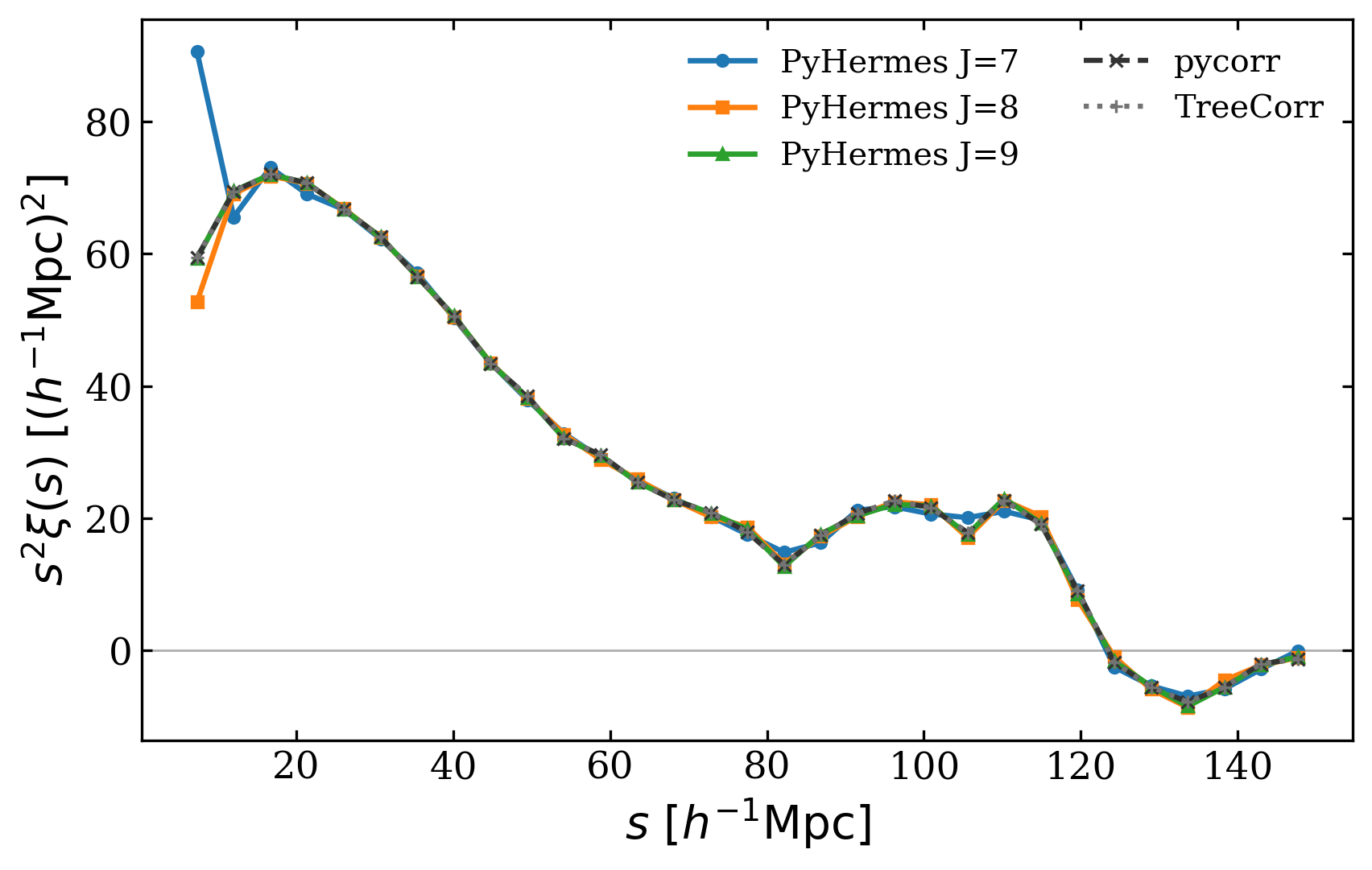}
    \end{minipage}
    \caption{Two complementary checks of the 2PCF window language.
    Left: one-dimensional binning-window choices measured from the same reconstructed
    halo field. The thin shell is compared with finite-thickness shells and
    Gaussian-shell binning windows, showing how the binning definition itself
    controls the radial smoothing of the 2PCF signal. Right:
    comparison of the isotropic 2PCF measured by \textsf{PyHermes},
    \textsf{pycorr}/\textsf{Corrfunc}, and \textsf{TreeCorr} using the same
    radial-bin edges. All curves are plotted at the common arithmetic bin centres
    \(s_i\), with ordinates \(s_i^2\xi_i\). \textsf{PyHermes} implements each
    bin as a volume-normalised difference of two spherical top-hat windows. The
    progression from \(J=7\) to \(J=9\) reduces the small-scale residuals
    relative to pair counting, while the large-scale curves remain in good
    agreement.}
    \label{fig:twopcf_binning_window_comparison}
\end{figure*}

The \textsf{PyHermes} and reference pair-counting measurements use the
same sharp radial bins, so their differences directly quantify the
effect of finite MRA resolution. At \(J=7\), the nominal grid spacing exceeds the \(4.68\,h^{-1}\mathrm{Mpc}\) bin width, producing visible
deviations, particularly at small separations. The offset near
\(s\simeq110\,h^{-1}\mathrm{Mpc}\) is likewise consistent with
insufficient resolution of the locally curved BAO feature.

At \(J=8\), the measurements agree well with pair counting over most of
the separation range, although the first bin,
\(5\leq s<9.68\,h^{-1}\mathrm{Mpc}\), retains a relative residual of
approximately \(-11.3\%\) with respect to
\textsf{pycorr}/\textsf{Corrfunc}. At \(J=9\), the agreement improves
markedly, especially in the first three bins, where the maximum absolute
relative residual is only approximately \(0.21\%\). All curves are evaluated at the same bin
centres and multiplied by the same \(s_i^2\) factor for display, so
these plotting choices do not affect the quoted residuals in the
measured \(\xi_i\).

At finite \(J\), the particle field is represented by its projection
onto the MRA approximation space; increasing \(J\) progressively
recovers the original Dirac-delta distribution. The MRA resolution must
therefore be sufficiently finer than the binning scale, although a grid
spacing smaller than the bin width is not by itself a sufficient
accuracy criterion. The binning and any deliberately applied smoothing
belong to the definition of the observable and must be included in the
theoretical modelling. By contrast, the \(J\)-dependence is a numerical
truncation effect that should be controlled through convergence tests,
rather than absorbed into the observational window.

\vskip0.1cm 
\item \textbf{Circular ring}: 
\vskip0.1cm 

For anisotropic statistics in redshift space, the binning window can be
constructed as a separable convolution of a one-dimensional kernel along the
line of sight and a two-dimensional kernel in the transverse plane. Owing to
the axial symmetry about the line of sight, the transverse kernel can adopt
various geometries, including a circular ring, disk, cylindrical shell, or
finite-thickness cylinder. Different window geometries perform different
averages over transverse and line-of-sight separations, leading to distinct
responses in Fourier space and different sensitivities to scale-dependent
clustering signals and measurement uncertainties.

The following example demonstrates two complementary levels of flexibility
provided by \textsf{PyHermes}. First, users can directly define custom window
kernels in Fourier space. Second, compatible kernels can be combined through
simple multiplication, providing a natural implementation of the separable
window construction described above. Specifically, as illustrated in
Fig.~\ref{fig:gauss_ring}, one user-defined kernel represents a circular ring
in the transverse plane, another introduces Gaussian smoothing in the
transverse direction, and a third specifies the line-of-sight translation
kernel. The product of the transverse ring kernel and the line-of-sight kernel
recovers the standard three-dimensional ring window, while an additional
multiplication by the Gaussian kernel produces the corresponding
Gaussian-smoothed ring window.

\vskip0.2cm
\par\noindent
\begin{lstlisting}[style=pyhermescode,
                   basicstyle=\ttfamily]
import numpy as np
from numba import njit
from pyhermes.utils.special_functions import jn_numba

@njit
def kernel_ring_2d(ki, kj, kk, R):
    k = (ki * ki + kj * kj) ** 0.5
    q = 2.0 * np.pi * k * R
    return jn_numba(0, q)

@njit
def kernel_gauss_2d(ki, kj, kk, sigma):
    k = (ki * ki + kj * kj) ** 0.5
    q = 2.0 * np.pi * k * sigma
    return np.exp(-0.5 * q * q)

@njit
def kernel_double_delta_1d(ki, kj, kk, H):
    return np.cos(2.0 * np.pi * kk * H)

R, H, sigma = 10, 20, 5           # transverse and LOS separations, transverse Gaussian width
W_ring_2d = WindowFunc({"func": kernel_ring_2d, "len_args": {"R": R}}, DeltaField.sfc_info)
W_gauss_2d = WindowFunc({"func": kernel_gauss_2d, "len_args": {"sigma": sigma}}, DeltaField.sfc_info)
W_los_1d = WindowFunc({"func": kernel_double_delta_1d, "len_args": {"H": H}}, DeltaField.sfc_info)

W_ring_3d = W_ring_2d * W_los_1d                         # standard ring
W_gauss_ring_3d = W_ring_3d * W_gauss_2d      # Gaussian-blurred ring
xi_smu_ring = ((DeltaField @ W_ring_3d) * DeltaField).as_array().mean()
xi_smu_gauss_ring = ((DeltaField @ W_gauss_ring_3d) * DeltaField).as_array().mean()
\end{lstlisting}
\par

Here \texttt{kernel\_ring\_2d} depends only on the transverse Fourier
wavenumber $k_\perp$ and defines a circular ring window in the transverse
plane. \texttt{kernel\_gauss\_2d} introduces Gaussian smoothing in the
transverse direction, while \texttt{kernel\_double\_delta\_1d} supplies the real, even response
of symmetric line-of-sight displacements at $\pm s_\parallel$,
$\cos(k_\parallel s_\parallel)$. Their products,
\texttt{W\_ring\_3d} and \texttt{W\_gauss\_ring\_3d}, are supplied to the same
2PCF estimator to obtain \texttt{xi\_smu\_ring} and
\texttt{xi\_smu\_gauss\_ring}, respectively. This example is not intended to
replace the built-in \texttt{ring} window, but rather to demonstrate that
separable Fourier-space components can be flexibly combined to construct new
window kernels without modifying the underlying 2PCF estimator.

Figure~\ref{fig:rsd_gaussian_ring} compares the redshift-space 2PCF measured
with two different transverse window functions. The left panel adopts the
standard infinitesimally thin ring window, which performs an average over a
fixed transverse separation $s_\perp$ while retaining the full line-of-sight
dependence. The right panel replaces this sharp transverse selection with a
Gaussian-smoothed ring window with $\sigma_\perp=5\,h^{-1}\mathrm{Mpc}$. Both
measurements are obtained using the same reconstructed density field and the
same 2PCF estimator; the only difference is the choice of the window kernel.

\begin{figure*}
    \centering
    \includegraphics[width=\textwidth]{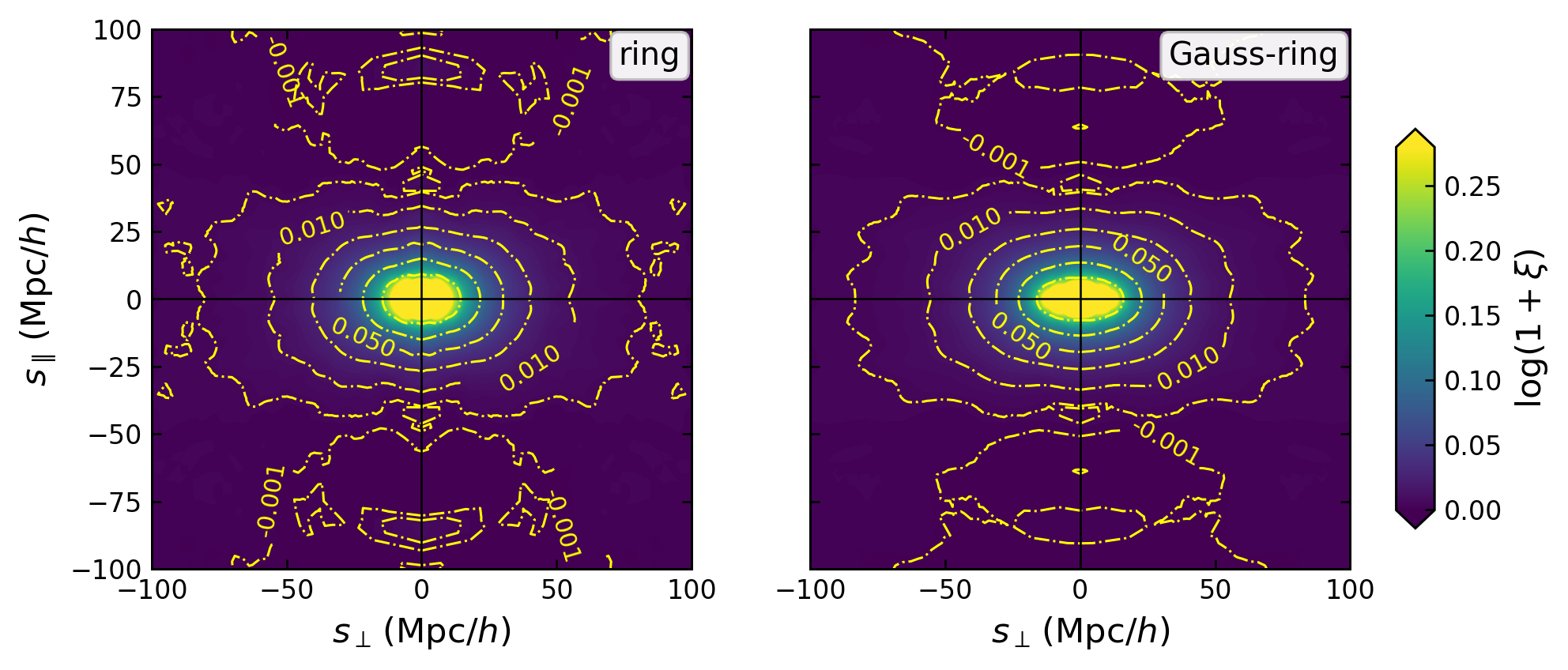}
    \caption{Redshift-space 2PCF measured with two different transverse window 
         choices. The left panel uses the standard thin ring window in the
         $(s_\perp,s_\parallel)$ plane. The right panel replaces the transverse ring selection with a Gaussian-smoothed ring window, where the transverse smoothing scale is $\sigma_\perp=5\,h^{-1}\mathrm{Mpc}$ while keeping the line-of-sight kernel unchanged. Both measurements are performed on the same
         \textsf{Quijote} halo field and show $\log(1+\xi)$.
         The comparison illustrates that \textsf{PyHermes} can modify the spatial
         weighting and binning geometry through flexible window functions without
         changing the underlying correlation estimator.}
    \label{fig:rsd_gaussian_ring}
\end{figure*}

The two measurements exhibit the same overall large-scale anisotropic
clustering pattern, demonstrating that the underlying redshift-space
distortion signal is preserved under the modified window operation. In
particular, the characteristic enhancement along the line of sight associated
with the Kaiser effect and the suppression of small-scale transverse
fluctuations remain clearly visible in both cases. Meanwhile, the
Gaussian-smoothed ring produces a smoother correlation field, with reduced
small-scale oscillatory features compared with the sharp ring window. This
difference originates from the Fourier-space response of the window function:
the Gaussian factor suppresses high-$k_\perp$ modes and therefore removes
small-scale transverse fluctuations while leaving the large-scale modes
almost unchanged.

This comparison illustrates the flexibility of the \textsf{Hermes} framework:
different clustering observables can be generated by modifying the window
function rather than redesigning the correlation estimator. The choice of
window therefore provides an additional degree of freedom to balance angular
resolution, noise suppression, and sensitivity to specific physical scales in
redshift-space clustering analyses.

The explicit constructions above demonstrate the underlying mechanism of
window composition. In typical applications, users do not need to manipulate
individual windowed fields or construct their products explicitly. Instead,
the same operations are automatically handled by the high-level task interface
of \textsf{PyHermes}:

\par\noindent
\begin{lstlisting}[style=pyhermescode,
                   basicstyle=\ttfamily]
task = Corr_2PCF()
task.sfc_field = "xxx_sfc.pkl"                  # SFC field path, or an SFCField object
task.random = "uniform"                         # "uniform", an SFCField object, or an SFC field path
task.binning_window = "shell"                   # separation-bin window, e.g. "shell" or "ring"
task.sampling["s"] = np.linspace(0.0, 180.0, 46) # radial samples
# task.sampling["mu"] = np.linspace(0.0, 1.0, 51) # optional angular samples for xi(s, mu)
task.products = "xi"                            # or ["dd", "dr", "rr", "xi"]
task.threads = 8
task.pair_reduction = "spectral"
task.spectral_window = "auto"
task.reuse_input_fields = True
task.spectral_fft = "shared"
task.memory_strategy = "speed"
task.run()
\end{lstlisting}

Here \texttt{sfc\_field} provides the reconstructed data density field.
Setting \texttt{random="uniform"} adopts the analytic constant-density field
corresponding to a uniform random catalogue, thereby avoiding the construction
of an explicit random field. Alternatively, users can provide an
\texttt{SFCField} reconstructed from an actual random catalogue when survey
geometry, angular masks, or selection effects need to be incorporated. The
\texttt{binning\_window} argument defines the geometry and scale dependence of
the separation kernel, while \texttt{products} controls whether the task returns
intermediate count-level quantities (e.g., $DD$, $DR$, and $RR$) or the final
dimensionless correlation function.

The task option \texttt{pair\_reduction} selects the convolution or spectral
formulation; \texttt{spectral\_window="auto"} selects the automatic window
planning used by the local pair interface. Built-in radial windows and
exactly grid-axis-aligned separable windows admit structural acceleration;
other windows retain the applicable generic or convolution path.
\texttt{reuse\_input\_fields=True} reuses compatible task preparation, while
\texttt{spectral\_fft="shared"} prepares source FFTs and pair powers on the
MPI root and broadcasts the powers. The local \texttt{pair\_with} interface
performs no MPI world collectives. These four task options are independent
and default, respectively, to \texttt{"convolution"},
\texttt{"materialized"}, \texttt{False}, and \texttt{"per\_rank"}; they must
be enabled explicitly. The spectral task path uses
\texttt{memory\_strategy="speed"}, which is the default; the
\texttt{"memory"} strategy retains the original convolution workflow.

\end{itemize}

This task-level interface provides the standard workflow for production 2PCF
measurements, while the lower-level window interface allows users to construct
custom separation kernels and develop alternative estimator variants. A direct
consequence of the field--window formulation is that 2PCF multipole moments can
be incorporated naturally as generalised window operations.

In conventional configuration-space analyses, an anisotropic correlation
function $\xi(s,\mu)$ is first measured in bins of separation and orientation,
after which the resulting two-dimensional correlation function is projected
onto Legendre multipole moments. This procedure introduces an additional
discretization choice: excessively fine $\mu$ binning increases statistical
noise, whereas overly coarse binning reduces angular resolution. In the
\textsf{Hermes} framework, the Legendre multipole kernel is incorporated
directly into the separation window function. Consequently, individual
multipole moments can be measured through the corresponding field--window
operation without explicitly constructing the intermediate $\mu$-binned
correlation function.

\subsubsection{Weighted fields and marked 2PCFs}
\label{subsec:weighted_marked_fields}

Weighted and marked statistics are obtained by changing the amplitude assigned to
each catalogue object before applying the same window convolutions and products
of window-filtered fields. In \textsf{PyHermes}, catalogue weights and
particle-carried field values are treated as separate factors. A generic weighted
field can therefore be written schematically as
\begin{equation}
F_q(\mathbf{x})=
\sum_i \underbrace{w_i^{\mathrm{c}}}_{\mathrm{completeness}}\times \underbrace{w_i^{\mathrm{opt}}}_{\mathrm{optimal\,weight}} \times \underbrace{m_i}_{\mathrm{mark}}\times \underbrace{q_i}_{\mathrm{physical\,field}}\,\delta_{\mathrm{D}}^3(\mathbf{x}-\mathbf{x}_i).
\label{eq:weighted_field}
\end{equation}
Here we distinguish between the weights, marks, and physical amplitudes assigned to each catalogue object. The factor $w_i^{\mathrm{c}}$ denotes a completeness weight that corrects for observational effects, such as redshift failures, fibre collisions, the survey mask, angular systematics, or luminosity-dependent selection. Some observational corrections may instead require pair- or triplet-dependent weights. The factor $w_i^{\mathrm{opt}}$ represents an analysis-dependent optimal weight. Examples include the minimum-variance FKP weighting scheme \citep{Feldman1994} and its bias-dependent PVP generalisation \citep{Percival2004}; multi-tracer weights that reduce sample-variance limitations by exploiting the relative clustering of tracers with different biases \citep{Seljak2009,McDonaldSeljak2009,Hamaus2012,Abramo2016}; and redshift-dependent weights designed to compress the redshift evolution of the BAO and RSD signals while retaining maximal cosmological information \citep{Zhu2015,Zhu2016,Ruggeri2017}. The quantity $m_i$ is a statistical mark designed, for example, to enhance or suppress environmental dependence, as discussed in Sec.~\ref{sec:pair-counting}. Finally, $q_i$ denotes a physical quantity carried by the object, such as its mass, a velocity component, angular momentum, or another intrinsic property.

The present convolution-based implementation in \textsf{PyHermes} naturally incorporates all factorisable weights through the weighted density field. Pair-dependent weights that are functions of separation can be absorbed into the convolution kernels. More general non-separable pair weights break the translation-invariant convolution structure and require bin-wise decomposition, low-rank approximations, or explicit pair-based corrections.

In the projection task, these choices are supplied either through catalogue
files or directly as in-memory arrays. The same interface therefore covers
ordinary catalogues, marked catalogues, and physical weighted fields:
\vskip0.2cm
\par\noindent
\begin{lstlisting}[style=pyhermescode,
                   basicstyle=\ttfamily]
task = SFCProjection()        # catalogue-to-field projection task
task.particle_pos = pos       # optional in-memory particle positions
task.catalog_weight = w      # optional catalogue or selection weights
task.field_value = q          # optional mark or physical value, e.g. velocity or mass
Q_field = task.run()              # reconstructed weighted field
\end{lstlisting}

The following example implements the power-law mark defined in Eq.~\eqref{eq:power-law-mark}. We first smooth the raw halo number-density field with a spherical window to estimate the local density at each halo position. The marked halo field is then constructed by assigning the corresponding mark to \texttt{field\_value}, after which the standard 2PCF driver can be applied without further modification. Since the marked correlation function is normalised by the corresponding unmarked pair counts, an overall rescaling of the mark cancels out; therefore, the mark does not need to be divided by its sample mean before projection.

\vskip0.2cm
\par\noindent
\begin{lstlisting}[style=pyhermescode,
                   basicstyle=\ttfamily]
# Estimate the local halo density around each object.
count_task = SFCProjection()
count_task.particle_pos = halo_pos
count_task.weight_normalization = "raw" # keep the count-field amplitude
D_h = count_task.run()

# Smooth the count field and read the local density at halo positions.
V_m = 4.0 * np.pi * R_m**3 / 3.0
W_m = WindowFunc({"type": "sphere", "len_args": {"R": R_m}}, D_h.sfc_info, threads=8)
n_m = (D_h @ W_m).field_density_at_pos(halo_pos, value_unit="physical") # local number density
Delta = np.clip((n_m - 1.0 / V_m) / nbar_h, 0.0, None) # remove self-count and normalise

# Project the power-law marked catalogue.
m = (Delta + epsilon) ** alpha
mark_task = SFCProjection()
mark_task.particle_pos = halo_pos
mark_task.field_value = m
D_m = mark_task.run()

# Measure the marked isotropic 2PCF.
corr = Corr_2PCF()
corr.sfc_field = D_m
corr.random = "uniform"
corr.binning_window = "shell"
corr.sampling["s"] = s_values
corr.products = "xi"
xi_mark = corr.run()
\end{lstlisting}

\subsubsection{Conventional 3PCF algorithm using a Monte Carlo method}

Conventional NPCF estimation can be viewed as a Monte Carlo evaluation of the spatial average of $N$-tuple counts, obtained by averaging over translations and rotations of a given $N$-point configuration. Following \citet{Yue2024}, the binning scheme for NPCFs does not need to be explicitly tied to a specific polyhedral configuration. Instead, it can be implemented by assigning a binning volume to each vertex, which is mathematically equivalent to convolving the density field at each vertex with a corresponding window function.

For the 3PCF, \citet{Yue2024} introduced a dual-/triplet-sphere-counting algorithm using either spherical top-hat or spherical Gaussian windows. In the dual-sphere scheme, the primary vertex is fixed on each data point, whereas in the triplet-sphere scheme, the primary vertex is randomly sampled within the survey volume. Both schemes provide an effective translational average. In addition, $N_{\mathrm{rot}}$ random rotations of the $N$-point configuration are generated around each primary vertex to estimate the rotational average. This Monte Carlo approach belongs to the class of particle-based summation algorithms described in Sec.~\ref{sec:particle-based-summation}.

At the task level, this 3PCF algorithm is specified by the SFC fields, the triangle configuration characterised by two side lengths ($R_1=r_{12}$ and $R_2=r_{13}$) and the included angle $\theta$, and the choice of the primary-vertex scheme. The latter determines whether the primary vertex is fixed on data points or randomly sampled within the survey volume, corresponding to the dual-sphere and triplet-sphere counting schemes, respectively.

\vskip0.2cm
\par\noindent\begin{lstlisting}[style=pyhermescode,
                   basicstyle=\ttfamily]
task = Corr_3PCF()
task.sfc_field = "xxx_sfc.pkl"        # shared SFC field path, or an SFCField object
task.random = "uniform"               # "uniform", an SFCField object, or an SFC field path
smooth_vertex = {"type": "sphere", "len_args": {"R": 5}}  # optional smoothing on each vertex
task.window2 = smooth_vertex             # window for the second vertex
task.window3 = smooth_vertex          # window for the third vertex
task.r12, task.r13 = 20.0, 40.0       # two fixed triangle side lengths
task.theta["n_theta"] = 20            # number of sampling angles between the two sides
task.n_rot = 1000                     # number of random rotations per primary vertex
task.center = "particle"              # or "box_random"
task.products = "Q"                   # or ["ddd", "zeta", "Q"]
task.run()
\end{lstlisting}
Since the same binning window is usually applied to each vertex of a given triangle, the default implementation uses a single \texttt{sfc\_field} as input. More generally, the three vertices can be assigned independent windows by replacing \texttt{sfc\_field} with three filtered fields, \texttt{sfc\_field1}, \texttt{sfc\_field2}, and \texttt{sfc\_field3}, obtained by filtering the density field with \texttt{window1}, \texttt{window2}, and \texttt{window3}, respectively.

The argument \texttt{random} specifies the random catalogue used for normalisation. It can either be generated from a prescribed mean density or supplied as a predefined reference sample. In the latter case, the corresponding SFC fields of the random catalogue must also be constructed. When \texttt{center="particle"} is selected, the primary vertices are taken from the data catalogue stored in \texttt{sfc\_field1}. In contrast, for \texttt{center="box\_random"}, the primary vertices are randomly sampled throughout the entire volume.

\begin{figure*}
    \centering
    \includegraphics[width=0.94\textwidth]{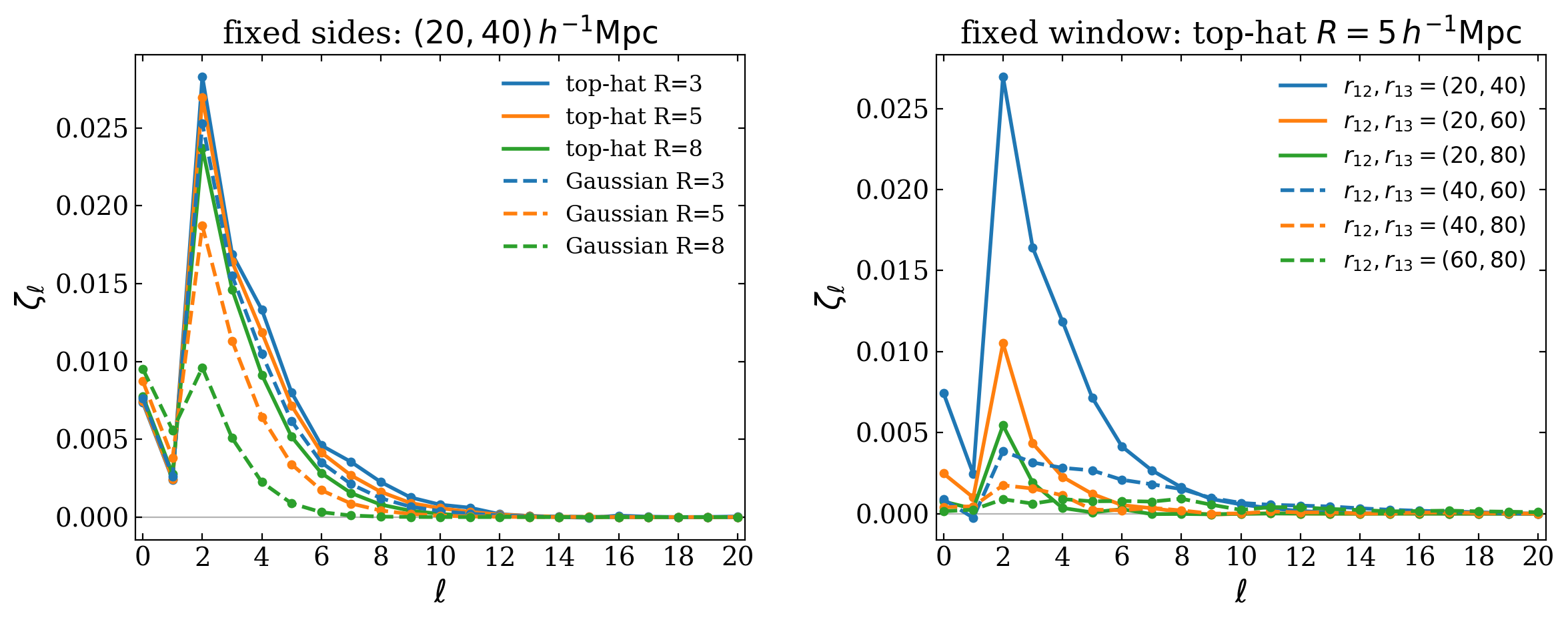}
    \caption{3PCF multipole spectra measured from the \textsf{Quijote} halo field at
    fixed field resolution $J=8$ with $\ell_{\max}=20$. Left: fixed side lengths
    $r_{12}=20\,h^{-1}\mathrm{Mpc}$ and $r_{13}=40\,h^{-1}\mathrm{Mpc}$, comparing
    spherical top-hat smoothing windows (solid lines) and Gaussian smoothing
    windows (dashed lines) with matched radii. Right: fixed top-hat smoothing
    radius $R=5\,h^{-1}\mathrm{Mpc}$, comparing several triangle side-length pairs;
    dashed curves indicate the larger-side configurations.}
\label{fig:threepcf_multipole_diagnostics}
\end{figure*}

\subsubsection{3PCF multipoles}

In the context of the multipole decomposition of the 3PCF, the angular dependence is encoded in the multipole indices, while the multipole coefficients are functions only of the two radial separations $R_1=r_{12}$ and $R_2=r_{13}$. Following the spherical-harmonic window formalism introduced in Section~\ref{sec:multipole-decomposition} and \citet{Ju2026}, \textsf{PyHermes} constructs the corresponding multipole fields for each $(\ell,m)$ and $(R_1,R_2)$ combination through the \texttt{legendre\_multipole} kernels defined in Eq.~(\ref{eq:iFFT-multipoleSFC}).

In the implementation, the negative-$m$ modes are obtained from the
complex-conjugation symmetry of spherical harmonics, so that only the
$m\geq0$ multipole fields need to be explicitly constructed. The convolution
stage is distributed through MPI and accelerated by \textsf{Numba} threading on
the CPU. The subsequent contractions and summations over the filtered
$(\ell,m)$ fields can be performed either using multithreaded CPU kernels or
CUDA kernels on GPUs. The CPU implementation provides a portable backend for
systems without compatible GPUs, while the GPU implementation significantly
accelerates the final multipole-field contractions. In both cases, the
calculation is reformulated from direct triplet enumeration into scalable
operations involving window convolutions and algebraic combinations of
multipole-decomposed fields.

The following boxed code snippet demonstrates the task-level interface of the
3PCF multipole estimator implemented in \textsf{PyHermes}. Following the
window-based formulation introduced above, the bins at the two
secondary vertices are specified through independent binning windows. The
\texttt{sampling} dictionary defines the triangle configuration through the
side-length parameters, while each \texttt{mapping} connects a sampled side length
to the radial coordinate of its associated window kernel. The output array
\texttt{zeta\_l} contains the connected 3PCF multipole moments obtained from the
corresponding products of filtered fields.

\vskip0.2cm
\par\noindent
\begin{lstlisting}[style=pyhermescode,
                   basicstyle=\ttfamily]
task = Corr_3PCF_Multipole()
task.sfc_field = "xxx_sfc.pkl"        # shared SFC field path, or an SFCField object
task.random = "uniform"               # "uniform", an SFCField object, or an SFC field path
task.window = {"type": "sphere", "len_args": {"R": 5}}  # smoothing for all vertices
task.binning_window12 = {"type": "shell", "mapping": {"R": "r12"}}
task.binning_window13 = {"type": "shell", "mapping": {"R": "r13"}}
task.sampling = {"r12": 20.0, "r13": 40.0}  # triangle side lengths
task.l_max = 20                       # highest multipole order
task.execution_mode = "pair_mpi"      # distribute multipole modes over MPI-rank pairs
task.summation_backend = "gpu"        # "gpu" or "cpu" for the final grid summation
task.gpu_threads_per_block = (8, 8, 8)  # CUDA block shape; GPU backend only
task.products = "zeta_l"              # or count-level multipole products
task.run()
\end{lstlisting}
\par

For the grid-based method, only the \texttt{summation\_backend} option differs
between the CPU and GPU workflows; the MPI-distributed CPU convolution stage is identical in both
implementations. The parameter \texttt{gpu\_threads\_per\_block} is used only
when \texttt{summation\_backend} is \texttt{"gpu"} and is ignored for the CPU backend.
Therefore, the choice of backend affects only the execution strategy of the
final multipole-field contraction and does not modify the estimator or the
underlying window-field formulation. Representative timings comparing the two
backends are presented in Section~\ref{subsec:parallel_performance}.

For particle-based summation (Section~\ref{sec:particle-based-summation}), set
\texttt{summation\_method} to \texttt{"particle"} and
\texttt{summation\_backend} to \texttt{"cpu"}. It reconstructs the two harmonic fields at
the primary catalogue positions and sums their weighted products. The primary
catalogue must remain accessible and unsmoothed: omit \texttt{window} and
\texttt{window1}, and apply endpoint smoothing through \texttt{window2} and
\texttt{window3}. Its finite-$J$ discretization can differ from the grid
contraction.

Figure~\ref{fig:threepcf_multipole_diagnostics} illustrates two practical choices in the multipole 3PCF workflow at a fixed field resolution $J=8$ and $\ell_{\max}=20$. The left panel fixes the triangle configuration with $r_{12}=20\,h^{-1}\mathrm{Mpc}$ and $r_{13}=40\,h^{-1}\mathrm{Mpc}$, and compares spherical top-hat and Gaussian windows with different smoothing radii. The right panel fixes the top-hat smoothing scale at $R=5\,h^{-1}\mathrm{Mpc}$ and varies the triangle side lengths. In both cases, the multipole signal is dominated by the lowest orders, while the detailed amplitude and the relative contribution of higher multipoles depend on the smoothing scale and triangle configuration. This illustrates the flexibility of the window-based estimator: the multipole decomposition provides a compact representation of the angular three-point information, while the choice of window functions allows one to regulate the effective scale dependence and suppress contributions from unresolved small-scale structures.

\subsubsection{Differential windows and Newtonian-potential construction}
\label{subsec:differential_weighted_fields}

Beyond the smoothing, binning, and multipole windows discussed above, \textsf{PyHermes} provides Fourier-space windows for applying mathematical operators to physical fields. These include directional derivatives and the Laplacian. Directional-derivative windows can be combined to obtain gradients, divergence, and curl. The package also provides the inverse-Laplacian part of the Newtonian-potential operator, which is combined with the Poisson prefactor below to construct the gravitational potential.

In this framework, a differential operator is treated as a window kernel acting on the field in Fourier space. For example, a smoothed directional derivative can be written as

\begin{equation}
    \mathrm{FT}\left[\partial_\alpha f_s(\mathbf{x})\right]
    =
    \mathrm{i} k_\alpha \widehat W_s(\mathbf{k})\widehat f(\mathbf{k}),
\end{equation}
where \(\widehat W_s(\mathbf{k})\) is an optional smoothing kernel. Accordingly, the built-in \texttt{directional\_derivative} and
\texttt{laplacian} windows can be combined with the smoothing and binning
windows introduced above. For example, smoothing followed by differentiation
is implemented by successively applying a smoothing window and a differential window to the same reconstructed field.

The same formalism naturally extends to inverse operators. The
\texttt{inverse\_laplacian} window applies the Fourier-space kernel
$\widehat{W}_{\nabla^{-2}}(\mathbf{k})=-{1}/{|\mathbf{k}|^2}$
for non-zero modes, while the zero mode is explicitly removed. Applying this
operator to a density field \(\rho\) yields \(\nabla^{-2}\rho\), the spatial
part of the gravitational potential through the Poisson equation. Equivalently,
the corresponding Fourier-space and real-space forms of the gravitational Green's function are
\begin{equation}
    \widehat{W}_{\mathrm{grav}}(\mathbf{k})=4\pi\widehat{W}_{\nabla^{-2}}(\mathbf{k})=-\frac{4\pi}{|\mathbf{k}|^2},\quad  W_{\mathrm{grav}}(\mathbf{x})=-\frac{1}{|\mathbf{x}|}.
\end{equation}
After the appropriate cosmological prefactor is included, this operation gives the
gravitational potential field. Directional differential windows can then
produce the acceleration field and higher-order derivatives, such as the
rank-2 tidal tensor. Likewise, the velocity divergence field is obtained by composing
directional-derivative windows with the individual velocity components. The
following example demonstrates smoothing, differentiation, inverse-Laplacian
operations, and velocity-divergence construction within \textsf{PyHermes}.

\vskip0.2cm
\par\noindent
\begin{lstlisting}[style=pyhermescode,
                   basicstyle=\ttfamily]
smooth = WindowFunc({"type": "gaussian", "len_args": {"R": 10}}, D.sfc_info, threads=8)  # 10 Mpc/h Gaussian smoothing
los_x = {"nx": 1.0, "ny": 0.0, "nz": 0.0}       # x direction
los_y = {"nx": 0.0, "ny": 1.0, "nz": 0.0}       # y direction
los_z = {"nx": 0.0, "ny": 0.0, "nz": 1.0}       # z direction
Dx = WindowFunc({"type": "directional_derivative", "los_args": los_x}, D.sfc_info, threads=8)     # x derivative
Dy = WindowFunc({"type": "directional_derivative", "los_args": los_y}, D.sfc_info, threads=8)     # y derivative
Dz = WindowFunc({"type": "directional_derivative", "los_args": los_z}, D.sfc_info, threads=8)     # z derivative
Lap = WindowFunc({"type": "laplacian"}, D.sfc_info, threads=8)
InvLap = WindowFunc({"type": "inverse_laplacian"}, dm_delta.sfc_info, threads=8)  # inverse-Laplacian part of potential operator

D_smooth = D @ smooth                  # smoothed reconstructed field
grad_x = D @ smooth @ Dx               # derivative after smoothing
lap_D = D @ smooth @ Lap               # Laplacian after smoothing
phi = (dm_delta @ smooth @ InvLap) * poisson_scale  # dimensionless potential Phi/c^2
g_x = -(c_light**2 / a) * (phi @ Dx)            # one component of acceleration
vx_dx = vx @ smooth @ Dx                        # smoothed velocity derivative
vy_dy = vy @ smooth @ Dy                        # smoothed velocity derivative
vz_dz = vz @ smooth @ Dz                        # smoothed velocity derivative
v_div = vx_dx + vy_dy + vz_dz                   # velocity divergence
\end{lstlisting}
\par

\begin{figure*}
    \centering
    \includegraphics[width=0.98\textwidth]{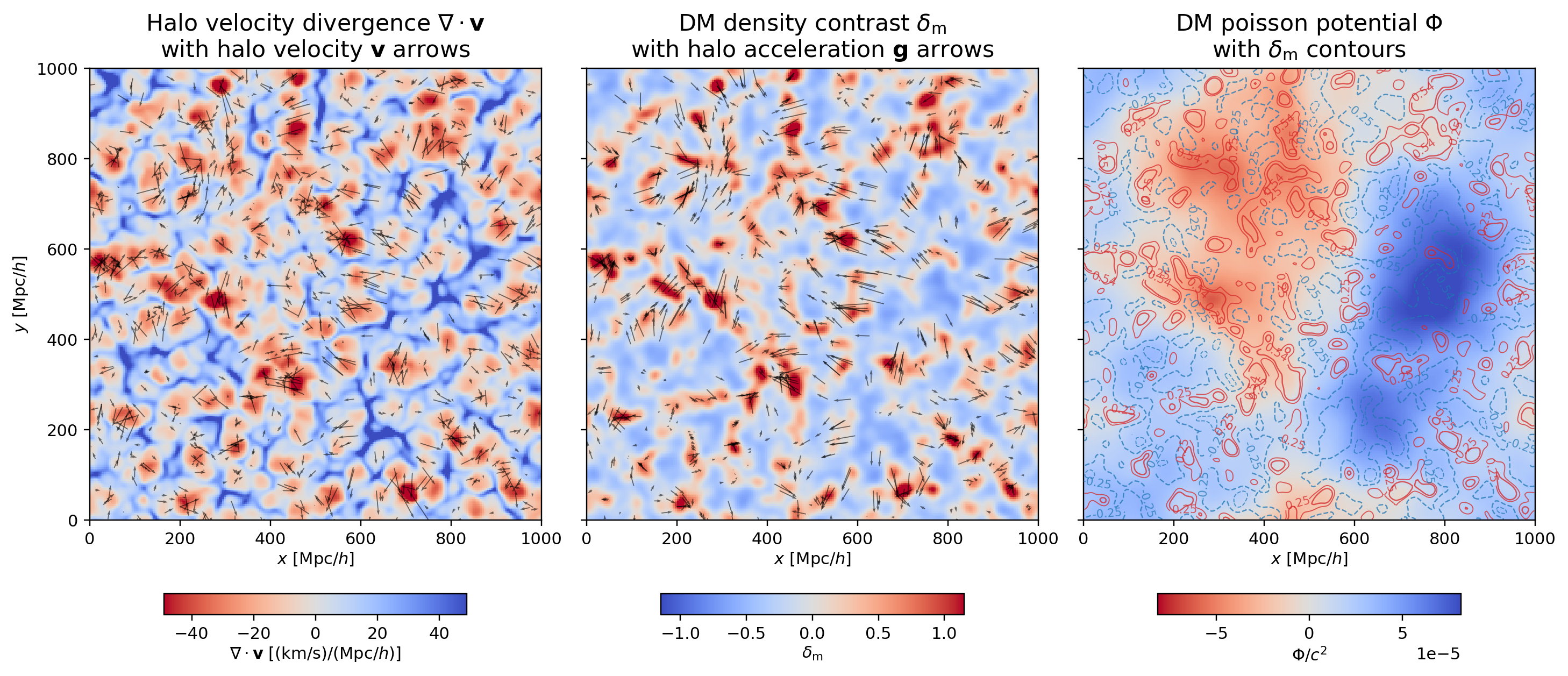}
    \caption{Weighted-field and Newtonian-potential diagnostics on the same
    two-dimensional slice. Left: halo velocity divergence
    \(\nabla\cdot\mathbf{v}\) reconstructed from velocity-weighted fields, with
    transverse halo velocity arrows overlaid. Middle: smoothed dark-matter
    density contrast \(\delta_{\mathrm{m}}\), with halo-position acceleration
    \(\mathbf{g}\) arrows obtained by applying the inverse-Laplacian component and
    the Poisson prefactor to \(\delta_{\mathrm{m}}\) to form \(\Phi/c^2\), followed by
    differential windows. Right: the resulting
    Newtonian potential \(\Phi/c^2\), with \(\delta_{\mathrm{m}}\) contours overlaid.
    The halo and dark-matter fields are Gaussian-smoothed with
    \(R=10\,h^{-1}\mathrm{Mpc}\) before these differential and potential
    operations are applied.
    The colour maps are oriented so that underdense regions appear blue
    and overdense regions are red; for \(\nabla\cdot\mathbf{v}\) and \(\Phi\), this
    orientation follows the signs expected from the linear continuity and
    Poisson relations.
    The similar large-scale patterns in the left and middle panels illustrate
    the linear continuity and gravitational-flow connections, while the
    potential panel highlights the long-range, low-pass character of the
    Newtonian-potential kernel.}
    \label{fig:weighted_velocity_gravity_overlay}
\end{figure*}

\begin{figure*}
    \centering
    \includegraphics[width=0.94\textwidth]{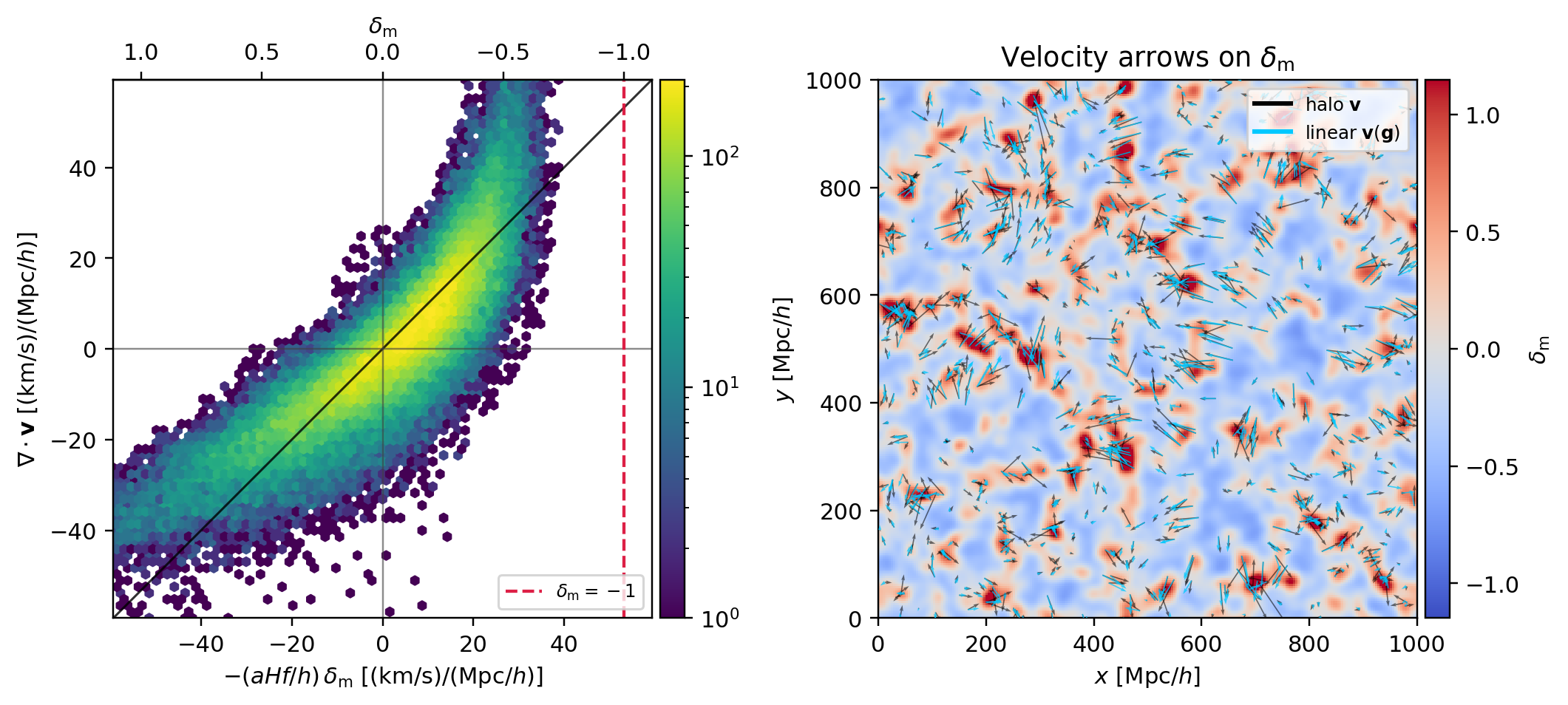}
    \caption{Linear-theory checks for the weighted-field and Newtonian-potential
    construction. Left: point-by-point comparison between the halo velocity
    divergence and the dark-matter prediction in
    Eq.~(\ref{eq:linear_theory}), with the diagonal showing
    equality and the top axis giving the corresponding \(\delta_{\mathrm{m}}\)
    value. The dashed vertical line marks \(\delta_{\mathrm{m}}=-1\). Right: the
    same slice as in Fig.~\ref{fig:weighted_velocity_gravity_overlay}, with
    halo velocities shown in black and the linear velocity inferred from
    Eq.~(\ref{eq:linear_theory}) shown in cyan on top of the
    \(\delta_{\mathrm{m}}\) background. The broad correlation and arrow alignment
    show that the differential windows and Newtonian-potential construction recover the
    expected large-scale continuity and gravity--velocity trends, while the
    common \(R=10\,h^{-1}\mathrm{Mpc}\) Gaussian smoothing keeps the comparison on
    large scales. The residual scatter reflects nonlinear motions, halo bias, and sampling
    noise.}
    \label{fig:weighted_velocity_linear_checks}
\end{figure*}

Figures~\ref{fig:weighted_velocity_gravity_overlay} and~\ref{fig:weighted_velocity_linear_checks} demonstrate this workflow using
\textsf{Quijote} fields. Both figures adopt the same Gaussian smoothing scale,
$R=10\,h^{-1}\mathrm{Mpc}$, for the halo and dark-matter fields. Figure ~\ref{fig:weighted_velocity_gravity_overlay} compares the halo velocity divergence, the smoothed dark-matter
density contrast, and the gravitational potential reconstructed through the
Poisson equation on the same slice, with halo velocity and acceleration
vectors overlaid.

In linear theory, the corresponding large-scale relations are
\begin{equation}\label{eq:linear_theory}
\delta_{\mathrm{m}} = -\frac{1}{a}\nabla\cdot\mathbf{v}_{\mathrm{lin}},\quad
\mathbf{v}_{\mathrm{lin}}
= \frac{2f(z)}{3\Omega_{\mathrm{m}}(z)H(z)}\mathbf{g}.
\end{equation}
Here \(a\) is the scale factor, \(H(z)\) the Hubble expansion rate,
\(f(z)=d\ln D/d\ln a\) the linear growth rate, and \(\Omega_{\mathrm{m}}(z)\)
the matter density parameter. The quantities \(\delta_{\mathrm{m}}\),
\(\mathbf{v}_{\mathrm{lin}}\), and \(\mathbf{g}\) denote the matter density
contrast, linear peculiar velocity, and peculiar gravitational acceleration.
Figure ~\ref{fig:weighted_velocity_linear_checks}  directly tests these two linear-theory predictions. As
expected, the agreement is primarily observed on large scales, since haloes are
biased tracers and the measured fields contain nonlinear velocity contributions,
finite smoothing effects, and sampling noise.

These examples illustrate the main practical advantage of the operator-window
framework: density, velocity, potential, and acceleration fields can all be
constructed by changing the weights and Fourier-space operators applied to the
same reconstructed density field.

It is worth noting that, within the \textsf{MRA} framework, spatial
differential operators admit an alternative real-space implementation in
addition to the Fourier-space representation introduced above. Specifically,
the SFCs of operator-derived fields, including gradient and higher-order
derivative fields, can be directly constructed from the SFCs of the original
density field through derivative connection coefficients associated with the
scaling-function basis \citep{Beylkin1991}. A brief description of this
real-space differentiation scheme is given in Appendix~B.

\subsection{Parallel execution and performance considerations}
\label{subsec:parallel_performance}

This section provides a compact summary of the computational cost for
representative production runs; detailed optimisation and tuning strategies are
described in the public documentation. \textsf{PyHermes} combines
process-level MPI parallelism with thread-level numerical kernels accelerated
by \textsf{Numba}. Depending on the specific task, independent sampling
points, random rotations, or multipole modes are distributed among MPI ranks,
while local numerical operations are executed using the configured number of
threads per rank. For the 3PCF multipole workflow, the final contractions and
grid summations over multipole fields can additionally be performed either on
the CPU or on a CUDA-capable GPU. These parallel strategies optimise the
execution of the same field--window operations without changing the underlying
estimator. In the timings reported below, $N_{\mathrm{MPI}}\times N_{\mathrm{th}}$
denotes the number of MPI ranks multiplied by the number of threads per rank.

The CPU benchmarks were performed on compute nodes equipped with two AMD EPYC
9754 processors, each providing 128 physical cores, resulting in a total of 256
physical CPU cores with one hardware thread per core. GPU benchmarks were
performed using a single NVIDIA GeForce RTX 4090 GPU running with CUDA 12.4.

The reported timings should be interpreted as representative reference values
rather than machine-independent performance predictions. Absolute wall times
depend on hardware characteristics, including processor and GPU architectures,
memory bandwidth, software environments, and process placement strategies.
Moreover, repeated runs on the same system may exhibit small variations due to
dynamic system conditions and background activity. Therefore, the more robust
metrics are the scaling trends with respect to task parameters and the relative
performance comparisons obtained under identical computational settings.

The core of the \textsf{PyHermes} algorithm is the efficient reconstruction of
physical fields on the MRA representation. The computational domain is
characterised by the scale-dilation level \(J\), which determines the number of
the MRA lattice, \(N_{\mathrm{grid}}=2^{3J},\)
in the three-dimensional representation. Increasing \(J\) by one therefore
increases the number of field coefficients by approximately a factor of eight.
In MPI runs, some field arrays and temporary convolution buffers are replicated
within individual ranks, so the total host-memory footprint depends not only on
the field resolution but also on the number of MPI processes. At high
resolution, using fewer MPI ranks with more threads per rank can reduce memory
duplication while maintaining efficient thread-level parallelism.

The second key ingredient is a library of window functions that encode the
spatial selection and filtering operations required by different statistical
estimators, including one-point, two-point, and higher-order statistics.
Window-filtered fields are constructed through Fourier-space convolution using
FFTs, with a computational cost of
\(\mathscr{O}\!\left(N_{\mathrm{w}} N_{\mathrm{FFT}}\log N_{\mathrm{FFT}}\right),\)
where \(N_{\mathrm{w}}\) denotes the number of applied window kernels and \(N_{\mathrm{FFT}}\)
is the number of FFT grid cells. In typical applications, the FFT grid is
chosen to match the MRA representation, such that
\(N_{\mathrm{FFT}}=N_{\mathrm{grid}}\). Therefore, once the continuous field has been
reconstructed, the cost of applying different statistical filters depends
primarily on the number of window operations rather than on the number of
objects in the original catalogue.

An important feature of this construction is that the computational cost of
field reconstruction and subsequent window operations is controlled by the
number of reconstructed field coefficients and applied kernels, rather than by
the number of particles or galaxies \(N_{\mathrm{g}}\). Therefore, once the catalogue has
been projected onto the MRA representation, the same reconstructed field can be
reused for multiple statistics without repeating particle-level counting.
This advantage becomes increasingly significant for large data sets, where
direct particle-based pair-, triplet-, and higher-order $N$-tuple counting
methods become prohibitively expensive due to their combinatorial scaling.

In the benchmark tests, we consider three workflows: the anisotropic 2PCF in redshift space, the conventional 3PCF based on Monte Carlo sampling, and the 3PCF multipole estimator. Except for the 3PCF multipole benchmarks, all tests are performed at resolution $J=8$.

\begin{figure*}
    \centering
    \includegraphics[width=0.96\textwidth]{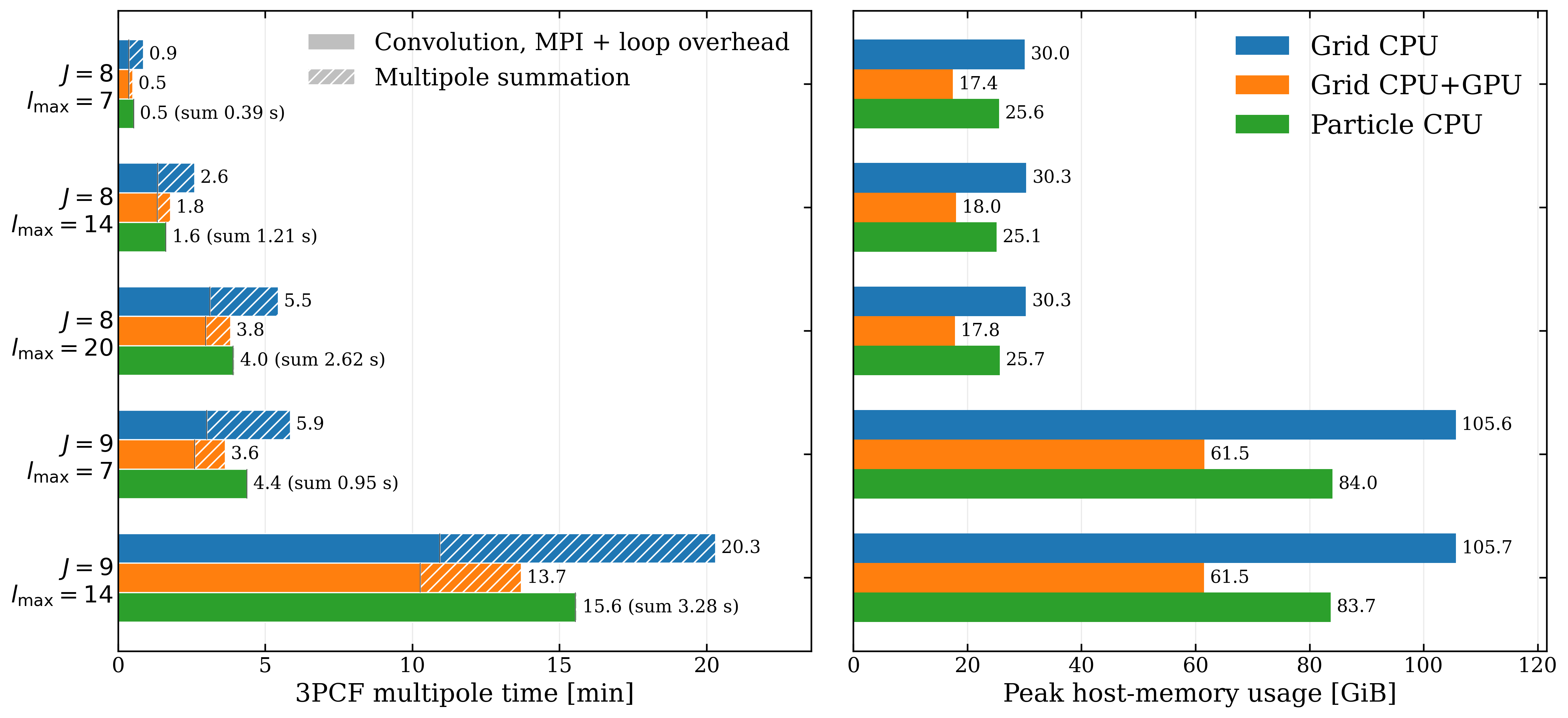}
    \caption{Runtime and host-memory usage of the grid-based CPU, hybrid CPU--GPU, and particle-based CPU implementations of the 3PCF multipole estimator for the \textsf{Quijote} halo example. All benchmarks use \((r_{12},r_{13})=(20,40)\,h^{-1}\mathrm{Mpc}\), with spherical top-hat smoothing of radius \(R=5\,h^{-1}\mathrm{Mpc}\) applied at the two secondary vertices and no additional smoothing at the primary vertex. Left: main-loop wall time, excluding data preparation, initialisation, and output. Hatched portions indicate the final-summation time, evaluated as the maximum cumulative time over all MPI ranks, while the solid portions show the remainder of the loop time, dominated by window convolutions and MPI communication. Labels give the total runtime in minutes; the particle-based summation times are additionally reported in seconds. Right: peak host-memory usage in GiB, measured using the \textsf{Slurm} batch-step \texttt{MaxRSS} and excluding GPU device memory. The \(J=8\) and \(J=9\) configurations use \(24\times4\) and \(12\times8\) MPI ranks~\(\times\)~threads, respectively. Each CPU--GPU run additionally uses one NVIDIA RTX~4090. Each reported configuration corresponds to a single benchmark run.}
    \label{fig:threepcf_multipole_cpu_gpu_performance}
\end{figure*}

\begin{itemize}

\item \textbf{Anisotropic 2PCF:}
For anisotropic two-point statistics, either
$\xi(s,\mu)$ parametrised by the pair separation $s$ and the cosine of the
line-of-sight angle $\mu$, or
$\xi(r_\perp,r_\parallel)$ parametrised by the transverse and line-of-sight
separations, the leading computational cost is determined by the number of
sampling points in the two-dimensional parameter space. Specifically,
$N_{\mathrm{samp}}=N_s\,N_\mu$ for $\xi(s,\mu)$ and
$N_{\mathrm{samp}}=N_{r_\perp}\,N_{r_\parallel}$ for
$\xi(r_\perp,r_\parallel)$. For a fixed field resolution $J$, the computational
complexity scales as
\(
\mathscr{O}(N_{\mathrm{samp}}N_{\mathrm{grid}}\log N_{\mathrm{grid}}),
\)
while the memory footprint is dominated by the stored SFC array, FFT work arrays, and temporary convolution buffers.

\item \textbf{Conventional 3PCF workflow:}
The dominant runtime is controlled by the number of translation samples,
$N_{\mathrm{trans}}$, and the number of random rotations, $N_{\mathrm{rot}}$, which
estimate the translational and rotational averages, respectively. The overall
computational cost therefore scales as
\(\mathscr{O}(N_{\mathrm{trans}}N_{\mathrm{rot}})\).
In the dual-sphere scheme (\texttt{center="particle"}),
$N_{\mathrm{trans}}$ is equal to the number of haloes in the catalogue. In the
triplet-sphere scheme (\texttt{center="box\_random"}), the primary vertices are
sampled uniformly throughout the survey volume, and $N_{\mathrm{trans}}$ is chosen
according to the required Monte Carlo accuracy. The reconstructed fields and
temporary convolution buffers provide the baseline memory footprint. Large
centre samples and their rank-local work arrays can nevertheless make an
additional contribution, so the measured peak memory also depends on the
centre strategy, sample size, and batching configuration.

\item \textbf{3PCF multipole workflow:}
The multipole estimator has a different computational cost driver from the
conventional Monte Carlo workflow. After the angular decomposition, the 3PCF
multipoles, $\zeta_\ell(r_1,r_2)$, depend only on the two radial separations
$\{r_1,r_2\}$. For each multipole order $\ell$, only the non-negative-$m$
fields are explicitly constructed, while the negative-$m$ contributions are
recovered from the conjugation symmetry of spherical harmonics. For a maximum
multipole order $\ell_{\max}$, the total number of independent multipole
windows is
\( N_{\ell_{\max}}={(\ell_{\max}+1)(\ell_{\max}+2)}/2,\)
which leads to the computational cost scaling $\mathscr{O}(N_{\ell_{\max}}N_{\mathrm{grid}}\log N_{\mathrm{grid}})$.
In practice, these multipole window functions are generated on demand rather than stored
simultaneously in memory. Consequently, increasing $\ell_{\max}$ primarily
increases the runtime through the number of multipole-window evaluations,
whereas the peak memory usage at fixed resolution $J$ remains nearly constant.
On the other hand, increasing $J$ enlarges the number of grid coefficients by a
factor of eight, leading to the dominant memory growth. As discussed above,
this increase can be partially mitigated in distributed runs by reducing the
number of MPI ranks and increasing the number of threads per rank, as
illustrated by the $J=9$ benchmarks in
Table~\ref{tab:pyhermes_performance_summary} and
Fig.~\ref{fig:threepcf_multipole_cpu_gpu_performance}.

\end{itemize}

Table~\ref{tab:pyhermes_performance_summary} summarises a compact set of
representative benchmark points for the main estimator categories. In these
tests, the random contribution is evaluated analytically assuming a constant
random density, and therefore no explicit random catalogue or random field is
constructed. Benchmarks using an explicit random catalogue generally require
additional memory and computational time. The associated overhead depends on the
specific estimator and window geometry, and we leave these implementation-dependent
details to the public documentation. The results in the table should
therefore be interpreted as a validation of the scaling behaviours discussed
above, rather than as an exhaustive performance survey over all possible task
configurations.

\begin{table*}
    \centering
    \caption{Runtime and host-memory usage of the representative benchmark jobs. Memory is reported as the peak resident set size in GiB, measured using the \textsf{Slurm} batch-step \texttt{MaxRSS}. For (DD) and (DDD), the reported main-loop runtimes are the minima of three independent fresh-process measurements and exclude input preparation, initial FFTs, and final post-processing. The corresponding average times are obtained by dividing the total loop time by \(n_{\mathrm{s}}n_\mu\) and \(N_{\mathrm{cen}}N_{\mathrm{rot}}N_\theta\), respectively. The (DD) row benchmarks the anisotropic 2PCF sampling loop using an analytic normalisation for a uniform random catalogue; its memory usage is measured in a separate run with the same configuration. The (DDD) rows benchmark the pure triplet calculation and exclude the additional pair measurements required to construct (Q). In the average-time column, \texttt{/sample} denotes the time per one sampled \((s,\mu)\) bin, \texttt{/comb.} one centre\,\&\,rotation--orientation combination, and \texttt{/(m)} one explicitly evaluated non-negative-(m) multipole mode. For a given \(\ell_{\max}\), the number of such modes is
\(N_{\mathrm{m}}=\sum_{\ell=0}^{\ell_{\max}}(\ell+1)
=(\ell_{\max}+1)(\ell_{\max}+2)/2\).
All 3PCF benchmarks use \(r_{12}=20\,h^{-1}\mathrm{Mpc}\) and
\(r_{13}=40\,h^{-1}\mathrm{Mpc}\). The \texttt{Parallel} column lists the CPU configuration as MPI ranks~\(\times\)~threads. The five 3PCF-multipole rows correspond to GPU-backend runs, which additionally use a CUDA thread-block configuration of \((8,8,8)\), or 512 GPU threads per block, for the final summation. The corresponding CPU-backend results are compared in Fig.~\ref{fig:threepcf_multipole_cpu_gpu_performance}.
}
    \label{tab:pyhermes_performance_summary}
    \footnotesize
    \renewcommand{\arraystretch}{1.12}
    \setlength{\tabcolsep}{3pt}
    \begin{tabularx}{\textwidth}{@{}>{\centering\arraybackslash}p{0.14\textwidth}>{\raggedright\arraybackslash}Xcc>{\centering\arraybackslash}p{0.11\textwidth}>{\centering\arraybackslash}p{0.12\textwidth}>{\centering\arraybackslash}p{0.085\textwidth}@{}}
        \toprule
        Product & Representative setup & $J$ & Parallel & Main-loop time & Average time & Memory \\
        \midrule
        $DD(s,\mu)$ & Ring binning windows, $n_{\mathrm{s}}=46$, $n_\mu=51$ & 8 & $8\times4$ & $6.87\,\mathrm{s}$ & $2.93\,\mathrm{ms}$/sample & $4.37\,\mathrm{GiB}$ \\
        \midrule
        \multirow{2}{=}{\centering \shortstack{$DDD$\\$(\theta;r_{12},r_{13})$}} & \begin{tabular}[c]{@{}l@{}}Particle centres\\$N_{\mathrm{cen}}=406{,}728$, $N_{\mathrm{rot}}=1000$, $N_\theta=20$\end{tabular} & 8 & $16\times8$ & $78.1\,\mathrm{s}$ & $9.60\,\mathrm{ns}$/comb. & $7.75\,\mathrm{GiB}$ \\
        & \begin{tabular}[c]{@{}l@{}}Box-random centres\\$N_{\mathrm{cen}}=8.0\times10^6$, $N_{\mathrm{rot}}=200$, $N_\theta=20$\end{tabular} & 8 & $16\times8$ & $294\,\mathrm{s}$ & $9.19\,\mathrm{ns}$/comb. & $9.91\,\mathrm{GiB}$ \\
        \midrule
        \multirow{5}{=}{\centering \shortstack{$DDD_\ell$\\$(r_{12},r_{13})$}} & $\ell_{\max}=7$ ($N_{\mathrm{m}}=36$) & 8 & $24\times4$ & $30\,\mathrm{s}$ & $0.84\,\mathrm{s}$/$m$ & $17.4\,\mathrm{GiB}$ \\
        & $\ell_{\max}=14$ ($N_{\mathrm{m}}=120$) & 8 & $24\times4$ & $107\,\mathrm{s}$ & $0.89\,\mathrm{s}$/$m$ & $18.0\,\mathrm{GiB}$ \\
        & $\ell_{\max}=20$ ($N_{\mathrm{m}}=231$) & 8 & $24\times4$ & $230\,\mathrm{s}$ & $0.99\,\mathrm{s}$/$m$ & $17.8\,\mathrm{GiB}$ \\
        & $\ell_{\max}=7$ ($N_{\mathrm{m}}=36$) & 9 & $12\times8$ & $219\,\mathrm{s}$ & $6.07\,\mathrm{s}$/$m$ & $61.5\,\mathrm{GiB}$ \\
        & $\ell_{\max}=14$ ($N_{\mathrm{m}}=120$) & 9 & $12\times8$ & $822\,\mathrm{s}$ & $6.85\,\mathrm{s}$/$m$ & $61.5\,\mathrm{GiB}$ \\
        \bottomrule
    \end{tabularx}
\end{table*}

The comparison in Fig.~\ref{fig:threepcf_multipole_cpu_gpu_performance}
demonstrates that the 3PCF multipole estimator can be efficiently executed
with either the CPU or GPU backend. Across the five benchmark configurations,
offloading the final multipole-field summation from the CPU to the GPU reduces
the cost of this stage by a factor of approximately $4.1$--$4.7$. Since the
preceding window convolutions and MPI communication remain CPU-side operations,
the acceleration of the complete multipole workflow is more moderate, with an
overall speedup of approximately $1.8$--$1.9$ in these runs.

In the CPU implementation with particle-based summation, the final contraction is evaluated only at the catalogue positions, rather than over the entire MRA lattice. For the sparse halo sample considered here, this summation requires at most a few seconds. Window convolutions and MPI communication consequently dominate the end-to-end runtime, so reducing the summation cost yields only a modest overall speed-up. The particle-based CPU implementation therefore does not consistently outperform the grid-based GPU implementation across all configurations.

This separation of convolution and contraction stages also allows the same estimator to operate efficiently across different hardware environments. Therefore, the CPU backend provides a portable implementation without requiring
specialised hardware, while GPU acceleration becomes increasingly advantageous when the final grid contraction represents a significant fraction of the total computational cost.

The memory comparison reveals two distinct scaling behaviours. At fixed $J$,
the peak host-memory usage is governed primarily by the field resolution rather
than by $\ell_{\max}$, because the multipole windows are generated and processed
sequentially rather than stored simultaneously. At $J=8$, the CPU measurements
span $19.1$--$19.4\,\mathrm{GiB}$ and the GPU measurements span
$17.4$--$18.0\,\mathrm{GiB}$. At $J=9$, the CPU measurements span
$65.9$--$66.9\,\mathrm{GiB}$, while both GPU measurements are approximately
$61.5\,\mathrm{GiB}$. The GPU runs therefore use about $5.6$--$9.1\%$ less host
memory in these measurements. This comparison does not include GPU device
memory, however, and should not be interpreted as a measurement of the total
CPU-plus-GPU memory footprint.

In contrast, increasing the field resolution from $J=8$ to $J=9$ increases the number of MRA coefficients per field by a factor of eight, making each reconstructed field and associated temporary grid buffer substantially larger. For this benchmark, the parallel configuration changes from $24\times4$ to $12\times8$: reducing the number of MPI ranks decreases the duplication of rank-local field arrays, while increasing the number of threads per rank maintains the total number of CPU threads at 96. This rank--thread redistribution limits the measured host-memory increase to approximately a factor of $3.4$--$3.5$ while retaining most of the available computational parallelism, which becomes increasingly important for high-$J$ calculations.

GPU acceleration of the FFT can be implemented using standard GPU FFT libraries, but its net benefit depends on data movement within the complete workflow. Offloading only the convolution stage may require transfers of large three-dimensional MRA coefficient arrays and window kernels between host and device memory. The potential savings therefore need to be assessed together with these transfer costs.

A fully GPU-resident workflow would avoid repeated host--device transfers, but would require the active fields, window arrays, and FFT workspaces to fit in GPU memory. Otherwise, the current CPU--MPI implementation can distribute independent convolution tasks across ranks, while each rank retains the full arrays needed for its local FFTs. The present hybrid design performs the MPI-parallel window convolutions on
the CPU and offloads the final MRA-grid summation to the GPU. Keeping the complete field--window pipeline resident on the GPU is a possible future optimisation when sufficient device memory is available.

\subsection{2PCF and 3PCF comparisons with established codes}
\label{sec:external-code-comparison}

We further validate the accuracy and computational performance of
\textsf{PyHermes} through direct comparisons with established
correlation-function codes. For the isotropic 2PCF, we compare with
\textsf{pycorr}/\textsf{Corrfunc} \citep{Sinha2020} and
\textsf{TreeCorr} \citep{Jarvis2004}; for the 3PCF multipoles, we compare
with \textsf{HITOMI}\footnote{\url{https://github.com/naonori/hitomi}} \citep{Sugiyama2023}. The 2PCF benchmark primarily demonstrates the
different scaling with catalogue size, whereas the 3PCF benchmark
provides a joint comparison of the measured multipoles, end-to-end
runtime, and peak host-memory consumption.

\begin{figure*}
    \centering
    \includegraphics[width=0.95\textwidth]{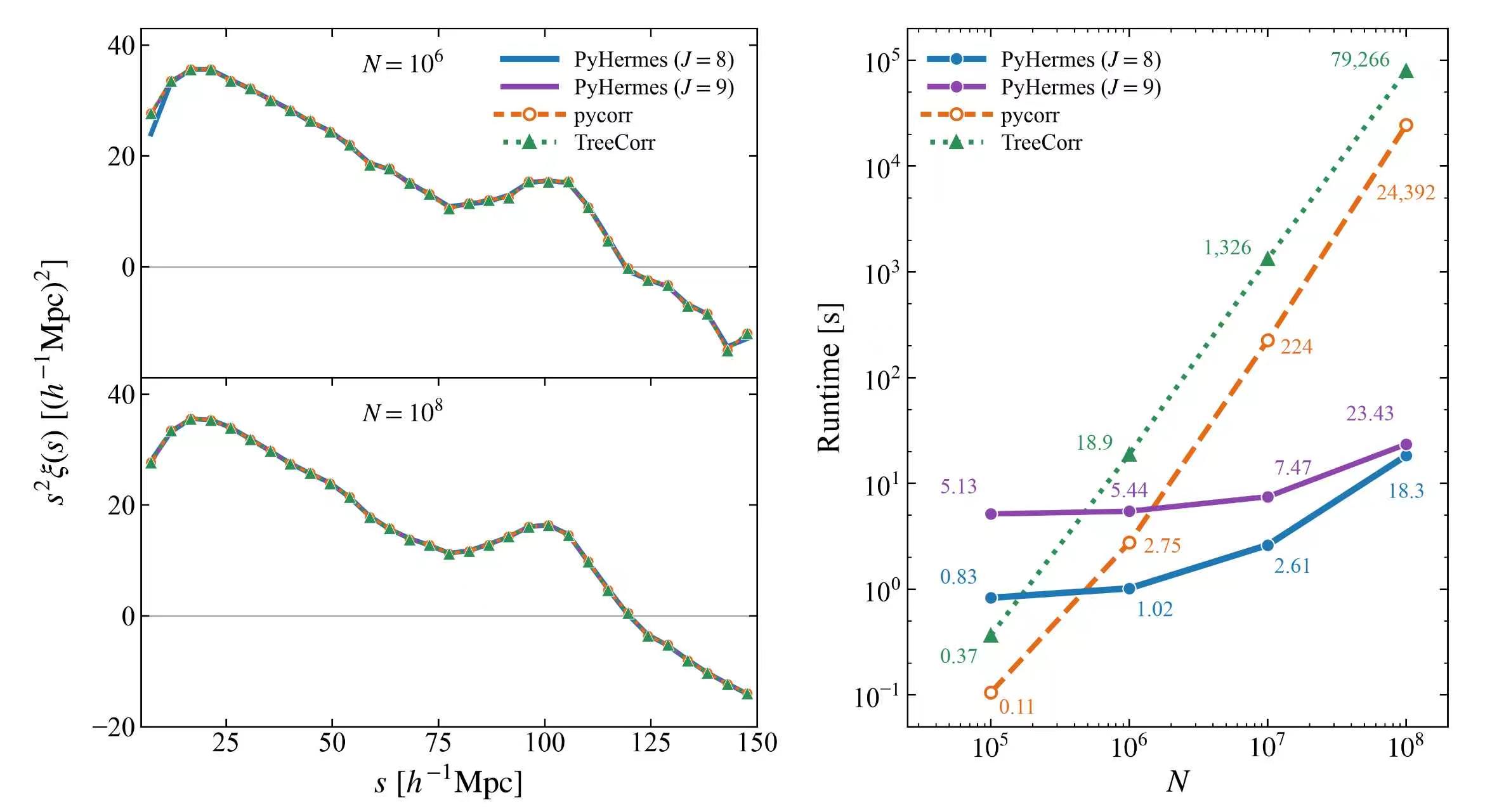}
    \caption{Comparison of isotropic 2PCF measurements and runtimes using dark-matter samples drawn from the fiducial \textsf{Quijote} simulation. Left:
\(s^2\xi(s)\) measured from catalogues containing \(N=10^6\) particles
(top) and \(N=10^8\) particles (bottom). The \textsf{PyHermes} results
at \(J=8\) and \(J=9\) are compared with
\textsf{pycorr}/\textsf{Corrfunc} and \textsf{TreeCorr}, using the same
radial bins. Right: wall-clock runtime as a function of particle number;
the annotated values are in seconds. All timing measurements use one MPI
process with 16 threads.}
    \label{fig_2pcf_code_comparison}
\end{figure*}

Figure~\ref{fig_2pcf_code_comparison} compares both the accuracy and
wall-clock performance of the isotropic 2PCF measurements. The left
panels show that the \textsf{PyHermes} results at \(J=8\) and \(J=9\)
agree closely with those from \textsf{pycorr}/\textsf{Corrfunc} and
\textsf{TreeCorr} for both \(N=10^6\) and \(N=10^8\). The right panel
shows the runtime as a function of particle number, with all calculations
using one MPI process and 16 threads. At \(N=10^5\), the fixed MRA and
FFT costs make \textsf{PyHermes} slower than the pair-counting codes:
the runtimes are \(0.83\) and \(5.13\) seconds for \(J=8\) and \(J=9\),
respectively, compared with \(0.105\) seconds for
\textsf{pycorr}/\textsf{Corrfunc} and \(0.37\) seconds for
\textsf{TreeCorr}. The scaling becomes increasingly favourable as \(N\)
grows. At \(N=10^7\), the runtimes are \(2.61\) and \(7.47\) seconds
for \textsf{PyHermes} at \(J=8\) and \(J=9\), compared with \(224\)
and \(1326\) seconds for \textsf{pycorr}/\textsf{Corrfunc} and
\textsf{TreeCorr}. At \(N=10^8\), the corresponding values are
\(18.3\), \(23.43\), \(24\,392\), and \(79\,266\) seconds. Thus, at
\(J=8\), \textsf{PyHermes} is approximately \(1.3\times10^3\) times
faster than \textsf{pycorr}/\textsf{Corrfunc} and
\(4.3\times10^3\) times faster than \textsf{TreeCorr} for the largest
catalogue.

The different scaling follows from the field-based formulation of
\textsf{PyHermes}. Once the scaling-function coefficients of the density
field have been constructed, the core 2PCF calculation depends primarily
on the number of MRA lattice elements,
\(N_{\rm grid}=2^{3J}\), rather than directly on the number of particles.
Only one forward FFT of the original coefficient array is required,
after which the configuration-space pair sum is replaced by
Fourier-space contractions with the separation-window kernels, with the
equivalence guaranteed by Parseval's theorem. At fixed \(J\), the total
cost is approximately $O(N)+O\!\left(N_{\rm grid}\log N_{\rm grid}\right)$,
where the first term describes the projection of the particle catalogue
onto the MRA basis. The larger fixed cost at \(J=9\) reflects the increase
in the number of lattice elements, but its dependence on \(N\) remains
weak. By contrast, the cost of conventional pair counting is governed
by the number of candidate particle pairs. Although cell and tree
searches reduce the prefactor, the
\textsf{pycorr}/\textsf{Corrfunc} and \textsf{TreeCorr} runtimes approach
quadratic growth at the large-\(N\) end of the benchmark. The fixed MRA
overhead is therefore disadvantageous for small catalogues, whereas the
field-based formulation becomes increasingly efficient for the large
particle samples targeted by HERMES.

\begin{figure*}
    \centering
    \includegraphics[width=0.96\textwidth]{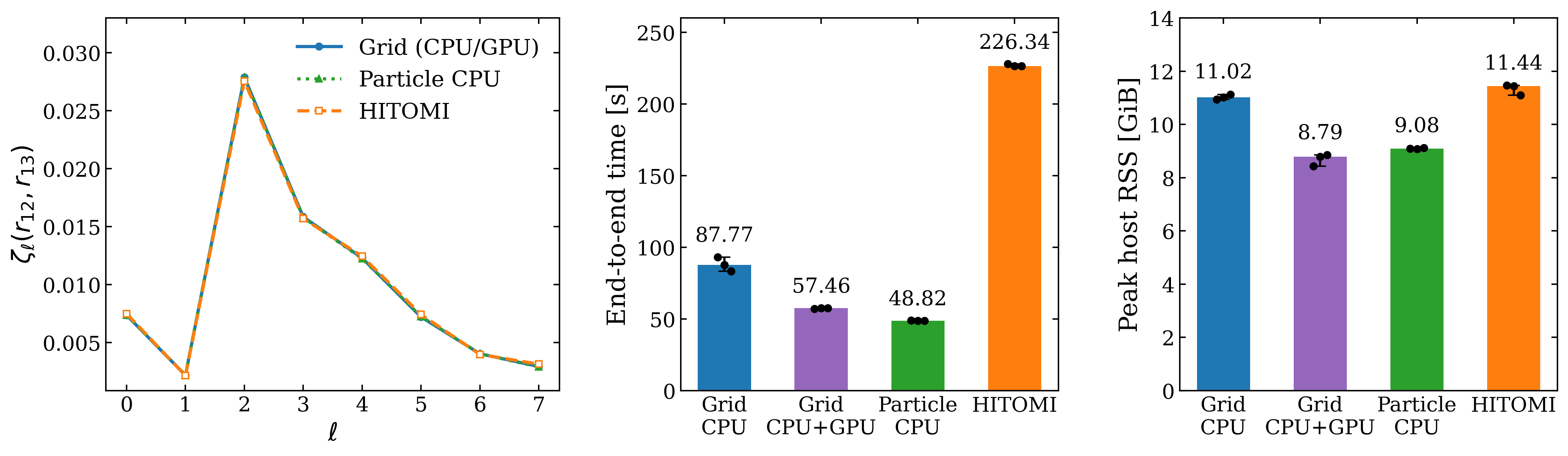}
    \caption{Comparison of \textsf{PyHermes} and \textsf{HITOMI} for the same
    \textsf{Quijote} halo example and window configuration as
    Fig.~\ref{fig:threepcf_multipole_cpu_gpu_performance}, with $J=8$ and
    $\ell_{\max}=7$. Left: measured multipole coefficients; the PyHermes
    grid CPU and GPU results coincide numerically and are shown as one
    curve. Centre: end-to-end wall time, including input preparation,
    start-up, calculation, and output. Right: peak host RSS in GiB,
    sampled from the process tree at $0.5$-s intervals, excluding GPU
    device memory.
    Bars show the medians of three runs, black points show all samples,
    and error bars span the minimum and maximum. All implementations use
    $64$ physical CPU cores: PyHermes uses $8$ MPI ranks with $8$ threads
    each, while HITOMI uses $8$ concurrent multipole processes with $8$
    FFTW threads each. GPU runs additionally use one RTX~4090.}
    \label{fig:pyhermes_hitomi_3pcf_comparison}
\end{figure*}

Figure~\ref{fig:pyhermes_hitomi_3pcf_comparison} presents the corresponding
3PCF comparison for the \textsf{Quijote} halo example, using the same
window configuration as Figure~\ref{fig:threepcf_multipole_cpu_gpu_performance},
with \(J=8\) and \(\ell_{\max}=7\). Here, the timings are medians of three
end-to-end runs on 64 CPU cores, whereas
Fig.~\ref{fig:threepcf_multipole_cpu_gpu_performance} reports single-run
main-loop timings on 96 CPU cores. The multipole coefficients measured
with the \textsf{PyHermes} grid-based CPU and GPU implementations,
the particle-based CPU implementation, and \textsf{HITOMI} agree
closely over the full multipole range. The CPU and GPU grid results
coincide numerically and are therefore shown as a single curve. This
agreement provides an external validation of both the grid-based and
particle-based \textsf{PyHermes} 3PCF estimators and confirms that the
different computational backends preserve the same statistic and
normalization.

The median end-to-end runtimes are \(87.77\) seconds for the
\textsf{PyHermes} grid CPU calculation, \(57.46\) seconds for the grid
CPU+GPU calculation, \(48.82\) seconds for the particle-centred CPU
calculation, and \(226.34\) seconds for \textsf{HITOMI}. Relative to
\textsf{HITOMI}, these correspond to speed-ups of approximately
\(2.6\), \(3.9\), and \(4.6\), respectively. The GPU accelerates the
grid-based calculation by a factor of approximately \(1.5\), even
though the reported end-to-end time also includes input preparation,
initialization, and output, which are not accelerated by the GPU. For
this particular catalogue and window configuration, the
particle-based CPU implementation gives the shortest total runtime.

The peak host-memory consumptions are \(11.02\), \(8.79\), \(9.08\),
and \(11.44\) GiB for the grid CPU, grid CPU+GPU, particle-based CPU,
and \textsf{HITOMI} calculations, respectively. The memory requirements
are therefore comparable, with all implementations remaining close to
\(9\)--\(12\) GiB. The GPU value refers only to host resident-set size
and excludes device memory. Taken together, the two benchmarks show
that \textsf{PyHermes} reproduces established 2PCF and 3PCF
measurements while providing increasingly favourable computational
scaling for large catalogues.

\section{Summary and Concluding Remarks}
\label{sec:Conclusions}

In this work, we have presented \textsf{Hermes}, a unified framework for cosmic
statistics based on multiresolution density reconstruction, window
convolution, and products of window-filtered fields. The central idea is to
replace the conventional view of each statistic as an independent catalogue
counting algorithm with a common field-based representation. A discrete
catalogue is first reconstructed as a continuous density field in a compact
scaling-function basis. Statistical measurements are then obtained by applying
different window operators and forming products of the resulting filtered
fields.

This formulation provides a general language that connects a wide range of
statistics. Conventional counts-in-cells measurements, isotropic and anisotropic
two-point correlation functions, higher-order correlation functions, and
multipole decompositions correspond to particular choices of window functions.
In this view, a spherical shell, a redshift-space ring, a disk or cylindrical
window, a smoothing filter, a high-pass kernel, or a spherical-harmonic
multipole projector are not separate estimators, but different operators acting
on the same reconstructed field. This perspective naturally extends the
traditional binning concept and allows customised statistics to be introduced
by modifying the window kernel rather than redesigning the counting algorithm.

A key aspect of the framework is the separation between the reconstructed field
and the statistical estimator. The field reconstruction is performed once,
while subsequent measurements reuse the same multiresolution representation.
This separation also clarifies the distinction between count-level quantities
constructed from data and random fields and the final connected statistics
obtained after random normalisation or subtraction. As a result, standard
estimators such as the \textsf{Landy--Szalay} 2PCF and higher-order correlation
estimators can be naturally expressed through the same products of filtered fields.

We have introduced \textsf{PyHermes}, an open-source \textsf{Python} implementation of this
framework. The current implementation combines multiresolution field
reconstruction, FFT-based window convolution, MPI and thread-level
parallelisation, and CUDA acceleration for computationally intensive
high-order contractions. The software supports a broad range of measurements,
including one-point statistics, isotropic and redshift-space two-point
correlation functions, marked and weighted statistics, standard and multipole
three-point correlation functions, and Fourier-space operator windows. The
latter provide a unified way to construct derived physical fields, such as
fields describing velocity divergence, gravitational potential, and acceleration by
composing differential and inverse-Laplacian operators with the same
reconstructed density fields.

Using halo catalogues from the \textsf{Quijote} simulations, we have demonstrated the
practical capability of the framework. The reconstructed density fields show
the expected convergence with increasing multiresolution level. The isotropic
2PCF measurements agree with conventional pair-counting codes once the field
resolution is sufficient to resolve the adopted bin scale. Redshift-space
examples demonstrate that anisotropic statistics can be constructed using
different geometric windows, while marked statistics and multipole 3PCFs
illustrate the flexibility of applying non-standard filters and products of
multiple filtered fields. We have also shown that physical diagnostics involving density,
velocity, potential, and acceleration fields can be generated without
introducing separate reconstruction pipelines.

The computational advantage of \textsf{Hermes} lies not simply in replacing specialised
pair-counting algorithms for a single conventional statistic. For standard
small catalogues, dedicated pair counters may remain faster because they are
optimised for a specific measurement. Instead, \textsf{Hermes} is designed for the
regime where many statistics, complex window geometries, or high-order
correlations are required from large data sets. After the initial field
reconstruction, the computational cost is primarily controlled by the field
resolution, the number of applied windows, and the complexity of the required
products of filtered fields, rather than directly by the number of particle pairs or
triplets. This makes the framework particularly suitable for future surveys
and simulations containing millions to billions of tracers.

The current implementation provides a foundation for further extensions. Future developments will include more general survey geometries, additional higher-order statistics, optimised weighting schemes, and improved hardware acceleration. By providing a common field--window language for cosmic statistics, \textsf{Hermes} offers a flexible and scalable approach for extracting information from increasingly large and complex large-scale-structure data sets.

Particularly, a principal direction of our future work is to extend the field–window formulation to 3PCF and, more generally, NPCF in redshift space, initially retaining the fixed-line-of-sight approximation. In this extension, the full spherical rotational symmetry of the isotropic estimator is reduced to the residual axial symmetry about the line of sight. The existing MRA convolution machinery can still be used to construct the required spherical-harmonic filtered fields, whereas the final field contractions must be generalized to the symmetry-allowed angular-mode combinations. For the 3PCF, these modes may be labelled by \((\ell_1,\ell_2,m)\) and subsequently recoupled into a covariant TripoSH representation; the same construction extends to general NPCFs through the corresponding polypolar basis. A local-line-of-sight treatment for realistic wide-angle galaxy surveys and light-cone simulations constitutes a further extension beyond this fixed-axis formulation. Because the local line of sight varies with the primary position, the angular kernels become position dependent and the global translation invariance used by the current FFT-based convolution is broken. Local-line-of-sight multipoles are therefore conceptually compatible with the field–window framework, but cannot be obtained by simply replacing the fixed line-of-sight direction in the present kernels. Possible approaches include an end-point or Yamamoto-type approximation, position-dependent rotations of the harmonic coefficients, or a fully position-dependent TripoSH/polypolar treatment. The implementation and validation of the redshift-space 3PCF and NPCFs, together with their eventual local-line-of-sight generalization, will be presented in future work.

Beyond the methodological and computational developments presented here, \textsf{PyHermes} is designed to enable precision cosmological analyses with the next generation of large-scale-structure surveys. Future spectroscopic and photometric surveys, including DESI, Euclid, the Nancy Grace Roman Space Telescope, and the Vera C. Rubin Observatory, will provide galaxy catalogues containing unprecedented numbers of tracers over increasingly large cosmic volumes. Extracting the full cosmological information from these surveys will require not only accurate theoretical modelling, but also efficient and flexible statistical estimators capable of combining multiple probes beyond
the conventional power spectrum and 2PCF. The unified window-function
formulation of \textsf{PyHermes} provides a natural framework for such analyses, allowing higher-order correlations, marked statistics, density-split statistics, and operator-based observables to be measured within the same field representation. These capabilities open the possibility of exploiting non-Gaussian information from nonlinear structure formation to improve constraints on dark energy, neutrino masses, primordial non-Gaussianity, galaxy bias, and possible deviations from general relativity. In this sense, \textsf{PyHermes} serves not only as a high-performance computational tool, but also as a flexible statistical infrastructure for future precision cosmology.

\section*{Acknowledgements}

This work is supported by the Major Program of the National Natural Science Foundation of China (NSFC) through grant 12595314. FLL is supported by the NSFC Key Program through grant 12533002. ZWS is supported by the National SKA Program of China (grant Nos. 2025SKA0150100 and 2025SKA0150104), the NSFC through grant 12173102, and the China Manned Space Program through grant CMS-CSST-2025-A10. YC acknowledges the support of the UK Royal Society through a University Research Fellowship.

We thank Zhiqi Huang, Ming Li, Xiaodong Li, Weipeng Lin, Jun Pan, Yu Yu for fruitful discussions and constructive comments. 

We thank the developers of the public scientific \textsf{Python} ecosystem on which
\textsf{PyHermes} is built. The numerical demonstrations use data products from the
\textsf{Quijote} simulation suite. For the purpose of open access, the author has applied a Creative Commons Attribution (CC BY) licence to any Author Accepted Manuscript version arising from this submission.

\section*{Data Availability}

The software release associated with this paper is \textsf{PyHermes}
v1.2.0\footnote{\url{https://github.com/SYSUSPA-Projects/PyHermes/releases/tag/v1.2.0}}.
The source code, documentation, configuration examples, benchmark scripts,
and notebooks used to generate the demonstration figures are publicly
available through the \textsf{PyHermes} project website at
\url{https://pyhermes.astroslacker.com}. The numerical examples presented in this paper use particle and halo
catalogues from the \textsf{Quijote} simulation suite
(\url{https://quijote-simulations.readthedocs.io/en/latest/}) and catalogues
from the \textsf{Kun} simulation suite
(\url{https://jiutian.sjtu.edu.cn/}). The derived data products required to
reproduce the figures can be generated using the scripts and notebooks
distributed with \textsf{PyHermes}. Additional data supporting the results
and analyses presented in this paper will be made available by the authors
upon reasonable request.

\bibliographystyle{mnras}
\bibliography{hermes}

@ARTICLE{2025SCPMA.6880402G,
       author = {{Gong}, Yan and {Miao}, Haitao and {Zhou}, Xingchen and {Xiong}, Qi and {Song}, Yingxiao and {Jiang}, Yuer and {Wang}, Minglin and {Yan}, Junhui and {Wu}, Beichen and {Deng}, Furen and {Chen}, Xuelei and {Fan}, Zuhui and {Jing}, Yipeng and {Yang}, Xiaohu and {Zhan}, Hu},
        title = "{Future cosmology: New physics and opportunity from the China Space Station Telescope (CSST)}",
      journal = {Science China Physics, Mechanics \& Astronomy},
         year = 2025,
        month = aug,
       volume = {68},
       number = {8},
          eid = {280402},
        pages = {280402},
          doi = {10.1007/s11433-025-2646-2},
archivePrefix = {arXiv},
       eprint = {2501.15023},
 primaryClass = {astro-ph.CO},
       adsurl = {https://ui.adsabs.harvard.edu/abs/2025SCPMA..6880402G}
}

@ARTICLE{2026SCPMA.6939501C,
       author = {{CSST Collaboration} and {Gong}, Yan and {Miao}, Haitao and {Zhan}, Hu and {Li}, Zhao-Yu and {Shangguan}, Jinyi and {Li}, Haining and {Liu}, Chao and {Chen}, Xuefei and {Yuan}, Haibo and {Zhou}, Jilin and {Liu}, Hui-Gen and {Yu}, Cong and {Ji}, Jianghui and {Qi}, Zhaoxiang and {Liu}, Jiacheng and {Dai}, Zigao and {Wang}, Xiaofeng and {Zheng}, Zhenya and {Hao}, Lei and {Dou}, Jiangpei and {Ao}, Yiping and {Lin}, Zhenhui and {Zhang}, Kun and {Wang}, Wei and {Sun}, Guotong and {Li}, Ran and {Li}, Guoliang and {Xu}, Youhua and {Li}, Xinfeng and {Li}, Shengyang and {Wu}, Peng and {Zhang}, Jiuxing and {Wang}, Bo and {Bai}, Jinming and {Cai}, Yi-Fu and {Cai}, Zheng and {Cao}, Jie and {Chan}, Kwan Chuen and {Chang}, Jin and {Chen}, Xiaodian and {Chen}, Xuelei and {Chen}, Yuqin and {Chen}, Yun and {Cui}, Wei and {Dong}, Subo and {Du}, Pu and {Duan}, Wenying and {Fan}, Junhui and {Fan}, LuLu and {Fan}, Zhou and {Fan}, Zuhui and {Fang}, Taotao and {Fu}, Jianning and {Fu}, Liping and {Fu}, Zhensen and {Gao}, Jian and {Gu}, Shenghong and {Gu}, Yidong and {Guo}, Qi and {Han}, Zhanwen and {Hu}, Bin and {Huang}, Zhiqi and {Ho}, Luis C. and {Jiang}, Linhua and {Jiang}, Ning and {Jing}, Yipeng and {Kang}, Xi and {Kong}, Xu and {Li}, Cheng and {Li}, Chengyuan and {Li}, Di and {Li}, Jing and {Li}, Nan and {Li}, Yang A. and {Liao}, Shilong and {Lin}, Weipeng and {Liu}, Fengshan and {Liu}, Jifeng and {Liu}, Xiangkun and {Liu}, Zhuokai and {Mao}, Ruiqing and {Mao}, Shude and {Meng}, Xianmin and {Pang}, Xiaoying and {Peng}, Xiyan and {Peng}, Yingjie and {Shan}, Huanyuan and {Shen}, Juntai and {Shen}, Shiyin and {Shen}, Zhiqiang and {Shi}, Sheng-Cai and {Shi}, Yong and {Tan}, Siyuan and {Tian}, Hao and {Wang}, Jianmin and {Wang}, Jun-Xian and {Wang}, Xin and {Wang}, Yuting and {Wu}, Hong and {Wu}, Jingwen and {Wu}, Xuebing and {Xu}, Chun and {Xue}, Xiang-Xiang and {Xue}, Yongquan and {Yang}, Ji and {Yang}, Xiaohu and {Yao}, Qijun and {Yuan}, Fangting and {Yuan}, Zhen and {Zhang}, Jun and {Zhang}, Pengjie and {Zhang}, Tianmeng and {Zhang}, Wei and {Zhang}, Xin and {Zhao}, Gang and {Zhao}, Gongbo and {Zhong}, Hongen and {Zhong}, Jing and {Zhou}, Liyong and {Zhu}, Wei and {Zu}, Ying},
        title = "{Introduction to the Chinese Space Station Survey Telescope (CSST)}",
      journal = {Science China Physics, Mechanics \& Astronomy},
         year = 2026,
        month = jan,
       volume = {69},
       number = {3},
          eid = {239501},
        pages = {239501},
          doi = {10.1007/s11433-025-2809-0},
archivePrefix = {arXiv},
       eprint = {2507.04618},
 primaryClass = {astro-ph.IM},
       adsurl = {https://ui.adsabs.harvard.edu/abs/2026SCPMA..6939501C}
}

@article{Abramo2016,
  author        = {Abramo, L. Raul and Secco, Lucas F. and Loureiro, Arthur},
  title         = {Fourier Analysis of Multitracer Cosmological Surveys},
  journal       = {Monthly Notices of the Royal Astronomical Society},
  volume        = {455},
  number        = {4},
  pages         = {3871--3889},
  year          = {2016},
  doi           = {10.1093/mnras/stv2588},
  eprint        = {1505.04106},
  archivePrefix = {arXiv},
  primaryClass  = {astro-ph.CO}
}

@ARTICLE{Achucarro2022,
       author = {{Ach{\'u}carro}, Ana and {Biagetti}, Matteo and {Braglia}, Matteo and {Cabass}, Giovanni and {Caldwell}, Robert and {Castorina}, Emanuele and {Chen}, Xingang and {Coulton}, William and {Flauger}, Raphael and {Fumagalli}, Jacopo and {Ivanov}, Mikhail M. and {Lee}, Hayden and {Maleknejad}, Azadeh and {Meerburg}, P. Daniel and {Moradinezhad Dizgah}, Azadeh and {Palma}, Gonzalo A. and {Pimentel}, Guilherme L. and {Renaux-Petel}, S{\'e}bastien and {Wallisch}, Benjamin and {Wandelt}, Benjamin D. and {Witkowski}, Lukas T. and {Kimmy Wu}, W.~L.},
        title = "{Inflation: Theory and Observations}",
      journal = {arXiv e-prints},
         year = 2022,
        month = mar,
          eid = {arXiv:2203.08128},
        pages = {arXiv:2203.08128},
          doi = {10.48550/arXiv.2203.08128},
archivePrefix = {arXiv},
       eprint = {2203.08128},
 primaryClass = {astro-ph.CO},
       adsurl = {https://ui.adsabs.harvard.edu/abs/2022arXiv220308128A}
}

@ARTICLE{Agarwal2021,
       author = {{Agarwal}, Nishant and {Desjacques}, Vincent and {Jeong}, Donghui and {Schmidt}, Fabian},
        title = "{Information content in the redshift-space galaxy power spectrum and bispectrum}",
      journal = {Journal of Cosmology and Astroparticle Physics},
         year = 2021,
        month = mar,
       volume = {2021},
       number = {3},
          eid = {021},
        pages = {021},
          doi = {10.1088/1475-7516/2021/03/021},
archivePrefix = {arXiv},
       eprint = {2007.04340},
 primaryClass = {astro-ph.CO},
       adsurl = {https://ui.adsabs.harvard.edu/abs/2021JCAP...03..021A}
}

@book{Aitchison1957,
	author = {Aitchison J., Brown J. A. C.},
	publisher = { Cambridge: Cambridge Univ. Press},
	title = {The Lognormal Distribution},
	year = {1957}}

@ARTICLE{Akeson2019,
       author = {{Akeson}, Rachel and {Armus}, Lee and {Bachelet}, Etienne and {Bailey}, Vanessa and {Bartusek}, Lisa and {Bellini}, Andrea and {Benford}, Dominic and {Bennett}, David and {Bhattacharya}, Aparna and {Bohlin}, Ralph and {Boyer}, Martha and {Bozza}, Valerio and {Bryden}, Geoffrey and {Calchi Novati}, Sebastiano and {Carpenter}, Kenneth and {Casertano}, Stefano and {Choi}, Ami and {Content}, David and {Dayal}, Pratika and {Dressler}, Alan and {Dor{\'e}}, Olivier and {Fall}, S. Michael and {Fan}, Xiaohui and {Fang}, Xiao and {Filippenko}, Alexei and {Finkelstein}, Steven and {Foley}, Ryan and {Furlanetto}, Steven and {Kalirai}, Jason and {Gaudi}, B. Scott and {Gilbert}, Karoline and {Girard}, Julien and {Grady}, Kevin and {Greene}, Jenny and {Guhathakurta}, Puragra and {Heinrich}, Chen and {Hemmati}, Shoubaneh and {Hendel}, David and {Henderson}, Calen and {Henning}, Thomas and {Hirata}, Christopher and {Ho}, Shirley and {Huff}, Eric and {Hutter}, Anne and {Jansen}, Rolf and {Jha}, Saurabh and {Johnson}, Samson and {Jones}, David and {Kasdin}, Jeremy and {Kelly}, Patrick and {Kirshner}, Robert and {Koekemoer}, Anton and {Kruk}, Jeffrey and {Lewis}, Nikole and {Macintosh}, Bruce and {Madau}, Piero and {Malhotra}, Sangeeta and {Mandel}, Kaisey and {Massara}, Elena and {Masters}, Daniel and {McEnery}, Julie and {McQuinn}, Kristen and {Melchior}, Peter and {Melton}, Mark and {Mennesson}, Bertrand and {Peeples}, Molly and {Penny}, Matthew and {Perlmutter}, Saul and {Pisani}, Alice and {Plazas}, Andr{\'e}s and {Poleski}, Radek and {Postman}, Marc and {Ranc}, Cl{\'e}ment and {Rauscher}, Bernard and {Rest}, Armin and {Roberge}, Aki and {Robertson}, Brant and {Rodney}, Steven and {Rhoads}, James and {Rhodes}, Jason and {Ryan}, Russell, Jr. and {Sahu}, Kailash and {Sand}, David and {Scolnic}, Dan and {Seth}, Anil and {Shvartzvald}, Yossi and {Siellez}, Karelle and {Smith}, Arfon and {Spergel}, David and {Stassun}, Keivan and {Street}, Rachel and {Strolger}, Louis-Gregory and {Szalay}, Alexander and {Trauger}, John and {Troxel}, M.~A. and {Turnbull}, Margaret and {van der Marel}, Roeland and {von der Linden}, Anja and {Wang}, Yun and {Weinberg}, David and {Williams}, Benjamin and {Windhorst}, Rogier and {Wollack}, Edward and {Wu}, Hao-Yi and {Yee}, Jennifer and {Zimmerman}, Neil},
        title = "{The Wide Field Infrared Survey Telescope: 100 Hubbles for the 2020s}",
      journal = {arXiv e-prints},
         year = 2019,
        month = feb,
          eid = {arXiv:1902.05569},
        pages = {arXiv:1902.05569},
          doi = {10.48550/arXiv.1902.05569},
archivePrefix = {arXiv},
       eprint = {1902.05569},
 primaryClass = {astro-ph.IM},
       adsurl = {https://ui.adsabs.harvard.edu/abs/2019arXiv190205569A}
}

@article{Alam2017,
    author = {Alam, Shadab and Ata, Metin and Bailey, Stephen and Beutler, Florian and Bizyaev, Dmitry and Blazek, Jonathan A. and Bolton, Adam S. and Brownstein, Joel R. and Burden, Angela and Chuang, Chia-Hsun and Comparat, Johan and Cuesta, Antonio J. and Dawson, Kyle S. and Eisenstein, Daniel J. and Escoffier, Stephanie and Gil-Marín, Héctor and Grieb, Jan Niklas and Hand, Nick and Ho, Shirley and Kinemuchi, Karen and Kirkby, David and Kitaura, Francisco and Malanushenko, Elena and Malanushenko, Viktor and Maraston, Claudia and McBride, Cameron K. and Nichol, Robert C. and Olmstead, Matthew D. and Oravetz, Daniel and Padmanabhan, Nikhil and Palanque-Delabrouille, Nathalie and Pan, Kaike and Pellejero-Ibanez, Marcos and Percival, Will J. and Petitjean, Patrick and Prada, Francisco and Price-Whelan, Adrian M. and Reid, Beth A. and Rodríguez-Torres, Sergio A. and Roe, Natalie A. and Ross, Ashley J. and Ross, Nicholas P. and Rossi, Graziano and Rubiño-Martín, Jose Alberto and Saito, Shun and Salazar-Albornoz, Salvador and Samushia, Lado and Sánchez, Ariel G. and Satpathy, Siddharth and Schlegel, David J. and Schneider, Donald P. and Scóccola, Claudia G. and Seo, Hee-Jong and Sheldon, Erin S. and Simmons, Audrey and Slosar, Anže and Strauss, Michael A. and Swanson, Molly E. C. and Thomas, Daniel and Tinker, Jeremy L. and Tojeiro, Rita and Magaña, Mariana Vargas and Vazquez, Jose Alberto and Verde, Licia and Wake, David A. and Wang, Yuting and Weinberg, David H. and White, Martin and Wood-Vasey, W. Michael and Yèche, Christophe and Zehavi, Idit and Zhai, Zhongxu and Zhao, Gong-Bo},
    title = "{The clustering of galaxies in the completed SDSS-III Baryon Oscillation Spectroscopic Survey: cosmological analysis of the DR12 galaxy sample}",
    journal = {Monthly Notices of the Royal Astronomical Society},
    volume = {470},
    number = {3},
    pages = {2617-2652},
    year = {2017},
    month = {03},
    issn = {0035-8711},
    doi = {10.1093/mnras/stx721},
    url = {https://doi.org/10.1093/mnras/stx721},
    eprint = {https://academic.oup.com/mnras/article-pdf/470/3/2617/18315003/stx721.pdf},
}

@ARTICLE{Alonso2012,
       author = {{Alonso}, David},
        title = "{CUTE solutions for two-point correlation functions from large cosmological datasets}",
      journal = {arXiv e-prints},
         year = 2012,
        month = oct,
          eid = {arXiv:1210.1833},
        pages = {arXiv:1210.1833},
          doi = {10.48550/arXiv.1210.1833},
archivePrefix = {arXiv},
       eprint = {1210.1833},
 primaryClass = {astro-ph.IM},
       adsurl = {https://ui.adsabs.harvard.edu/abs/2012arXiv1210.1833A}
}

@article{Anbajagane2023, 
year = {2023}, 
title = {{Beyond the 3rd moment: a practical study of using lensing convergence CDFs for cosmology with DES Y3}}, 
author = {Anbajagane, D and Chang, C and Banerjee, A and Abel, T and Gatti, M and Ajani, V and Alarcon, A and Amon, A and Baxter, E J and Bechtol, K and Becker, M R and Bernstein, G M and Campos, A and Rosell, A Carnero and Kind, M Carrasco and Chen, R and Choi, A and Davis, C and DeRose, J and Diehl, H T and Dodelson, S and Doux, C and Drlica-Wagner, A and Eckert, K and Elvin-Poole, J and Everett, S and Ferté, A and Gruen, D and Gruendl, R A and Harrison, I and Hartley, W G and Huff, E M and Jain, B and Jarvis, M and Jeffrey, N and Kacprzak, T and Kokron, N and Kuropatkin, N and Leget, P-F and MacCrann, N and McCullough, J and Myles, J and Navarro-Alsina, A and Pandey, S and Prat, J and Raveri, M and Rollins, R P and Roodman, A and Rykoff, E S and Sánchez, C and Secco, L F and Sevilla-Noarbe, I and Sheldon, E and Shin, T and Troxel, M A and Tutusaus, I and Whiteway, L and Yanny, B and Yin, B and Zhang, Y and Abbott, T M C and Allam, S and Aguena, M and Alves, O and Andrade-Oliveira, F and Annis, J and Bacon, D and Blazek, J and Brooks, D and Cawthon, R and Costa, L N da and Pereira, M E S and Davis, T M and Desai, S and Doel, P and Ferrero, I and Frieman, J and Giannini, G and Gutierrez, G and Hinton, S R and Hollowood, D L and Honscheid, K and James, D J and Kuehn, K and Lahav, O and Marshall, J L and Mena-Fernández, J and Menanteau, F and Miquel, R and Palmese, A and Pieres, A and Malagón, A A Plazas and Reil, K and Sanchez, E and Smith, M and Swanson, M E C and Tarle, G and Wiseman, P and Collaboration), (DES}, 
journal = {Monthly Notices of the Royal Astronomical Society},
issn = {0035-8711}, 
doi = {10.1093/mnras/stad3118}, 
pages = {5530--5554}, 
number = {4}, 
volume = {526}, 
}

@article{Anderson2014,
    author = {Anderson, Lauren and Aubourg, Éric and Bailey, Stephen and Beutler, Florian and Bhardwaj, Vaishali and Blanton, Michael and Bolton, Adam S. and Brinkmann, J. and Brownstein, Joel R. and Burden, Angela and Chuang, Chia-Hsun and Cuesta, Antonio J. and Dawson, Kyle S. and Eisenstein, Daniel J. and Escoffier, Stephanie and Gunn, James E. and Guo, Hong and Ho, Shirley and Honscheid, Klaus and Howlett, Cullan and Kirkby, David and Lupton, Robert H. and Manera, Marc and Maraston, Claudia and McBride, Cameron K. and Mena, Olga and Montesano, Francesco and Nichol, Robert C. and Nuza, Sebastián E. and Olmstead, Matthew D. and Padmanabhan, Nikhil and Palanque-Delabrouille, Nathalie and Parejko, John and Percival, Will J. and Petitjean, Patrick and Prada, Francisco and Price-Whelan, Adrian M. and Reid, Beth and Roe, Natalie A. and Ross, Ashley J. and Ross, Nicholas P. and Sabiu, Cristiano G. and Saito, Shun and Samushia, Lado and Sánchez, Ariel G. and Schlegel, David J. and Schneider, Donald P. and Scoccola, Claudia G. and Seo, Hee-Jong and Skibba, Ramin A. and Strauss, Michael A. and Swanson, Molly E. C. and Thomas, Daniel and Tinker, Jeremy L. and Tojeiro, Rita and Magaña, Mariana Vargas and Verde, Licia and Wake, David A. and Weaver, Benjamin A. and Weinberg, David H. and White, Martin and Xu, Xiaoying and Yèche, Christophe and Zehavi, Idit and Zhao, Gong-Bo},
    title = "{The clustering of galaxies in the SDSS-III Baryon Oscillation Spectroscopic Survey: baryon acoustic oscillations in the Data Releases 10 and 11 Galaxy samples}",
    journal = {Monthly Notices of the Royal Astronomical Society},
    volume = {441},
    number = {1},
    pages = {24-62},
    year = {2014},
    month = {04},
    issn = {0035-8711},
    doi = {10.1093/mnras/stu523},
    url = {https://doi.org/10.1093/mnras/stu523},
    eprint = {https://academic.oup.com/mnras/article-pdf/441/1/24/3007885/stu523.pdf},
}

@article{baldauf2020,
  title={Effective Field Theory of Large-Scale Structure},
  author={Baldauf, Tobias},
  journal={Effective Field Theory in Particle Physics and Cosmology: Lecture Notes of the Les Houches Summer School: Volume 108, July 2017},
  volume={108},
  pages={415},
  year={2020},
  publisher={Oxford University Press, USA}
}

@article{Beisbart2000, 
year = {2000}, 
title = {{Luminosity- and Morphology-dependent Clustering of Galaxies}}, 
author = {Beisbart, Claus and Kerscher, Martin}, 
journal = {The Astrophysical Journal},
issn = {0004-637X}, 
doi = {10.1086/317788}, 
eprint = {astro-ph/0003358}, 
pages = {6--25}, 
number = {1}, 
volume = {545}
}

@article{Bernardeau2002,
title = {Large-scale structure of the Universe and cosmological perturbation theory},
journal = {Physics Reports},
volume = {367},
number = {1},
pages = {1-248},
year = {2002},
issn = {0370-1573},
doi = {https://doi.org/10.1016/S0370-1573(02)00135-7},
url = {https://www.sciencedirect.com/science/article/pii/S0370157302001357},
author = {F. Bernardeau and S. Colombi and E. Gaztañaga and R. Scoccimarro}
}

@article{Bernardeau2014,
  title = {Statistics of cosmic density profiles from perturbation theory},
  author = {Bernardeau, Francis and Pichon, Christophe and Codis, Sandrine},
  journal = {Physical Review D},
  volume = {90},
  issue = {10},
  pages = {103519},
  numpages = {23},
  year = {2014},
  month = {Nov},
  publisher = {American Physical Society},
  doi = {10.1103/PhysRevD.90.103519},
  url = {https://link.aps.org/doi/10.1103/PhysRevD.90.103519}
}

@ARTICLE{BernardeauReimberg2016,
       author = {{Bernardeau}, Francis and {Reimberg}, Paulo},
        title = "{A large deviation principle at play in large-scale structure cosmology}",
      journal = {Physical Review D},
         year = 2016,
        month = sep,
       volume = {94},
       number = {6},
          eid = {063520},
        pages = {063520},
          doi = {10.1103/PhysRevD.94.063520},
archivePrefix = {arXiv},
       eprint = {1511.08641},
 primaryClass = {astro-ph.CO},
       adsurl = {https://ui.adsabs.harvard.edu/abs/2016PhRvD..94f3520B}
}

@article{Beylkin1991,
  author  = {Beylkin, G. and Coifman, R. and Rokhlin, V.},
  title   = {On the representation of operators in bases of compactly supported wavelets},
  journal = {Communications on Pure and Applied Mathematics},
  volume  = {44},
  pages   = {141--183},
  year    = {1991}
}

@article{Beylkin1992,
  author  = {Beylkin, Gregory},
  title   = {On the Representation of Operators in Bases of Compactly Supported Wavelets},
  journal = {SIAM Journal on Numerical Analysis},
  volume  = {29},
  number  = {6},
  pages   = {1716--1740},
  year    = {1992},
  doi     = {10.1137/0729097}
}

@article{Brown2022, 
year = {2022}, 
title = {{ConKer: An algorithm for evaluating correlations of arbitrary order}}, 
author = {Brown, Z. and Mishtaku, G. and Demina, R.}, 
journal = {Astronomy \& Astrophysics},
issn = {0004-6361}, 
doi = {10.1051/0004-6361/202141917}, 
eprint = {2108.00015}, 
pages = {A129}, 
volume = {667}, 
}

@article{Burger2023, 
year = {2023}, 
title = {{KiDS-1000 cosmology: Constraints from density split statistics}}, 
author = {Burger, Pierre A. and Friedrich, Oliver and Harnois-Déraps, Joachim and Schneider, Peter and Asgari, Marika and Bilicki, Maciej and Hildebrandt, Hendrik and Wright, Angus H. and Castro, Tiago and Dolag, Klaus and Heymans, Catherine and Joachimi, Benjamin and Kuijken, Konrad and Martinet, Nicolas and Shan, HuanYuan and Tröster, Tilman}, 
journal = {Astronomy \& Astrophysics},
issn = {0004-6361}, 
doi = {10.1051/0004-6361/202244673}, 
eprint = {2208.02171}, 
pages = {A69}, 
volume = {669}, 
}

@article{Cahn_2023,
doi = {10.1088/1751-8121/acdfc4},
url = {https://doi.org/10.1088/1751-8121/acdfc4},
year = {2023},
month = {jul},
publisher = {IOP Publishing},
volume = {56},
number = {32},
pages = {325204},
author = {Cahn, Robert N and Slepian, Zachary},
title = {Isotropic N-point basis functions and their properties},
journal = {Journal of Physics A: Mathematical and Theoretical}
}

@article{carron2011, 
year = {2011}, 
title = {{ON THE INCOMPLETENESS OF THE MOMENT AND CORRELATION FUNCTION HIERARCHY AS PROBES OF THE LOGNORMAL FIELD}}, 
author = {Carron, Julien}, 
journal = {The Astrophysical Journal},
issn = {0004-637X}, 
doi = {10.1088/0004-637x/738/1/86}, 
eprint = {1105.4467}, 
pages = {86}, 
number = {1}, 
volume = {738}, 
}

@ARTICLE{CarronNeyrinck2012,
       author = {{Carron}, Julien and {Neyrinck}, Mark C.},
        title = "{On the Inadequacy of N-point Correlation Functions to Describe Nonlinear Cosmological Fields: Explicit Examples and Connection to Simulations}",
      journal = {The Astrophysical Journal},
         year = 2012,
        month = may,
       volume = {750},
       number = {1},
          eid = {28},
        pages = {28},
          doi = {10.1088/0004-637X/750/1/28},
archivePrefix = {arXiv},
       eprint = {1201.1444},
 primaryClass = {astro-ph.CO},
       adsurl = {https://ui.adsabs.harvard.edu/abs/2012ApJ...750...28C}
}

@article{Cataneo2022, 
year = {2022}, 
title = {{The matter density PDF for modified gravity and dark energy with Large Deviations Theory}}, 
author = {Cataneo, Matteo and Uhlemann, Cora and Arnold, Christian and Gough, Alex and Li, Baojiu and Heymans, Catherine}, 
journal = {Monthly Notices of the Royal Astronomical Society},
issn = {0035-8711}, 
doi = {10.1093/mnras/stac904}, 
eprint = {2109.02636}, 
pages = {1623--1641}, 
number = {2}, 
volume = {513}, 
}

@article{Chellino_2026,
doi = {10.1088/1475-7516/2026/06/021},
url = {https://doi.org/10.1088/1475-7516/2026/06/021},
year = {2026},
month = {jun},
publisher = {IOP Publishing},
volume = {2026},
number = {06},
pages = {021},
author = {Chellino, Jessica and Greco, Alessandro and May, Simon and Slepian, Zachary},
title = {Analytic model for covariance matrices of the 2-, 3-, and 4-Point Correlation Functions in the Gaussian random field approximation},
journal = {Journal of Cosmology and Astroparticle Physics}
}

@article{Chen2025EmulatorI,
  author  = {Chen, Zhao and Yu, Yu and Han, Jiaxin and Jing, Yipeng},
  title   = {{CSST} cosmological emulator {I}: Matter power spectrum emulation with one percent accuracy to \(k = 10\,h\,{\rm Mpc}^{-1}\)},
  journal = {Science China Physics, Mechanics \& Astronomy},
  year    = {2025},
  volume  = {68},
  number  = {8},
  pages   = {289512},
  doi     = {10.1007/s11433-025-2671-0},
  eprint  = {2502.11160},
  archivePrefix = {arXiv},
  primaryClass  = {astro-ph.CO}
}

@ARTICLE{Chen2025EmulatorII,
       author = {{Chen}, Zhao and {Yu}, Yu},
        title = "{CSST cosmological emulator II: Generalized accurate halo mass function emulation}",
      journal = {Science China Physics, Mechanics \& Astronomy},
         year = {2025},
       volume = {68},
        pages = {109513},
          doi = {10.1007/s11433-025-2764-x},
archivePrefix = {arXiv},
       eprint = {2506.09688},
 primaryClass = {astro-ph.CO}
}

@article{Clerkin2016, 
year = {2016}, 
title = {{Testing the lognormality of the galaxy and weak lensing convergence distributions from Dark Energy Survey maps}}, 
author = {Clerkin, L and Kirk, D and Manera, M and Lahav, O and Abdalla, F and Amara, A and Bacon, D and Chang, C and Gaztañaga, E and Hawken, A and Jain, B and Joachimi, B and Vikram, V and Abbott, T and Allam, S and Armstrong, R and Benoit-Lévy, A and Bernstein, G M and Bernstein, R A and Bertin, E and Brooks, D and Burke, D L and Rosell, A Carnero and Kind, M Carrasco and Crocce, M and Cunha, C E and D'Andrea, C B and Costa, L N da and Desai, S and Diehl, H T and Dietrich, J P and Eifler, T F and Evrard, A E and Flaugher, B and Fosalba, P and Frieman, J and Gerdes, D W and Gruen, D and Gruendl, R A and Gutierrez, G and Honscheid, K and James, D J and Kent, S and Kuehn, K and Kuropatkin, N and Lima, M and Melchior, P and Miquel, R and Nord, B and Plazas, A A and Romer, A K and Roodman, A and Sanchez, E and Schubnell, M and Sevilla-Noarbe, I and Smith, R C and Soares-Santos, M and Sobreira, F and Suchyta, E and Swanson, M E C and Tarle, G and Walker, A R}, 
journal = {Monthly Notices of the Royal Astronomical Society},
issn = {0035-8711}, 
doi = {10.1093/mnras/stw2106}, 
eprint = {1605.02036}, 
pages = {1444--1461}, 
number = {2}, 
volume = {466}
}

@article{coles1991, 
year = {1991}, 
title = {{A lognormal model for the cosmological mass distribution}}, 
author = {Coles, Peter and Jones, Bernard}, 
journal = {Monthly Notices of the Royal Astronomical Society},
issn = {0035-8711}, 
doi = {10.1093/mnras/248.1.1}, 
pages = {1--13}, 
number = {1}, 
volume = {248}, 
}

@article{Colombi2009, 
year = {2009}, 
title = {{Accurate estimators of power spectra in N‐body simulations}}, 
author = {Colombi, Stéphane and Jaffe, Andrew and Novikov, Dmitri and Pichon, Christophe}, 
journal = {Monthly Notices of the Royal Astronomical Society},
issn = {0035-8711}, 
doi = {10.1111/j.1365-2966.2008.14176.x}, 
eprint = {0811.0313}, 
pages = {511--526}, 
number = {2}, 
volume = {393}
}

@article{Connell2016,
    author = {O'Connell, Ross and Eisenstein, Daniel and Vargas, Mariana and Ho, Shirley and Padmanabhan, Nikhil},
    title = "{Large covariance matrices: smooth models from the two-point correlation function}",
    journal = {Monthly Notices of the Royal Astronomical Society},
    volume = {462},
    number = {3},
    pages = {2681-2694},
    year = {2016},
    month = {07},
    issn = {0035-8711},
    doi = {10.1093/mnras/stw1821},
    url = {https://doi.org/10.1093/mnras/stw1821},
    eprint = {https://academic.oup.com/mnras/article-pdf/462/3/2681/8010260/stw1821.pdf},
}

@ARTICLE{Connell2019,
       author = {{O'Connell}, Ross and {Eisenstein}, Daniel J.},
        title = "{Large covariance matrices: accurate models without mocks}",
      journal = {Monthly Notices of the Royal Astronomical Society},
         year = 2019,
        month = aug,
       volume = {487},
       number = {2},
        pages = {2701-2717},
          doi = {10.1093/mnras/stz1359},
archivePrefix = {arXiv},
       eprint = {1808.05978},
 primaryClass = {astro-ph.CO},
       adsurl = {https://ui.adsabs.harvard.edu/abs/2019MNRAS.487.2701O}
}

@article{Cooray2002,
title = {Halo models of large scale structure},
journal = {Physics Reports},
volume = {372},
number = {1},
pages = {1-129},
year = {2002},
issn = {0370-1573},
doi = {https://doi.org/10.1016/S0370-1573(02)00276-4},
url = {https://www.sciencedirect.com/science/article/pii/S0370157302002764},
author = {Asantha Cooray and Ravi Sheth}
}

@article{Cowell2024,
  author        = {Cowell, Jessica A. and Alonso, David and Liu, Jia},
  title         = {Optimizing Marked Power Spectra for Cosmology},
  journal       = {Monthly Notices of the Royal Astronomical Society},
  year          = {2024},
  volume        = {535},
  number        = {4},
  pages         = {3129--3140},
  doi           = {10.1093/mnras/stae2492},
  archivePrefix = {arXiv},
  eprint        = {2409.05695},
  primaryClass  = {astro-ph.CO}
}

@ARTICLE{Cui2008,
       author = {{Cui}, Weiguang and {Liu}, Lei and {Yang}, Xiaohu and {Wang}, Yu and {Feng}, Longlong and {Springel}, Volker},
        title = "{An Ideal Mass Assignment Scheme for Measuring the Power Spectrum with Fast Fourier Transforms}",
      journal = {The Astrophysical Journal},
         year = 2008,
        month = nov,
       volume = {687},
       number = {2},
        pages = {738-744},
          doi = {10.1086/592079},
archivePrefix = {arXiv},
       eprint = {0804.0070},
 primaryClass = {astro-ph},
       adsurl = {https://ui.adsabs.harvard.edu/abs/2008ApJ...687..738C}
}

@book{Daubechies1992,
	author = {Daubechies, I.},
	publisher = {Philadelphia: SIAM},
	title = {Ten Lectures on Wavelets},
	year = {1992}}

@misc{DESI2016,
      title={The DESI Experiment Part I: Science,Targeting, and Survey Design}, 
      author={{DESI Collaboration}},
      year={2016},
      eprint={1611.00036},
      archivePrefix={arXiv},
      primaryClass={astro-ph.IM}
}

@article{DESI2024VI_BAO, 
  year     = {2025}, 
  title    = {{DESI} 2024 {VI}: cosmological constraints from the measurements of baryon acoustic oscillations}, 
  author   = {Adame, A G and Aguilar, J and Ahlen, S and Alam, S and Alexander, D M and Alvarez, M and Alves, O and Anand, A and Andrade, U and Armengaud, E and Avila, S and Aviles, A and Awan, H and Bahr-Kalus, B and Bailey, S and Baltay, C and Bault, A and Behera, J and {BenZvi}, S and Bera, A and Beutler, F and Bianchi, D and Blake, C and Blum, R and Brieden, S and Brodzeller, A and Brooks, D and Buckley-Geer, E and Burtin, E and Calderon, R and Canning, R and Rosell, A Carnero and Cereskaite, R and Cervantes-Cota, J L and Chabanier, S and Chaussidon, E and Chaves-Montero, J and Chen, S and Chen, X and Claybaugh, T and Cole, S and Cuceu, A and Davis, T M and Dawson, K and Macorra, A de la and Mattia, A de and Deiosso, N and Dey, A and Dey, B and Ding, Z and Doel, P and Edelstein, J and Eftekharzadeh, S and Eisenstein, D J and Elliott, A and Fagrelius, P and Fanning, K and Ferraro, S and Ereza, J and Findlay, N and Flaugher, B and Font-Ribera, A and Forero-Sánchez, D and Forero-Romero, J E and Frenk, C S and Garcia-Quintero, C and Gaztañaga, E and Gil-Marín, H and Gontcho, S Gontcho A and Gonzalez-Morales, A X and Gonzalez-Perez, V and Gordon, C and Green, D and Gruen, D and Gsponer, R and Gutierrez, G and Guy, J and Hadzhiyska, B and Hahn, C and Hanif, M M S and Herrera-Alcantar, H K and Honscheid, K and Howlett, C and Huterer, D and Iršič, V and Ishak, M and Juneau, S and Karaçaylı, N G and Kehoe, R and Kent, S and Kirkby, D and Kremin, A and Krolewski, A and Lai, Y and Lan, T -W and Landriau, M and Lang, D and Lasker, J and Goff, J M Le and Guillou, L Le and Leauthaud, A and Levi, M E and Li, T S and Linder, E and Lodha, K and Magneville, C and Manera, M and Margala, D and Martini, P and Maus, M and {McDonald}, P and Medina-Varela, L and Meisner, A and Mena-Fernández, J and Miquel, R and Moon, J and Moore, S and Moustakas, J and Mudur, N and Mueller, E and Muñoz-Gutiérrez, A and Myers, A D and Nadathur, S and Napolitano, L and Neveux, R and Newman, J A and Nguyen, N M and Nie, J and Niz, G and Noriega, H E and Padmanabhan, N and Paillas, E and Palanque-Delabrouille, N and Pan, J and Penmetsa, S and Percival, W J and Pieri, M M and Pinon, M and Poppett, C and Porredon, A and Prada, F and Pérez-Fernández, A and Pérez-Ràfols, I and Rabinowitz, D and Raichoor, A and Ramírez-Pérez, C and Ramirez-Solano, S and Ravoux, C and Rashkovetskyi, M and Rezaie, M and Rich, J and Rocher, A and Rockosi, C and Roe, N A and Rosado-Marin, A and Ross, A J and Rossi, G and Ruggeri, R and Ruhlmann-Kleider, V and Samushia, L and Sanchez, E and Saulder, C and Schlafly, E F and Schlegel, D and Schubnell, M and Seo, H and Shafieloo, A and Sharples, R and Silber, J and Slosar, A and Smith, A and Sprayberry, D and Tan, T and Tarlé, G and Taylor, P and Trusov, S and Ureña-López, L A and Vaisakh, R and Valcin, D and Valdes, F and Vargas-Magaña, M and Verde, L and Walther, M and Wang, B and Wang, M S and Weaver, B A and Weaverdyck, N and Wechsler, R H and Weinberg, D H and White, M and Yu, J and Yu, Y and Yuan, S and Yèche, C and Zaborowski, E A and Zarrouk, P and Zhang, H and Zhao, C and Zhao, R and Zhou, R and Zhuang, T and Zou, H}, 
  journal  = {Journal of Cosmology and Astroparticle Physics}, 
  doi      = {10.1088/1475-7516/2025/02/021}, 
  eprint   = {2404.03002}, 
  pages    = {021}, 
  number   = {02}, 
  volume   = {2025}
}

@article{DESI2024VII_cosmo, 
  year     = {2025}, 
  title    = {{DESI} 2024 {VII}: cosmological constraints from the full-shape modeling of clustering measurements}, 
  author   = {Adame, A.G. and Aguilar, J. and Ahlen, S. and Alam, S. and Alexander, D.M. and Prieto, C. Allende and Alvarez, M. and Alves, O. and Anand, A. and Andrade, U. and Armengaud, E. and Avila, S. and Aviles, A. and Awan, H. and Bahr-Kalus, B. and Bailey, S. and Baltay, C. and Bault, A. and Behera, J. and {BenZvi}, S. and Beutler, F. and Bianchi, D. and Blake, C. and Blum, R. and Bonici, M. and Brieden, S. and Brodzeller, A. and Brooks, D. and Buckley-Geer, E. and Burtin, E. and Calderon, R. and Canning, R. and Rosell, A. Carnero and Cereskaite, R. and Cervantes-Cota, J.L. and Chabanier, S. and Chaussidon, E. and Chaves-Montero, J. and Chebat, D. and Chen, S. and Chen, X. and Claybaugh, T. and Cole, S. and Cuceu, A. and Davis, T.M. and Dawson, K. and Macorra, A. de la and Mattia, A. de and Deiosso, N. and Dey, A. and Dey, B. and Ding, Z. and Doel, P. and Edelstein, J. and Eftekharzadeh, S. and Eisenstein, D.J. and Elbers, W. and Elliott, A. and Fagrelius, P. and Fanning, K. and Ferraro, S. and Ereza, J. and Findlay, N. and Flaugher, B. and Font-Ribera, A. and Forero-Sánchez, D. and Forero-Romero, J.E. and Frenk, C.S. and Garcia-Quintero, C. and Garrison, L.H. and Gaztañaga, E. and Gil-Marín, H. and Gontcho, S.Gontcho A. and Gonzalez-Morales, A.X. and Gonzalez-Perez, V. and Gordon, C. and Green, D. and Gruen, D. and Gsponer, R. and Gutierrez, G. and Guy, J. and Hadzhiyska, B. and Hahn, C. and Hanif, M.M.S. and Herrera-Alcantar, H.K. and Honscheid, K. and Howlett, C. and Huterer, D. and Iršič, V. and Ishak, M. and Joyce, R. and Juneau, S. and Karaçaylı, N.G. and Kehoe, R. and Kent, S. and Kirkby, D. and Kong, H. and Koposov, S.E. and Kremin, A. and Krolewski, A. and Lahav, O. and Lai, Y. and Lan, T.-W. and Landriau, M. and Lang, D. and Lasker, J. and Goff, J.M. Le and Guillou, L. Le and Leauthaud, A. and Levi, M.E. and Li, T.S. and Lodha, K. and Magneville, C. and Manera, M. and Margala, D. and Martini, P. and Matthewson, W. and Maus, M. and {McDonald}, P. and Medina-Varela, L. and Meisner, A. and Mena-Fernández, J. and Miquel, R. and Moon, J. and Moore, S. and Moustakas, J. and Mudur, N. and Mueller, E. and Muñoz-Gutiérrez, A. and Myers, A.D. and Nadathur, S. and Napolitano, L. and Neveux, R. and Newman, J.A. and Nguyen, N.M. and Nie, J. and Niz, G. and Noriega, H.E. and Padmanabhan, N. and Paillas, E. and Palanque-Delabrouille, N. and Pan, J. and Penmetsa, S. and Percival, W.J. and Pieri, M.M. and Pinon, M. and Poppett, C. and Porredon, A. and Prada, F. and Pérez-Fernández, A. and Pérez-Ràfols, I. and Rabinowitz, D. and Raichoor, A. and Ramírez-Pérez, C. and Ramirez-Solano, S. and Rashkovetskyi, M. and Ravoux, C. and Rezaie, M. and Rich, J. and Rocher, A. and Rockosi, C. and Roe, N.A. and Rosado-Marin, A. and Ross, A.J. and Rossi, G. and Ruggeri, R. and Ruhlmann-Kleider, V. and Samushia, L. and Sanchez, E. and Saulder, C. and Schlafly, E.F. and Schlegel, D. and Schubnell, M. and Seo, H. and Shafieloo, A. and Sharples, R. and Silber, J. and Slosar, A. and Smith, A. and Sprayberry, D. and Tan, T. and Tarlé, G. and Taylor, P. and Trusov, S. and Vaisakh, R. and Valcin, D. and Valdes, F. and Valogiannis, G. and Vargas-Magaña, M. and Verde, L. and Walther, M. and Wang, B. and Wang, M.S. and Weaver, B.A. and Weaverdyck, N. and Wechsler, R.H. and Weinberg, D.H. and White, M. and Wilson, M.J. and Yi, L. and Yu, J. and Yu, Y. and Yuan, S. and Yèche, C. and Zaborowski, E.A. and Zarrouk, P. and Zhang, H. and Zhao, C. and Zhao, R. and Zhou, R. and Zhuang, T. and Zou, H. and collaboration, The {DESI}}, 
  journal  = {Journal of Cosmology and Astroparticle Physics}, 
  doi      = {10.1088/1475-7516/2025/07/028}, 
  eprint   = {2411.12022}, 
  pages    = {028}, 
  number   = {07}, 
  volume   = {2025}
}

@article{DESIBispectrum2025,
  author        = {{Novell-Masot}, S. and {Gil-Mar{\'i}n}, H. and {Verde}, L.
                   and others},
  title         = {Full-Shape Analysis of the Power Spectrum and Bispectrum of
                   {DESI} {DR1} {LRG} and {QSO} Samples},
  journal       = {Journal of Cosmology and Astroparticle Physics},
  year          = {2025},
  volume        = {2025},
  number        = {06},
  pages         = {005},
  doi           = {10.1088/1475-7516/2025/06/005},
  archivePrefix = {arXiv},
  eprint        = {2503.09714},
  primaryClass  = {astro-ph.CO}
}

@article{DESIDR2BAOValidation,
  author        = {Andrade, U. and others},
  title         = {Validation of the {DESI} {DR2} Measurements of Baryon
                   Acoustic Oscillations from Galaxies and Quasars},
  journal       = {Physical Review D},
  year          = {2025},
  volume        = {112},
  number        = {8},
  pages         = {083512},
  doi           = {10.1103/kdys-w8vl},
  archivePrefix = {arXiv},
  eprint        = {2503.14742},
  primaryClass  = {astro-ph.CO}
}

@article{DESIDR2Cosmo,
  title = {DESI DR2 results. II. Measurements of baryon acoustic oscillations and cosmological constraints},
  author = {Abdul Karim, M. and Aguilar, J. and Ahlen, S. and Alam, S. and Allen, L. and Prieto, C. Allende and Alves, O. and Anand, A. and Andrade, U. and Armengaud, E. and Aviles, A. and Bailey, S. and Baltay, C. and Bansal, P. and Bault, A. and Behera, J. and BenZvi, S. and Bianchi, D. and Blake, C. and Brieden, S. and Brodzeller, A. and Brooks, D. and Buckley-Geer, E. and Burtin, E. and Calderon, R. and Canning, R. and Rosell, A. Carnero and Carrilho, P. and Casas, L. and Castander, F. J. and Charles, M. and Chaussidon, E. and Chaves-Montero, J. and Chebat, D. and Chen, X. and Claybaugh, T. and Cole, S. and Cooper, A. P. and Cuceu, A. and Dawson, K. S. and de la Macorra, A. and de Mattia, A. and Deiosso, N. and Della Costa, J. and Demina, R. and Dey, A. and Dey, B. and Ding, Z. and Doel, P. and Edelstein, J. and Eisenstein, D. J. and Elbers, W. and Fagrelius, P. and Fanning, K. and Fern{\'a}ndez-Garc{\'{\i}}a, E. and Ferraro, S. and Font-Ribera, A. and Forero-Romero, J. E. and Frenk, C. S. and Garcia-Quintero, C. and Garrison, L. H. and Gazta{\~n}aga, E. and Gil-Mar{\'{\i}}n, H. and Gontcho, S. Gontcho A. and Gonzalez, D. and Gonzalez-Morales, A. X. and Gordon, C. and Green, D. and Gutierrez, G. and Guy, J. and Hadzhiyska, B. and Hahn, C. and He, S. and Herbold, M. and Herrera-Alcantar, H. K. and Ho, M.-F. and Honscheid, K. and Howlett, C. and Huterer, D. and Ishak, M. and Juneau, S. and Kamble, N. V. and Kara{\ifmmode \mbox{\c{c}}\else \c{c}\fi{}}ayl��, N. G. and Kehoe, R. and Kent, S. and Kim, A. G. and Kirkby, D. and Kisner, T. and Koposov, S. E. and Kremin, A. and Krolewski, A. and Lahav, O. and Lamman, C. and Landriau, M. and Lang, D. and Lasker, J. and Le Goff, J. M. and Le Guillou, L. and Leauthaud, A. and Levi, M. E. and Li, Q. and Li, T. S. and Lodha, K. and Lokken, M. and Lozano-Rodr{\'{\i}}guez, F. and Magneville, C. and Manera, M. and Martini, P. and Matthewson, W. L. and Meisner, A. and Mena-Fern{\'a}ndez, J. and Menegas, A. and Mergulh{\~a}o, T. and Miquel, R. and Moustakas, J. and Mu{\~n}oz-Guti{\'e}rrez, A. and Mu{\~n}oz-Santos, D. and Myers, A. D. and Nadathur, S. and Naidoo, K. and Napolitano, L. and Newman, J. A. and Niz, G. and Noriega, H. E. and Paillas, E. and Palanque-Delabrouille, N. and Pan, J. and Peacock, J. A. and Ibanez, M. P. and Percival, W. J. and P{\'e}rez-Fern{\'a}ndez, A. and P{\'e}rez-R{\`a}fols, I. and Pieri, M. M. and Poppett, C. and Prada, F. and Rabinowitz, D. and Raichoor, A. and Ram{\'{\i}}rez-P{\'e}rez, C. and Rashkovetskyi, M. and Ravoux, C. and Rich, J. and Rocher, A. and Rockosi, C. and Rohlf, J. and Rom{\'a}n-Herrera, J. O. and Ross, A. J. and Rossi, G. and Ruggeri, R. and Ruhlmann-Kleider, V. and Samushia, L. and Sanchez, E. and Sanders, N. and Schlegel, D. and Schubnell, M. and Seo, H. and Shafieloo, A. and Sharples, R. and Silber, J. and Sinigaglia, F. and Sprayberry, D. and Tan, T. and Tarl{\'e}, G. and Taylor, P. and Turner, W. and Ure{\~n}a-L{\'o}pez, L. A. and Vaisakh, R. and Valdes, F. and Valogiannis, G. and Vargas-Maga{\~n}a, M. and Verde, L. and Walther, M. and Weaver, B. A. and Weinberg, D. H. and White, M. and Wolfson, M. and Y{\`e}che, C. and Yu, J. and Zaborowski, E. A. and Zarrouk, P. and Zhai, Z. and Zhang, H. and Zhao, C. and Zhao, G. B. and Zhou, R. and Zou, H.},
  collaboration = {DESI Collaboration},
  journal = {Physical Review D},
  volume = {112},
  issue = {8},
  pages = {083515},
  numpages = {40},
  year = {2025},
  month = {Oct},
  publisher = {American Physical Society},
  doi = {10.1103/tr6y-kpc6},
  url = {https://link.aps.org/doi/10.1103/tr6y-kpc6}
}

@article{DESIDR2DarkEnergy,
  author        = {Lodha, K. and others},
  title         = {Extended Dark Energy Analysis Using {DESI} {DR2} {BAO}
                   Measurements},
  journal       = {Physical Review D},
  year          = {2025},
  volume        = {112},
  number        = {8},
  pages         = {083511},
  doi           = {10.1103/w4c6-1r5j},
  archivePrefix = {arXiv},
  eprint        = {2503.14743},
  primaryClass  = {astro-ph.CO}
}

@article{DESIDR2DLACatalog,
  author        = {Brodzeller, A. and others},
  title         = {Construction of the Damped Lyman Alpha Absorber Catalog for
                   {DESI} {DR2} Lyman Alpha {BAO}},
  journal       = {Physical Review D},
  year          = {2025},
  volume        = {112},
  number        = {8},
  pages         = {083510},
  doi           = {10.1103/wxyv-46kb},
  archivePrefix = {arXiv},
  eprint        = {2503.14740},
  primaryClass  = {astro-ph.CO}
}

@article{DESIDR2LyAValidation,
  author        = {Casas, L. and others},
  title         = {Validation of the {DESI} {DR2} Lyman Alpha {BAO} Analysis
                   Using Synthetic Datasets},
  journal       = {Physical Review D},
  year          = {2026},
  volume        = {113},
  number        = {2},
  pages         = {023520},
  doi           = {10.1103/fvgh-kswf},
  archivePrefix = {arXiv},
  eprint        = {2503.14741},
  primaryClass  = {astro-ph.IM}
}

@article{DESIDR2Neutrinos,
  author        = {Elbers, W. and others},
  title         = {Constraints on Neutrino Physics from {DESI} {DR2} {BAO} and
                   {DR1} Full Shape},
  journal       = {Physical Review D},
  year          = {2025},
  volume        = {112},
  number        = {8},
  pages         = {083513},
  doi           = {10.1103/w9pk-xsk7},
  archivePrefix = {arXiv},
  eprint        = {2503.14744},
  primaryClass  = {astro-ph.CO}
}

@article{Desjacques2010,
doi = {10.1088/0264-9381/27/12/124011},
url = {https://dx.doi.org/10.1088/0264-9381/27/12/124011},
year = {2010},
month = {may},
publisher = {},
volume = {27},
number = {12},
pages = {124011},
author = {V Desjacques and U Seljak},
title = {Primordial non-Gaussianity from the large-scale structure},
journal = {Classical and Quantum Gravity}
}

@article{Desjacques2018, 
year = {2018}, 
title = {{Large-scale galaxy bias}}, 
author = {Desjacques, Vincent and Jeong, Donghui and Schmidt, Fabian}, 
journal = {Physics Reports}, 
issn = {0370-1573}, 
doi = {10.1016/j.physrep.2017.12.002}, 
eprint = {1611.09787}, 
pages = {1--193}, 
volume = {733}, 
}

@article{Dodelson2013,
  author        = {Dodelson, Scott and Schneider, Michael D.},
  title         = {The Effect of Covariance Estimator Error on
                   Cosmological Parameter Constraints},
  journal       = {Physical Review D},
  year          = {2013},
  volume        = {88},
  number        = {6},
  pages         = {063537},
  doi           = {10.1103/PhysRevD.88.063537},
  eprint        = {1304.2593},
  archivePrefix = {arXiv}
}

@article{Donoso2019, 
year = {2019}, 
title = {{GUNDAM: a toolkit for fast spatial correlation functions in galaxy surveys}}, 
author = {Donoso, E}, 
journal = {Monthly Notices of the Royal Astronomical Society},
issn = {0035-8711}, 
doi = {10.1093/mnras/stz1469}, 
pages = {2824--2835}, 
number = {2}, 
volume = {487}, 
}

@ARTICLE{Eisenstein2005,
       author = {{Eisenstein}, Daniel J. and {Zehavi}, Idit and {Hogg}, David W. and {Scoccimarro}, Roman and {Blanton}, Michael R. and {Nichol}, Robert C. and {Scranton}, Ryan and {Seo}, Hee-Jong and {Tegmark}, Max and {Zheng}, Zheng and {Anderson}, Scott F. and {Annis}, Jim and {Bahcall}, Neta and {Brinkmann}, Jon and {Burles}, Scott and {Castander}, Francisco J. and {Connolly}, Andrew and {Csabai}, Istvan and {Doi}, Mamoru and {Fukugita}, Masataka and {Frieman}, Joshua A. and {Glazebrook}, Karl and {Gunn}, James E. and {Hendry}, John S. and {Hennessy}, Gregory and {Ivezi{\'c}}, Zeljko and {Kent}, Stephen and {Knapp}, Gillian R. and {Lin}, Huan and {Loh}, Yeong-Shang and {Lupton}, Robert H. and {Margon}, Bruce and {McKay}, Timothy A. and {Meiksin}, Avery and {Munn}, Jeffery A. and {Pope}, Adrian and {Richmond}, Michael W. and {Schlegel}, David and {Schneider}, Donald P. and {Shimasaku}, Kazuhiro and {Stoughton}, Christopher and {Strauss}, Michael A. and {SubbaRao}, Mark and {Szalay}, Alexander S. and {Szapudi}, Istv{\'a}n and {Tucker}, Douglas L. and {Yanny}, Brian and {York}, Donald G.},
        title = "{Detection of the Baryon Acoustic Peak in the Large-Scale Correlation Function of SDSS Luminous Red Galaxies}",
      journal = {The Astrophysical Journal},
         year = 2005,
        month = nov,
       volume = {633},
       number = {2},
        pages = {560-574},
          doi = {10.1086/466512},
archivePrefix = {arXiv},
       eprint = {astro-ph/0501171},
 primaryClass = {astro-ph},
       adsurl = {https://ui.adsabs.harvard.edu/abs/2005ApJ...633..560E}
}

@book{Fang1998,
		author = {Fang, L.Z. and Thews, R. L.},
	publisher = {Singapore: World Scientific},
	title = {Wavelets in Physics},
	year = {1998}}

@article{fang2000, 
year = {2000}, 
title = {{Measuring the Galaxy Power Spectrum and Scale-Scale Correlations with Multiresolution-decomposed Covariance. I. Method}}, 
author = {Fang, Li-Zhi and Feng, Long-long}, 
journal = {The Astrophysical Journal},
issn = {0004-637X}, 
doi = {10.1086/309207}, 
eprint = {astro-ph/0003259}, 
pages = {5--19}, 
number = {1}, 
volume = {539}, 
}

@article{Feldman1994,
  author  = {Feldman, Hume A. and Kaiser, Nick and Peacock, John A.},
  title   = {Power-spectrum analysis of three-dimensional redshift surveys},
  journal = {The Astrophysical Journal},
  volume  = {426},
  pages   = {23--37},
  year    = {1994},
  doi     = {10.1086/174036}
}

@article{Feng2007, 
year = {2007}, 
title = {{The Beylkin‐Cramer Summation Rule and a New Fast Algorithm of Cosmic Statistics for Large Data Sets}}, 
author = {Feng, Long-Long}, 
journal = {The Astrophysical Journal},
issn = {0004-637X}, 
doi = {10.1086/511024}, 
eprint = {astro-ph/0512167}, 
pages = {25--35}, 
number = {1}, 
volume = {658}, 
}

@ARTICLE{Friedrich2020,
       author = {{Friedrich}, Oliver and {Uhlemann}, Cora and {Villaescusa-Navarro}, Francisco and {Baldauf}, Tobias and {Manera}, Marc and {Nishimichi}, Takahiro},
        title = "{Primordial non-Gaussianity without tails - how to measure f$_{NL}$ with the bulk of the density PDF}",
      journal = {Monthly Notices of the Royal Astronomical Society},
         year = 2020,
        month = oct,
       volume = {498},
       number = {1},
        pages = {464-483},
          doi = {10.1093/mnras/staa2160},
archivePrefix = {arXiv},
       eprint = {1912.06621},
 primaryClass = {astro-ph.CO},
       adsurl = {https://ui.adsabs.harvard.edu/abs/2020MNRAS.498..464F}
}

@inproceedings{Friesen2017, 
author = {Friesen, Brian and Patwary, Md. Mostofa Ali and Austin, Brian and Satish, Nadathur and Slepian, Zachary and Sundaram, Narayanan and Bard, Deborah and Eisenstein, Daniel J. and Deslippe, Jack and Dubey, Pradeep and Prabhat}, 
title = {Galactos: computing the anisotropic 3-point correlation function for 2 billion galaxies}, 
year = {2017}, 
isbn = {9781450351140}, 
publisher = {Association for Computing Machinery}, 
address = {New York, NY, USA}, 
url = {https://doi.org/10.1145/3126908.3126927}, 
doi = {10.1145/3126908.3126927}, 
booktitle = {Proceedings of the International Conference for High Performance Computing, Networking, Storage and Analysis}, articleno = {20}, 
numpages = {11}, 
location = {Denver, Colorado}, 
series = {SC '17} 
}

@ARTICLE{Fry1993,
       author = {{Fry}, J.~N. and {Gaztanaga}, Enrique},
        title = "{Biasing and Hierarchical Statistics in Large-Scale Structure}",
      journal = {The Astrophysical Journal},
         year = 1993,
        month = aug,
       volume = {413},
        pages = {447},
          doi = {10.1086/173015},
archivePrefix = {arXiv},
       eprint = {astro-ph/9302009},
 primaryClass = {astro-ph},
       adsurl = {https://ui.adsabs.harvard.edu/abs/1993ApJ...413..447F}
}

@ARTICLE{Gagrani2017,
       author = {{Gagrani}, Praful and {Samushia}, Lado},
        title = "{Information Content of the Angular Multipoles of Redshift-Space Galaxy Bispectrum}",
      journal = {Monthly Notices of the Royal Astronomical Society},
         year = 2017,
        month = may,
       volume = {467},
       number = {1},
        pages = {928-935},
          doi = {10.1093/mnras/stx135},
archivePrefix = {arXiv},
       eprint = {1610.03488},
 primaryClass = {astro-ph.CO},
       adsurl = {https://ui.adsabs.harvard.edu/abs/2017MNRAS.467..928G}
}

@ARTICLE{Gaztanaga1994,
       author = {{Gaztanaga}, Enrique and {Frieman}, Joshua A.},
        title = "{Bias and High-Order Galaxy Correlation Functions in the APM Galaxy Survey}",
      journal = {The Astrophysical Journal Letters},
         year = 1994,
        month = dec,
       volume = {437},
        pages = {L13},
          doi = {10.1086/187671},
archivePrefix = {arXiv},
       eprint = {astro-ph/9407079},
 primaryClass = {astro-ph},
       adsurl = {https://ui.adsabs.harvard.edu/abs/1994ApJ...437L..13G}
}

@article{gaztanaga2005,
  title={The three-point function in large-scale structure: redshift distortions and galaxy bias},
  author={Gaztanaga, Enrique and Scoccimarro, R},
  journal={Monthly Notices of the Royal Astronomical Society},
  volume={361},
  number={3},
  pages={824--836},
  year={2005},
  publisher={Blackwell Science Ltd Oxford, UK}
}

@INPROCEEDINGS{Gray2004,
       author = {{Gray}, A.~G. and {Moore}, A.~W. and {Nichol}, R.~C. and {Connolly}, A.~J. and {Genovese}, C. and {Wasserman}, L.},
        title = "{Multi-Tree Methods for Statistics on Very Large Datasets in Astronomy}",
    booktitle = {Astronomical Data Analysis Software and Systems (ADASS) XIII},
         year = 2004,
       editor = {{Ochsenbein}, Francois and {Allen}, Mark G. and {Egret}, Daniel},
       series = {Astronomical Society of the Pacific Conference Series},
       volume = {314},
        month = jul,
        pages = {249},
          doi = {10.48550/arXiv.astro-ph/0401121},
archivePrefix = {arXiv},
       eprint = {astro-ph/0401121},
 primaryClass = {astro-ph},
       adsurl = {https://ui.adsabs.harvard.edu/abs/2004ASPC..314..249G}
}

@article{Gruen2018, 
year = {2018}, 
title = {{Density split statistics: Cosmological constraints from counts and lensing in cells in DES Y1 and SDSS data}}, 
author = {Gruen, D and Friedrich, O and Krause, E and DeRose, J and Cawthon, R and Davis, C and Elvin-Poole, J and Rykoff, E  S and Wechsler, R  H and Alarcon, A and Bernstein, G  M and Blazek, J and Chang, C and Clampitt, J and Crocce, M and Vicente, J De and Gatti, M and Gill, M  S  S and Hartley, W  G and Hilbert, S and Hoyle, B and Jain, B and Jarvis, M and Lahav, O and MacCrann, N and McClintock, T and Prat, J and Rollins, R  P and Ross, A  J and Rozo, E and Samuroff, S and Sánchez, C and Sheldon, E and Troxel, M  A and Zuntz, J and Abbott, T  M  C and Abdalla, F  B and Allam, S and Annis, J and Bechtol, K and Benoit-Lévy, A and Bertin, E and Bridle, S  L and Brooks, D and Buckley-Geer, E and Rosell, A Carnero and Kind, M Carrasco and Carretero, J and Cunha, C  E and D’Andrea, C  B and Costa, L  N da and Desai, S and Diehl, H  T and Dietrich, J  P and Doel, P and Drlica-Wagner, A and Fernandez, E and Flaugher, B and Fosalba, P and Frieman, J and García-Bellido, J and Gaztanaga, E and Giannantonio, T and Gruendl, R  A and Gschwend, J and Gutierrez, G and Honscheid, K and James, D  J and Jeltema, T and Kuehn, K and Kuropatkin, N and Lima, M and March, M and Marshall, J  L and Martini, P and Melchior, P and Menanteau, F and Miquel, R and Mohr, J  J and Plazas, A  A and Roodman, A and Sanchez, E and Scarpine, V and Schubnell, M and Sevilla-Noarbe, I and Smith, M and Smith, R  C and Soares-Santos, M and Sobreira, F and Swanson, M  E  C and Tarle, G and Thomas, D and Vikram, V and Walker, A  R and Weller, J and Zhang, Y and Collaboration, DES}, 
journal = {Physical Review D},
issn = {2470-0010}, 
doi = {10.1103/physrevd.98.023507}, 
eprint = {1710.05045}, 
pages = {023507}, 
number = {2}, 
volume = {98}
}

@ARTICLE{Guo2014,
       author = {{Guo}, Hong and {Li}, Cheng and {Jing}, Y.~P. and {B{\"o}rner}, Gerhard},
        title = "{Stellar Mass and Color Dependence of the Three-point Correlation Function of Galaxies in the Local Universe}",
      journal = {The Astrophysical Journal},
         year = 2014,
        month = jan,
       volume = {780},
       number = {2},
          eid = {139},
        pages = {139},
          doi = {10.1088/0004-637X/780/2/139},
archivePrefix = {arXiv},
       eprint = {1303.2609},
 primaryClass = {astro-ph.CO},
       adsurl = {https://ui.adsabs.harvard.edu/abs/2014ApJ...780..139G}
}

@article{Hamaus2012,
  author        = {Hamaus, Nico and Seljak, Uro{\v{s}} and Desjacques, Vincent},
  title         = {Optimal Weighting in Galaxy Surveys: Application to Redshift-Space Distortions},
  journal       = {Physical Review D},
  volume        = {86},
  number        = {10},
  pages         = {103513},
  year          = {2012},
  doi           = {10.1103/PhysRevD.86.103513},
  eprint        = {1207.1102},
  archivePrefix = {arXiv},
  primaryClass  = {astro-ph.CO}
}

@article{Hand2018, 
year = {2018}, 
title = {{nbodykit: An Open-source, Massively Parallel Toolkit for Large-scale Structure}}, 
author = {Hand, Nick and Feng, Yu and Beutler, Florian and Li, Yin and Modi, Chirag and Seljak, Uroš and Slepian, Zachary}, 
journal = {The Astronomical Journal},
doi = {10.3847/1538-3881/aadae0}, 
eprint = {1712.05834}, 
pages = {160}, 
number = {4}, 
volume = {156}, 
}

@article{Hartlap2007,
  author        = {Hartlap, J. and Simon, P. and Schneider, P.},
  title         = {Why Your Model Parameter Confidences Might Be Too
                   Optimistic. Unbiased Estimation of the Inverse
                   Covariance Matrix},
  journal       = {Astronomy \& Astrophysics},
  year          = {2007},
  volume        = {464},
  number        = {1},
  pages         = {399--404},
  doi           = {10.1051/0004-6361:20066170},
  archivePrefix = {arXiv},
  eprint        = {astro-ph/0608064}
}

@article{Hearin2017, 
year = {2017}, 
title = {{Forward Modeling of Large-scale Structure: An Open-source Approach with Halotools}}, 
author = {Hearin, Andrew P. and Campbell, Duncan and Tollerud, Erik and Behroozi, Peter and Diemer, Benedikt and Goldbaum, Nathan J. and Jennings, Elise and Leauthaud, Alexie and Mao, Yao-Yuan and More, Surhud and Parejko, John and Sinha, Manodeep and Sipöcz, Brigitta and Zentner, Andrew}, 
journal = {The Astronomical Journal},
issn = {0004-6256}, 
doi = {10.3847/1538-3881/aa859f}, 
eprint = {1606.04106}, 
pages = {190}, 
number = {5}, 
volume = {154}, 
}

@book{Hockney1988,
  author    = {Hockney, Roger W. and Eastwood, James W.},
  title     = {Computer Simulation Using Particles},
  edition   = {1},
  publisher = {Taylor \& Francis},
  year      = {1988},
  address   = {Boca Raton, FL},
  isbn      = {978-1560321195}
}

@article{Hou2022,
  title = {Analytic Gaussian covariance matrices for galaxy $N$-point correlation functions},
  author = {Hou, Jiamin and Cahn, Robert N. and Philcox, Oliver H. E. and Slepian, Zachary},
  journal = {Physical Review D},
  volume = {106},
  issue = {4},
  pages = {043515},
  numpages = {34},
  year = {2022},
  month = {Aug},
  publisher = {American Physical Society},
  doi = {10.1103/PhysRevD.106.043515},
  url = {https://link.aps.org/doi/10.1103/PhysRevD.106.043515}
}

@article{Hurtado2017, 
year = {2017}, 
title = {{The best fit for the observed galaxy counts-in-cell distribution function}}, 
author = {Hurtado-Gil, Lluís and Martínez, Vicent J. and Arnalte-Mur, Pablo and Pons-Bordería, María-Jesús and Pareja-Flores, Cristóbal and Paredes, Silvestre}, 
journal = {Astronomy \& Astrophysics},
issn = {0004-6361}, 
doi = {10.1051/0004-6361/201629097}, 
eprint = {1703.01087}, 
pages = {A40}, 
volume = {601}, 
}

@article{Ivanov2020,
doi = {10.1088/1475-7516/2020/05/042},
url = {https://dx.doi.org/10.1088/1475-7516/2020/05/042},
year = {2020},
month = {may},
publisher = {},
volume = {2020},
number = {05},
pages = {042},
author = {Mikhail M. Ivanov and Marko Simonović and Matias Zaldarriaga},
title = {Cosmological parameters from the BOSS galaxy power spectrum},
journal = {Journal of Cosmology and Astroparticle Physics}
}

@ARTICLE{Ivezic2019,
       author = {{Ivezi{\'c}}, {\v{Z}}eljko and {Kahn}, Steven M. and {Tyson}, J. Anthony and {Abel}, Bob and {Acosta}, Emily and {Allsman}, Robyn and {Alonso}, David and {AlSayyad}, Yusra and {Anderson}, Scott F. and {Andrew}, John and {Angel}, James Roger P. and {Angeli}, George Z. and {Ansari}, Reza and {Antilogus}, Pierre and {Araujo}, Constanza and {Armstrong}, Robert and {Arndt}, Kirk T. and {Astier}, Pierre and {Aubourg}, {\'E}ric and {Auza}, Nicole and {Axelrod}, Tim S. and {Bard}, Deborah J. and {Barr}, Jeff D. and {Barrau}, Aurelian and {Bartlett}, James G. and {Bauer}, Amanda E. and {Bauman}, Brian J. and {Baumont}, Sylvain and {Bechtol}, Ellen and {Bechtol}, Keith and {Becker}, Andrew C. and {Becla}, Jacek and {Beldica}, Cristina and {Bellavia}, Steve and {Bianco}, Federica B. and {Biswas}, Rahul and {Blanc}, Guillaume and {Blazek}, Jonathan and {Blandford}, Roger D. and {Bloom}, Josh S. and {Bogart}, Joanne and {Bond}, Tim W. and {Booth}, Michael T. and {Borgland}, Anders W. and {Borne}, Kirk and {Bosch}, James F. and {Boutigny}, Dominique and {Brackett}, Craig A. and {Bradshaw}, Andrew and {Brandt}, William Nielsen and {Brown}, Michael E. and {Bullock}, James S. and {Burchat}, Patricia and {Burke}, David L. and {Cagnoli}, Gianpietro and {Calabrese}, Daniel and {Callahan}, Shawn and {Callen}, Alice L. and {Carlin}, Jeffrey L. and {Carlson}, Erin L. and {Chandrasekharan}, Srinivasan and {Charles-Emerson}, Glenaver and {Chesley}, Steve and {Cheu}, Elliott C. and {Chiang}, Hsin-Fang and {Chiang}, James and {Chirino}, Carol and {Chow}, Derek and {Ciardi}, David R. and {Claver}, Charles F. and {Cohen-Tanugi}, Johann and {Cockrum}, Joseph J. and {Coles}, Rebecca and {Connolly}, Andrew J. and {Cook}, Kem H. and {Cooray}, Asantha and {Covey}, Kevin R. and {Cribbs}, Chris and {Cui}, Wei and {Cutri}, Roc and {Daly}, Philip N. and {Daniel}, Scott F. and {Daruich}, Felipe and {Daubard}, Guillaume and {Daues}, Greg and {Dawson}, William and {Delgado}, Francisco and {Dellapenna}, Alfred and {de Peyster}, Robert and {de Val-Borro}, Miguel and {Digel}, Seth W. and {Doherty}, Peter and {Dubois}, Richard and {Dubois-Felsmann}, Gregory P. and {Durech}, Josef and {Economou}, Frossie and {Eifler}, Tim and {Eracleous}, Michael and {Emmons}, Benjamin L. and {Fausti Neto}, Angelo and {Ferguson}, Henry and {Figueroa}, Enrique and {Fisher-Levine}, Merlin and {Focke}, Warren and {Foss}, Michael D. and {Frank}, James and {Freemon}, Michael D. and {Gangler}, Emmanuel and {Gawiser}, Eric and {Geary}, John C. and {Gee}, Perry and {Geha}, Marla and {Gessner}, Charles J.~B. and {Gibson}, Robert R. and {Gilmore}, D. Kirk and {Glanzman}, Thomas and {Glick}, William and {Goldina}, Tatiana and {Goldstein}, Daniel A. and {Goodenow}, Iain and {Graham}, Melissa L. and {Gressler}, William J. and {Gris}, Philippe and {Guy}, Leanne P. and {Guyonnet}, Augustin and {Haller}, Gunther and {Harris}, Ron and {Hascall}, Patrick A. and {Haupt}, Justine and {Hernandez}, Fabio and {Herrmann}, Sven and {Hileman}, Edward and {Hoblitt}, Joshua and {Hodgson}, John A. and {Hogan}, Craig and {Howard}, James D. and {Huang}, Dajun and {Huffer}, Michael E. and {Ingraham}, Patrick and {Innes}, Walter R. and {Jacoby}, Suzanne H. and {Jain}, Bhuvnesh and {Jammes}, Fabrice and {Jee}, M. James and {Jenness}, Tim and {Jernigan}, Garrett and {Jevremovi{\'c}}, Darko and {Johns}, Kenneth and {Johnson}, Anthony S. and {Johnson}, Margaret W.~G. and {Jones}, R. Lynne and {Juramy-Gilles}, Claire and {Juri{\'c}}, Mario and {Kalirai}, Jason S. and {Kallivayalil}, Nitya J. and {Kalmbach}, Bryce and {Kantor}, Jeffrey P. and {Karst}, Pierre and {Kasliwal}, Mansi M. and {Kelly}, Heather and {Kessler}, Richard and {Kinnison}, Veronica and {Kirkby}, David and {Knox}, Lloyd and {Kotov}, Ivan V. and {Krabbendam}, Victor L. and {Krughoff}, K. Simon and {Kub{\'a}nek}, Petr and {Kuczewski}, John and {Kulkarni}, Shri and {Ku}, John and {Kurita}, Nadine R. and {Lage}, Craig S. and {Lambert}, Ron and {Lange}, Travis and {Langton}, J. Brian and {Le Guillou}, Laurent and {Levine}, Deborah and {Liang}, Ming and {Lim}, Kian-Tat and {Lintott}, Chris J. and {Long}, Kevin E. and {Lopez}, Margaux and {Lotz}, Paul J. and {Lupton}, Robert H. and {Lust}, Nate B. and {MacArthur}, Lauren A. and {Mahabal}, Ashish and {Mandelbaum}, Rachel and {Markiewicz}, Thomas W. and {Marsh}, Darren S. and {Marshall}, Philip J. and {Marshall}, Stuart and {May}, Morgan and {McKercher}, Robert and {McQueen}, Michelle and {Meyers}, Joshua and {Migliore}, Myriam and {Miller}, Michelle and {Mills}, David J. and {Miraval}, Connor and {Moeyens}, Joachim and {Moolekamp}, Fred E. and {Monet}, David G. and {Moniez}, Marc and {Monkewitz}, Serge and {Montgomery}, Christopher and {Morrison}, Christopher B. and {Mueller}, Fritz and {Muller}, Gary P. and {Mu{\~n}oz Arancibia}, Freddy and {Neill}, Douglas R. and {Newbry}, Scott P. and {Nief}, Jean-Yves and {Nomerotski}, Andrei and {Nordby}, Martin and {O'Connor}, Paul and {Oliver}, John and {Olivier}, Scot S. and {Olsen}, Knut and {O'Mullane}, William and {Ortiz}, Sandra and {Osier}, Shawn and {Owen}, Russell E. and {Pain}, Reynald and {Palecek}, Paul E. and {Parejko}, John K. and {Parsons}, James B. and {Pease}, Nathan M. and {Peterson}, J. Matt and {Peterson}, John R. and {Petravick}, Donald L. and {Libby Petrick}, M.~E. and {Petry}, Cathy E. and {Pierfederici}, Francesco and {Pietrowicz}, Stephen and {Pike}, Rob and {Pinto}, Philip A. and {Plante}, Raymond and {Plate}, Stephen and {Plutchak}, Joel P. and {Price}, Paul A. and {Prouza}, Michael and {Radeka}, Veljko and {Rajagopal}, Jayadev and {Rasmussen}, Andrew P. and {Regnault}, Nicolas and {Reil}, Kevin A. and {Reiss}, David J. and {Reuter}, Michael A. and {Ridgway}, Stephen T. and {Riot}, Vincent J. and {Ritz}, Steve and {Robinson}, Sean and {Roby}, William and {Roodman}, Aaron and {Rosing}, Wayne and {Roucelle}, Cecille and {Rumore}, Matthew R. and {Russo}, Stefano and {Saha}, Abhijit and {Sassolas}, Benoit and {Schalk}, Terry L. and {Schellart}, Pim and {Schindler}, Rafe H. and {Schmidt}, Samuel and {Schneider}, Donald P. and {Schneider}, Michael D. and {Schoening}, William and {Schumacher}, German and {Schwamb}, Megan E. and {Sebag}, Jacques and {Selvy}, Brian and {Sembroski}, Glenn H. and {Seppala}, Lynn G. and {Serio}, Andrew and {Serrano}, Eduardo and {Shaw}, Richard A. and {Shipsey}, Ian and {Sick}, Jonathan and {Silvestri}, Nicole and {Slater}, Colin T. and {Smith}, J. Allyn and {Smith}, R. Chris and {Sobhani}, Shahram and {Soldahl}, Christine and {Storrie-Lombardi}, Lisa and {Stover}, Edward and {Strauss}, Michael A. and {Street}, Rachel A. and {Stubbs}, Christopher W. and {Sullivan}, Ian S. and {Sweeney}, Donald and {Swinbank}, John D. and {Szalay}, Alexander and {Takacs}, Peter and {Tether}, Stephen A. and {Thaler}, Jon J. and {Thayer}, John Gregg and {Thomas}, Sandrine and {Thornton}, Adam J. and {Thukral}, Vaikunth and {Tice}, Jeffrey and {Trilling}, David E. and {Turri}, Max and {Van Berg}, Richard and {Vanden Berk}, Daniel and {Vetter}, Kurt and {Virieux}, Francoise and {Vucina}, Tomislav and {Wahl}, William and {Walkowicz}, Lucianne and {Walsh}, Brian and {Walter}, Christopher W. and {Wang}, Daniel L. and {Wang}, Shin-Yawn and {Warner}, Michael and {Wiecha}, Oliver and {Willman}, Beth and {Winters}, Scott E. and {Wittman}, David and {Wolff}, Sidney C. and {Wood-Vasey}, W. Michael and {Wu}, Xiuqin and {Xin}, Bo and {Yoachim}, Peter and {Zhan}, Hu},
        title = "{LSST: From Science Drivers to Reference Design and Anticipated Data Products}",
      journal = {The Astrophysical Journal},
         year = 2019,
        month = mar,
       volume = {873},
       number = {2},
          eid = {111},
        pages = {111},
          doi = {10.3847/1538-4357/ab042c},
archivePrefix = {arXiv},
       eprint = {0805.2366},
 primaryClass = {astro-ph},
       adsurl = {https://ui.adsabs.harvard.edu/abs/2019ApJ...873..111I}
}

@ARTICLE{Jarvis2004,
       author = {{Jarvis}, M. and {Bernstein}, G. and {Jain}, B.},
        title = "{The skewness of the aperture mass statistic}",
      journal = {Monthly Notices of the Royal Astronomical Society},
         year = 2004,
        month = jul,
       volume = {352},
       number = {1},
        pages = {338-352},
          doi = {10.1111/j.1365-2966.2004.07926.x},
archivePrefix = {arXiv},
       eprint = {astro-ph/0307393},
 primaryClass = {astro-ph},
       adsurl = {https://ui.adsabs.harvard.edu/abs/2004MNRAS.352..338J}
}

@ARTICLE{Jing1998,
       author = {{Jing}, Y.~P. and {B{\"o}rner}, G.},
        title = "{The Three-Point Correlation Function of Galaxies Determined from the Las Campanas Redshift Survey}",
      journal = {The Astrophysical Journal},
         year = 1998,
        month = aug,
       volume = {503},
       number = {1},
        pages = {37-47},
          doi = {10.1086/305997},
archivePrefix = {arXiv},
       eprint = {astro-ph/9802011},
 primaryClass = {astro-ph},
       adsurl = {https://ui.adsabs.harvard.edu/abs/1998ApJ...503...37J}
}

@article{Jing2004,
doi = {10.1086/383343},
url = {https://dx.doi.org/10.1086/383343},
year = {2004},
month = {may},
publisher = {},
volume = {607},
number = {1},
pages = {140},
author = {Y. P. Jing and G. Börner},
title = {The Three-Point Correlation Function of Galaxies Determined from the Two-Degree Field Galaxy Redshift Survey},
journal = {The Astrophysical Journal}
}

@ARTICLE{Jiutian,
       author = {{Han}, Jiaxin and {Li}, Ming and {Jiang}, Wenkang and
                 {Chen}, Zhao and {Wang}, Huiyuan and {Wei}, Chengliang and
                 {He}, Feihong and {He}, Jianhua and {Zhang}, Jiajun and
                 {Liu}, Yu and {Cui}, Weiguang and {Gu}, Yizhou and
                 {Guo}, Qi and {Jing}, Yipeng and {Kang}, Xi and
                 {Li}, Guoliang and {Luo}, Xiong and {Luo}, Yu and
                 {Pei}, Wenxiang and {Qiu}, Yisheng and {Tan}, Zhenlin and
                 {Xie}, Lizhi and {Yang}, Xiaohu and {Yu}, Hao-Ran and
                 {Yu}, Yu and {Zhou}, Jiale},
        title = "{The Jiutian simulations for the CSST extra-galactic surveys}",
      journal = {Science China Physics, Mechanics \& Astronomy},
         year = {2025},
       volume = {68},
        pages = {109511},
          doi = {10.1007/s11433-025-2712-1},
archivePrefix = {arXiv},
       eprint = {2503.21368},
 primaryClass = {astro-ph.CO}
}

@article{joachimi2017,
  title={Non-linear shrinkage estimation of large-scale structure covariance},
  author={Joachimi, Benjamin},
  journal={Monthly Notices of the Royal Astronomical Society},
  volume={466},
  number={1},
  pages={L83--L87},
  year={2017},
  publisher={Oxford University Press}
}

@article{Ju2026,
    author = {Ju, Wenjie and Feng, Longlong and Huang, Zhiqi and Sun, Xin and Zhu, Weishan},
    title = {An optimal in situ multipole algorithm for the isotropic three-point correlation function},
    journal = {Monthly Notices of the Royal Astronomical Society},
    volume = {546},
    number = {1},
    pages = {staf2275},
    year = {2026},
    month = {02},
    issn = {0035-8711},
    doi = {10.1093/mnras/staf2275},
    url = {https://doi.org/10.1093/mnras/staf2275},
    eprint = {https://academic.oup.com/mnras/article-pdf/546/1/staf2275/66132035/staf2275.pdf},
}

@article{Karim2023, 
year = {2023}, 
title = {{On the impact of the galaxy window function on cosmological parameter estimation}}, 
author = {Karim, Tanveer and Rezaie, Mehdi and Singh, Sukhdeep and Eisenstein, Daniel}, 
journal = {Monthly Notices of the Royal Astronomical Society}, 
issn = {0035-8711}, 
doi = {10.1093/mnras/stad2210}, 
eprint = {2305.11956}, 
pages = {311--324}, 
number = {1}, 
volume = {525}, 
}

@article{Kayo2001,
  author        = {Kayo, Issha and Taruya, Atsushi and Suto, Yasushi},
  title         = {Probability Distribution Function of Cosmological Density
                   Fluctuations from a Gaussian Initial Condition:
                   Comparison of One-Point and Two-Point Lognormal Model
                   Predictions with {N}-Body Simulations},
  journal       = {The Astrophysical Journal},
  year          = {2001},
  volume        = {561},
  number        = {1},
  pages         = {22--34},
  doi           = {10.1086/323227},
  archivePrefix = {arXiv},
  eprint        = {astro-ph/0105218}
}

@article{Landy1993,
  author  = {Landy, Stephen D. and Szalay, Alexander S.},
  title   = {Bias and variance of angular correlation functions},
  journal = {The Astrophysical Journal},
  year    = {1993},
  volume  = {412},
  number  = {1},
  pages   = {64--71},
  doi     = {10.1086/172900}
}

@misc{laureijs2011,
      title={Euclid Definition Study Report}, 
      author={R. Laureijs and J. Amiaux and S. Arduini and J. -L. Auguères and J. Brinchmann and R. Cole and M. Cropper and C. Dabin and L. Duvet and A. Ealet and B. Garilli and P. Gondoin and L. Guzzo and J. Hoar and H. Hoekstra and R. Holmes and T. Kitching and T. Maciaszek and Y. Mellier and F. Pasian and W. Percival and J. Rhodes and G. Saavedra Criado and M. Sauvage and R. Scaramella and L. Valenziano and S. Warren and R. Bender and F. Castander and A. Cimatti and O. Le Fèvre and H. Kurki-Suonio and M. Levi and P. Lilje and G. Meylan and R. Nichol and K. Pedersen and V. Popa and R. Rebolo Lopez and H. -W. Rix and H. Rottgering and W. Zeilinger and F. Grupp and P. Hudelot and R. Massey and M. Meneghetti and L. Miller and S. Paltani and S. Paulin-Henriksson and S. Pires and C. Saxton and T. Schrabback and G. Seidel and J. Walsh and N. Aghanim and L. Amendola and J. Bartlett and C. Baccigalupi and J. -P. Beaulieu and K. Benabed and J. -G. Cuby and D. Elbaz and P. Fosalba and G. Gavazzi and A. Helmi and I. Hook and M. Irwin and J. -P. Kneib and M. Kunz and F. Mannucci and L. Moscardini and C. Tao and R. Teyssier and J. Weller and G. Zamorani and M. R. Zapatero Osorio and O. Boulade and J. J. Foumond and A. Di Giorgio and P. Guttridge and A. James and M. Kemp and J. Martignac and A. Spencer and D. Walton and T. Blümchen and C. Bonoli and F. Bortoletto and C. Cerna and L. Corcione and C. Fabron and K. Jahnke and S. Ligori and F. Madrid and L. Martin and G. Morgante and T. Pamplona and E. Prieto and M. Riva and R. Toledo and M. Trifoglio and F. Zerbi and F. Abdalla and M. Douspis and C. Grenet and S. Borgani and R. Bouwens and F. Courbin and J. -M. Delouis and P. Dubath and A. Fontana and M. Frailis and A. Grazian and J. Koppenhöfer and O. Mansutti and M. Melchior and M. Mignoli and J. Mohr and C. Neissner and K. Noddle and M. Poncet and M. Scodeggio and S. Serrano and N. Shane and J. -L. Starck and C. Surace and A. Taylor and G. Verdoes-Kleijn and C. Vuerli and O. R. Williams and A. Zacchei and B. Altieri and I. Escudero Sanz and R. Kohley and T. Oosterbroek and P. Astier and D. Bacon and S. Bardelli and C. Baugh and F. Bellagamba and C. Benoist and D. Bianchi and A. Biviano and E. Branchini and C. Carbone and V. Cardone and D. Clements and S. Colombi and C. Conselice and G. Cresci and N. Deacon and J. Dunlop and C. Fedeli and F. Fontanot and P. Franzetti and C. Giocoli and J. Garcia-Bellido and J. Gow and A. Heavens and P. Hewett and C. Heymans and A. Holland and Z. Huang and O. Ilbert and B. Joachimi and E. Jennins and E. Kerins and A. Kiessling and D. Kirk and R. Kotak and O. Krause and O. Lahav and F. van Leeuwen and J. Lesgourgues and M. Lombardi and M. Magliocchetti and K. Maguire and E. Majerotto and R. Maoli and F. Marulli and S. Maurogordato and H. McCracken and R. McLure and A. Melchiorri and A. Merson and M. Moresco and M. Nonino and P. Norberg and J. Peacock and R. Pello and M. Penny and V. Pettorino and C. Di Porto and L. Pozzetti and C. Quercellini and M. Radovich and A. Rassat and N. Roche and S. Ronayette and E. Rossetti and B. Sartoris and P. Schneider and E. Semboloni and S. Serjeant and F. Simpson and C. Skordis and G. Smadja and S. Smartt and P. Spano and S. Spiro and M. Sullivan and A. Tilquin and R. Trotta and L. Verde and Y. Wang and G. Williger and G. Zhao and J. Zoubian and E. Zucca},
      year={2011},
      eprint={1110.3193},
      archivePrefix={arXiv},
      primaryClass={astro-ph.CO}
}

@article{Lee2019PyWaveletsAP,
  title={PyWavelets: A Python package for wavelet analysis},
  author={Gregory R. Lee and Ralf Gommers and Filip Wasilewski and Kai Wohlfahrt and Aaron O'Leary},
  journal={Journal of Open Source Software},
  year={2019},
  volume={4},
  pages={1237},
  url={https://api.semanticscholar.org/CorpusID:146106279}
}

@article{Leicht2018, 
year = {2018}, 
title = {{Extreme spheres: counts-in-cells for 21cm intensity mapping}}, 
author = {Leicht, Oliver and Uhlemann, Cora and Villaescusa-Navarro, Francisco and Codis, Sandrine and Hernquist, Lars and Genel, Shy}, 
journal = {Monthly Notices of the Royal Astronomical Society},
issn = {0035-8711}, 
doi = {10.1093/mnras/sty3469}, 
eprint = {1808.09968}, 
pages = {269--281}, 
number = {1}, 
volume = {484}
}

@INPROCEEDINGS{March2012,
  author={March, William B. and Czechowski, Kenneth and Dukhan, Marat and Benson, Thomas and Lee, Dongryeol and Connolly, Andrew J. and Vuduc, Richard and Chow, Edmond and Gray, Alexander G.},
  booktitle={SC '12: Proceedings of the International Conference on High Performance Computing, Networking, Storage and Analysis}, 
  title={Optimizing the computation of n-point correlations on large-scale astronomical data}, 
  year={2012},
  volume={},
  number={},
  pages={1-12},
  doi={10.1109/SC.2012.89}}

@article{Marulli2016, 
year = {2016}, 
title = {{CosmoBolognaLib: C++ libraries for cosmological calculations}}, 
author = {Marulli, F. and Veropalumbo, A. and Moresco, M.}, 
journal = {Astronomy and Computing}, 
issn = {2213-1337}, 
doi = {10.1016/j.ascom.2016.01.005}, 
eprint = {1511.00012}, 
pages = {35--42}, 
volume = {14}, 
}

@article{Massara2021, 
year = {2021}, 
title = {{Using the Marked Power Spectrum to Detect the Signature of Neutrinos in Large-Scale Structure}}, 
author = {Massara, Elena and Villaescusa-Navarro, Francisco and Ho, Shirley and Dalal, Neal and Spergel, David N.}, 
journal = {Physical Review Letters},
issn = {0031-9007}, 
doi = {10.1103/physrevlett.126.011301}, 
pmid = {33480786}, 
eprint = {2001.11024}, 
pages = {011301}, 
number = {1}, 
volume = {126}
}

@article{Massara2023,
  author        = {Massara, Elena and
                   Villaescusa-Navarro, Francisco and
                   Hahn, ChangHoon and
                   Abidi, Muntazir M. and
                   Eickenberg, Michael and
                   Ho, Shirley and
                   Lemos, Pablo and
                   Moradinezhad Dizgah, Azadeh and
                   R{\'e}galdo-Saint Blancard, Bruno},
  title         = {Cosmological Information in the Marked Power Spectrum
                   of the Galaxy Field},
  journal       = {The Astrophysical Journal},
  year          = {2023},
  volume        = {951},
  number        = {1},
  pages         = {70},
  doi           = {10.3847/1538-4357/acd44d},
  archivePrefix = {arXiv},
  eprint        = {2206.01709},
  primaryClass  = {astro-ph.CO}
}

@article{Matarrese2000, 
year = {2000}, 
title = {{The Abundance of High-Redshift Objects as a Probe of Non-Gaussian Initial Conditions}}, 
author = {Matarrese, Sabino and Verde, Licia and Jimenez, Raul}, 
journal = {The Astrophysical Journal},
issn = {0004-637X}, 
doi = {10.1086/309412}, 
eprint = {astro-ph/0001366}, 
pages = {10--24}, 
number = {1}, 
volume = {541}
}

@ARTICLE{McCarthyGould2024,
       author = {{McCarthy Gould}, Beth and {Castiblanco}, Lina and {Uhlemann}, Cora and {Friedrich}, Oliver},
        title = "{Cosmology on point: modelling spectroscopic tracer one-point statistics}",
      journal = {The Open Journal of Astrophysics},
         year = 2025,
        month = jan,
       volume = {8},
          doi = {10.33232/001c.127800},
archivePrefix = {arXiv},
       eprint = {2409.18182},
 primaryClass = {astro-ph.CO}
}

@article{McDonaldSeljak2009,
  author        = {McDonald, Patrick and Seljak, Uro{\v{s}}},
  title         = {How to Evade the Sample Variance Limit on Measurements of Redshift-Space Distortions},
  journal       = {Journal of Cosmology and Astroparticle Physics},
  volume        = {2009},
  number        = {10},
  pages         = {007},
  year          = {2009},
  doi           = {10.1088/1475-7516/2009/10/007},
  eprint        = {0810.0323},
  archivePrefix = {arXiv},
  primaryClass  = {astro-ph}
}

@INPROCEEDINGS{Moore2001,
       author = {{Moore}, Andrew W. and {Connolly}, Andy J. and {Genovese}, Chris and {Gray}, Alex and {Grone}, Larry and {Kanidoris}, Nick, II and {Nichol}, Robert C. and {Schneider}, Jeff and {Szalay}, Alex S. and {Szapudi}, Istvan and {Wasserman}, Larry},
        title = "{Fast Algorithms and Efficient Statistics: N-Point Correlation Functions}",
    booktitle = {Mining the Sky},
         year = 2001,
       editor = {{Banday}, Anthony J. and {Zaroubi}, Saleem and {Bartelmann}, Matthias},
        month = jan,
        pages = {71},
          doi = {10.1007/10849171_5},
archivePrefix = {arXiv},
       eprint = {astro-ph/0012333},
 primaryClass = {astro-ph},
       adsurl = {https://ui.adsabs.harvard.edu/abs/2001misk.conf...71M}
}

@ARTICLE{Neyrinck2006,
       author = {{Neyrinck}, Mark C. and {Szapudi}, Istv{\'a}n and {Rimes}, Christopher D.},
        title = "{Information content in the halo-model dark-matter power spectrum}",
      journal = {Monthly Notices of the Royal Astronomical Society},
         year = 2006,
        month = jul,
       volume = {370},
       number = {1},
        pages = {L66-L70},
          doi = {10.1111/j.1745-3933.2006.00190.x},
archivePrefix = {arXiv},
       eprint = {astro-ph/0604282},
 primaryClass = {astro-ph},
       adsurl = {https://ui.adsabs.harvard.edu/abs/2006MNRAS.370L..66N}
}

@ARTICLE{Nichol2006,
       author = {{Nichol}, R.~C. and {Sheth}, R.~K. and {Suto}, Y. and {Gray}, A.~J. and {Kayo}, I. and {Wechsler}, R.~H. and {Marin}, F. and {Kulkarni}, G. and {Blanton}, M. and {Connolly}, A.~J. and {Gardner}, J.~P. and {Jain}, B. and {Miller}, C.~J. and {Moore}, A.~W. and {Pope}, A. and {Pun}, J. and {Schneider}, D. and {Schneider}, J. and {Szalay}, A. and {Szapudi}, I. and {Zehavi}, I. and {Bahcall}, N.~A. and {Csabai}, I. and {Brinkmann}, J.},
        title = "{The effect of large-scale structure on the SDSS galaxy three-point correlation function}",
      journal = {Monthly Notices of the Royal Astronomical Society},
         year = 2006,
        month = jun,
       volume = {368},
       number = {4},
        pages = {1507-1514},
          doi = {10.1111/j.1365-2966.2006.10239.x},
archivePrefix = {arXiv},
       eprint = {astro-ph/0602548},
 primaryClass = {astro-ph},
       adsurl = {https://ui.adsabs.harvard.edu/abs/2006MNRAS.368.1507N}
}

@ARTICLE{Novell-Masot2023,
       author = {{Novell-Masot}, Sergi and {Gualdi}, Davide and {Gil-Mar{\'\i}n}, H{\'e}ctor and {Verde}, Licia},
        title = "{GEO-FPT: a model of the galaxy bispectrum at mildly non-linear scales}",
      journal = {Journal of Cosmology and Astroparticle Physics},
         year = 2023,
        month = nov,
       volume = {2023},
       number = {11},
          eid = {044},
        pages = {044},
          doi = {10.1088/1475-7516/2023/11/044},
archivePrefix = {arXiv},
       eprint = {2303.15510},
 primaryClass = {astro-ph.CO},
       adsurl = {https://ui.adsabs.harvard.edu/abs/2023JCAP...11..044N}
}

@article{Padmanabhan2016,
    author = {Padmanabhan, Nikhil and White, Martin and Zhou, Harrison H. and O'Connell, Ross},
    title = {Estimating sparse precision matrices},
    journal = {Monthly Notices of the Royal Astronomical Society},
    volume = {460},
    number = {2},
    pages = {1567-1576},
    year = {2016},
    month = {08},
    issn = {0035-8711},
    doi = {10.1093/mnras/stw1042},
    url = {https://doi.org/10.1093/mnras/stw1042},
    eprint = {https://academic.oup.com/mnras/article-pdf/460/2/1567/8115592/stw1042.pdf},
}

@article{Paillas2021,
    author = {Paillas, Enrique and Cai, Yan-Chuan and Padilla, Nelson and Sánchez, Ariel G},
    title = "{Redshift-space distortions with split densities}",
    journal = {Monthly Notices of the Royal Astronomical Society},
    volume = {505},
    number = {4},
    pages = {5731-5752},
    year = {2021},
    month = {06},
    issn = {0035-8711},
    doi = {10.1093/mnras/stab1654},
    url = {https://doi.org/10.1093/mnras/stab1654},
    eprint = {https://academic.oup.com/mnras/article-pdf/505/4/5731/38864518/stab1654.pdf},
}

@ARTICLE{Paillas2023a,
       author = {{Paillas}, Enrique and {Cuesta-Lazaro}, Carolina and {Percival}, Will J. and {Nadathur}, Seshadri and {Cai}, Yan-Chuan and {Yuan}, Sihan and {Beutler}, Florian and {de Mattia}, Arnaud and {Eisenstein}, Daniel and {Forero-Sanchez}, Daniel and {Padilla}, Nelson and {Pinon}, Mathilde and {Ruhlmann-Kleider}, Vanina and {S{\'a}nchez}, Ariel G. and {Valogiannis}, Georgios and {Zarrouk}, Pauline},
        title = "{Cosmological constraints from density-split clustering in the BOSS CMASS galaxy sample}",
      journal = {Monthly Notices of the Royal Astronomical Society},
         year = 2024,
        month = jun,
       volume = {531},
       number = {1},
        pages = {898--918},
          doi = {10.1093/mnras/stae1118},
archivePrefix = {arXiv},
       eprint = {2309.16541},
 primaryClass = {astro-ph.CO},
       adsurl = {https://ui.adsabs.harvard.edu/abs/2023arXiv230916541P}
}

@article{Paillas2023b,
    author = {Paillas, Enrique and Cuesta-Lazaro, Carolina and Zarrouk, Pauline and Cai, Yan-Chuan and Percival, Will J and Nadathur, Seshadri and Pinon, Mathilde and de Mattia, Arnaud and Beutler, Florian},
    title = "{Constraining νΛCDM with density-split clustering}",
    journal = {Monthly Notices of the Royal Astronomical Society},
    volume = {522},
    number = {1},
    pages = {606-625},
    year = {2023},
    month = {04},
    issn = {0035-8711},
    doi = {10.1093/mnras/stad1017},
    url = {https://doi.org/10.1093/mnras/stad1017},
    eprint = {https://academic.oup.com/mnras/article-pdf/522/1/606/49992302/stad1017.pdf},
}

@article{Pan2005, 
year = {2005}, 
title = {{The monopole moment of the three‐point correlation function of the two‐degree Field Galaxy Redshift Survey}}, 
author = {Pan, Jun and Szapudi, István}, 
journal = {Monthly Notices of the Royal Astronomical Society},
issn = {0035-8711}, 
doi = {10.1111/j.1365-2966.2005.09407.x}, 
eprint = {astro-ph/0505422}, 
pages = {1363--1370}, 
number = {4}, 
volume = {362}, 
}

@article{Paz2015,
    author = {Paz, Dante J. and Sánchez, Ariel G.},
    title = {Improving the precision matrix for precision cosmology},
    journal = {Monthly Notices of the Royal Astronomical Society},
    volume = {454},
    number = {4},
    pages = {4326-4334},
    year = {2015},
    month = {12},
    issn = {0035-8711},
    doi = {10.1093/mnras/stv2259},
    url = {https://doi.org/10.1093/mnras/stv2259},
    eprint = {https://academic.oup.com/mnras/article-pdf/454/4/4326/18505893/stv2259.pdf},
}

@article{Percival2004,
  author  = {Percival, Will J. and Verde, Licia and Peacock, John A.},
  title   = {Fourier analysis of luminosity-dependent galaxy clustering},
  journal = {Monthly Notices of the Royal Astronomical Society},
  volume  = {347},
  number  = {2},
  pages   = {645--653},
  year    = {2004},
  doi     = {10.1111/j.1365-2966.2004.07245.x}
}

@article{Percival2011,
    author = {Percival, Will J. and Reid, Beth A. and Eisenstein, Daniel J. and Bahcall, Neta A. and Budavari, Tamas and Frieman, Joshua A. and Fukugita, Masataka and Gunn, James E. and Ivezić, Željko and Knapp, Gillian R. and Kron, Richard G. and Loveday, Jon and Lupton, Robert H. and McKay, Timothy A. and Meiksin, Avery and Nichol, Robert C. and Pope, Adrian C. and Schlegel, David J. and Schneider, Donald P. and Spergel, David N. and Stoughton, Chris and Strauss, Michael A. and Szalay, Alexander S. and Tegmark, Max and Vogeley, Michael S. and Weinberg, David H. and York, Donald G. and Zehavi, Idit},
    title = "{Baryon acoustic oscillations in the Sloan Digital Sky Survey Data Release 7 galaxy sample}",
    journal = {Monthly Notices of the Royal Astronomical Society},
    volume = {401},
    number = {4},
    pages = {2148-2168},
    year = {2010},
    month = {01},
    issn = {0035-8711},
    doi = {10.1111/j.1365-2966.2009.15812.x},
    url = {https://doi.org/10.1111/j.1365-2966.2009.15812.x},
    eprint = {https://academic.oup.com/mnras/article-pdf/401/4/2148/3901461/mnras0401-2148.pdf},
}

@ARTICLE{Philcox2019,
       author = {{Philcox}, Oliver H.~E. and {Eisenstein}, Daniel J.},
        title = "{Estimating covariance matrices for two- and three-point correlation function moments in Arbitrary Survey Geometries}",
      journal = {Monthly Notices of the Royal Astronomical Society},
         year = 2019,
        month = dec,
       volume = {490},
       number = {4},
        pages = {5931-5951},
          doi = {10.1093/mnras/stz2896},
archivePrefix = {arXiv},
       eprint = {1910.04764},
 primaryClass = {astro-ph.CO},
       adsurl = {https://ui.adsabs.harvard.edu/abs/2019MNRAS.490.5931P}
}

@article{Philcox2020, 
year = {2020}, 
title = {{What does the marked power spectrum measure? Insights from perturbation theory}}, 
author = {Philcox, Oliver H. E. and Massara, Elena and Spergel, David N.}, 
journal = {Physical Review D},
issn = {2470-0010}, 
doi = {10.1103/physrevd.102.043516}, 
eprint = {2006.10055}, 
pages = {043516}, 
number = {4}, 
volume = {102}
}

@ARTICLE{Philcox2021a,
       author = {{Philcox}, Oliver H.~E.},
        title = "{A faster Fourier transform? Computing small-scale power spectra and bispectra for cosmological simulations in {\u{g}}�'{\textordfeminine}(N$^{2}$) time}",
      journal = {Monthly Notices of the Royal Astronomical Society},
         year = 2021,
        month = mar,
       volume = {501},
       number = {3},
        pages = {4004-4034},
          doi = {10.1093/mnras/staa3882},
archivePrefix = {arXiv},
       eprint = {2005.01739},
 primaryClass = {astro-ph.CO},
       adsurl = {https://ui.adsabs.harvard.edu/abs/2021MNRAS.501.4004P}
}

@article{Philcox2021b,
    author = {Philcox, Oliver H E and Slepian, Zachary and Hou, Jiamin and Warner, Craig and Cahn, Robert N and Eisenstein, Daniel J},
    title = "{encore: an O (Ng2) estimator for galaxy N-point correlation functions}",
    journal = {Monthly Notices of the Royal Astronomical Society},
    volume = {509},
    number = {2},
    pages = {2457-2481},
    year = {2021},
    month = {10},
    issn = {0035-8711},
    doi = {10.1093/mnras/stab3025},
    url = {https://doi.org/10.1093/mnras/stab3025},
    eprint = {https://academic.oup.com/mnras/article-pdf/509/2/2457/41227438/stab3025.pdf},
}

@article{Philcox2022, 
year = {2022}, 
title = {{Efficient computation of N-point correlation functions in D dimensions}}, 
author = {Philcox, Oliver H. E. and Slepian, Zachary}, 
journal = {Proceedings of the National Academy of Sciences}, 
issn = {0027-8424}, 
doi = {10.1073/pnas.2111366119}, 
pmid = {35939667}, 
pmcid = {PMC9388109}, 
eprint = {2106.10278}, 
pages = {e2111366119}, 
number = {33}, 
volume = {119}
}

@article{PolyBin3D,
  author        = {Philcox, Oliver H. E. and Fl{\"o}ss, Thomas},
  title         = {A Suite of Optimal and Efficient Power Spectrum and
                   Bispectrum Estimators for Large-Scale Structure Analyses},
  journal       = {Physical Review D},
  year          = {2025},
  volume        = {112},
  number        = {6},
  pages         = {063507},
  doi           = {10.1103/rr6c-cf3s},
  archivePrefix = {arXiv},
  eprint        = {2404.07249},
  primaryClass  = {astro-ph.CO}
}

@article{Pope2008,
    author = {Pope, Adrian C. and Szapudi, István},
    title = {Shrinkage estimation of the power spectrum covariance matrix},
    journal = {Monthly Notices of the Royal Astronomical Society},
    volume = {389},
    number = {2},
    pages = {766-774},
    year = {2008},
    month = {09},
    issn = {0035-8711},
    doi = {10.1111/j.1365-2966.2008.13561.x},
    url = {https://doi.org/10.1111/j.1365-2966.2008.13561.x},
    eprint = {https://academic.oup.com/mnras/article-pdf/389/2/766/2957244/mnras0389-0766.pdf},
}

@ARTICLE{Portillo2018,
       author = {{Portillo}, Stephen K.~N. and {Slepian}, Zachary and {Burkhart}, Blakesley and {Kahraman}, Sule and {Finkbeiner}, Douglas P.},
        title = "{Developing the 3-point Correlation Function for the Turbulent Interstellar Medium}",
      journal = {The Astrophysical Journal},
         year = 2018,
        month = aug,
       volume = {862},
       number = {2},
          eid = {119},
        pages = {119},
          doi = {10.3847/1538-4357/aacb80},
archivePrefix = {arXiv},
       eprint = {1711.09907},
 primaryClass = {astro-ph.CO},
       adsurl = {https://ui.adsabs.harvard.edu/abs/2018ApJ...862..119P}
}

@ARTICLE{Rashkovetskyi2023,
       author = {{Rashkovetskyi}, Michael and {Eisenstein}, Daniel J. and {Aguilar}, Jessica Nicole and {Brooks}, David and {Claybaugh}, Todd and {Cole}, Shaun and {Dawson}, Kyle and {de la Macorra}, Axel and {Doel}, Peter and {Fanning}, Kevin and {Font-Ribera}, Andreu and {Forero-Romero}, Jaime E. and {Gontcho A Gontcho}, Satya and {Hahn}, ChangHoon and {Honscheid}, Klaus and {Kehoe}, Robert and {Kisner}, Theodore and {Landriau}, Martin and {Levi}, Michael and {Manera}, Marc and {Miquel}, Ramon and {Moon}, Jeongin and {Nadathur}, Seshadri and {Nie}, Jundan and {Poppett}, Claire and {Ross}, Ashley J. and {Rossi}, Graziano and {Sanchez}, Eusebio and {Saulder}, Christoph and {Schubnell}, Michael and {Seo}, Hee-Jong and {Tarle}, Gregory and {Valcin}, David and {Weaver}, Benjamin Alan and {Zhao}, Cheng and {Zhou}, Zhimin and {Zou}, Hu},
        title = "{Validation of semi-analytical, semi-empirical covariance matrices for two-point correlation function for early DESI data}",
      journal = {Monthly Notices of the Royal Astronomical Society},
         year = 2023,
        month = sep,
       volume = {524},
       number = {3},
        pages = {3894-3911},
          doi = {10.1093/mnras/stad2078},
archivePrefix = {arXiv},
       eprint = {2306.06320},
 primaryClass = {astro-ph.CO},
       adsurl = {https://ui.adsabs.harvard.edu/abs/2023MNRAS.524.3894R}
}

@article{Repp2020, 
year = {2020}, 
title = {{Galaxy bias and σ8 from counts in cells from the SDSS main sample}}, 
author = {Repp, Andrew and Szapudi, István}, 
journal = {Monthly Notices of the Royal Astronomical Society},
issn = {1745-3925}, 
doi = {10.1093/mnrasl/slaa139}, 
eprint = {2006.01146}, 
pages = {L125--L129}, 
number = {1}, 
volume = {498}, 
}

@article{Rimes2005,
    author = {Rimes, Christopher D. and Hamilton, Andrew J. S.},
    title = "{Information content of the non-linear matter power spectrum}",
    journal = {Monthly Notices of the Royal Astronomical Society},
    volume = {360},
    number = {1},
    pages = {L82-L86},
    year = {2005},
    month = {06},
    issn = {1745-3925},
    doi = {10.1111/j.1745-3933.2005.00051.x},
    url = {https://doi.org/10.1111/j.1745-3933.2005.00051.x},
    eprint = {https://academic.oup.com/mnrasl/article-pdf/360/1/L82/54693321/mnrasl\_360\_1\_l82.pdf},
}

@ARTICLE{Rohin2018,
       author = {{Rohin}, Y.},
        title = "{correlcalc: A 'generic' recipe for calculation of two-point correlation function}",
      journal = {Astronomy and Computing},
         year = 2018,
        month = oct,
       volume = {25},
        pages = {149-158},
          doi = {10.1016/j.ascom.2018.09.011},
archivePrefix = {arXiv},
       eprint = {1710.01723},
 primaryClass = {astro-ph.CO},
       adsurl = {https://ui.adsabs.harvard.edu/abs/2018A&C....25..149R}
}

@article{Ruggeri2017,
  author        = {Ruggeri, Rossana and Percival, Will J.
                   and Gil-Mar{\'i}n, H{\'e}ctor and Zhu, Fangzhou
                   and Zhao, Gong-Bo and Wang, Yuting},
  title         = {Optimal Redshift Weighting for Redshift-Space Distortions},
  journal       = {Monthly Notices of the Royal Astronomical Society},
  volume        = {464},
  number        = {3},
  pages         = {2698--2707},
  year          = {2017},
  doi           = {10.1093/mnras/stw2422},
  eprint        = {1602.05195},
  archivePrefix = {arXiv},
  primaryClass  = {astro-ph.CO}
}

@article{Sabiu2019, 
year = {2019}, 
title = {{Graph Database Solution for Higher-order Spatial Statistics in the Era of Big Data}}, 
author = {Sabiu, Cristiano G. and Hoyle, Ben and Kim, Juhan and Li, Xiao-Dong}, 
journal = {The Astrophysical Journal Supplement Series},
issn = {0067-0049}, 
doi = {10.3847/1538-4365/ab22b5}, 
eprint = {1901.00296}, 
pages = {29}, 
number = {2}, 
volume = {242}
}

@ARTICLE{Samushia2021,
       author = {{Samushia}, Lado and {Slepian}, Zachary and {Villaescusa-Navarro}, Francisco},
        title = "{Information content of higher order galaxy correlation functions}",
      journal = {Monthly Notices of the Royal Astronomical Society},
         year = 2021,
        month = jul,
       volume = {505},
       number = {1},
        pages = {628-641},
          doi = {10.1093/mnras/stab1199},
archivePrefix = {arXiv},
       eprint = {2102.01696},
 primaryClass = {astro-ph.CO},
       adsurl = {https://ui.adsabs.harvard.edu/abs/2021MNRAS.505..628S}
}

@article{Scoccimarro1999,
doi = {10.1086/308059},
url = {https://dx.doi.org/10.1086/308059},
year = {1999},
month = {dec},
publisher = {},
volume = {527},
number = {1},
pages = {1},
author = {Román Scoccimarro and Matias Zaldarriaga and Lam Hui},
title = {Power Spectrum Correlations Induced by Nonlinear Clustering},
journal = {The Astrophysical Journal}
}

@ARTICLE{Scoccimarro2001,
       author = {{Scoccimarro}, Rom{\'a}n and {Feldman}, Hume A. and {Fry}, J.~N. and {Frieman}, Joshua A.},
        title = "{The Bispectrum of IRAS Redshift Catalogs}",
      journal = {The Astrophysical Journal},
         year = 2001,
        month = jan,
       volume = {546},
       number = {2},
        pages = {652-664},
          doi = {10.1086/318284},
archivePrefix = {arXiv},
       eprint = {astro-ph/0004087},
 primaryClass = {astro-ph},
       adsurl = {https://ui.adsabs.harvard.edu/abs/2001ApJ...546..652S}
}

@article{Seljak2009,
  author        = {Seljak, Uro{\v{s}}},
  title         = {Extracting Primordial Non-Gaussianity without Cosmic Variance},
  journal       = {Physical Review Letters},
  volume        = {102},
  number        = {2},
  pages         = {021302},
  year          = {2009},
  doi           = {10.1103/PhysRevLett.102.021302},
  eprint        = {0807.1770},
  archivePrefix = {arXiv},
  primaryClass  = {astro-ph}
}

@article{SellentinHeavens2016,
  author        = {Sellentin, Elena and Heavens, Alan F.},
  title         = {Parameter Inference with Estimated Covariance Matrices},
  journal       = {Monthly Notices of the Royal Astronomical Society},
  year          = {2016},
  volume        = {456},
  number        = {1},
  pages         = {L132--L136},
  doi           = {10.1093/mnrasl/slv190},
  archivePrefix = {arXiv},
  eprint        = {1511.05969},
  primaryClass  = {astro-ph.CO}
}

@article{Sheth2005, 
year = {2005}, 
title = {{The halo‐model description of marked statistics}}, 
author = {Sheth, Ravi K.}, 
journal = {Monthly Notices of the Royal Astronomical Society},
issn = {0035-8711}, 
doi = {10.1111/j.1365-2966.2005.09609.x}, 
eprint = {astro-ph/0511772}, 
pages = {796--806}, 
number = {3}, 
volume = {364}
}

@ARTICLE{Singh2021,
       author = {{Singh}, S.},
        title = "{improved MASTER for the LSS: fast and accurate analysis of the two-point power spectra and correlation functions.}",
      journal = {Monthly Notices of the Royal Astronomical Society},
         year = 2021,
        month = jan,
       volume = {508},
        pages = {1632-1651},
          doi = {10.48550/arXiv.2105.04548},
archivePrefix = {arXiv},
       eprint = {2105.04548},
 primaryClass = {astro-ph.CO},
       adsurl = {https://ui.adsabs.harvard.edu/abs/2021MNRAS.508.1632S}
}

@ARTICLE{Sinha2020,
       author = {{Sinha}, Manodeep and {Garrison}, Lehman H.},
        title = "{CORRFUNC - a suite of blazing fast correlation functions on the CPU}",
      journal = {Monthly Notices of the Royal Astronomical Society},
         year = 2020,
        month = jan,
       volume = {491},
       number = {2},
        pages = {3022-3041},
          doi = {10.1093/mnras/stz3157},
archivePrefix = {arXiv},
       eprint = {1911.03545},
 primaryClass = {astro-ph.CO},
       adsurl = {https://ui.adsabs.harvard.edu/abs/2020MNRAS.491.3022S}
}

@article{Skibba2006, 
year = {2006}, 
title = {{The luminosity‐weighted or ‘marked’ correlation function}}, 
author = {Skibba, Ramin and Sheth, Ravi K. and Connolly, Andrew J. and Scranton, Ryan}, 
journal = {Monthly Notices of the Royal Astronomical Society},
issn = {0035-8711}, 
doi = {10.1111/j.1365-2966.2006.10196.x}, 
eprint = {astro-ph/0512463}, 
pages = {68--76}, 
number = {1}, 
volume = {369}
}

@article{Skibba2013, 
year = {2013}, 
title = {{Measures of galaxy environment – II. Rank-ordered mark correlations}}, 
author = {Skibba, Ramin A. and Sheth, Ravi K. and Croton, Darren J. and Muldrew, Stuart I. and Abbas, Ummi and Pearce, Frazer R. and Shattow, Genevieve M.}, 
journal = {Monthly Notices of the Royal Astronomical Society},
issn = {0035-8711}, 
doi = {10.1093/mnras/sts349}, 
eprint = {1211.0287}, 
pages = {458--468}, 
number = {1}, 
volume = {429}
}

@ARTICLE{Slepian2015,
       author = {{Slepian}, Zachary and {Eisenstein}, Daniel J.},
        title = "{Computing the three-point correlation function of galaxies in O(N\^2) time}",
      journal = {Monthly Notices of the Royal Astronomical Society},
         year = 2015,
        month = dec,
       volume = {454},
       number = {4},
        pages = {4142-4158},
          doi = {10.1093/mnras/stv2119},
archivePrefix = {arXiv},
       eprint = {1506.02040},
 primaryClass = {astro-ph.CO},
       adsurl = {https://ui.adsabs.harvard.edu/abs/2015MNRAS.454.4142S}
}

@ARTICLE{Slepian2016,
       author = {{Slepian}, Zachary and {Eisenstein}, Daniel J.},
        title = "{Accelerating the two-point and three-point galaxy correlation functions using Fourier transforms}",
      journal = {Monthly Notices of the Royal Astronomical Society},
         year = 2016,
        month = jan,
       volume = {455},
       number = {1},
        pages = {L31-L35},
          doi = {10.1093/mnrasl/slv133},
archivePrefix = {arXiv},
       eprint = {1506.04746},
 primaryClass = {astro-ph.CO},
       adsurl = {https://ui.adsabs.harvard.edu/abs/2016MNRAS.455L..31S}
}

@ARTICLE{Slepian2017,
       author = {{Slepian}, Zachary and {Eisenstein}, Daniel J. and {Brownstein}, Joel R. and {Chuang}, Chia-Hsun and {Gil-Mar{\'\i}n}, H{\'e}ctor and {Ho}, Shirley and {Kitaura}, Francisco-Shu and {Percival}, Will J. and {Ross}, Ashley J. and {Rossi}, Graziano and {Seo}, Hee-Jong and {Slosar}, An{\v{z}}e and {Vargas-Maga{\~n}a}, Mariana},
        title = "{Detection of baryon acoustic oscillation features in the large-scale three-point correlation function of SDSS BOSS DR12 CMASS galaxies}",
      journal = {Monthly Notices of the Royal Astronomical Society},
         year = 2017,
        month = aug,
       volume = {469},
       number = {2},
        pages = {1738-1751},
          doi = {10.1093/mnras/stx488},
archivePrefix = {arXiv},
       eprint = {1607.06097},
 primaryClass = {astro-ph.CO},
       adsurl = {https://ui.adsabs.harvard.edu/abs/2017MNRAS.469.1738S}
}

@article{Slepian2018, 
year = {2018}, 
title = {{A practical computational method for the anisotropic redshift-space three-point correlation function}}, 
author = {Slepian, Zachary and Eisenstein, Daniel J}, 
journal = {Monthly Notices of the Royal Astronomical Society},
issn = {0035-8711}, 
doi = {10.1093/mnras/sty1063}, 
eprint = {1709.10150}, 
pages = {1468--1483}, 
number = {2}, 
volume = {478}
}

@ARTICLE{Sugiyama2023,
       author = {{Sugiyama}, Naonori S. and {Yamauchi}, Daisuke and {Kobayashi}, Tsutomu and {Fujita}, Tomohiro and {Arai}, Shun and {Hirano}, Shin'ichi and {Saito}, Shun and {Beutler}, Florian and {Seo}, Hee-Jong},
        title = "{New constraints on cosmological modified gravity theories from anisotropic three-point correlation functions of BOSS DR12 galaxies}",
      journal = {Monthly Notices of the Royal Astronomical Society},
         year = 2023,
        month = aug,
       volume = {523},
       number = {2},
        pages = {3133-3191},
          doi = {10.1093/mnras/stad1505},
archivePrefix = {arXiv},
       eprint = {2302.06808},
 primaryClass = {astro-ph.CO},
       adsurl = {https://ui.adsabs.harvard.edu/abs/2023MNRAS.523.3133S}
}

@article{Sunseri2023, 
year = {2023}, 
title = {{sarabande: 3/4 point correlation functions with fast Fourier transforms}}, 
author = {Sunseri, James and Slepian, Zachary and Portillo, Stephen and Hou, Jiamin and Kahraman, Sule and Finkbeiner, Douglas P}, 
journal = {RAS Techniques and Instruments}, 
doi = {10.1093/rasti/rzad003}, 
pages = {62--77}, 
number = {1}, 
volume = {2}, 
}

@ARTICLE{Szapudi1998,
       author = {{Szapudi}, Istv{\'a}n and {Szalay}, Alexander S.},
        title = "{A New Class of Estimators for the N-Point Correlations}",
      journal = {The Astrophysical Journal Letters},
         year = 1998,
        month = feb,
       volume = {494},
       number = {1},
        pages = {L41-L44},
          doi = {10.1086/311146},
       adsurl = {https://ui.adsabs.harvard.edu/abs/1998ApJ...494L..41S}
}

@ARTICLE{Szapudi2004,
       author = {{Szapudi}, Istv{\'a}n},
        title = "{Three-Point Statistics from a New Perspective}",
      journal = {The Astrophysical Journal Letters},
         year = 2004,
        month = apr,
       volume = {605},
       number = {2},
        pages = {L89-L92},
          doi = {10.1086/420894},
archivePrefix = {arXiv},
       eprint = {astro-ph/0404476},
 primaryClass = {astro-ph},
       adsurl = {https://ui.adsabs.harvard.edu/abs/2004ApJ...605L..89S}
}

@article{Taylor2013,
  author        = {Taylor, Andy and Joachimi, Benjamin and Kitching, Thomas},
  title         = {Putting the precision in precision cosmology:
                   How accurate should your data covariance matrix be?},
  journal       = {Monthly Notices of the Royal Astronomical Society},
  year          = {2013},
  volume        = {432},
  number        = {3},
  pages         = {1928--1946},
  doi           = {10.1093/mnras/stt270},
  eprint        = {1212.4359},
  archivePrefix = {arXiv}
}

@ARTICLE{Tegmark2006,
       author = {{Tegmark}, Max and {Eisenstein}, Daniel J. and {Strauss}, Michael A. and {Weinberg}, David H. and {Blanton}, Michael R. and {Frieman}, Joshua A. and {Fukugita}, Masataka and {Gunn}, James E. and {Hamilton}, Andrew J.~S. and {Knapp}, Gillian R. and {Nichol}, Robert C. and {Ostriker}, Jeremiah P. and {Padmanabhan}, Nikhil and {Percival}, Will J. and {Schlegel}, David J. and {Schneider}, Donald P. and {Scoccimarro}, Roman and {Seljak}, Uro{\v{s}} and {Seo}, Hee-Jong and {Swanson}, Molly and {Szalay}, Alexander S. and {Vogeley}, Michael S. and {Yoo}, Jaiyul and {Zehavi}, Idit and {Abazajian}, Kevork and {Anderson}, Scott F. and {Annis}, James and {Bahcall}, Neta A. and {Bassett}, Bruce and {Berlind}, Andreas and {Brinkmann}, Jon and {Budavari}, Tam{\'a}s and {Castander}, Francisco and {Connolly}, Andrew and {Csabai}, Istvan and {Doi}, Mamoru and {Finkbeiner}, Douglas P. and {Gillespie}, Bruce and {Glazebrook}, Karl and {Hennessy}, Gregory S. and {Hogg}, David W. and {Ivezi{\'c}}, {\v{Z}}eljko and {Jain}, Bhuvnesh and {Johnston}, David and {Kent}, Stephen and {Lamb}, Donald Q. and {Lee}, Brian C. and {Lin}, Huan and {Loveday}, Jon and {Lupton}, Robert H. and {Munn}, Jeffrey A. and {Pan}, Kaike and {Park}, Changbom and {Peoples}, John and {Pier}, Jeffrey R. and {Pope}, Adrian and {Richmond}, Michael and {Rockosi}, Constance and {Scranton}, Ryan and {Sheth}, Ravi K. and {Stebbins}, Albert and {Stoughton}, Christopher and {Szapudi}, Istv{\'a}n and {Tucker}, Douglas L. and {vanden Berk}, Daniel E. and {Yanny}, Brian and {York}, Donald G.},
        title = "{Cosmological constraints from the SDSS luminous red galaxies}",
      journal = {Physical Review D},
         year = 2006,
        month = dec,
       volume = {74},
       number = {12},
          eid = {123507},
        pages = {123507},
          doi = {10.1103/PhysRevD.74.123507},
archivePrefix = {arXiv},
       eprint = {astro-ph/0608632},
 primaryClass = {astro-ph},
       adsurl = {https://ui.adsabs.harvard.edu/abs/2006PhRvD..74l3507T}
}

@ARTICLE{Trusov2024,
       author = {{Trusov}, Svyatoslav and {Zarrouk}, Pauline and {Cole}, Shaun and {Norberg}, Peder and {Zhao}, Cheng and {Aguilar}, Jessica Nicole and {Ahlen}, Steven and {Brooks}, David and {de la Macorra}, Axel and {Doel}, Peter and {Font-Ribera}, Andreu and {Honscheid}, Klaus and {Kisner}, Theodore and {Landriau}, Martin and {Magneville}, Christophe and {Miquel}, Ramon and {Nie}, Jundan and {Poppett}, Claire and {Schubnell}, Michael and {Tarl{\'e}}, Gregory and {Zhou}, Zhimin},
        title = "{The two-point correlation function covariance with fewer mocks}",
      journal = {Monthly Notices of the Royal Astronomical Society},
         year = 2024,
        month = jan,
       volume = {527},
       number = {3},
        pages = {9048-9060},
          doi = {10.1093/mnras/stad3710},
archivePrefix = {arXiv},
       eprint = {2306.16332},
 primaryClass = {astro-ph.CO},
       adsurl = {https://ui.adsabs.harvard.edu/abs/2024MNRAS.527.9048T}
}

@article{Uhlemann2016,
    author = {Uhlemann, C. and Codis, S. and Kim, J. and Pichon, C. and Bernardeau, F. and Pogosyan, D. and Park, C. and L'Huillier, B.},
    title = "{Beyond Kaiser bias: mildly non-linear two-point statistics of densities in distant spheres}",
    journal = {Monthly Notices of the Royal Astronomical Society},
    volume = {466},
    number = {2},
    pages = {2067-2084},
    year = {2016},
    month = {12},
    issn = {0035-8711},
    doi = {10.1093/mnras/stw3221},
    url = {https://doi.org/10.1093/mnras/stw3221},
    eprint = {https://academic.oup.com/mnras/article-pdf/466/2/2067/10868206/stw3221.pdf},
}

@ARTICLE{Uhlemann2018,
       author = {{Uhlemann}, C. and {Pajer}, E. and {Pichon}, C. and {Nishimichi}, T. and {Codis}, S. and {Bernardeau}, F.},
        title = "{Hunting high and low: disentangling primordial and late-time non-Gaussianity with cosmic densities in spheres}",
      journal = {Monthly Notices of the Royal Astronomical Society},
         year = 2018,
        month = mar,
       volume = {474},
       number = {3},
        pages = {2853-2870},
          doi = {10.1093/mnras/stx2623},
archivePrefix = {arXiv},
       eprint = {1708.02206},
 primaryClass = {astro-ph.CO},
       adsurl = {https://ui.adsabs.harvard.edu/abs/2018MNRAS.474.2853U}
}

@ARTICLE{Valageas2002,
       author = {{Valageas}, P.},
        title = "{Dynamics of gravitational clustering. III. The quasi-linear regime for some non-Gaussian initial conditions}",
      journal = {Astronomy \& Astrophysics},
         year = 2002,
        month = feb,
       volume = {382},
        pages = {431-449},
          doi = {10.1051/0004-6361:20011675},
archivePrefix = {arXiv},
       eprint = {astro-ph/0107196},
 primaryClass = {astro-ph},
       adsurl = {https://ui.adsabs.harvard.edu/abs/2002A&A...382..431V}
}

@article{vanDaalen2011,
  author        = {van Daalen, Marcel P. and
                   Schaye, Joop and
                   Booth, C. M. and
                   Dalla Vecchia, Claudio},
  title         = {The Effects of Galaxy Formation on the Matter Power
                   Spectrum: A Challenge for Precision Cosmology},
  journal       = {Monthly Notices of the Royal Astronomical Society},
  year          = {2011},
  volume        = {415},
  number        = {4},
  pages         = {3649--3665},
  doi           = {10.1111/j.1365-2966.2011.18981.x},
  archivePrefix = {arXiv},
  eprint        = {1104.1174},
  primaryClass  = {astro-ph.CO}
}

@ARTICLE{villaescusa2020quijote,
       author = {{Villaescusa-Navarro}, Francisco and {Hahn}, ChangHoon and {Massara}, Elena and {Banerjee}, Arka and {Delgado}, Ana Maria and {Ramanah}, Doogesh Kodi and {Charnock}, Tom and {Giusarma}, Elena and {Li}, Yin and {Allys}, Erwan and {Brochard}, Antoine and {Uhlemann}, Cora and {Chiang}, Chi-Ting and {He}, Siyu and {Pisani}, Alice and {Obuljen}, Andrej and {Feng}, Yu and {Castorina}, Emanuele and {Contardo}, Gabriella and {Kreisch}, Christina D. and {Nicola}, Andrina and {Alsing}, Justin and {Scoccimarro}, Roman and {Verde}, Licia and {Viel}, Matteo and {Ho}, Shirley and {Mallat}, Stephane and {Wandelt}, Benjamin and {Spergel}, David N.},
        title = "{The Quijote Simulations}",
      journal = {The Astrophysical Journal Supplement Series},
         year = 2020,
        month = sep,
       volume = {250},
       number = {1},
          eid = {2},
        pages = {2},
          doi = {10.3847/1538-4365/ab9d82},
archivePrefix = {arXiv},
       eprint = {1909.05273},
 primaryClass = {astro-ph.CO},
       adsurl = {https://ui.adsabs.harvard.edu/abs/2020ApJS..250....2V}
}

@ARTICLE{Wang2007,
       author = {{Wang}, Yu and {Yang}, Xiaohu and {Mo}, H.~J. and {van den Bosch}, Frank C.},
        title = "{The Cross-Correlation between Galaxies of Different Luminosities and Colors}",
      journal = {The Astrophysical Journal},
         year = 2007,
        month = aug,
       volume = {664},
       number = {2},
        pages = {608-632},
          doi = {10.1086/519245},
archivePrefix = {arXiv},
       eprint = {astro-ph/0703253},
 primaryClass = {astro-ph},
       adsurl = {https://ui.adsabs.harvard.edu/abs/2007ApJ...664..608W}
}

@article{Wang2023Triumvirate,
  author  = {Wang, Mike Shengbo and Beutler, Florian and Sugiyama, Naonori S.},
  title   = {Triumvirate: A Python/C++ Package for Three-Point Clustering
             Measurements},
  journal = {Journal of Open Source Software},
  year    = {2023},
  volume  = {8},
  number  = {91},
  pages   = {5571},
  doi     = {10.21105/joss.05571}
}

@article{Wang2025Window,
  author        = {Wang, Mike Shengbo and Beutler, Florian and others},
  title         = {Window Convolution of the Galaxy Clustering Bispectrum},
  journal       = {Journal of Cosmology and Astroparticle Physics},
  year          = {2025},
  volume        = {2025},
  number        = {06},
  pages         = {031},
  doi           = {10.1088/1475-7516/2025/06/031},
  archivePrefix = {arXiv},
  eprint        = {2411.14947},
  primaryClass  = {astro-ph.CO}
}

@article{Wangyun2022A, 
year = {2022}, 
title = {{Continuous Wavelet Analysis of Matter Clustering Using the Gaussian-derived Wavelet}}, 
author = {Wang, Yun and Yang, Hua-Yu and He, Ping}, 
journal = {The Astrophysical Journal},
issn = {0004-637X}, 
doi = {10.3847/1538-4357/ac752c}, 
eprint = {2112.06114}, 
pages = {77}, 
number = {1}, 
volume = {934}, 
}

@article{Wangyun2022B, 
year = {2022}, 
title = {{Simultaneous Dependence of Matter Clustering on Scale and Environment}}, 
author = {Wang, Yun and He, Ping}, 
journal = {The Astrophysical Journal},
issn = {0004-637X}, 
doi = {10.3847/1538-4357/ac7a3d}, 
eprint = {2202.11964}, 
pages = {112}, 
number = {2}, 
volume = {934}, 
}

@article{White1979, 
year = {1979}, 
title = {{The hierarchy of correlation functions and its relation to other measures of galaxy clustering}}, 
author = {White, Simon D. M.}, 
journal = {Monthly Notices of the Royal Astronomical Society},
issn = {0035-8711}, 
doi = {10.1093/mnras/186.2.145}, 
pages = {145--154}, 
number = {2}, 
volume = {186}, 
}

@article{White2009, 
year = {2009}, 
title = {{Breaking halo occupation degeneracies with marked statistics}}, 
author = {White, Martin and Padmanabhan, Nikhil}, 
journal = {Monthly Notices of the Royal Astronomical Society},
issn = {0035-8711}, 
doi = {10.1111/j.1365-2966.2009.14732.x}, 
eprint = {0812.4288}, 
pages = {2381--2384}, 
number = {4}, 
volume = {395}
}

@article{White2016, 
year = {2016}, 
title = {{A marked correlation function for constraining modified gravity models}}, 
author = {White, Martin}, 
journal = {Journal of Cosmology and Astroparticle Physics},
doi = {10.1088/1475-7516/2016/11/057}, 
eprint = {1609.08632}, 
pages = {057--057}, 
number = {11}, 
volume = {2016}
}

@article{Wild2005, 
year = {2005}, 
title = {{The 2dF Galaxy Redshift Survey: stochastic relative biasing between galaxy populations}}, 
author = {Wild, Vivienne and Peacock, John A. and Lahav, Ofer and Conway, Edward and Maddox, Steve and Baldry, Ivan K. and Baugh, Carlton M. and Bland‐Hawthorn, Joss and Bridges, Terry and Cannon, Russell and Cole, Shaun and Colless, Matthew and Collins, Chris and Couch, Warrick and Dalton, Gavin and Propris, Roberto De and Driver, Simon P. and Efstathiou, George and Ellis, Richard S. and Frenk, Carlos S. and Glazebrook, Karl and Jackson, Carole and Lewis, Ian and Lumsden, Stuart and Madgwick, Darren and Norberg, Peder and Peterson, Bruce A. and Sutherland, Will and Taylor, Keith and Team, The 2dFGRS}, 
journal = {Monthly Notices of the Royal Astronomical Society},
issn = {0035-8711}, 
doi = {10.1111/j.1365-2966.2004.08447.x}, 
pages = {247--269}, 
number = {1}, 
volume = {356}, 
}

@article{Xiao2022, 
year = {2022}, 
title = {{Cosmological constraints from the density gradient weighted correlation function}}, 
author = {Xiao, Xiaoyuan and Yang, Yizhao and Luo, Xiaolin and Ding, Jiacheng and Huang, Zhiqi and Wang, Xin and Zheng, Yi and Sabiu, Cristiano G and Forero-Romero, Jaime and Miao, Haitao and Li, Xiao-Dong}, 
journal = {Monthly Notices of the Royal Astronomical Society},
issn = {0035-8711}, 
doi = {10.1093/mnras/stac879}, 
eprint = {2203.15986}, 
pages = {595--603}, 
number = {1}, 
volume = {513}, 
}

@article{xiao2025_AP, 
  year     = {2025}, 
  title    = {Tomographic Alcock-Paczynski test with marked correlation functions}, 
  author   = {Xiao, Liang and Lai, Limin and Jiang, Zhujun and Li, Xiao-Dong and Zhang, Le}, 
  journal  = {Physical Review D}, 
  issn     = {2470-0010}, 
  doi      = {10.1103/9gg5-wm7t}, 
  eprint   = {2504.20478}, 
  pages    = {123510}, 
  number   = {12}, 
  volume   = {112}
}

@article{Xiao2026,
year = {2026},
title = {{Cosmological constraints from neighbor-density-weighted marked correlation functions}},
author = {Xiao, Xu and Chen, Zhao and Yu, Yu and Li, Xiao-Dong and Huang, Yiqi and Zhang, Le},
journal = {arXiv e-prints},
doi = {10.48550/arxiv.2605.23367},
eprint = {2605.23367}
}

@article{yang2001a, 
year = {2001}, 
title = {{Measuring the Galaxy Power Spectrum with Multiresolution Decomposition. II. Diagonal and Off-Diagonal Power Spectra of the Las Campanas Redshift Survey Galaxies}}, 
author = {Yang, XiaoHu and Feng, Long-Long and Chu, YaoQuan and Fang, Li-Zhi}, 
journal = {The Astrophysical Journal},
issn = {0004-637X}, 
doi = {10.1086/320661}, 
eprint = {astro-ph/0102160}, 
pages = {1--13}, 
number = {1}, 
volume = {553}, 
}

@article{yang2001b, 
year = {2001}, 
title = {{Measuring the Galaxy Power Spectrum with Multiresolution Decomposition. III. Velocity Field Analysis}}, 
author = {Yang, Xiao Hu and Feng, Long-Long and Chu, Yao Quan and Fang, Li-Zhi}, 
journal = {The Astrophysical Journal},
issn = {0004-637X}, 
doi = {10.1086/323059}, 
eprint = {astro-ph/0107083}, 
pages = {549}, 
number = {2}, 
volume = {560}, 
}

@article{yang2002, 
year = {2002}, 
title = {{Measuring the Galaxy Power Spectrum with Multiresolution Decomposition. IV. Redshift Distortion}}, 
author = {Yang, XiaoHu and Feng, Long-Long and Chu, YaoQuan and Fang, Li-Zhi}, 
journal = {The Astrophysical Journal},
issn = {0004-637X}, 
doi = {10.1086/338274}, 
eprint = {astro-ph/0110530}, 
pages = {630--640}, 
number = {2}, 
volume = {566}, 
}

@article{Yang2020, 
year = {2020}, 
title = {{Using the Mark Weighted Correlation Functions to Improve the Constraints on Cosmological Parameters}}, 
author = {Yang, Yizhao and Miao, Haitao and Ma, Qinglin and Liu, Miaoxin and Sabiu, Cristiano G. and Forero-Romero, Jaime and Huang, Yuanzhu and Lai, Limin and Qian, Qiyue and Zheng, Yi and Li, Xiao-Dong}, 
journal = {The Astrophysical Journal},
issn = {0004-637X}, 
doi = {10.3847/1538-4357/aba35b}, 
eprint = {2007.03150}, 
pages = {6}, 
number = {1}, 
volume = {900}
}

@article{YangYB_2009,
doi = {10.1088/1674-4527/9/2/012},
url = {https://doi.org/10.1088/1674-4527/9/2/012},
year = {2009},
month = {feb},
publisher = {},
volume = {9},
number = {2},
pages = {227},
author = {Yi-Bin Yang and Long-Long Feng and Jun Pan and Xiao-Hu Yang},
title = {An optimal method for the power spectrum measurement},
journal = {Research in Astronomy and Astrophysics}
}

@article{Yue2024,
    author = {Yue, Shiyu and Feng, Longlong and Ju, Wenjie and Pan, Jun and Huang, Zhiqi and Fang, Feng and Li, Zhuoyang and Cai, Yan-Chuan and Zhu, Weishan},
    title = {Pair counting without binning – a new approach to correlation functions in clustering statistics},
    journal = {Monthly Notices of the Royal Astronomical Society},
    volume = {535},
    number = {4},
    pages = {3500-3516},
    year = {2024},
    month = {11},
    issn = {0035-8711},
    doi = {10.1093/mnras/stae2513},
    url = {https://doi.org/10.1093/mnras/stae2513},
    eprint = {https://academic.oup.com/mnras/article-pdf/535/4/3500/60925763/stae2513.pdf},
}

@article{zehavi2011,
  title={Galaxy clustering in the completed SDSS redshift survey: the dependence on color and luminosity},
  author={Zehavi, Idit and Zheng, Zheng and Weinberg, David H and Blanton, Michael R and Bahcall, Neta A and Berlind, Andreas A and Brinkmann, Jon and Frieman, Joshua A and Gunn, James E and Lupton, Robert H and others},
  journal={The Astrophysical Journal},
  volume={736},
  number={1},
  pages={59},
  year={2011},
  publisher={IOP Publishing}
}

@article{Zhang2005,
title = {Fast n-point correlation functions and three-point lensing application},
journal = {New Astronomy},
volume = {10},
number = {7},
pages = {569-590},
year = {2005},
issn = {1384-1076},
doi = {https://doi.org/10.1016/j.newast.2005.04.002},
url = {https://www.sciencedirect.com/science/article/pii/S1384107605000485},
author = {Lucy Liuxuan Zhang and Ue-Li Pen}
}

@article{Zhao2023, 
year = {2023}, 
title = {{Fast correlation function calculator}}, 
author = {Zhao, Cheng}, 
journal = {Astronomy \& Astrophysics},
issn = {0004-6361}, 
doi = {10.1051/0004-6361/202346015}, 
eprint = {2301.12557}, 
pages = {A83}, 
volume = {672}, 
}

@ARTICLE{Zhou2025EmulatorIII,
       author = {{Zhou}, Shuren and {Chen}, Zhao and {Yu}, Yu},
        title = "{CSST cosmological emulator III: Hybrid Lagrangian bias expansion emulation of galaxy clustering}",
      journal = {Science China Physics, Mechanics \& Astronomy},
         year = {2025},
       volume = {68},
        pages = {129512},
          doi = {10.1007/s11433-025-2755-x},
archivePrefix = {arXiv},
       eprint = {2506.04671},
 primaryClass = {astro-ph.CO}
}

@article{Zhu2015,
  author        = {Zhu, Fangzhou and Padmanabhan, Nikhil and White, Martin},
  title         = {Optimal Redshift Weighting for Baryon Acoustic Oscillations},
  journal       = {Monthly Notices of the Royal Astronomical Society},
  volume        = {451},
  number        = {1},
  pages         = {236--243},
  year          = {2015},
  doi           = {10.1093/mnras/stv964},
  eprint        = {1411.1424},
  archivePrefix = {arXiv},
  primaryClass  = {astro-ph.CO}
}

@article{Zhu2016,
  author        = {Zhu, Fangzhou and Padmanabhan, Nikhil and White, Martin
                   and Ross, Ashley J. and Zhao, Gong-Bo},
  title         = {Redshift Weights for Baryon Acoustic Oscillations:
                   Application to Mock Galaxy Catalogues},
  journal       = {Monthly Notices of the Royal Astronomical Society},
  volume        = {461},
  number        = {3},
  pages         = {2867--2878},
  year          = {2016},
  doi           = {10.1093/mnras/stw1515},
  eprint        = {1604.01050},
  archivePrefix = {arXiv},
  primaryClass  = {astro-ph.CO}
}

\appendix

\section{Built-in basic window-function library}
\label{app:window_functions}

\setcounter{table}{0}
\renewcommand{\thetable}{A\arabic{table}}
\renewcommand{\theHfigure}{appendix.\arabic{figure}}
\renewcommand{\theHtable}{appendix.\arabic{table}}

\begingroup
\setlength{\tabcolsep}{3pt}
\begin{longtable}{>{\centering\arraybackslash}m{0.28in} >{\raggedright\arraybackslash}m{1.17in} >{\centering\arraybackslash}m{3.1in} >{\centering\arraybackslash}m{2.05in}}
    \caption{\large\bf Analytic window kernels integrated into the \textsf{PyHermes} window-function library.}
    \label{tab:kernel}\\
    \hline
    \textbf{Item} & \textbf{Description} & \textbf{Window Function (Real \& Fourier Space)} & \textbf{Fourier Space Sketch} \\
    \hline
    \endfirsthead
    \multicolumn{4}{c}{\tablename\ \thetable\ -- \textit{Continued from previous page}}\\
    \hline
    \textbf{Item} & \textbf{Description} & \textbf{Window Function (Real \& Fourier Space)} & \textbf{Fourier Space Sketch} \\
    \hline
    \endhead
    \hline
    \multicolumn{4}{r}{\textit{Continued on next page}} \\
    \endfoot
    \hline
    \endlastfoot
        \hline
        (a) & Infinitesimal spherical shell; standard isotropic binning window. & $${W_{\mathrm{shell}}(\mathbf{x};R)=\frac{1}{4\pi R^2}\delta_{\mathrm{D}}(|\mathbf{x}|-R),}$$ $$\hat{W}_{\mathrm{shell}}(k;R)=\frac{\sin(kR)}{kR}$$ & \includegraphics[width=0.28\textwidth]{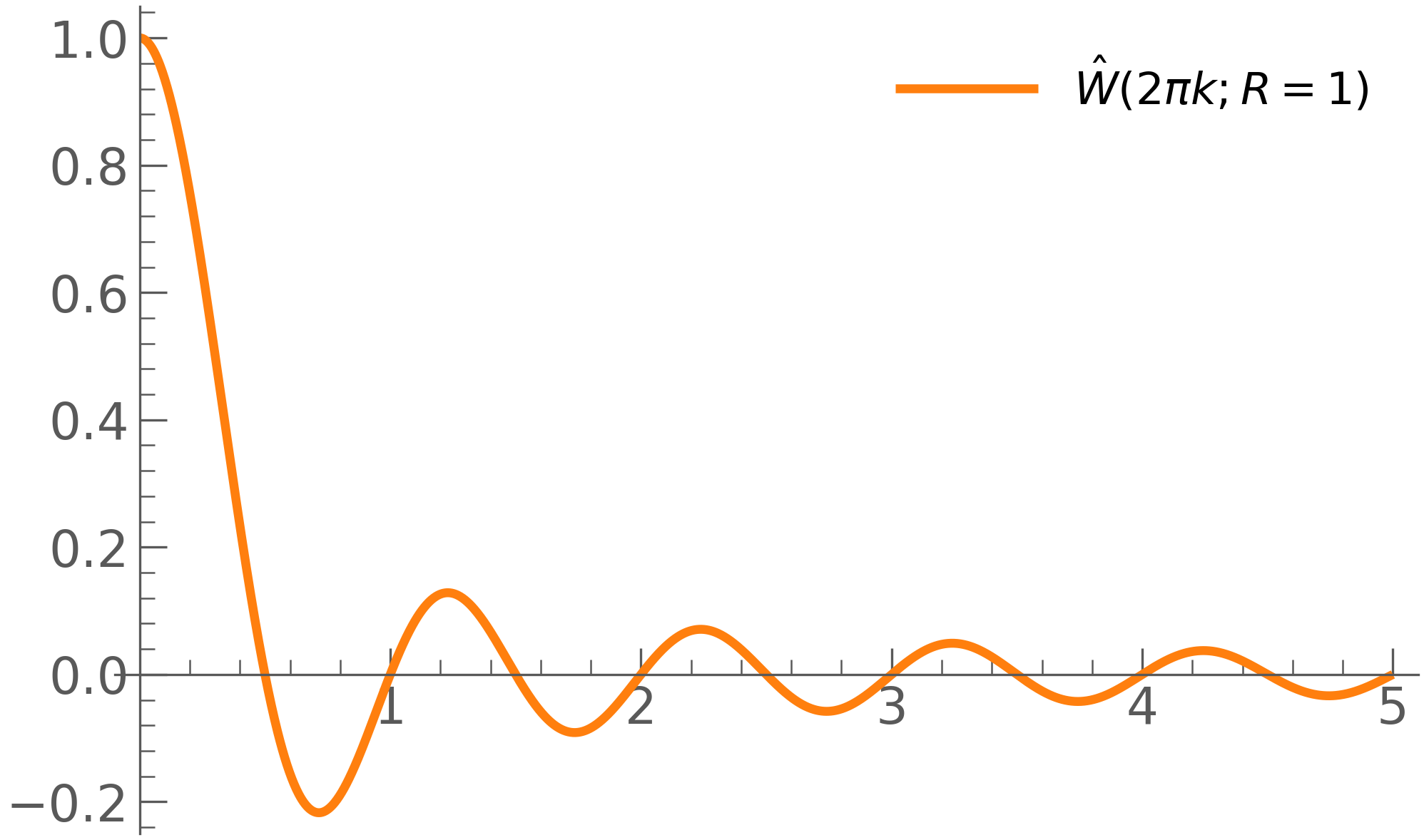} \\
        \hline
        (b) & Spherical top-hat; counts-in-spheres smoothing and variance window. &$${W_{\mathrm{sphere}}(\mathbf{x};R)=\frac{3}{4\pi R^3}\theta(R-|\mathbf{x}|),}$$ $$\hat{W}_{\mathrm{sphere}}(k;R)=\frac{3(\sin(kR)-kR\cos(kR))}{k^3R^3}$$ & \includegraphics[width=0.28\textwidth]{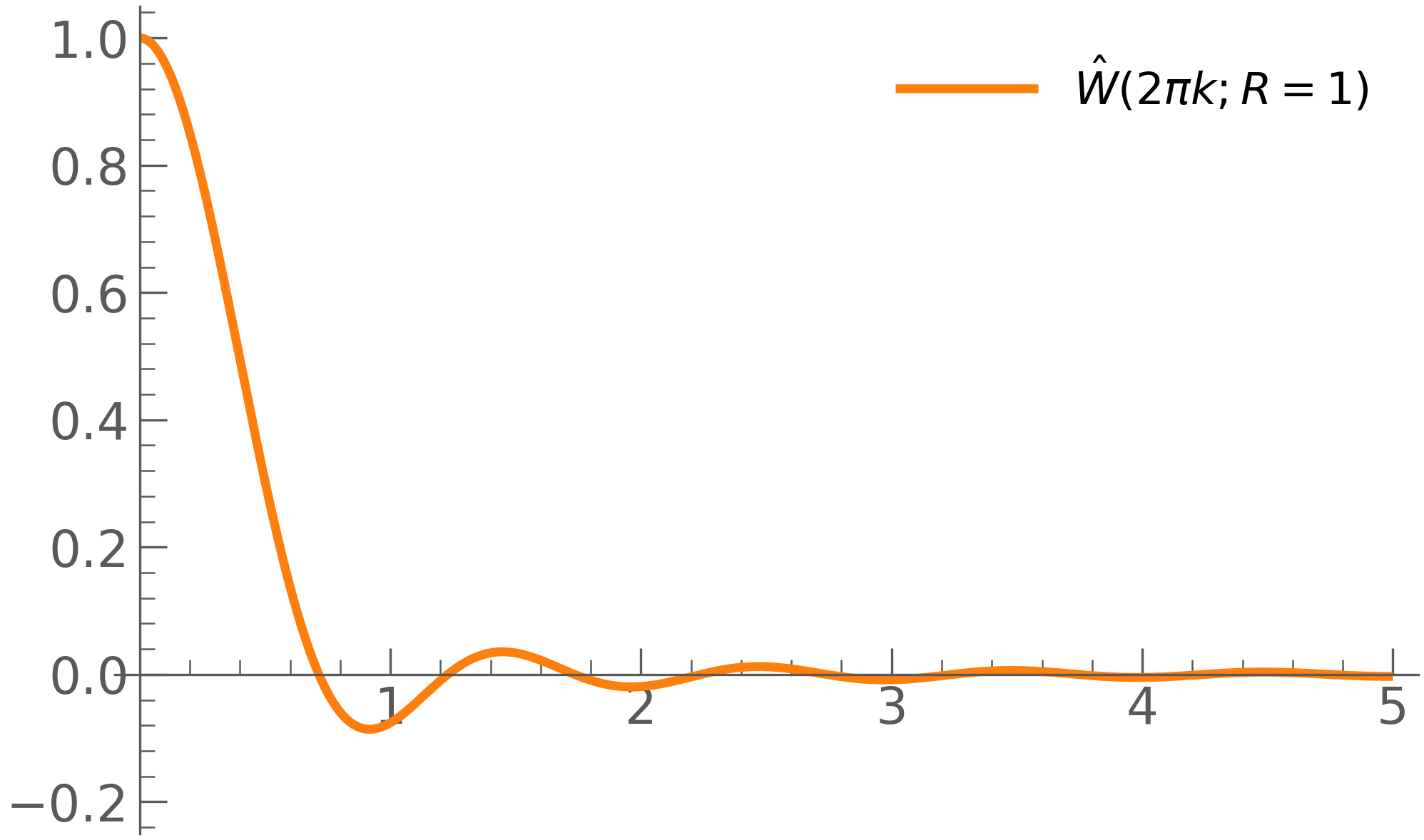} \\
        \hline
        (c) & Smooth low-pass filter with characteristic scale \(R\). & $${W_{\mathrm{Gauss}}(\mathbf{x};R)=\frac{1}{\sqrt{(2\pi)^3}R^3}\exp\left(-\frac{|\mathbf{x}|^2}{2R^2}\right),}$$ $$\hat{W}_{\mathrm{Gauss}}(k;R)=\exp\left(-\frac{1}{2}k^2R^2\right)$$ & \includegraphics[width=0.28\textwidth]{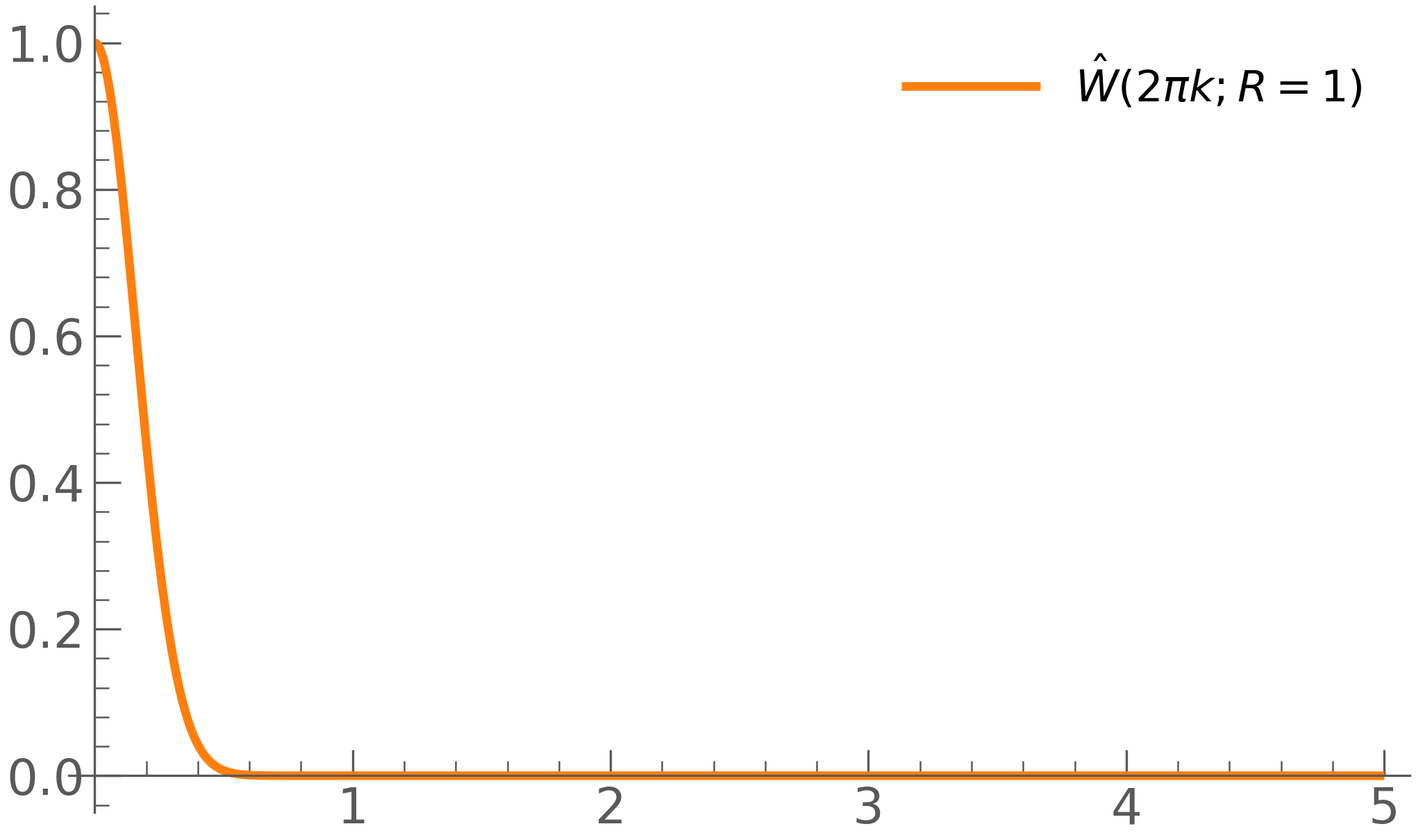} \\
        \hline
        (d) & Shell response smoothed by a Gaussian radial width \(R_g\) around \(R_s\). & $${W_{\mathrm{gsh}}(\mathbf{x};R_s,R_g)=\mathcal{F}^{-1}\{\hat{W}_{\mathrm{gsh}}\},}$$ $$\begin{aligned}\hat{W}_{\mathrm{gsh}}(k;R_s,R_g)&=\frac{R_g^2\cos(kR_s)+R_s^2\sin(kR_s)/(kR_s)}{R_s^2+R_g^2}\\
        &\quad\times\exp\left(-\frac{1}{2}k^2R_g^2\right)\end{aligned}$$ & \includegraphics[width=0.28\textwidth]{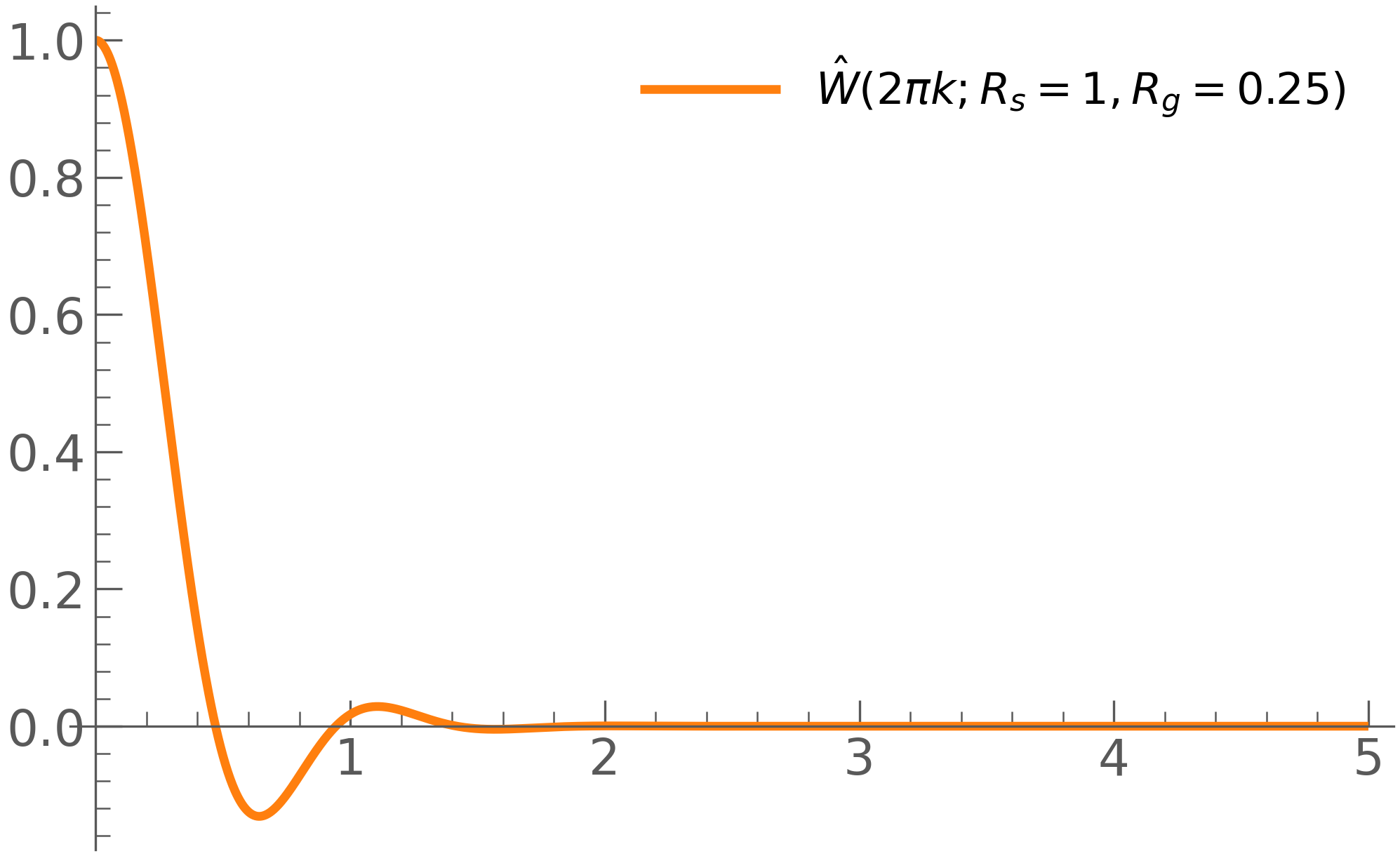} \\
        \hline
        (e) & Separable Cartesian top-hat for anisotropic box smoothing. & $$W_{\mathrm{cubic}}(\mathbf{x};\mathbf{L})=\prod_{i\in\{x,y,z\}}\frac{1}{L_i}\theta(L_i/2-|x_i|),$$ $$\hat{W}_{\mathrm{cubic}}(\mathbf{k};\mathbf{L})=\prod_{i\in\{x,y,z\}}\frac{\sin(k_iL_i/2)}{k_iL_i/2}$$  & \includegraphics[width=0.28\textwidth]{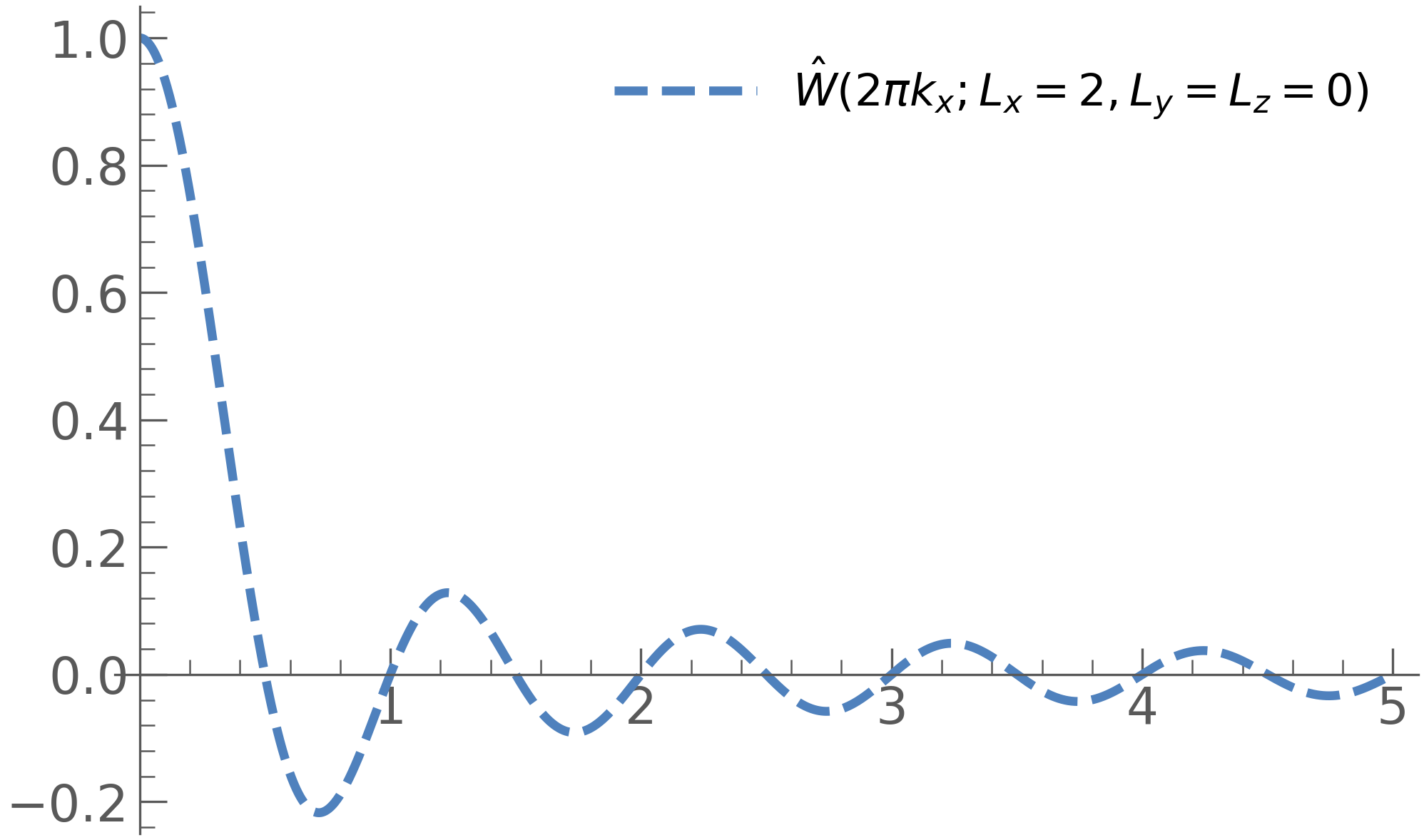} \\
        \hline
        (f) & Filled cylinder; averages over a transverse disk and a finite line-of-sight slab. & $$W_{\mathrm{cylinder}}(\rho,z;R,H)=\frac{\theta(R-\rho)}{\pi R^2}\frac{\theta(H-|z|)}{2H},$$ $$\hat{W}_{\mathrm{cylinder}}(k_{\perp},k_{\parallel};R,H)=\frac{2J_1(k_{\perp}R)}{k_{\perp}R}\frac{\sin(k_{\parallel}H)}{k_{\parallel}H}$$ & \includegraphics[width=0.28\textwidth]{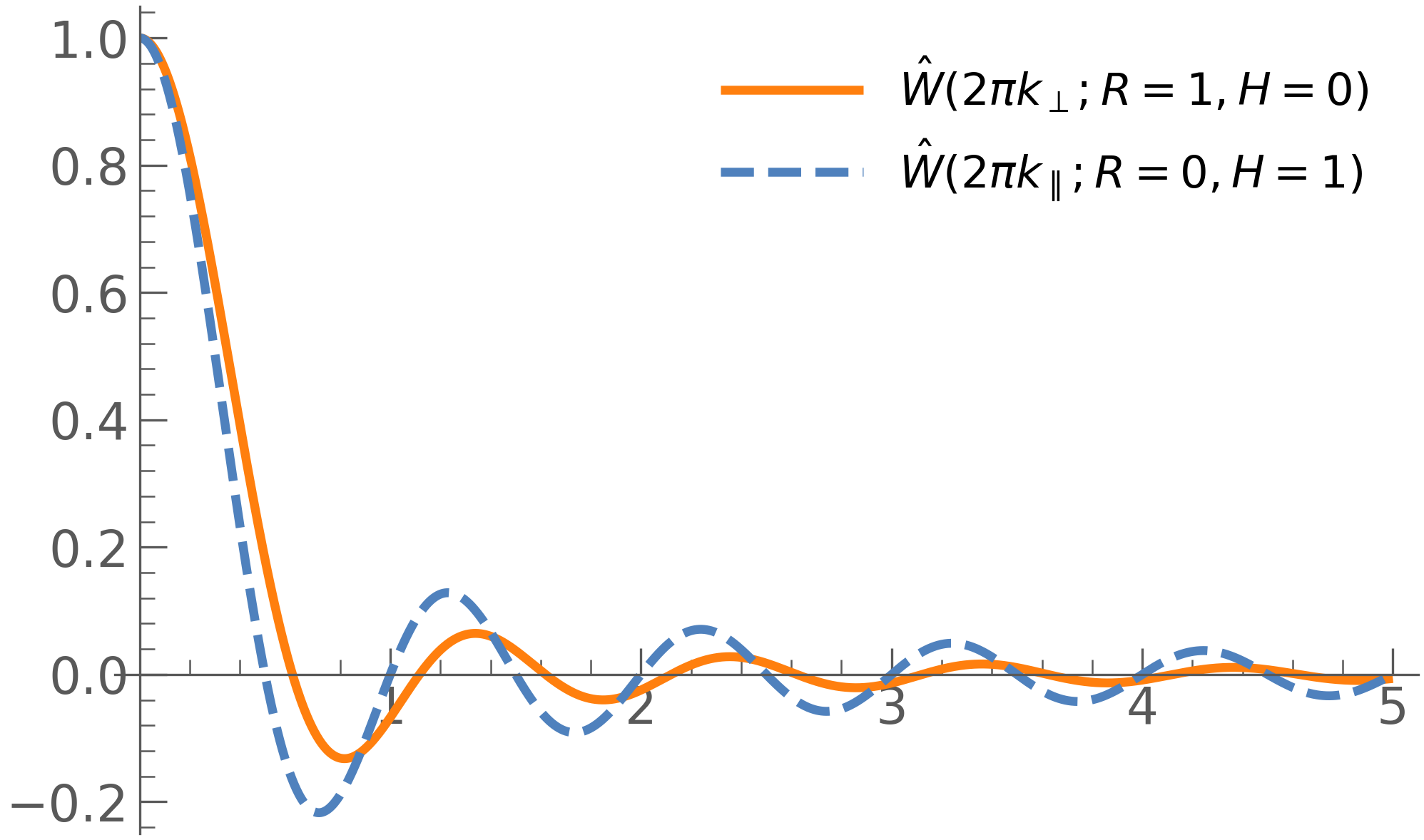} \\
        \hline
        (g) & Cylindrical side wall with finite line-of-sight extent. & $$W_{\mathrm{cylshell}}(\rho,z;R,H)=\frac{\delta_{\mathrm{D}}(\rho-R)}{2\pi R}\frac{\theta(H-|z|)}{2H},$$ $$\hat{W}_{\mathrm{cylshell}}(k_{\perp},k_{\parallel};R,H)=J_0(k_\perp R)\frac{\sin(k_{\parallel}H)}{k_{\parallel}H} $$ & \includegraphics[width=0.28\textwidth]{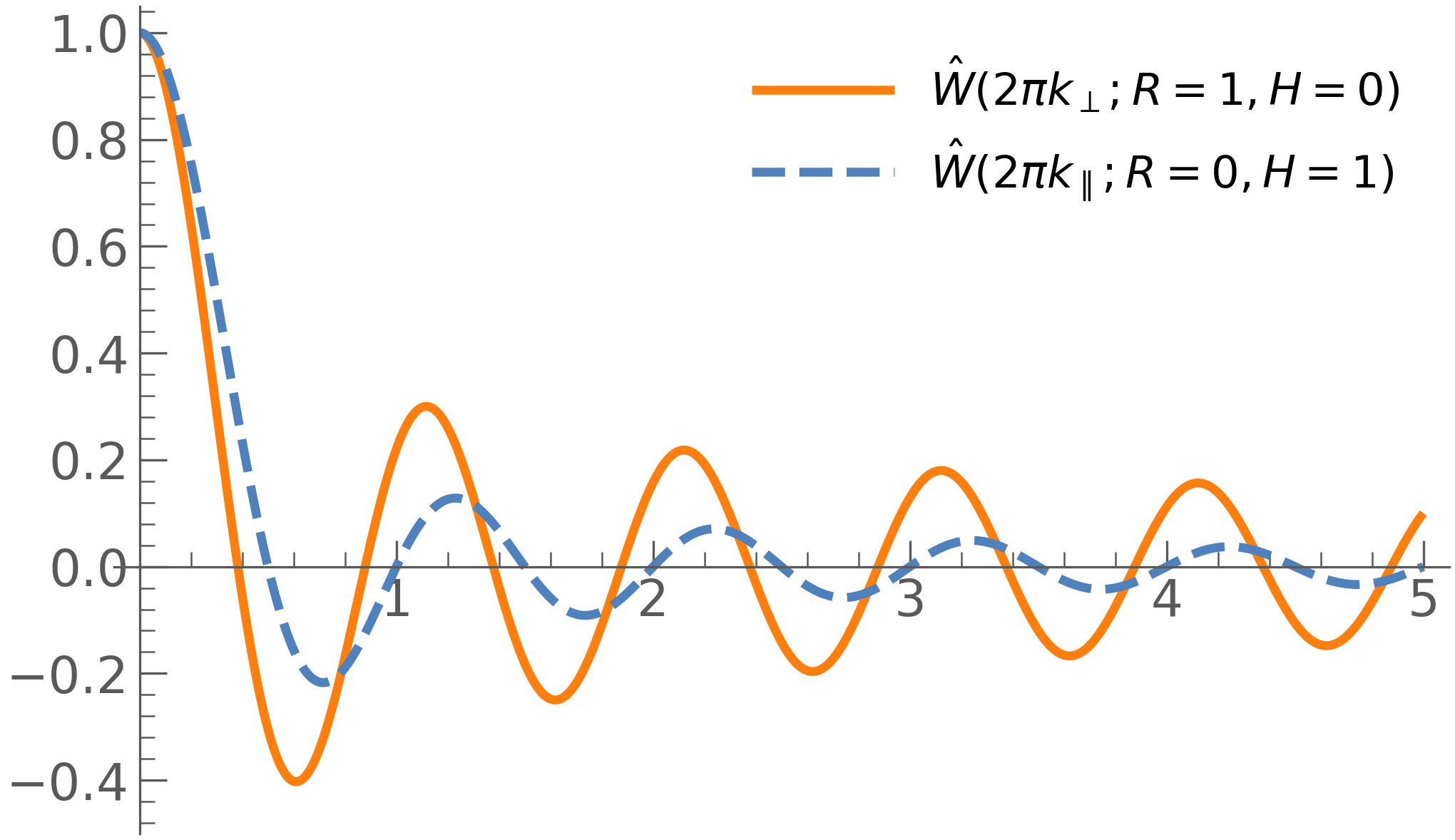} \\
        \hline
        (h) & Pair of transverse disks at fixed line-of-sight offset. & $$W_{\mathrm{disk}}(\rho,z;R,H)=\frac{\theta(R-\rho)}{\pi R^2}\frac{\delta_{\mathrm{D}}(|z|-H)}{2},$$ $$\hat{W}_{\mathrm{disk}}(k_{\perp},k_{\parallel};R,H)=\frac{2J_1(k_{\perp} R)}{k_{\perp} R}\cos(k_{\parallel}H)$$ & \includegraphics[width=0.28\textwidth]{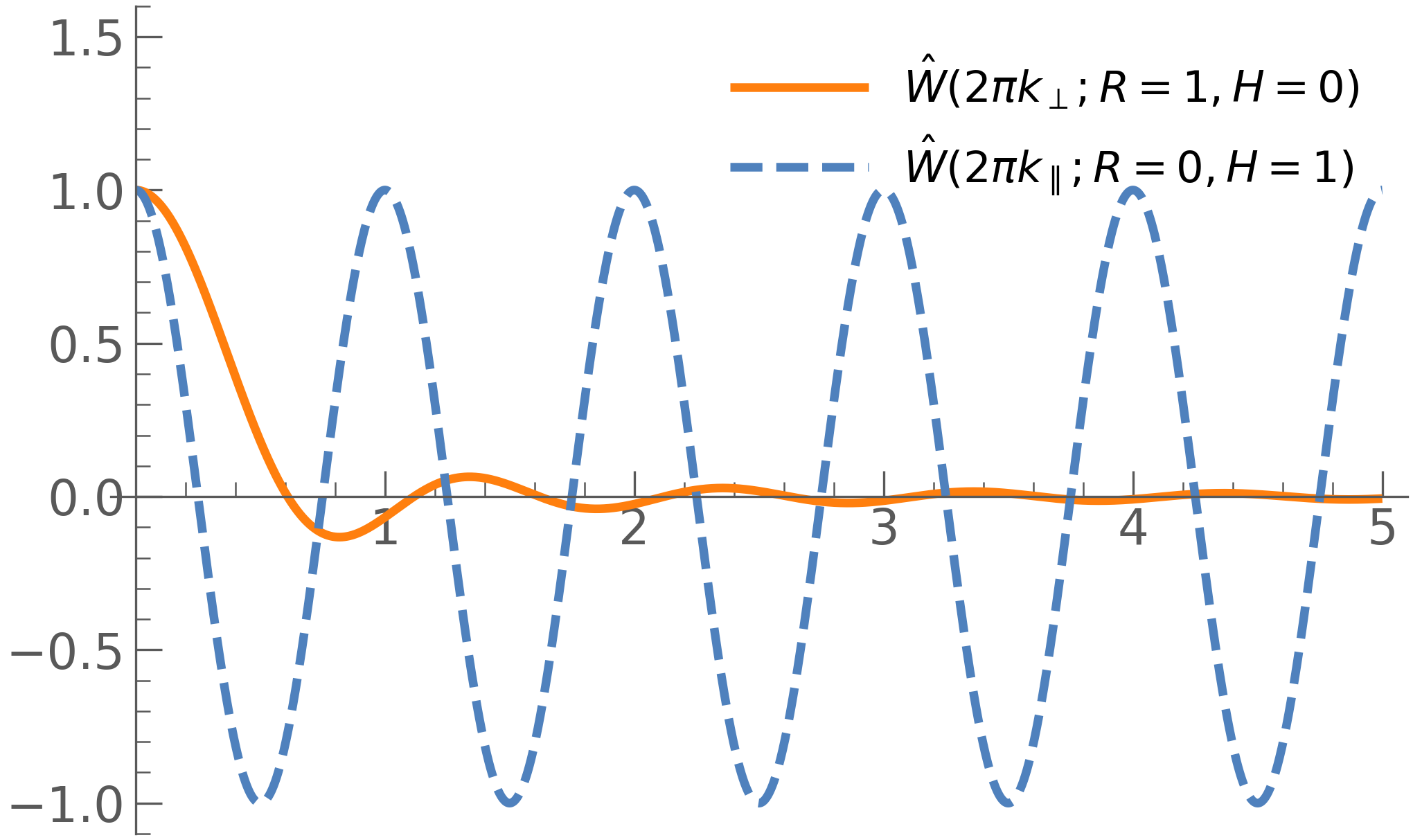} \\
        \hline
        (i) & Thin redshift-space ring; standard \((s,\mu)\) binning geometry. & $$W_{\mathrm{ring}}(\rho,z;R,H)=\frac{\delta_{\mathrm{D}}(\rho-R)}{2\pi R}\frac{\delta_{\mathrm{D}}(|z|-H)}{2},$$ $$\hat{W}_{\mathrm{ring}}(k_{\perp},k_{\parallel};R,H)=J_0(k_{\perp}R)\cos(k_{\parallel}H)$$ & \includegraphics[width=0.28\textwidth]{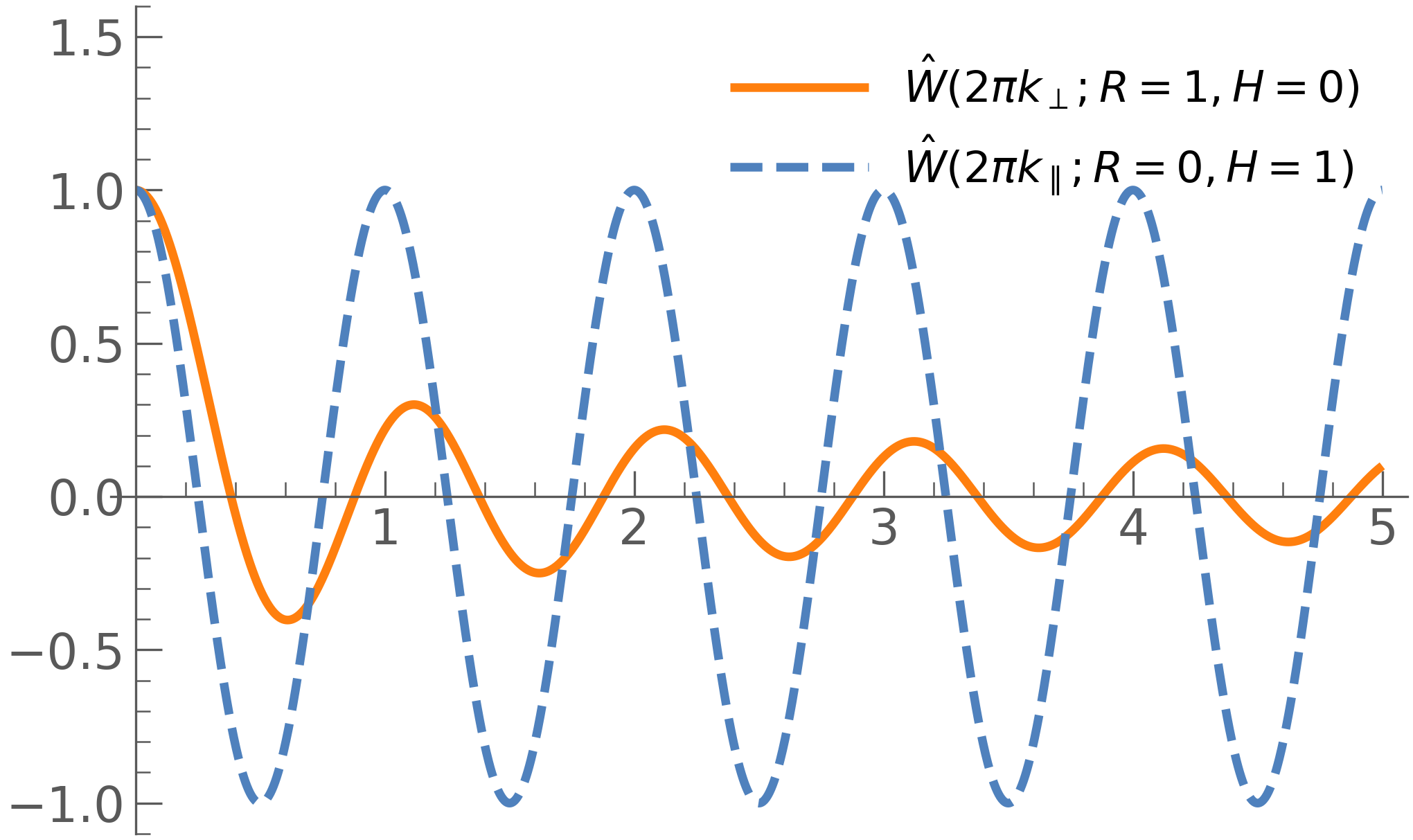} \\
        \hline
        (j) & One-dimensional cosine wavelet high-pass filter. & $$ W_{\mathrm{CW}}(x,R) = C_{\mathrm{CW}} R^{-1/2} F_{\mathrm{CW}}(x/R),$$ $$ \hat{W}_{\mathrm{CW}}(k,R) = (2\pi)^{1/2}C_{\mathrm{CW}} R^{1/2}G_{\mathrm{CW}}(kR),$$
        $$ F_{\mathrm{CW}}(x)=\bigl[(1-x^2)\cos x - x \sin x \bigr]\exp\bigl(\frac{1}{2}(1-x^2)\bigr),$$
        $$ G_{\mathrm{CW}}(x) = x\bigl[x\cosh x - \sinh x\bigr] \exp\bigl(-\frac{1}{2}x^2\bigr),$$
        $$ C_{\mathrm{CW}}=\frac{2\sqrt{2}}{\sqrt{1+5\mathrm{e}}\pi^{1/4}}$$
         & \includegraphics[width=0.28\textwidth]{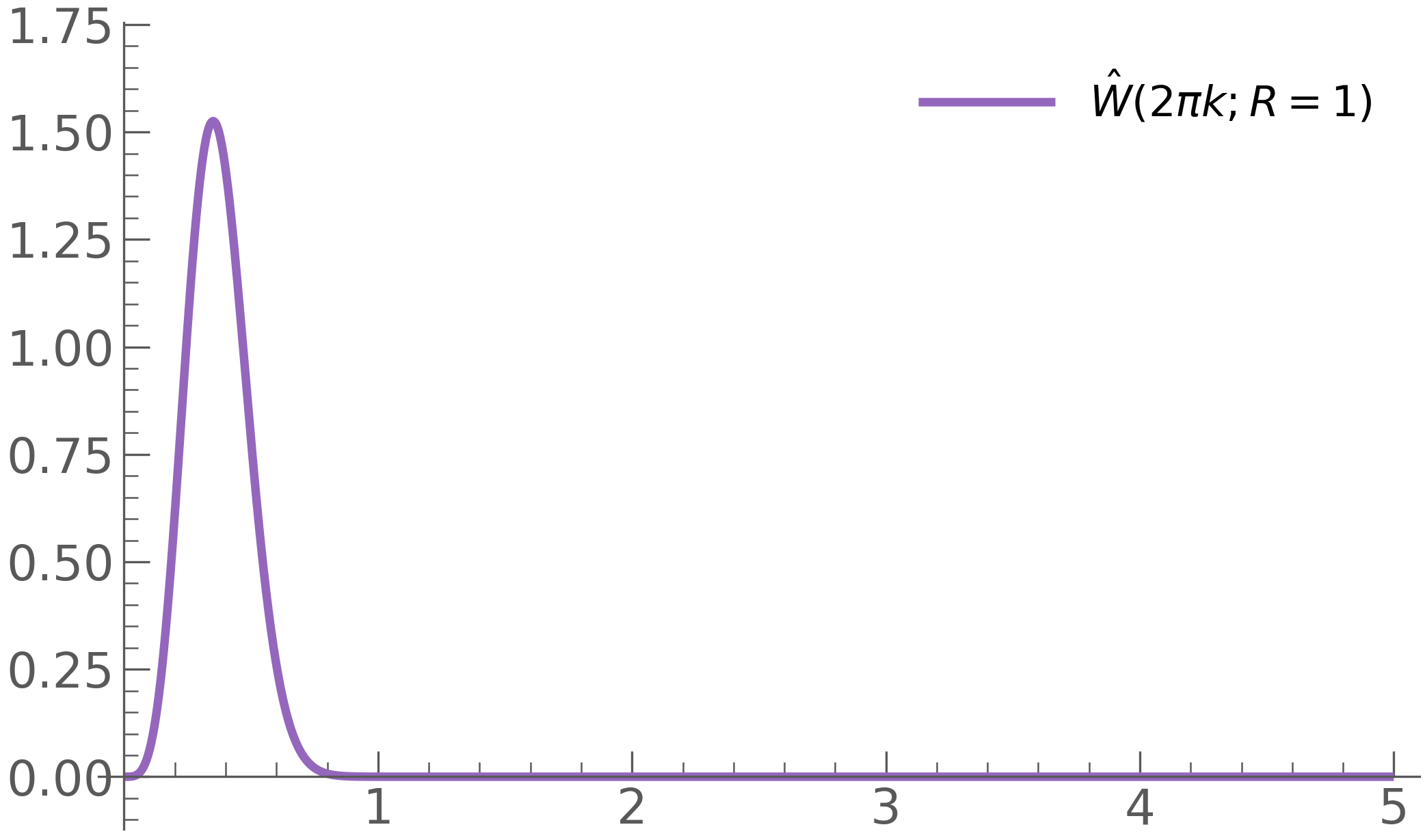} \\
        \hline
         (k) & Spherical cosine-wavelet high-pass filter. & $${W_{\mathrm{CWS}}(\mathbf{x},R) = C_{\mathrm{CWS}} R^{-3/2}F_{\mathrm{CWS}}(|\mathbf{x}|/R) }$$,$$ \hat{W}_{\mathrm{CWS}}(k,R) = (2\pi)^{3/2}C_{\mathrm{CWS}} R^{3/2}G_{\mathrm{CW}}(kR),$$
         $$F_{\mathrm{CWS}}(x)= \mathrm{e}^{\frac{1}{2}(1-x^2)}\Bigl[(4-x^2)\cos x +  2(1-x^2)\frac{\sin x}{x}\bigr] $$
        $$ C_{\mathrm{CWS}}=\frac{2\sqrt{2}}{\sqrt{9+55\mathrm{e}}\pi^{3/4}}$$
         & \includegraphics[width=0.28\textwidth]{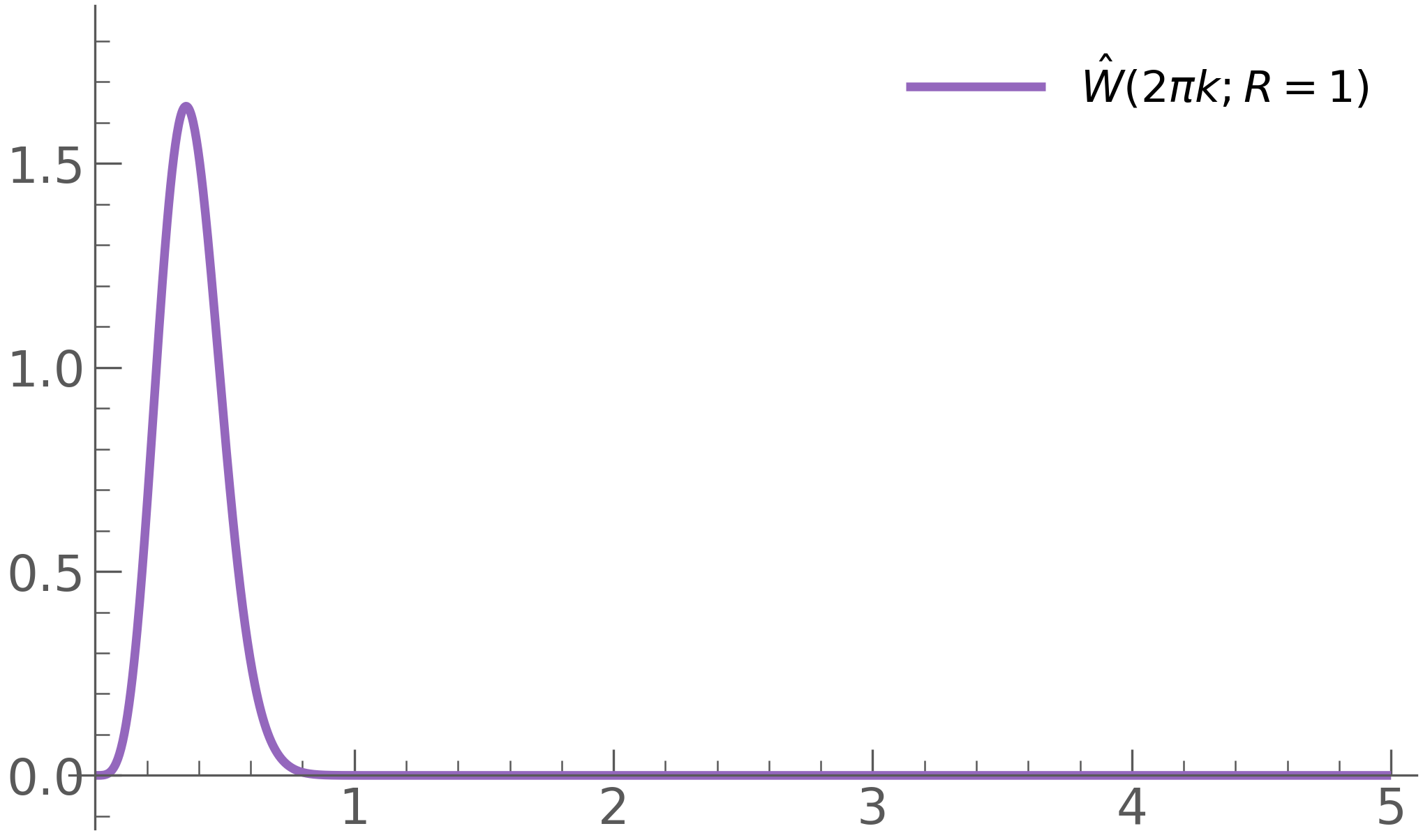} \\
        \hline
        (l) & Isotropic Gaussian-derivative wavelet; localised band-pass/high-pass response. & $$ {W_{\mathrm{GDW}}(\mathbf{x},R) =C_{\mathrm{GDW}} R^{-{3}/{2}}\Bigl(3-\frac{|\mathbf{x}|^2}{R^2}\Bigr)\exp\Bigl({-\frac{|\mathbf{x}|^2}{2R^2}}\Bigr),}$$ $$\hat{W}_{\mathrm{GDW}}(\mathbf{k},R) = C_{\mathrm{GDW}}^{(k)}R^{3/2}(kR)^2\exp\Bigl({-\frac{1}{2}k^2R^2}\Bigr)$$
        $$ C_{\mathrm{GDW}}^{(k)}=(2\pi)^{3/2}C_{\mathrm{GDW}},\quad C_{\mathrm{GDW}} = \frac{2}{\sqrt{15}\pi^{3/4}}$$
        & \includegraphics[width=0.28\textwidth]{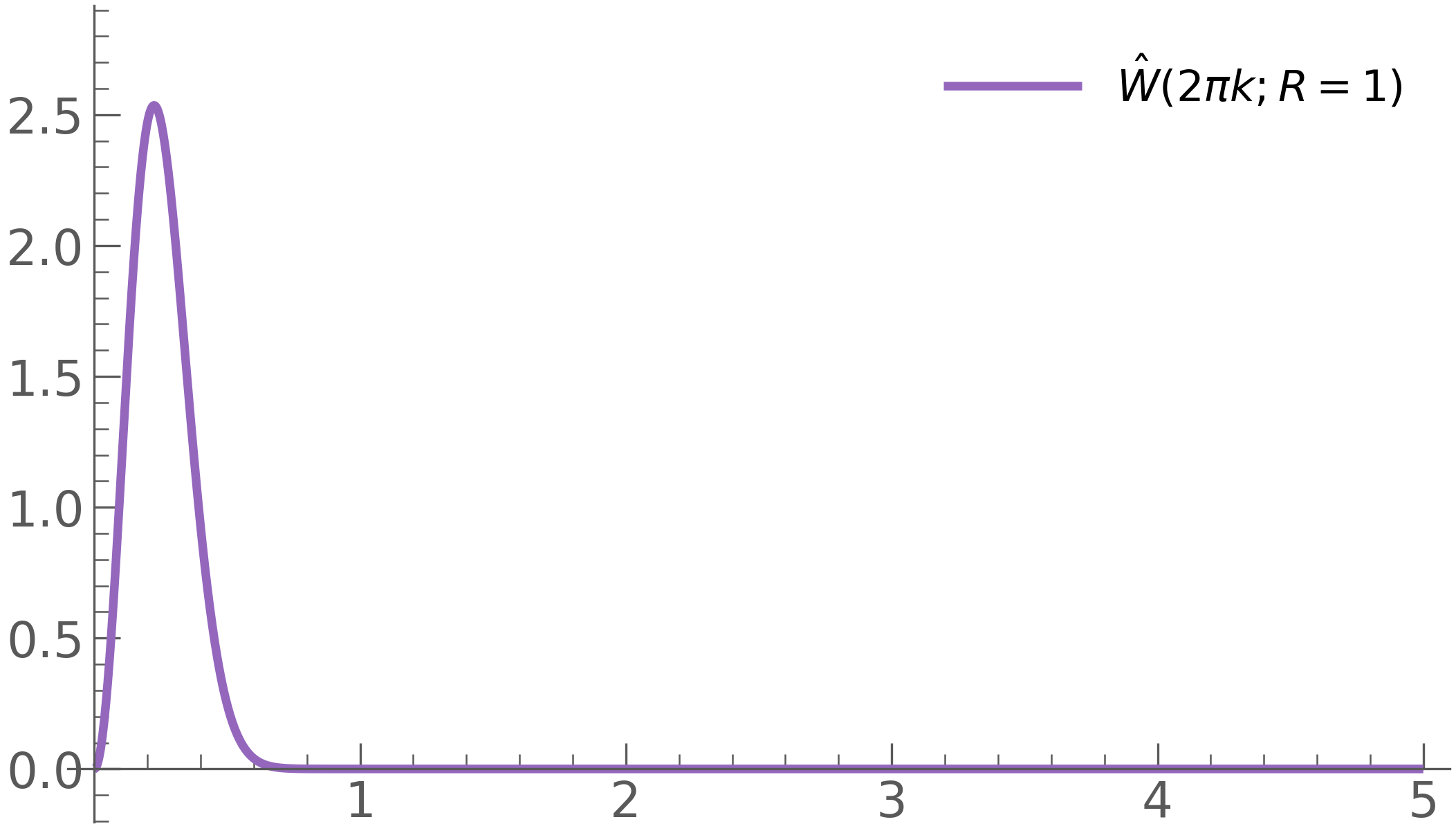} \\
        \hline
        (m) & Directional derivative operator along \(\hat{\mathbf{n}}\). & $${W_{\nabla_{\hat{\mathbf{n}}}}(\mathbf{x})=\hat{\mathbf{n}}\cdot\nabla,}$$ $$\hat{W}_{\nabla_{\hat{\mathbf{n}}}}(\mathbf{k})=\mathrm{i}\,\mathbf{k}\cdot\hat{\mathbf{n}}$$
        & \includegraphics[width=0.28\textwidth]{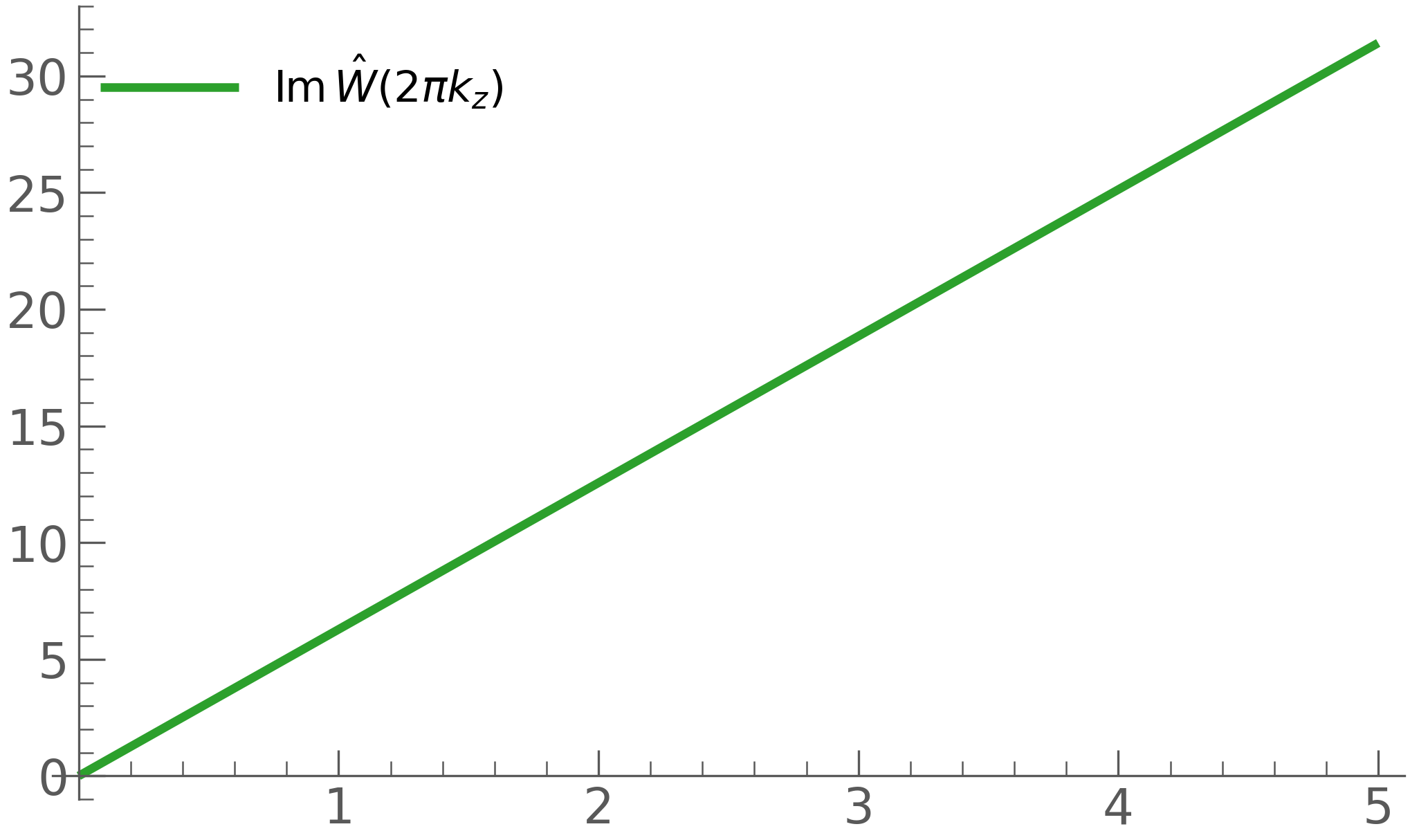} \\
        \hline
        (n) & Laplacian operator for second-derivative diagnostics. & $${W_{\nabla^2}(\mathbf{x})=\nabla^2,}$$ $$\hat{W}_{\nabla^2}(\mathbf{k})=-|\mathbf{k}|^2$$
        & \includegraphics[width=0.28\textwidth]{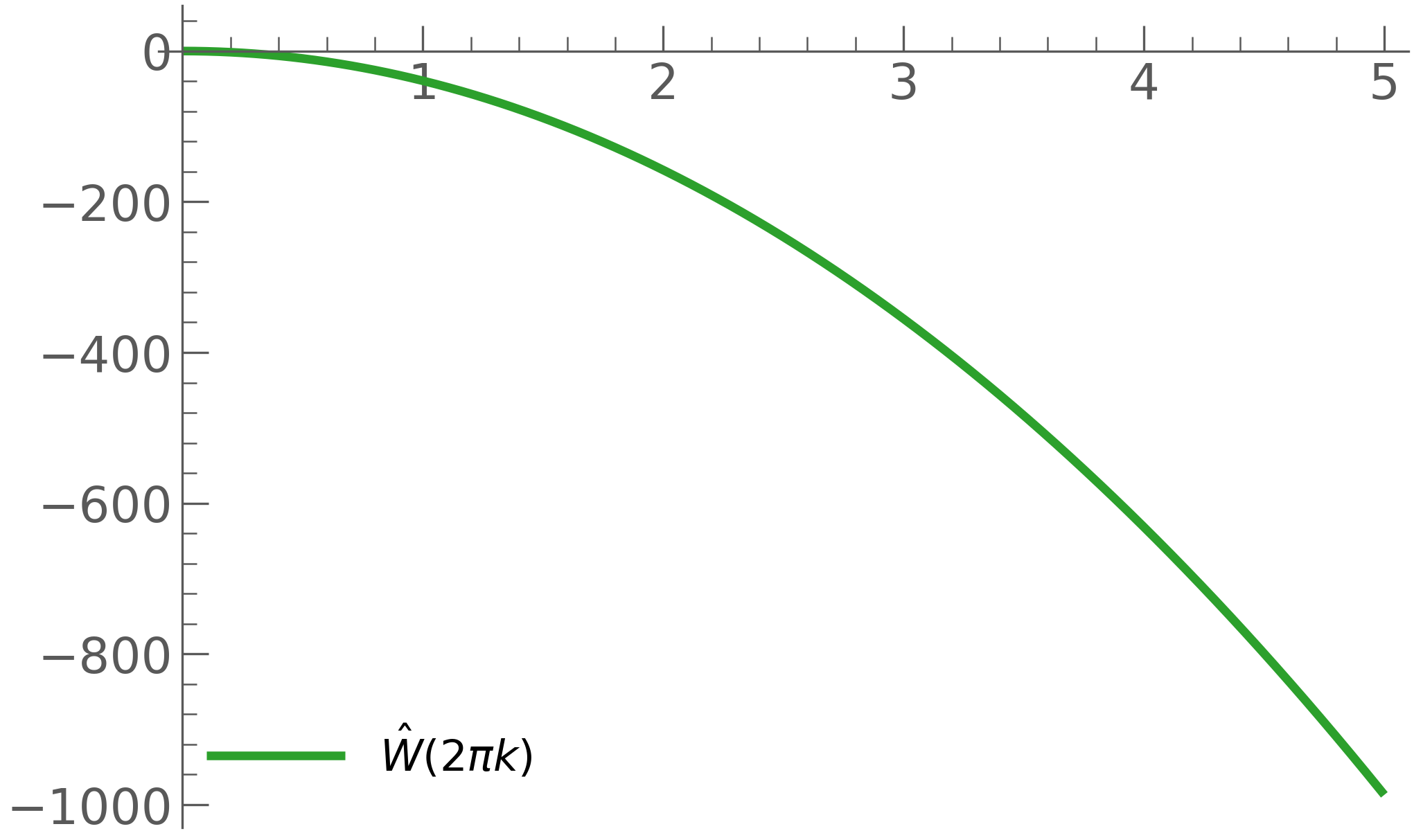} \\
        \hline
        (o) & Newtonian gravitational potential. & $$W_{\mathrm{grav}}(\mathbf{x})=-\frac{1}{|\mathbf{x}|}$$ $$\hat{W}_{\mathrm{grav}}(\mathbf{k})=-\frac{4\pi}{|\mathbf{k}|^{2}}\quad(\mathbf{k}\ne\mathbf{0})$$
        $$\hat{W}_{\mathrm{grav}}(\mathbf{0})=0$$
        & \includegraphics[width=0.28\textwidth]{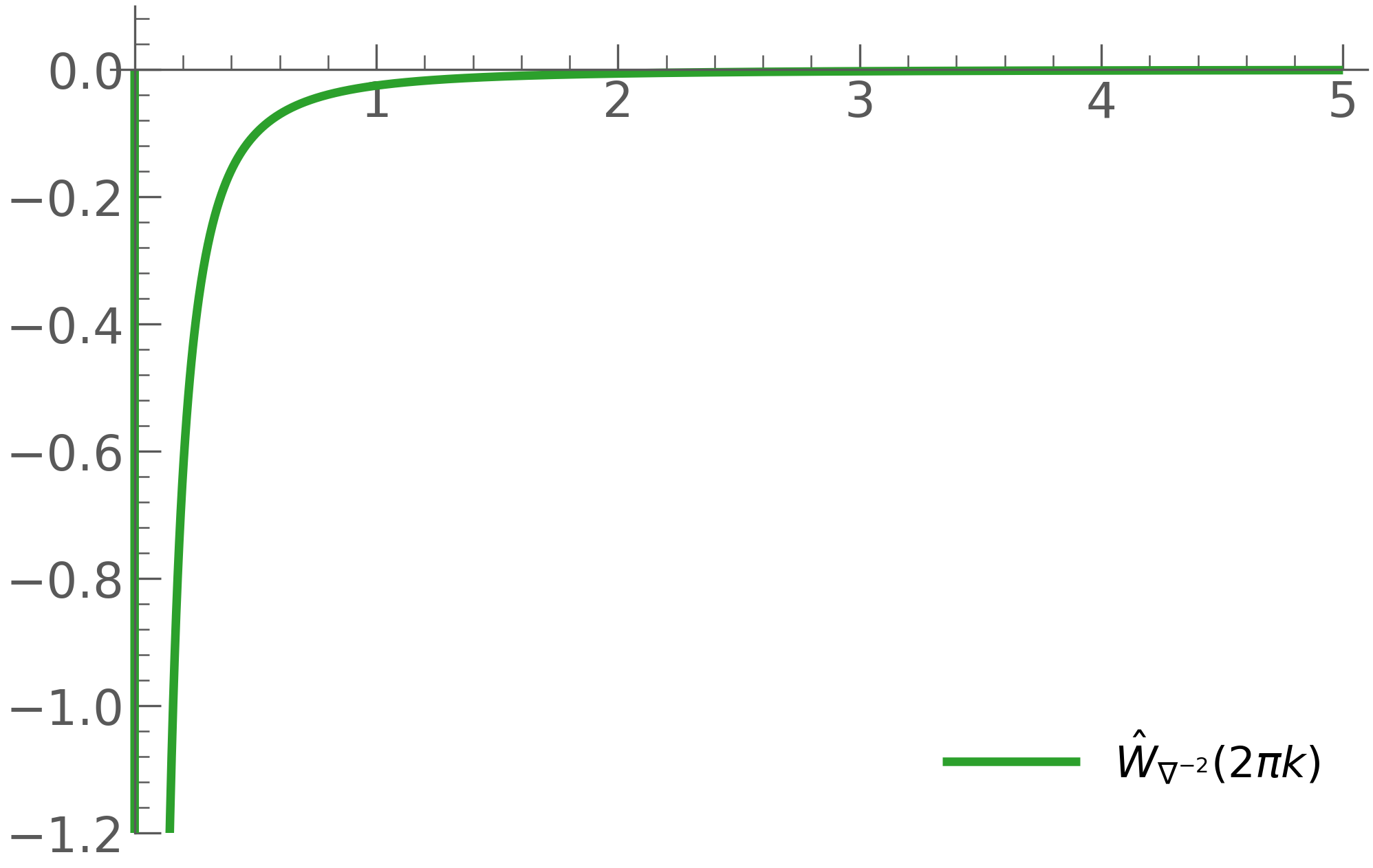} \\

        \hline
        \multicolumn{4}{p{6.6in}}{\textbf{Notes.} \noindent The Fourier expressions in Table~\ref{tab:kernel} use the same
non-unitary angular-wavenumber convention as the main text,
\(\hat W(\mathbf{k})=\int W(\mathbf{x})\mathrm{e}^{-\mathrm{i}\mathbf{k}\cdot\mathbf{x}}d^3\mathbf{x}\).
For the \textsf{PyHermes} line-of-sight windows, \(H\) denotes the line-of-sight
offset for \texttt{ring} and \texttt{disk}, and the half-height for
\texttt{cylshell} and \texttt{cylinder}; the tabulated kernels are the
real, even-in-\(z\) forms used by the code. The Fourier-space amplitudes of
the scale-normalised wavelet rows include the corresponding
\((2\pi)^{d/2}\) factors for this convention.

Rows (a)--(i) list the ordinary smoothing or binning windows. Rows (j)--(l) list band-pass, scale-normalised filters: the cosine wavelet (CW), its spherical counterpart (CWS), and the Gaussian-derivative wavelet (GDW). Rows (m)--(n) are differential operators, and row (o) is the Newtonian-potential kernel. The right column presents the corresponding Fourier-space kernels as functions of the angular wavenumber \(k\).} \\

\end{longtable}
\endgroup

\section{Differential operators in the MRA basis}
\label{app:differential_operator}

In the main text, differential operations are implemented through
Fourier-space operator windows, where a derivative is represented by the
corresponding Fourier kernel. This approach provides a flexible framework for
combining differential operations with smoothing and other window functions.
Alternatively, within the MRA framework, differential operators can be applied
directly to the reconstructed fields by differentiating the scaling basis
functions. This appendix briefly describes this basis-space approach.

At a fixed resolution level $j$, a one-dimensional field can be represented as
\begin{equation}
    f(x)=\sum_l f_l \phi_{j,l}(x),
\end{equation}
where
\begin{equation}
    \phi_{j,l}(x)=2^{j/2}\phi(2^j x-l),
\end{equation}
is the scaling function of the MRA basis. The derivative field is obtained by
acting the differential operator directly on the basis functions,
\begin{equation}
    \frac{df}{dx}
    =\sum_l f_l \frac{d}{dx}\phi_{j,l}(x).
\end{equation}

For compactly supported Daubechies scaling functions, the derivative of the
basis functions can be projected back onto the same scaling-function space
through derivative connection coefficients \citep{Beylkin1991},
\begin{equation}
    r_m=
    \int dx\,
    \phi(x-m)\frac{d\phi(x)}{dx}.
\end{equation}
The derivative operator at resolution level $j$ is then represented by the
matrix
\begin{equation}
    D^{(j)}_{ll'}
    =
    2^j r_{l'-l},
\end{equation}
which leads to the local convolution form
\begin{equation}
    \left(\frac{df}{dx}\right)_l
    =
    2^j\sum_m r_m f_{l+m}.
\end{equation}

The coefficients $r_m$ are determined entirely by the choice of the
Daubechies basis and are finite because of the compact support of the scaling
functions. For the Daubechies D4 (\texttt{db2} in \textsf{PyWavelets}) basis adopted in the current
implementation, the non-zero first-derivative connection coefficients are
\begin{equation}
    r_{\pm1}=\mp\frac{2}{3},
    \qquad
    r_{\pm2}=\pm\frac{1}{12},
    \qquad
    r_0=0,
\end{equation}
up to the sign convention adopted for the derivative operator. Therefore, the
first derivative can be evaluated through a compact stencil involving only
neighbouring MRA coefficients.

For the three-dimensional tensor-product scaling basis,
\begin{equation}
    \Phi_{\mathbf{l}}(\mathbf{x})
    =
    \phi_{j,l_x}(x)
    \phi_{j,l_y}(y)
    \phi_{j,l_z}(z),
\end{equation}
the derivative operator along the $x$ direction becomes
\begin{equation}
    D^x_{\mathbf{l},\mathbf{l}'}
    =
    2^j r_{l'_x-l_x}
    \delta_{l_y l'_y}
    \delta_{l_z l'_z},
\end{equation}
with analogous expressions for the $y$ and $z$ directions. Thus, differential
operations remain local stencil operations on the MRA grid.

The basis-space derivative and the Fourier-space operator window represent two
equivalent realisations of spatial differential operators. The Fourier-space
approach is more general for constructing composite filters, for example
combining derivatives with arbitrary smoothing windows. In contrast, the
connection-coefficient approach preserves the locality and sparsity of the MRA
representation and is advantageous for repeated differential operations, such
as constructing velocity gradients, tidal tensors, and other derived physical
fields.

The derivative operators in the \textsf{MRA} framework provide a natural
approach for constructing physical fields and their spatial derivatives
directly within the same reconstructed field representation. In particular, the derivative
connection coefficients allow gravitational forces to be evaluated from the
reconstructed density and potential fields without the conventional
mesh-to-particle force interpolation step used in
\textsf{particle--mesh} (\textsf{PM})
algorithms \citep{Hockney1988}. Combined with the simultaneous representation
of density, gravitational potential, and their derivatives in the same \textsf{MRA}
basis, this approach suggests a possible formulation of
\textsf{MRA}-based \textsf{PM} simulations, in which field reconstruction, Poisson solving,
and force evaluation are consistently performed within a unified
multiresolution representation.

Another important advantage arises from performing Fourier operations directly
on the \textsf{MRA} representation. Since the FFT is applied to the scaling-function
coefficients rather than to a conventionally interpolated mesh density field,
the resulting Fourier representation naturally preserves the multiresolution
structure of the basis functions. This construction can be viewed as a
generalised Fourier representation on a multiresolution basis, closely related
to wavelet-based operator representations and Fourier transforms beyond
uniform sampling schemes \citep{Beylkin1991,Beylkin1992}. Previous studies have
demonstrated that power-spectrum estimation based on such \textsf{MRA}
representations can effectively suppress aliasing contamination and maintain
high accuracy close to the Nyquist frequency \citep{Cui2008,YangYB_2009}.
When incorporated into cosmological \(N\)-body simulations, these properties
may offer a route to improving both the accuracy and computational
efficiency of \textsf{PM} schemes, particularly for high-resolution
simulations where force-interpolation errors and aliasing effects become
increasingly important.

\label{lastpage}
\end{document}